%% file: pcdr.tex
\documentclass[11pt,tightenlines,letterpaper,superscriptaddress,amsmath,amssymb,bibnotes,altaffilletter,graphics,nofootinbib]{revtex4-1}

\input{gen/header}        
\input{gen/abbreviations} 

\begin{document}
\raggedbottom
\pagenumbering{roman}

\title{The Large Enriched Germanium Experiment for Neutrinoless \BB\ Decay}
\subtitle{\Lthou\ Preconceptual Design Report}
\input{gen/authors-and-affiliations-2021-06-29}
\documentname{LEGEND-1000 Preconceptual Design Report}
\date{\today}

\mymaketitle
\myauthorpage
\tocless{\section*}{Acknowledgments}
\input{gen/Acknowledgments-2021-02-18}

\newpage
\tableofcontents
\newpage

\setcounter{page}{0}
\pagenumbering{arabic}
\import{sec_overview/}{sec_overview}
\clearpage
\import{sec_introduction/}{sec_introduction.tex}

\clearpage
\import{sec_ge/}{sec_ge}

\clearpage
\import{sec_legend-approach/}{sec_legend-approach.tex}

\clearpage\
\import{sec_technical/}{sec_technical}

\clearpage
\import{sec_facilities/}{sec_facilities}

\clearpage
\appendix

\import{sec_appendix/}{sec_specs}
\clearpage
\import{sec_appendix/}{sec_glossary}
\clearpage
\bibliography{references}

\end{document}

%% file: gen/header.tex
\usepackage[includeheadfoot,top=0.3in,bottom=0.7in,left=1in,right=1in]{geometry} 

\renewcommand{\footnotesize}{\fontsize{9pt}{10pt}\selectfont}

\renewcommand{\scriptsize}{\fontsize{8pt}{9pt}\selectfont}

\usepackage{titlesec}
\usepackage[normalem]{ulem}
\titleformat{\section}{\normalfont\fontsize{11pt}{12pt}\selectfont\bfseries\centering}{\thesection.}{1em}{\MakeUppercase}[\vspace{-1ex}]
\titleformat{\subsection}{\normalfont\fontsize{11pt}{12pt}\selectfont\bfseries\centering}{\thesubsection.}{1em}{}[\vspace{-1ex}]
\titleformat{\subsubsection}{\normalfont\it\centering}{\thesubsubsection.}{1em}{}[\vspace{-1ex}]
\titleformat{\paragraph}[hang]{\normalfont\normalsize\it}{}{0em}{\uline}
\titlespacing*{\paragraph}{0pt}{1.25ex plus 1ex minus .2ex}{0em}

\usepackage{fancyhdr}
\fancypagestyle{uniheader}
{
\fancyhf{}
\fancyfoot[C]{-\thepage-}
\fancyhead[L]{\footnotesize\thissubtitle \hfill
  \raisebox{1.ex-0.65\height}
  {\includegraphics[height=20pt]{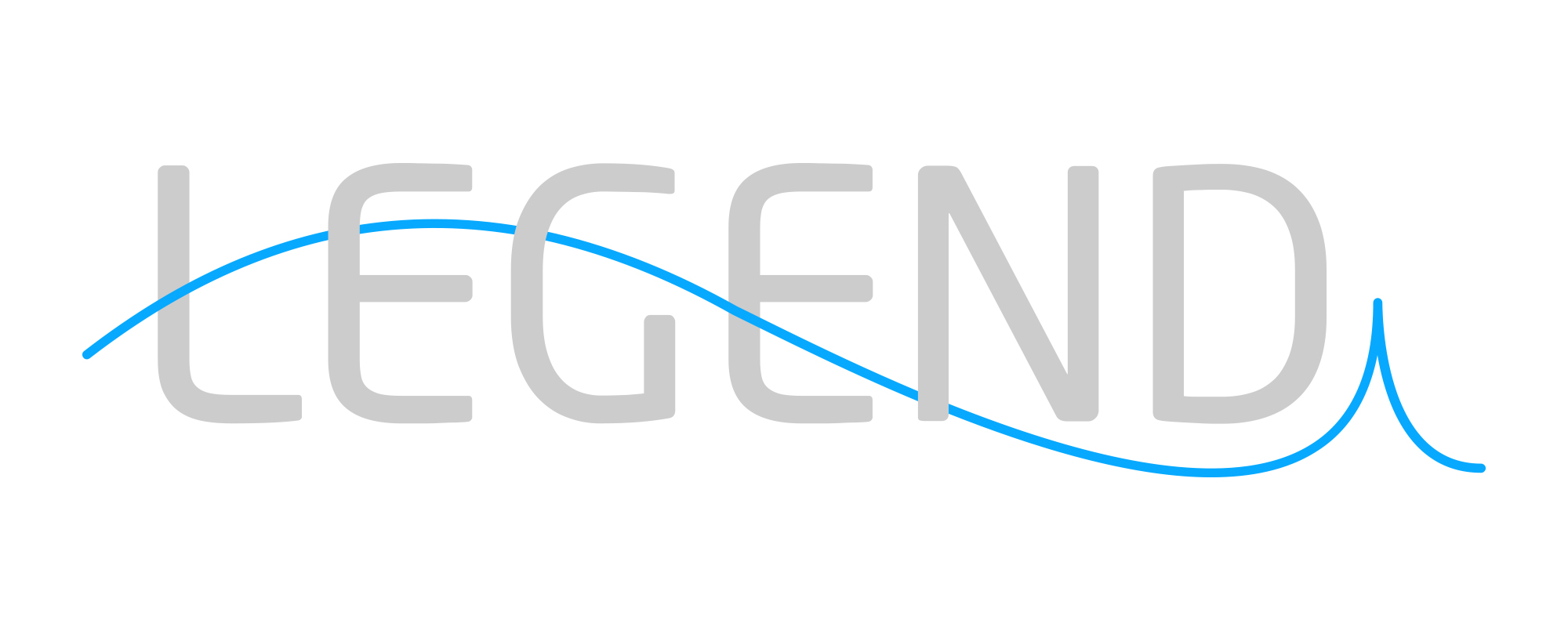}}
  }
  \renewcommand{\headrulewidth}{1pt}%
  \renewcommand{\headrule}{\hbox to\headwidth{%
    \color{legendblue}\leaders\hrule height \headrulewidth\hfill}}
}

\fancypagestyle{myheader}{%
  \fancyhf{}
  \fancyhead[C]{My header}
  \fancyfoot[C]{-\thepage-}
  \renewcommand{\headrulewidth}{2pt}
  \renewcommand{\headrule}{\hbox to\headwidth{%
    \color{red}\leaders\hrule height \headrulewidth\hfill}}
}

\newenvironment{myitemize}
{ \begin{itemize}
  \setlength{\itemsep}{0pt}
  \setlength{\parskip}{0pt}
  \setlength{\parsep}{0pt}   }
{ \end{itemize}         }
\newenvironment{myenumerate}
{ \begin{enumerate}
  \setlength{\itemsep}{0pt}
  \setlength{\parskip}{0pt}
  \setlength{\parsep}{0pt}   }
{ \end{enumerate}         }

\makeatletter
\renewcommand{\p@subsection}{\thesection.}
\renewcommand{\p@subsubsection}{\thesection.\thesubsection.}
\makeatother

\usepackage{float}
\usepackage[table]{xcolor}
\definecolor{blue}{RGB}{50,0,255}
\definecolor{BLUE}{RGB}{50,0,255}
\definecolor{orange}{RGB}{255,128,0}
\definecolor{beaublue}{rgb}{0.74, 0.83, 0.9}
\definecolor{bubbles}{rgb}{0.91, 1.0, 1.0}
\definecolor{columbiablue}{rgb}{0.61, 0.87, 1.0}
\definecolor{electricblue}{rgb}{0.49, 0.98, 1.0}
\definecolor{cyan}{rgb}{0.0, 1.0, 1.0}
\definecolor{lightcornflowerblue}{rgb}{0.6, 0.81, 0.93}
\colorlet{blueh}{blue!30}
\definecolor{legendblue}{RGB}{  7, 169, 255}
\definecolor{legendgrey}{RGB}{204, 204, 204}
\definecolor{legenddarkblue}{RGB}{  26, 42, 91}

\usepackage{soul}

\newcommand{\new}[1] {#1}

\makeatletter 
\def\Dated@name{Date: }%
\makeatother

\makeatletter 
 \def\CT@@do@color{%
 \global\let\CT@do@color\relax
  \@tempdima\wd\z@
  \advance\@tempdima\@tempdimb
  \advance\@tempdima\@tempdimc
  \advance\@tempdimb\tabcolsep
  \advance\@tempdimc\tabcolsep
  \advance\@tempdima2\tabcolsep
  \kern-\@tempdimb
  \leaders\vrule
  \hskip\@tempdima\@plus  1fill
  \kern-\@tempdimc
  \hskip-\wd\z@ \@plus -1fill }
\makeatother

\newlength\tw
\usepackage{tabularx}

\usepackage{longtable}
\newcolumntype{Y}{>{\raggedright\arraybackslash}X}
\newcolumntype{Z}{>{\centering\arraybackslash}X}
\newcolumntype{U}{>{\raggedleft\arraybackslash}X}
\newcolumntype{L}[1]{>{\raggedright\arraybackslash}p{#1}}
\newcolumntype{R}[1]{>{\raggedleft\arraybackslash}p{#1}}
\newcolumntype{C}[1]{>{\centering\arraybackslash}p{#1}}
\usepackage{multirow}

\usepackage{makecell}

\makeatletter
\newcommand{\subtitle}[1]{\gdef\@subtitle{#1}}
\def\thissubtitle{\@subtitle}
\newcommand{\documentname}[1]{\gdef\@documentname{#1}}
\makeatother

\makeatletter
\def\my@affil@script#1#2#3#4{%
 \@ifnum{#1=\z@}{}{%
  \begingroup
   \frontmatter@affiliationfont
   \footnotesize
   \ignorespaces#3%
   \@if@empty{#4}{}{\frontmatter@footnote{#4}}%
   \par
  \endgroup
 }%
}%
\makeatother

\makeatletter
\def\mymaketitle{
 \begingroup
 \thispagestyle{titlepage}
  {\raggedleft
  \includegraphics[trim={0 0.7in 0 2.4in},width=0.45\columnwidth, keepaspectratio=true]
  {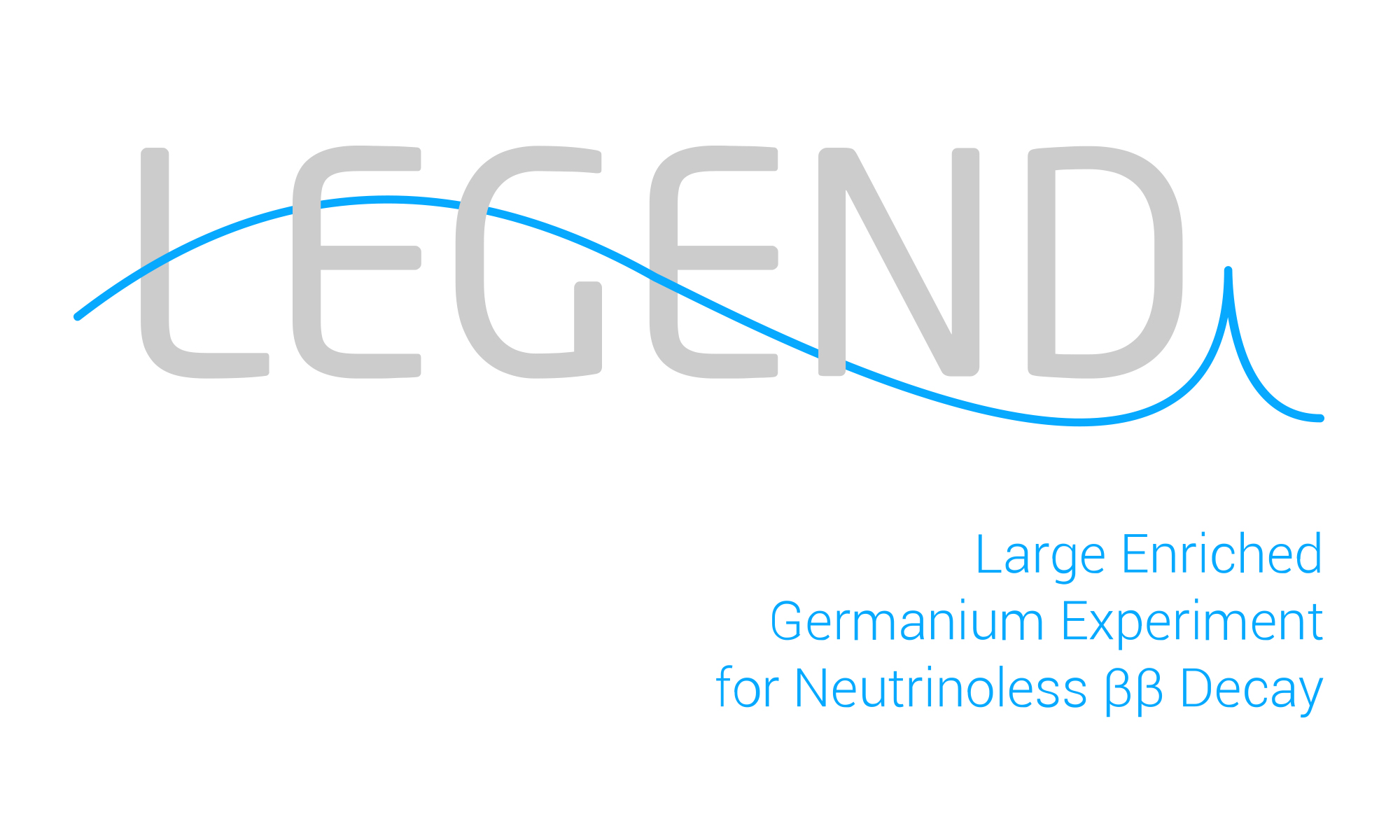}\\} 
  {\raggedright
  {\vspace{1.0in}\Huge{\@title \vspace{0.4in} \\  \@subtitle}\par} 
  \vspace{\fill}
  {\Large\@documentname}\\
  {\vspace{0.1in}\@date}
  }
  \newpage
 \endgroup
}%
\makeatother

\newcommand{\nocontentsline}[3]{}
\newcommand{\tocless}[2]{\bgroup\let\addcontentsline=\nocontentsline#1{#2}\egroup}

\makeatletter
\def\myauthorpage{
  \begingroup
  \tocless{\section*}{LEGEND Collaboration}
  \@author@finish
  \setlength{\parindent}{0cm}
  \ltx@footnote@pop
  \def\@mpfn{mpfootnote}%
  \def\thempfn{\thempfootnote}%
  \c@mpfootnote\z@
  \let\@makefnmark\frontmatter@makefnmark

  \let\@author@present\@author@present@script
  \let\AU@temp\@empty
  \@tempcnta\z@
  \let\AF@opr \@gobble
  \def\AU@opr{\@author@count\@tempcnta}%
  \def\CO@opr{\@collaboration@count\AU@temp\@tempcnta}%
  \@AAC@list
  \expandafter\CO@opr\@author@cleared

  \let\AF@opr \@affilID@def
  \let\AU@opr \@author@present
  \def\CO@opr{\@collaboration@present\AU@temp}%
  \set@listcomma@list\AU@temp
  \@AAC@list
    \vspace{0.1in}
  \let\AF@opr \@gobble
  \let\AFF@opr \@affil@script
  \@AFF@list
  \newpage
 \endgroup
}
\makeatother

\usepackage{lineno}
\usepackage{graphicx}
\usepackage{amsmath,amssymb,bbold,bm}
\usepackage{epstopdf}
\usepackage[table]{xcolor}
\usepackage{lipsum}
\usepackage{calc}
\usepackage{import}
\usepackage{upgreek}
\usepackage{hyperref}
\usepackage{bold-extra}
\usepackage{wasysym}
\usepackage[caption=false]{subfig} 
\usepackage{rotating} 
\usepackage[utf8]{inputenc}
\usepackage{currfile}
\usepackage{units} 
\usepackage{textcomp} 



%% file: gen/abbreviations.tex
\newcommand{\mee}       {\ensuremath{m_{\beta\beta}}}

\newcommand{\onbb}      {\ensuremath{0\nu\beta\beta}}
\newcommand{\BBz}       {\ensuremath{0\nu\beta\beta}}

\newcommand{\nnbb}      {\ensuremath{2\nu\beta\beta}}
\newcommand{\BBt}       {\ensuremath{2\nu\beta\beta}}
\newcommand{\BB}        {\ensuremath{\beta\beta}}
\newcommand{\Mz}        {\ensuremath{M_{0\nu}}}
\newcommand{\Mcont}        {\ensuremath{M_{0\nu}^{cont}}}

\newcommand{\MzG}       {\ensuremath{M^{(0\nu)}_{\textrm{GT}}}}
\newcommand{\MzF}       {\ensuremath{M^{(0\nu)}_{\textrm{F}}}}
\newcommand{\MzT}       {\ensuremath{M^{(0\nu)}_{\textrm{T}}}}

\newcommand{\Gz}        {\ensuremath{G_{0\nu}}}

\newcommand{\gA}        {\ensuremath{g_{A}}}
\newcommand{\gV}        {\ensuremath{g_{V}}}
\newcommand{\qval}      {\ensuremath{Q_{\beta\beta}}}
\newcommand{\Qbb}       {\ensuremath{Q_{\beta\beta}}}
\newcommand{\thalf}     {\ensuremath{T_{1/2}}}

\newcommand{\Tz}        {\ensuremath{T^{0\nu}_{1/2}}}

\newcommand{\powten}[1] {$10^{#1}$}
\newcommand{\cpowten}[2]{$#1\times10^{#2}$}
\newcommand{\cpKkgy}    {cts/(keV\,kg\,yr)}
\newcommand{\cpFty}     {cts/(FWHM\,t\,yr)}
\newcommand{\ctsper}    {cts/(keV\,kg\,yr)}

\newcommand{\kgyr}      {kg$\,$yr}
\newcommand{\tyr}       {t$\,$yr}
\newcommand{\mubq}      {$\upmu$Bq}

\newcommand{\mus}       {$\upmu$s}

\newcommand{\bgfwhm}      {less than $0.025$\,\cpFty}
\newcommand{\mbgfwhm}      {$<0.025$\,\cpFty}
\newcommand{\bgkev}       {less than \cpowten{1}{-5}\,\cpKkgy}
\newcommand{\mbgkev}       {\cpowten{<1}{-5}\,\cpKkgy}

\newcommand{\BGprojfwhm}      {$0.023\pm\nobreak0.012$\,\cpFty}
\newcommand{\BGprojkev}       {\cpowten{	9.1	^{+4.9}	_{-6.3}	}{-	6	}\,\cpKkgy}

\newcommand{\nuc}[2]    {$^{#1}$\textrm{#2}} 
\newcommand{\gesix}     {\nuc{76}{Ge}}
\newcommand{\geVII}        {\nuc{77}{Ge}}
\newcommand{\geVIIm}        {\nuc{77\textrm{m}}{Ge}}
\newcommand{\geVIIpm}        {\nuc{77\textrm{(m)}}{Ge}}
\newcommand{\U}        {\nuc{238}{U}}
\newcommand{\Th}        {\nuc{232}{Th}}

\newcommand{\ppc}       {PPC} 
\newcommand{\icpc}       {ICPC} 
\newcommand{\bege}      {BEGe}

\newcommand{\nPlus}     {n$^+$} 
\newcommand{\pPlus}     {p$^+$} 
\newcommand{\lar}       {LAr}
\newcommand{\UGLAr}	{UGLAr}

\newcommand{\lnn}       {LN$_2$}
\newcommand{\nt}       {N$_2$}

\newcommand{\LEG}       {LEGEND}
\newcommand{\Ltwo}      {{\LEG-200}}
\newcommand{\Lthou}     {{\LEG-1000}}
\newcommand{\Lk}        {{\LEG-1000}}

\newcommand{\MJ}        {\textsc{Majorana}}
\newcommand{\DEM}       {\textsc{Demonstrator}}
\newcommand{\MJD}       {\textsc{Majorana Demonstrator}}
\newcommand{\Gerda}     {\textsc{Gerda}}
\newcommand{\ptwo}      {{Phase\,II}}

\newcommand{\Lngs}      {LNGS}
\newcommand{\SL}	{SNOLAB}

\newcommand{\Geant}     {Geant4}
\newcommand{\GF}        {Geant4}
\newcommand{\MaGe}      {\textsc{MaGe}}
\newcommand{\Siggen}    {\texttt{siggen}}

\newcommand{\be}        {\begin{equation}}
\newcommand{\ee}        {\end{equation}}

\newcommand{\ultem}     {Ultem\textsuperscript{\tiny\texttrademark}}
\newcommand{\suprasil}  {Suprasil\textsuperscript{\tiny\textregistered}}



%% file: gen/authors-and-affiliations-2021-06-29.tex
\newcommand{\UNM}{Department of Physics and Astronomy, University of New Mexico, Albuquerque, NM 87131, USA}
\newcommand{\LAquila}{Department of Physical and Chemical Sciences University of L'Aquila, L'Aquila, Italy}
\newcommand{\GSSI}{Gran Sasso Science Institute, L'Aquila, Italy}
\newcommand{\LNGS}{Istituto Nazionale di Fisica Nucleare, Laboratori Nazionali del Gran Sasso, Assergi (AQ), Italy}
\newcommand{\UTAustin}{Department of Physics, University of Texas at Austin, Austin, TX 78712, USA}
\newcommand{\LBNL}{Institute for Nuclear and Particle Astrophysics and Nuclear Science Division, Lawrence Berkeley National Laboratory, Berkeley, CA 94720, USA}
\newcommand{\UCBPH}{Department of Physics, University of California, Berkeley, CA, 94720, USA}
\newcommand{\UCBNE}{Department of Nuclear Engineering, University of California, Berkeley, CA, 94720, USA}
\newcommand{\IKZ}{Leibniz Institute for Crystal Growth, Berlin, Germany}
\newcommand{\IU}{Department of Physics, Indiana University, Bloomington, IN 47405, USA}
\newcommand{\Bratislava}{Department of Nuclear Physics and Biophysics, Comenius University, Bratislava, Slovakia}
\newcommand{\SFU}{Department of Chemistry, Simon Fraser University, Burnaby, British Columbia, Canada}
\newcommand{\LNS}{Istituto Nazionale di Fisica Nucleare, Laboratori Nazionali del Sud, Catania, Italy}
\newcommand{\UNC}{Department of Physics and Astronomy, University of North Carolina, Chapel Hill, NC 27514, USA}
\newcommand{\Warwick}{Department of Physics, University of Warwick, Coventry, United Kingdom}
\newcommand{\TUNL}{Triangle Universities Nuclear Laboratory, Durham, NC 27708, USA}
\newcommand{\Duke}{Department of Physics, Duke University, Durham, NC 27708, USA}
\newcommand{\USC}{Department of Physics and Astronomy, University of South Carolina, Columbia, SC 29208, USA}
\newcommand{\Jag}{Institute of Physics, Jagiellonian University, Cracow, Poland}
\newcommand{\Dresden}{Technische Universit\"{a}t Dresden, Dresden, Germany}
\newcommand{\JINR}{Joint Institute for Nuclear Research, Dubna, Russia}
\newcommand{\INRRAS}{Institute for Nuclear Research of the Russian Academy of Sciences, Moscow, Russia}
\newcommand{\Geel}{European Commission, Joint Research Centre, Directorate for Nuclear Safety \& Security, Geel, Belgium}
\newcommand{\MPIK}{Max-Planck-Institut f\"{u}r Kernphysik, Heidelberg, Germany}
\newcommand{\Queens}{Department of Physics, Engineering Physics \& Astronomy, Queen's University, Kingston, Ontario, Canada}
\newcommand{\UTK}{Department of Physics and Astronomy, University of Tennessee, Knoxville, TN 37916, USA}
\newcommand{\Lancaster}{Department of Physics, Lancaster University, Lancaster, United Kingdom}
\newcommand{\Liverpool}{University of Liverpool, Liverpool, United Kingdom}
\newcommand{\UCL}{University College London, London, United Kingdom}
\newcommand{\LANL}{Los Alamos National Laboratory, Los Alamos, NM 87545, USA}
\newcommand{\MILB}{Istituto Nazionale di Fisica Nucleare, Milano Biocca, Milano, Italy}
\newcommand{\MILC}{Milano Univ. and Milano Istituto Nazionale di Fisica Nucleare, Milano, Italy}
\newcommand{\NRCKI}{National Research Centre ``Kurchatov Institute'', Moscow, Russia}
\newcommand{\MEPhI}{National Research Nuclear University MEPhI (Moscow Engineering Physics Institute), 115409 Moscow, Russia}
\newcommand{\MPP}{Max-Planck-Institut f\"{u}r Physik, M\"{u}nchen, Germany}
\newcommand{\TUMPhy}{Physik-Department E15, Technische Universit\"{a}t, M\"{u}nchen, Germany}
\newcommand{\TUMClu}{Excellence Cluster Universe, Technische Universit\"{a}t M\"{u}nchen, M\"{u}nchen, Germany}
\newcommand{\ORNL}{Oak Ridge National Laboratory, Oak Ridge, TN 37830, USA}
\newcommand{\PadovaUniv}{Dipartimento di Fisica e Astronomia dell'Universita' di Padova, Italy}
\newcommand{\PadovaINFN}{Padova Istituto Nazionale di Fisica Nucleare, Padova, Italy}
\newcommand{\CTU}{Czech Technical University, Institute of Experimental and Applied Physics, CZ-12800 Prague, Czech Republic}
\newcommand{\Williams}{Department of Physics, Williams College, Williamstown, MA 01267, USA}
\newcommand{\Roma}{Roma Tre University and INFN Roma Tre, Rome, Italy}
\newcommand{\TTU}{Tennessee Tech University, Cookeville, TN 38505, USA}
\newcommand{\NCSU}{Department of Physics, North Carolina State University, Raleigh, NC 27607, USA}
\newcommand{\SDSMT}{South Dakota School of Mines and Technology, Rapid City, SD, 57701, USA}
\newcommand{\UW}{Center for Experimental Nuclear Physics and Astrophysics, and Department of Physics, University of Washington, Seattle, WA 98195, USA}
\newcommand{\Tuebingen}{University T\"{u}bingen, T\"{u}bingen, Germany}
\newcommand{\USD}{Department of Physics, University of South Dakota, Vermillion, SD 57069, USA}
\newcommand{\UZH}{Physik-Institut, University of Z\"{u}rich, Z\"{u}rich, Switzerland}
\newcommand{\Regina}{Department of Physics, University of Regina, Regina, Saskatchewan, Canada} 
\newcommand{\IPPFDD}{Leibniz-Institute of Polymer Research Dresden e.V., Dresden, Germany}

\affiliation{\LBNL}
\affiliation{\MPP}
\affiliation{\UCL}
\affiliation{\SFU}
\affiliation{\UZH}
\affiliation{\USC}
\affiliation{\ORNL}
\affiliation{\UTAustin}
\affiliation{\NRCKI}
\affiliation{\LNGS}
\affiliation{\MPIK}
\affiliation{\INRRAS}
\affiliation{\Duke}
\affiliation{\TUNL}
\affiliation{\USD}
\affiliation{\Roma}
\affiliation{\UNC}
\affiliation{\PadovaUniv}
\affiliation{\PadovaINFN}
\affiliation{\NCSU}
\affiliation{\MEPhI}
\affiliation{\UW}
\affiliation{\TUMPhy}
\affiliation{\Liverpool}
\affiliation{\UNM}
\affiliation{\MILB}
\affiliation{\SDSMT}
\affiliation{\LANL}
\affiliation{\LAquila}
\affiliation{\GSSI}
\affiliation{\UTK}
\affiliation{\CTU}
\affiliation{\IU}
\affiliation{\JINR}
\affiliation{\Lancaster}
\affiliation{\MILC}
\affiliation{\IKZ}
\affiliation{\Regina}
\affiliation{\Jag}
\affiliation{\Dresden}
\affiliation{\Geel}
\affiliation{\Tuebingen}
\affiliation{\Bratislava}
\affiliation{\Williams}
\affiliation{\TTU}
\affiliation{\Queens}
\affiliation{\Warwick}
\affiliation{\UCBPH}
\affiliation{\IPPFDD}
\affiliation{\UCBNE}

\author{N.~Abgrall}\affiliation{\LBNL}
\author{I.~Abt}\affiliation{\MPP}
\author{M.~Agostini}\affiliation{\UCL}
\author{A.~Alexander}\affiliation{\UCL}
\author{C.~Andreoiu}\affiliation{\SFU}
\author{G.R.~Araujo}\affiliation{\UZH}
\author{F.T.~Avignone III}\affiliation{\USC}\affiliation{\ORNL}
\author{W.~Bae}\affiliation{\UTAustin}
\author{A.~Bakalyarov}\affiliation{\NRCKI}
\author{M.~Balata}\affiliation{\LNGS}
% ...

%% file: gen/Acknowledgments-2021-02-18.tex
This material is based upon work supported by the U.S.~Department of Energy, Office of Science, Office of Nuclear Physics under Federal Prime Agreements
DE-AC02-05CH11231,  	
DE-AC05-00OR22725, 	
LANLEM77, 			
and under award numbers
DE-FG02-97ER41020, 	
DE-FG02-97ER41033, 	
DE-FG02-97ER41041, 	
DE-FG02-97ER41042, 	
DE-SC0017594,		
DE-SC0012612, DE-SC0018060, and		
DE-SC0014445. 		
We acknowledge support from the Nuclear Precision Measurements program of the Division of Physics of the National Science Foundation through grant numbers
NSF PHY-1812374,			
NSF PHY-1812356,		
NSF PHY-1812409,			
and from the Office of International Science and Engineering of the National Science Foundation through grant number
NSF OISE 1743790.			
We gratefully acknowledge the support of the U.S.~Department of Energy through the LANL, ORNL and LBNL Laboratory Directed Research and Development (LDRD) Programs for this work. 	
This research was supported in part by the Excellence Cluster ORIGINS (EXC-2094 : 390783311) and the SFB1258 		
which are funded by the Deutsche Forschungsgemeinschaft (DFG, German Research Foundation). 	
We acknowledge the support of the German Federal Ministry for Education and Research (BMBF) through grant number 05A2020. 	
and the Max Planck Society (MPG). 	
This work is supported in part by the European Research Council (ERC) under the European Union's Horizon 2020 research and innovation programme (Grant agreement No.~786430 - GemX).	
We gratefully acknowledge the Italian Istituto Nazionale di Fisica Nucleare (INFN), 		
the Polish National Science Centre (NCN, grant number UMO-2020/37/B/ST2/03905), the Polish Ministry
of Science and Higher Education (MNiSW, grant number DIR/WK/2018/08), 		
the Czech Republic Ministry of Education, Youth and Sports CZ.02.1.01/0.0/0.0/16\_019/0000766 and LM2018107,	
the Slovak Research and Development Agency, grant No.~15-0576,	
and the Swiss National Science Foundation (SNF).		
This project has received funding /support from the European Union's Horizon 2020 research and innovation programme under the Marie Sk\l{}odowska-Curie grant agreement No 860881-HIDDeN. 	
This work has been supported by the Science and Technology Facilities Council, part of U.K.~Research and Innovation  (grant numbers ST/T002042/1 and ST/T004169/1).	
We acknowledge support from the Russian Foundation for Basic Research (RFBR), grant No.~15-02-02919, 	
and from the Institute of Nuclear Physics and Technology of National Research Nuclear University ``Moscow Engineering Physics Institute'' and by the Ministry of Science and Higher Education of the Russian Federation, Project ``Fundamental properties of elementary particles and cosmology'' No 0723-2020-0041.	
We acknowledge the support of the Natural Sciences and Engineering Research Council of Canada, funding reference number SAPIN-2017-00023, 	
and from the Canada Foundation for Innovation John R.~Evans Leaders Fund.  
This research used resources provided by
National Energy Research Scientific Computing Center, a U.S.~Department of Energy Office of Science User Facility under Contract No.~DE-AC02-05CH11231,  	
and the Oak Ridge Leadership Computing Facility at Oak Ridge National Laboratory.	
The collaboration thanks the directors and the staff of the Laboratori Nazionali del Gran Sasso for their continuous strong support of the LEGEND experiment.		
We thank our hosts and colleagues at the Sanford Underground Research Facility for their support. 		

%% file: sec_overview/sec_overview.tex
\section{Executive Summary}\label{sec:overview}

\textbf{Objective:}
We propose the construction of \Lk, the ton-scale Large Enriched Germanium Experiment for Neutrinoless $\beta \beta$  Decay. This international experiment is designed to answer one of the highest priority questions in fundamental physics.
It consists of 1000 kg of Ge detectors enriched to more than 90\% in the \nuc{76}{Ge} isotope of interest operated in a liquid argon active shield at a deep underground laboratory.
The experiment is designed to achieve a discovery potential that covers  the inverted-ordering neutrino mass scale region. The baseline design assumes installation in the SNOLAB cryopit. A similar experimental setup could also be realized at the alternative LNGS site.

\textbf{Vision and Mission:}
Neutrinos have been at the forefront of discovery in particle physics for
decades, and the study of their properties drove the conception of the weak
interaction and modern quantum field theories.
Still unanswered is the important fundamental question of whether the
neutrino is identical to its anti-particle, i.e., a Majorana particle, a
property connected to the origin of its mass.
Majorana neutrinos are naturally predicted by many extensions of the Standard
Model (SM). They are also predicted by leptogenesis, a leading model to account for
the predominance of matter over antimatter in the Universe.

The only known, feasible probe of the Majorana nature of the neutrino is
neutrinoless double-beta (\BBz) decay, an as yet unobserved radioactive transition.
The discovery of \BBz\ decay would prove unambiguously not only that new
lepton-number-violating physics exists but also that it is connected to the
mysterious origin of the neutrino's mass.
There is a long history of searches for \BBz\ decay. At present, the most
stringent constraints are set by experiments using the isotopes
\nuc{76}{Ge}, 
\nuc{130}{Te}, 
and \nuc{136}{Xe}, 
with lower bounds on the decay half-life surpassing \powten{26}\,years.
For minimal extensions of the SM, these limits constrain the effective Majorana
neutrino mass \mee\ at the scale of 100\,meV.

\Lk\ is designed to probe \BBz\ decay with a 99.7\% CL discovery sensitivity, defined to be a 50\% chance of measuring a signal of at least 3$\sigma$ significance, in the \nuc{76}{Ge} half-life beyond $10^{28}$\,years, corresponding to a \mee\ upper limit in the range of 9--21 meV in 10 yr of live time.  By combining the lowest background levels and the best energy
resolution in the field, \Lk\ will perform a quasi-background-free search and
can make an unambiguous discovery of \BBz\ decay with just a handful of counts
at the \BBz\ decay $Q$ value (\qval).
The experiment uses an unambiguous signature for the events of interest: a fully
contained event at a very specific energy (2039 keV) with a distinctive signal
shape that indicates a single-interaction event topology in the bulk of a Ge
detector and with no accompanying signals from other detectors.
\Lk's discovery sensitivity covers the inverted-ordering neutrino mass scale. It
also probes the next order of magnitude for the normal ordering and
other exchange mechanisms.

 \textbf{Experimental Advantages:}
 Germanium is a leading material for \BBz\ decay searches:
 \begin{myitemize}\vspace{-1em}
 \item Germanium detectors achieve the best energy resolution of any \onbb\
 decay experiment, while also providing detailed information on the event topology.
 \item It can be isotopically enriched to greater than 90\% in \nuc{76}{Ge} in sufficient
 quantities and converted into high-purity Ge detectors with high yield.
 \item Nearly all \nuc{76}{Ge} decays occur in active detector regions that do
 not require self-shielding to eliminate background, resulting in high detection
 efficiency.
 \item Germanium detectors have undetectably low \Th- and \U-chain internal
 contamination and no known background source produces a peak in the vicinity
 of \qval.
 \item   Experiments based on \nuc{76}{Ge} (\Gerda\ and \MJD) have achieved the
 lowest background of any \onbb\ decay experiment when normalized to energy
 resolution and operate in a quasi-background-free regime, with no
 contamination from two-neutrino double-beta (\BBt) decays.
 \emph{Thus, \BBz\ decay events would create a lone, sharp peak in the energy spectrum,
 visible to the eye.  The extraction of a \onbb\ decay signal does not rely on
 background modeling and so has negligible systematic uncertainty.}
 \end{myitemize}

\textbf{Key Innovations:}
 While Ge detector technology is mature and proven, it continues to be improved.
 \Lk\ builds upon major breakthroughs achieved in the current-generation Ge-based
 searches, combining excellent energy resolution with the most effective
 background-mitigating practices employed in \Gerda\ and the \MJD.
 The immersion of Ge detectors in a LAr scintillating medium by \Gerda\ has
 shown that backgrounds can be greatly suppressed. The dominant backgrounds in
 \Gerda, other than \nuc{42}{Ar}, can be reduced using \MJD's materials and
 improved energy resolution. The use of underground-sourced liquid argon
 (UGLAr) can
 drastically reduce the \nuc{42}{Ar} background. \Lk\ is informed by the design of a
 200-kg phase, \Ltwo, which is using new inverted-coaxial
 point-contact (\icpc) Ge detectors with more than a factor of two greater mass
 per crystal over previous experiments and excellent energy resolution. Combining these innovations leads to an achievable \Lk\ background goal of \bgkev.

\textbf{\Lthou\ Baseline Design:}
The \Lk\ experiment utilizes the demonstrated low background and excellent
energy performance of high-purity p-type, \icpc\ Ge semiconductor detectors,
enriched to more than 90\% in \nuc{76}{Ge}.
The background rejection power of \icpc\ detectors begins with their superb
energy resolution, demonstrated to have a full-width at half-maximum (FWHM)
resolution of 0.12\% (0.05\% $\sigma$) at \Qbb.
Pulse shape analysis of the signal distinguishes bulk \BBz\
decay energy depositions from surface events and backgrounds from $\gamma$
rays with multiple interaction sites. The granular nature of the Ge detector array
allows rejection of background events that span multiple detectors.
Finally, background interactions external to the Ge detectors are identified by
LAr scintillation light.

About 400 \icpc\ detectors with an average mass of 2.6~kg each are distributed among
four 250-kg modules to allow independent operation and phased commissioning. In
each module, the detectors are arranged into 14 vertical strings, are supported by
ultra-clean materials and read out using ultra-low-background ASIC-based electronics.
The detector strings are immersed in radiopure UGLAr, reduced in
the \nuc{42}{Ar} isotope and contained within an electroformed copper reentrant
tube. Each of the four UGLAr modules is surrounded by LAr
produced from atmospheric Ar, contained within a vacuum-insulated cryostat.
The LAr volumes are instrumented with
an active veto system comprising optical fibers read out by Si
photomultipliers.  The cryostat is enveloped by a water tank providing
additional shielding. The baseline design assumes installation in the SNOLAB cryopit.

%% file: sec_introduction/sec_introduction.tex
\section{Introduction and Science Program}\label{sec:science}
\input{sec_introduction/subsec_A-Lepton-viol}
\input{sec_introduction/subsec_B-double-beta}
\input{sec_introduction/subsec_C-theory-uncer}
\input{sec_introduction/subsec_D-norm-order}
\input{sec_introduction/subsec_G-experimental-context}

%% file: sec_introduction/subsec_A-Lepton-viol.tex
\subsection{Global Symmetries of the Standard Model and Lepton Number Violation}\label{subsec:LeptonViol}

In the Standard Model (SM)~\cite{Weinberg:2018apv} of particle physics,
lepton flavor and total lepton number, $L$, are both conserved
quantities~\cite{deGouvea:2013zba}. The existence of neutrino oscillation
indicates that flavor lepton number is not conserved. The quantity $L$, on the
other hand, is associated with the SM's $U(1)_L$ symmetry and remains a
conserved quantity at the classical level in the SM.

This empirically observed symmetry, like that associated with baryon number,
$B$, is accidental in the SM~\cite{Rodejohann:2011mu}.  In Grand Unified
Theories (GUTs), quarks and leptons are grouped in multiplets, and thus $B$ and
$L$ are not expected to be conserved.  In the SM, the combination $B-L$ is
conserved, but it is usually broken at some scale in GUTs.  Furthermore, there
is an excess of baryons over anti-baryons in the Universe.  The Big Bang
presumably created equal numbers of each, suggesting that some $B$- or
$L$-violating process must have subsequently generated the asymmetry.
The SM contains in principle all the ingredients required for the baryon asymmetry, but it quantitatively predicts a much smaller asymmetry than is  observed~\cite{tHooft:1976rip,Davidson:2008bu}.
Therefore, new physics is required.

Neutrino oscillations prove that neutrinos have mass, but they provide
information only on the differences between their masses squared. Precise measurements
of beta-decay kinematics constrain the neutrino mass to be much smaller than
that of their charged-lepton counterparts. The seesaw mechanism provides an
explanation for this very low neutrino mass and the observation of  left-handed
neutrinos  without their right-handed counterparts. In addition, the seesaw
mechanism predicts neutrinos to be their own antiparticles, i.e., Majorana particles.
Light Majorana neutrinos, along with their heavy partners, could be harbingers of leptogenesis as an explanation of the baryon-antibaryon imbalance~\cite{Fukugita:1986hr,Davidson:2008bu}. Furthermore, the Majorana or Dirac nature of neutrinos directly impacts how their mass would be incorporated as extensions to the SM.

The question of the Majorana or Dirac nature of neutrinos and whether $L$ is violated is an experimental one; theory requires guidance from measurements. Since the answer is critical to many physics questions, searches for lepton number violation are extremely well motivated and a range of potential experimental probes exist. The most sensitive experimental test for lepton number violation and Majorana neutrinos is the search for neutrinoless double-beta (\BBz) decay~\cite{Haxton:1985am,Davidson:2008bu,Rodejohann:2011mu}.

%% file: sec_introduction/subsec_B-double-beta.tex
\subsection{Neutrinoless Double-Beta Decay}\label{subsec:BBintro}

Double-beta decay is a transition between nuclei of the same atomic mass number, $A$, that
changes the nuclear charge, $Z$, by two units through the emission of light particles. For nuclei close to the valley of stability with even numbers of both neutrons and protons,  beta decays are often energetically forbidden or highly suppressed, leaving double-beta decay as the only potential decay mode.  The transformation can occur by two-neutrino double-beta (\BBt) decay,%
\be
(A,Z) \rightarrow (A,Z+2) + e^- + e^- + \bar{\nu}_e + \bar{\nu}_e ,
\ee%
\noindent conserving lepton number.
This \BBt\ decay is allowed within the SM as a rare, second-order process and has been observed in 11 nuclei~\cite{Barabash:2020nck}.  In contrast, \BBz\ decay,%
\be
(A,Z) \rightarrow (A,Z+2) + e^- + e^- ,
\ee
violates lepton number by two units and is forbidden in standard electroweak theory. There have been a great number of reviews on the topic of \BBz\ decay over the years, and we refer the reader to Refs.~\cite{Schwingenheuer:2012zs,Cremonesi:2012av,Elliott:2014iha,Engel:2016xgb,Henning:2016fad,Barabash:2019suz,Dolinski:2019nrj}.

The two decay modes are distinguishable by the spectrum of the summed energies of the two outgoing electrons.
For the \BBt\ decay mode, the summed electron kinetic energy is reduced by the total energy of the outgoing antineutrinos and therefore
displays a continuous spectrum with a broad maximum around one third of the endpoint energy.
In contrast, the \BBz\ decay mode exhibits a monoenergetic line at the decay $Q$ value (\qval), as
the electrons carry the full available energy.

The \BBz\ decay mode can be mediated by various mechanisms. One requiring minimal new
physics is the exchange of a light Majorana neutrino interacting via standard,
left-handed V-A weak currents. The half-life (\Tz) can be expressed
as~\cite{Cirigliano:2018yza}
\begin{eqnarray}
\label{Eqn:HalfLife}
(\Tz) ^{-1}= \Gz \, \gA^4 \left(\Mz + \dfrac{g_{\nu}^{NN} m_{\pi}^2}{\gA^2} \Mcont \right)^2 \mee^2
\end{eqnarray}
where \Gz\ is the phase-space integral and \gA\ is the axial-vector coupling constant.
The nuclear matrix element \Mz\ is composed of Gamow-Teller (GT), Fermi (F) and tensor (T) components as follows:
\begin{eqnarray}
\label{eqn:Mz}
\mbox{\Mz} = \mbox{\MzG} - \left( \frac{\mbox{\gV}}{\mbox{\gA}} \right)^2 \mbox{\MzF}+\mbox{\MzT}.
\end{eqnarray}
\Mcont\ is a recently identified
contact operator that comes with its own hadronic coupling $g_{\nu}^{NN}$
normalized by the pion mass $m_{\pi}$~\cite{Cirigliano:2018hja}.
The \mee\ term is the effective Majorana neutrino mass that captures the physics of the
light-neutrinos that mediate the decay. It is a coherent sum of the neutrino parameters:
\begin{equation}
  \label{eqn:Mbb}
  \mee = \left| \sum_{i=1}^{3}U_{ei}^2 m_i \right|
\end{equation}
where $U_{ei}$ are the
elements of the neutrino mixing matrix, which include the Dirac and Majorana CP-phases, and $m_i$ are the neutrino mass eigenvalues~\cite{Zyla:2020zbs}.

\begin{figure}
\includegraphics[trim={7mm 0 7mm 0}, clip=false, width=1.0\columnwidth]{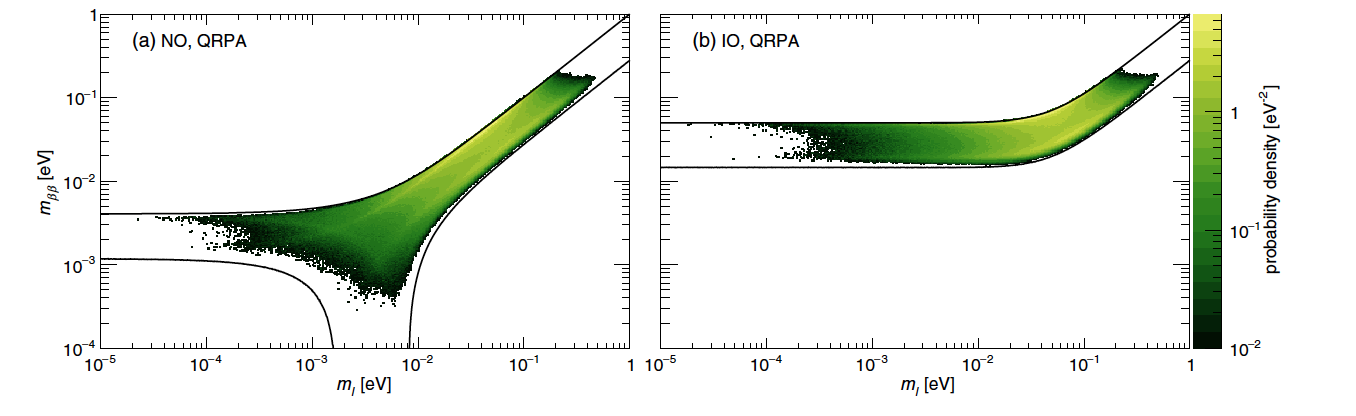}
\caption{\label{fig:benatoposterior} Parameter space for the effective Majorana
neutrino mass \mee\ as a function of the lightest neutrino mass
$m_l$ for (left) the normal mass ordering (NO) and (right) the inverted
ordering (IO). The solid lines show the allowed regions for all possible
CP-phases and assume 3$\sigma$ intervals of the neutrino oscillation
observables (NuFIT~\cite{Esteban:2016qun}). The color scale denotes
marginalized posterior distributions for \mee\ and $m_l$,
obtained by combining Eqn.~\eqref{eqn:Mbb} with neutrino oscillation measurements,
and assuming the absence of mechanisms that drive $m_l$ or \mee\ to 0.
Figure taken from Ref.~\cite{Agostini:2017jim}.
}
\end{figure}

Figure~\ref{fig:benatoposterior} displays the range of \mee\ as a function of the
lightest neutrino mass $m_l$: $m_l = m_1$ for normal ordering and $m_l = m_3$ for inverted
ordering. The width of the bands is primarily due to the uncertainty on the Majorana phases, for which no experimental information is available.
In the normal ordering, the unconstrained Majorana phases can conspire to give
vanishing values of \mee. However, in the inverted ordering, \mee\ is
constrained from below.
Using the central values and uncertainties of the latest neutrino-oscillation
data~\cite{Zyla:2020zbs}, the minimum value of \mee\ for the inverted ordering corresponds to $18.4\pm1.3$~meV~\cite{Agostini:2021kba}.

%% file: sec_introduction/subsec_C-theory-uncer.tex
\subsection{Nuclear Physics Uncertainties}\label{subsec:TheoryUnc}

The conversion of an experimental \BBz\ decay half-life into
a value for \mee\ requires inputs from nuclear theory, as shown by
Eqn.~\eqref{Eqn:HalfLife}.
The uncertainty in the phase space factor is small, and the two available
calculations for Ge
agree very well: $2.36\times10^{-15}$/yr~\cite{Kotila:2012zza} and
$2.37\times10^{-15}$/yr~\cite{Mirea:2015nsl}. Conversely,
state-of-the-art nuclear matrix element calculations can differ by more than a
factor of two as shown in Table~\ref{tab:MatrixElements}.  A world-wide effort
to reduce these uncertainties is ongoing within the nuclear theory community,
with significant advancements made in recent years.
While this effort bears fruit, following the recommendation by the U.S. Nuclear Science Advisory Committee's
subcommittee on neutrinoless double-beta decay~\cite{McKeown2014}, we continue to quote our \mee\ sensitivity
providing a range of values that extends from the conversion obtained with the
largest and smallest matrix elements available among four primary calculation methods, using
an unquenched $g_A$ value of 1.27 and no contribution from the contact term.
Uncertainties associated with the possible quenching of the axial vector coupling
and with the contact term are not
explicitly included. Their impacts are discussed below.

\begin{table}
\begin{center}
  \caption{A summary of the state-of-the-art matrix element calculations for
  \gesix\ used in this document. Very recently, a first ab-initio calculation has
  been reported for a few light $\beta\beta$ nuclei~\cite{Belley:2020ejd},
  resulting in a value of $2.14 \pm 0.09$ for \nuc{76}{Ge}. This calculation
  should be treated differently from the others, in particular because it
  incorporates the physics responsible for quenching.  It will also take some
  time before this theoretically more accurate method can be applied to
  higher-mass nuclei, allowing for comparison.  Hence it is not included in the
  collection of values we use to calculate the effective Majorana mass.}
  \label{tab:MatrixElements}
\begin{tabularx}{0.95\tw}{XX }
\rowcolor{legendgrey}
\thead[l]{Framework} &
\thead[l]{Nuclear Matrix Element Values (\Mz)} \\\hline
Nuclear Shell Model &
3.37\,-\,3.57~\cite{Horoi:2015tkc},
2.89\,-\,3.07~\cite{Menendez:2017fdf},
         2.66~\cite{Coraggio:2020hwx}\\
Quasiparticle Random Phase Approximation &
         5.09~\cite{Mustonen:2013zu},
         5.26~\cite{Hyvarinen:2015bda},
         4.85~\cite{Simkovic:2018hiq},
3.12\,-\,3.40~\cite{Fang:2018tui} \\
Interacting Boson Model&
         4.68~\cite{Barea:2015kwa}\\
Energy Density Functional&
         4.60~\cite{Rodriguez:2010mn,Rodriguez:2012rv},
         5.55~\cite{Vaquero:2014dna},
         6.04~\cite{Song:2017ktj} \\
Range&
         2.66 -- 6.04 \\
\end{tabularx}
\end{center}
\vspace{-5mm}
\end{table}

\paragraph{Quenching}
It is well-known that in the case of single $\beta$ decay, many-body
calculations yield decay rates that significantly exceed measurements, but
this disagreement can be alleviated by ``quenching'' \gA\ by roughly
25\%~\cite{Towner:1987zz}.
Equation~\eqref{Eqn:HalfLife} shows the overall \gA$^4$ dependence on the \BBz\
decay rate, while Eqn.~\eqref{eqn:Mz} shows its explicit factor in the Fermi
matrix element contribution to \Mz.
Simply scaling \Mz\ as \gA$^2$, quenching would introduce a 40\%
effect on \Mz. Since the Fermi contribution is sizable, however, the values quoted by
Refs.~\cite{Simkovic:2013qiy} and \cite{Hyvarinen:2015bda} for \nuc{76}{Ge}, for
example, are closer to 20--30\%; for Ref.~\cite{Vaquero:2014dna}, the value
is only a 10\% effect.  Since the decay rate depends on $M^2_{0\nu}$, a 25\% quenched
matrix element would alter the decay rate by about 44\%.

Some authors have claimed that the quenching could be very large (see the summary in
Ref.~\cite{Ejiri:2015wna}) based on the fact that large quenching is seen in \BBt\ decay.
Unlike \BBt\ decay, however, \BBz\ decay is a process that includes all
multipoles of the intermediate states. Muon capture, which is similar in this
regard, requires little if any quenching~\cite{Suzuki:2018aey}; recent
neutrino cross-section calculations also seem to not require
quenching~\cite{Lovato:2020kba}.

The theory is advancing, and investigations using effective field theory
indicate that quenching in single $\beta$ decay likely arises from the
combination of two-nucleon weak currents and strong correlations in the
nucleus that are omitted in traditional
calculations~\cite{Pastore:2017uwc,Gysbers:2019uyb}. New ab-initio \Mz\
calculations such as in Ref.~\cite{Belley:2020ejd} attempt to explicitly include all
of this typically omitted physics and indicate a suppression of \Mz\ in
$^{76}$Ge indeed by about 25\%, albeit with still large uncertainty.
The picture is converging, and it is
likely that this uncertainty can be resolved in the near future.

\paragraph{The contact term from effective field theory}
It has recently been recognized that a leading order, short-range contribution
has previously been ignored in calculations of \BBz\ decay transition
operators~\cite{Cirigliano:2018hja,Cirigliano:2019vdj}.
This contact term represents the effects of heavy mesons and quark/gluon
physics that can be excited by the exchanged neutrino when its energy is above
about a GeV.  Effective field theory
and ab-initio nuclear structure provide a scheme for estimating how large
the coefficient of such a term in the double-beta operator should be,
and recent work in this direction~\cite{Wirth:2021pij} indicates that this term
is potentially tens of percent of \Mz\ in magnitude with the same sign, leading
to enhancement of the decay rate.
This would offset some of the possible reduction in decay rate expected due to
quenching.  Lattice quantum-chromodynamics calculations are also being pursued to compute
this essential contribution~\cite{Davoudi:2020ngi}.

%% file: sec_introduction/subsec_D-norm-order.tex
\subsection{Discovery Opportunities for Next Generation Experiments}\label{subsec:NormOrd}

There is no theoretical motivation to favor one neutrino mass ordering over another.
Long-baseline neutrino experiments (NOvA~\cite{Acero:2019ksn} and T2K~\cite{Abe:2019vii}) are currently the most sensitive probes we
have to test the mass ordering. Their results had favored the normal ordering
for a few years, but interpretations of their latest higher-statistics data sets are now pushing in
the opposite direction~\cite{Esteban:2020cvm,Kelly:2020fkv}.
Global fits that include long- and short-baseline data
currently show a mild preference for normal ordering, but the statistical
significance is too low to draw any conclusion at the
moment~\cite{Esteban:2020cvm,Capozzi:2020add,deSalas:2020pgw}. It seems unlikely that the mass ordering will be
established in this decade, before data from future experiments such as
JUNO~\cite{Lu:2021lfe}, KM3NeT~\cite{KM3NeT:2021ozk}, DUNE~\cite{DUNE:2020lwj}, and Hyper-Kamiokande~\cite{Abe:2018uyc} become available.

Future \BBz\ decay experiments probing \mee\ values down to 18.4\,meV will
be able to test the inverted ordering parameter space~\cite{Agostini:2021kba}. Reaching such a sensitivity will offer exciting discovery opportunities even when assuming normal ordering. The current best bounds on \mee\ are at the level of 160--180\,meV assuming
less favorable nuclear matrix element calculations.
Reaching 18.4\,meV means probing 80--90\% of the currently allowed range for the normal
ordering.
Bayesian analyses~\cite{Agostini:2017jim, Caldwell:2017mqu} have tried to
quantify the discovery probability of future experiments using all experimental information available.
The posterior distributions for \mee\ and the lightest neutrino mass $m_l$ are typically not uniform (see Fig.~\ref{fig:benatoposterior}).
Low \mee\ values are disfavored as they require a fine tuning of the Majorana
phases whose values are random in many reasonable scenarios. This leads to
discovery probabilities of up to 50\% even when assuming the normal ordering.
Moreover, cosmological data suggests non-vanishing discovery opportunities even assuming the less favorable combination of Majorana phases~\cite{Agostini:2020oiv}.

A discovery of \BBz\ decay with an \mee\ value close to the current experimental limits would also create discovery expectations for cosmology and tritium $\beta$-decay experiments.
Cosmological observations of the cosmic microwave background, baryon acoustic oscillations, and Lyman-$\alpha$ forest are sensitive to neutrino properties, in particular to the sum of the neutrino mass eigenstates~\cite{Lahav:2019bbc, Zyla:2020zbs}. Precision measurements of the tritium $\beta$-decay kinematics are instead sensitive to the incoherent sum of neutrino masses. Both mass quantities will be measured with increasing precision in the years to come, with the tritium $\beta$-decay experiment KATRIN~\cite{Aker:2021exx} continuing data-taking, to be followed up by Project 8~\cite{Project8:2017nal}, and planned cosmological observations~\cite{Dvorkin:2019jgs} coming online.
The three-Majorana-neutrino model predicts correlations among these quantities and \mee, offering a unique opportunity to pin down the model parameters or prove the existence of additional \BBz\ decay mechanisms.

If one considers alternative mediation mechanisms beyond the light, left-handed neutrino exchange paradigm, the parameter space for discovery opens up
considerably. The existence of a sterile neutrino, for example, could swap the
order in which \BBz\ decay probes neutrino masses, with the normal ordering
being potentially fully probed by ton-scale experiments and the
inverted ordering allowing for vanishing decay rates~\cite{Rodejohann:2011mu}.
For more general new physics, a ``master formula'' has been developed
in Ref.~\cite{Cirigliano:2018yza} for computing
contributions to \BBz\ decay for Beyond-SM Lagrangian contributions up to
dimension nine. In the case of the exchange of heavy particles, the \BBz\ decay rate scales
generically as an inverse power of the energy scale at which new physics
appears. From this perspective, any improvement in sensitivity in \BBz\ decay experiments
probes ever higher energy scales. This includes energy scales that can be probed
in complementary searches at particle colliders as
well as energy scales that are not accessible by current accelerator technologies
(see Ref.~\cite{Peng:2015haa}).

\new{In addition to the crucial search for the lepton-number-violating \BBz\ decay, future \BBz\ decay experiments will have broad physics programs.
The larger mass and lower backgrounds of \Lk\ extends its reach for other Beyond Standard Model (BSM) searches well beyond that of the \MJD\ and \Gerda\ experiments. Table~\ref{tab:BSM} lists some of these other BSM physics accessible to \Lk.
A potentially-significant background for these searches is \nuc{39}{Ar} decay in the LAr, but this risk has been mitigated by the use of underground-sourced Ar. This brings the \nuc{39}{Ar} background rate in \Lk\ below 100\,keV more than an order of magnitude lower than the background rate achieved in the \MJD.
}

\begin{table}[t]
\begin{center}
\caption{\new{A non-exhaustive listing of recent and proposed BSM physics searches by Ge-based experiments.}}
\label{tab:BSM}
\begin{tabularx}{0.95\tw}{Y Y C{0.15\tw} C{0.2\tw}}
\rowcolor{legendgrey}
\thead{Physics} &
\thead{Signature} &
\thead{Energy} &
\thead{Experiment}\\
\rowcolor{legendgrey}
&
&
\thead{Range} &
\\
\hline
Bosonic dark matter	&
Peak at DM mass &
$<1$ MeV &
\MJ \cite{Abgrall:2016tnn}, \Gerda\
\cite{GERDA:2020emj} \\
Electron decay &
Peak at 11.8 keV &
$\sim10$~keV &
\MJ\
\cite{Abgrall:2016tnn} \\
Pauli exclusion principle violation	&
Peak at 10.6~keV &
$\sim10$~keV &
\MJ\
\cite{Abgrall:2016tnn} \\
Solar axions &
Peaked spectra, daily modulation &
$<10$~keV&
\MJ
\cite{Abgrall:2016tnn,Xu:2016tap} \\
Majoron emission &
\nnbb\ spectral distortion &
$<$ \Qbb &
\Gerda\
\cite{Agostini:2015nwa} \\
Exotic fermions &
\nnbb\ spectral distortion &
$<$ \Qbb &
(proposed)
\cite{Bolton:2020ncv, Agostini:2020cpz} \\
Lorentz violation &
\nnbb\ spectral distortion &
$<$ \Qbb &
(proposed)
\cite{Diaz:2013saa, Diaz:2013ywa, CUPID:2019kto}\\
Exotic currents in \nnbb\ decay &
\nnbb\ spectral distortion &
$<$ \Qbb &
(proposed)
\cite{Deppisch:2020mxv} \\
Time-dependent \nnbb\ decay rate &
Modulation of \nnbb\ spectrum &
$<$ \Qbb &
(proposed)
\cite{NEMO-3:2020mcq} \\
WIMP and related searches &
Exponential excess, annual modulation&
$<10$~keV &
CDEX
\cite{Liu:2019kzq}\\
Baryon decay &
Timing coincidence &
$>10$ MeV &
\MJ\
\cite{Alvis:2018pne} \\
Fractionally charged cosmic-rays &
Straight tracks&
few keV &
\MJ\
\cite{Alvis:2018yte}\\
Fermionic dark matter &
Nuclear recoil/deexcitation &
$<$ few MeV &
(proposed)
\cite{Dror:2019dib}\\
Inelastic boosted dark matter &
Positron production &
$<$ few MeV &
(proposed)
\cite{COSINE-100:2018ged}\\
BSM physics in Ar &
Features in Ar veto spectrum &
ECEC in $^{36}$Ar  &
\Gerda\
\cite{Agostini:2016rsa}\\
\end{tabularx}
\end{center}
\end{table}%

\new{The detection signatures of these new physics phenomena align well with the strengths of \Lk. New BSM particles such as dark matter WIMPs or axions could interact within the detectors, creating peaks in the energy spectrum or an accumulation of events at low energy.
Exotic fermions and bosons could also be created by double-beta decays, deforming the energy distribution of the two electrons ejected by the nucleus~\cite{Blum:2018ljv}. Other BSM searches include electron decay~\cite{Piscicchia:2015beq} and fractionally charged cosmic-rays~\cite{Graesser:2021vkr}.
\Lk\ could also scrutinize the recently reported excess in the XENON1T low-energy electron recoil spectrum~\cite{Aprile:2020tmw}.
These science opportunities complement the already exciting physics program of \Lk.}

%% file: sec_introduction/subsec_G-experimental-context.tex
\subsection{The Worldwide Program in Experimental \BBz\ Decay}\label{subsec:ExptProg}

At the end of the past century, the \BBz\ decay panorama was defined
by the Heidelberg-Moscow~\cite{KlapdorKleingrothaus:2000sn} and IGEX~\cite{Aalseth:2002rf} experiments,  both of which used
high-purity Ge detectors made from material isotopically enriched
in \nuc{76}{Ge} to around 86\%.
Finding no signal, these two experiments set leading limits on the \BBz\ decay half-life.
While part of the Heidelberg-Moscow collaboration~\cite{KlapdorKleingrothaus:2004wj} later claimed evidence for a signal,
there were various inconsistencies in the methodology~\cite{Schwingenheuer:2012zs}, and the claim is now refuted by present experiments.
In the initial decade of the new century,  CUORICINO~\cite{Andreotti:2010vj} and
NEMO-3~\cite{Arnold:2015wpy} provided bounds on the half-lives of new isotopes: \nuc{130}{Te}, \nuc{100}{Mo}, and \nuc{82}{Se}.
Although the results were no more restrictive than the previous Ge experiments, they were harbingers of technology to come. CUORICINO was the precursor of the CUORE and CUPID programs, and NEMO-3 developed a technique to study the various kinematic variables in \BBz\ decay.

At present, the landscape is shaped by the results of five experiments. The
\Gerda~\cite{Agostini:2020xta},
\MJD\ \cite{Aalseth:2017btx,Alvis:2019sil},
EXO-200~\cite{Anton:2019wmi}, KamLAND-Zen~\cite{KamLAND-Zen:2016pfg} and CUORE~\cite{Adams:2019jhp,Adams:2021rbc} experiments provide the strongest constraints on \BBz\ decay.
\Gerda, located at the Laboratori Nazionali del Gran Sasso (\Lngs) in Italy, operated an array of high-purity Ge detectors
enriched in the isotope \nuc{76}{Ge}, immersed
in a LAr active shield (see Sect.~\ref{subsec:gerda}).
Similarly,  the \MJ\ \DEM, located at the Sanford Underground Research Facility (SURF) in the USA, investigated the \BBz\ decay of \nuc{76}{Ge} with a Ge detector array deployed in vacuum
cryostats with shielding produced from ultra-pure Cu and Pb (see Sect.~\ref{subsec:mjd}).
The EXO-200 experiment, located at the Waste Isolation Pilot Plant facility in the USA, operated a  liquid Xe time-projection-chamber (TPC) enriched in the isotope  \nuc{136}{Xe}.
KamLAND-Zen, located in the Kamioka mine in Japan,
is a conversion of the neutrino experiment KamLAND into an
apparatus capable of studying \BBz\ decay by dissolving enriched Xe gas
in organic scintillator.
CUORE, located at \Lngs, searches for the  \BBz\ decay of \nuc{130}{Te}. CUORE operates an array
of natural isotopic abundance TeO$_2$ bolometers at 10 mK in a specially designed cryostat.

\begin{table}
\begin{center}
  \caption{Comparison of lower half-life limits \Tz\ (90\% CL) and corresponding upper Majorana neutrino
  mass \mee\ limits for the present-generation experiments. The range of \mee\ upper limits are from each collaboration's choice of multiple matrix elements.}
  \label{tab:present_exp}
\begin{tabularx}{0.95\tw}{Z Z Z Z Z }
\rowcolor{legendgrey}
\thead{Experiment} &
\thead{Isotope} &
\thead{Exposure  $[$kg-yr$]$} &
\thead{\Tz   $[$10$^{25}$ yr$]$} &
\thead{\mee  $[$meV$]$}
 \\\hline
\Gerda~\cite{Agostini:2020xta}     						&  \nuc{76}{Ge} & 127.2   & 18   & $79-180$ \\
\MJ~\cite{Alvis:2019sil}     							&  \nuc{76}{Ge} & 26   & 2.7  & $200-433$ \\
KamLAND-Zen~\cite{KamLAND-Zen:2016pfg}			& \nuc{136}{Xe} & 594  & 10.7 & $61-165$ \\
EXO-200~\cite{Anton:2019wmi}         					& \nuc{136}{Xe} & 234.1  & 3.5  & $93-286$ \\
CUORE~\cite{Adams:2021rbc}      					& \nuc{130}{Te} & 1038.4   & 2.2  & $90-305$ \\
\end{tabularx}
\end{center}
\vspace{-5mm}
\end{table}

These five experiments explored the quasi-degenerate neutrino mass spectrum with no evidence for a positive signal,
and Table~\ref{tab:present_exp}  shows their most
recent results.
The lower limits on \Tz\ can be
converted to upper limits on \mee\ assuming
light neutrino exchange as the dominant mechanism (see Eqn.~\eqref{Eqn:HalfLife}).
In Table~\ref{tab:present_exp}
the upper limits on \mee\ quoted were derived by the collaborations. There is some uniformity in their respective derivations, as choices of the values for  \gA\,=\,1.27, the phase space factors~\cite{Kotila:2012zza}, and the set of nuclear matrix elements (discussed in a recent review~\cite{Engel:2016xgb}) are fairly standard.

\begin{figure}
  \includegraphics[width=0.6\tw]{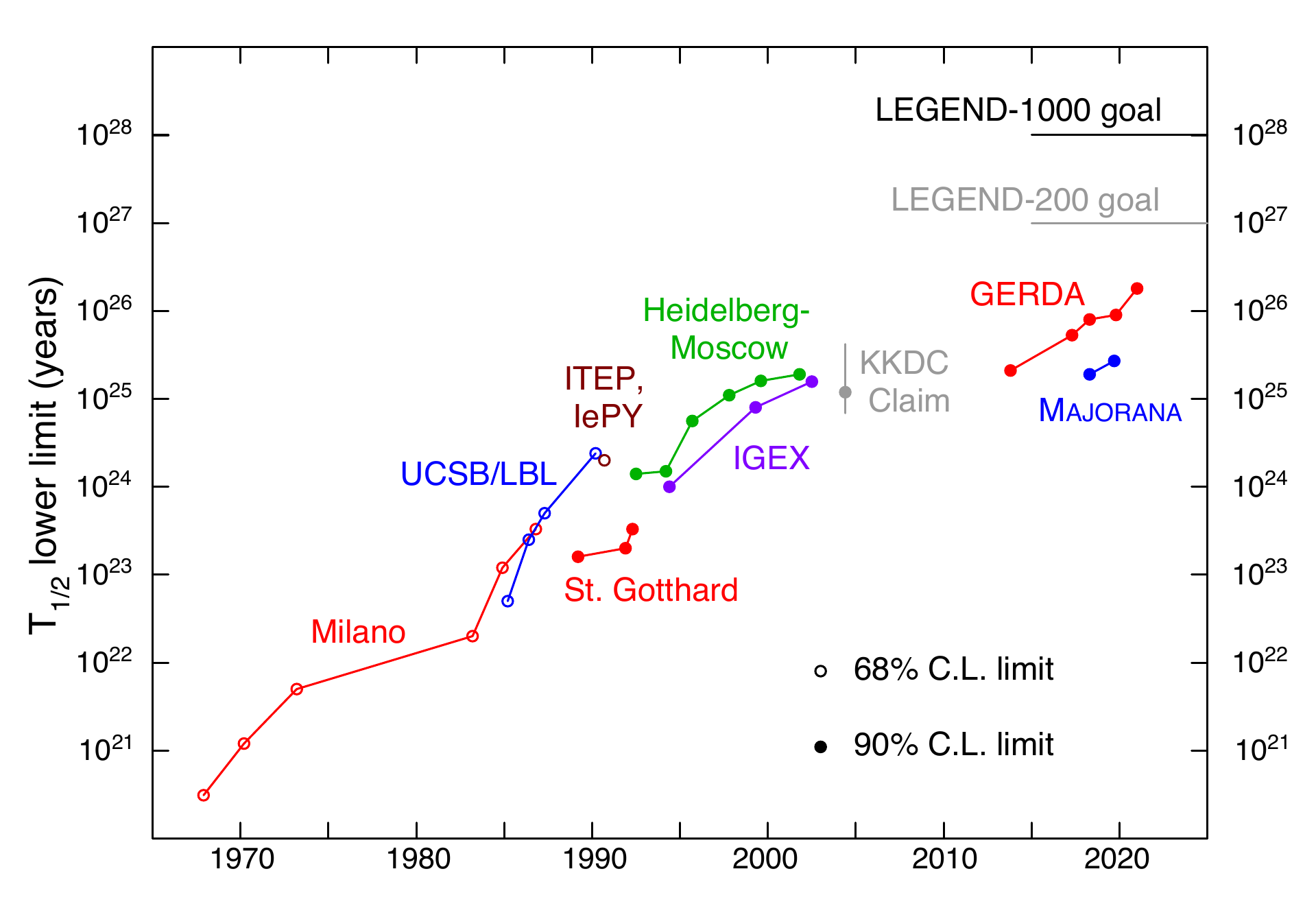}
  \caption{A historical timeline of Ge-based \BBz\ decay experiments and their reported half-life lower limit. The projections for the LEGEND program indicate the future prospects for Ge-based experiments. References to the named experiments can be found in Ref.~\cite{Avignone:2019phg}.}
  \label{fig:GeHistory}
\end{figure}

The next-generation of \BBz\ decay experiments aims to probe
 the inverted mass ordering~\cite{Agostini:2021kba,nEXO:2021ujk,CUPID:2019imh}.
It is clear from Table~\ref{tab:present_exp} that \nuc{76}{Ge} experiments reach
competitive sensitivities with much smaller exposures than the other
technologies. This competitiveness derives from Ge-based experiments'  high
detection efficiency, excellent energy resolution, and operation in a quasi-background-free regime.
As a result, experiments using \nuc{76}{Ge} have been historically impactful in \BBz\ decay~\cite{Avignone:2019phg}; a past timeline of Ge experiments and future projections for LEGEND are shown in Fig.~\ref{fig:GeHistory}.

%% file: sec_ge/sec_ge.tex
\section{Innovation Toward LEGEND-1000}

\input{subsec_exec}

\subsection{Ge Detectors and \BBz\ Decay Searches}
\input{subsec_I-Ge-choice.tex}

\input{subsec_gerda.tex}
\input{subsec_mjd.tex}

\subsection{\Ltwo}\label{subsec:L200}
\input{subsec_H-L200.tex}

%% file: sec_ge/subsec_exec.tex
\Lk\ combines the fundamental strengths of Ge-based \BBz\ decay searches with
significant innovations in Ge-detector technology, active shielding, clean materials, and
high-resolution electronics developed and demonstrated by members of the \Gerda, \MJ, and \Ltwo\
experiments.
The \LEG\ collaboration was formed to bring together the technical expertise and leadership from both the \MJ\ and \Gerda\ collaborations, as well as add new members to strengthen core capabilities.
Its members bring a broad range of deep technical expertise and a history of innovation developed through the design and operation of the \MJD, \Gerda, and \Ltwo\ experiments.  The existing strengths of Ge detectors, described next, provide a basis for their competitive standing in the search for \BBz\ decay; technical advancements have resulted in larger detectors with better energy resolution, operating within a scintillating medium for enhanced background suppression, cleaner materials, and low-noise electronics to expand the overall reach of the technology.

%% file: sec_ge/subsec_I-Ge-choice.tex
\label{subsec:GeChoice}

Germanium detectors have many advantages that make them well suited for
\BBz\ decay searches. As a result, they have historically
provided some of the most sensitive limits on this process~\cite{Avignone:2019phg}.
The excellent energy resolution of Ge detectors is of paramount importance for a
\BBz\ decay search, as energy is the only observable that is truly both necessary and
sufficient for a discovery.
In addition, Ge detectors are essentially solid-state TPCs,
allowing for accurate event reconstruction and a clear
topological discrimination between background and signal events through pulse shape discrimination (PSD) techniques.

\begin{figure}[tb]
  \includegraphics[height=0.35\columnwidth]{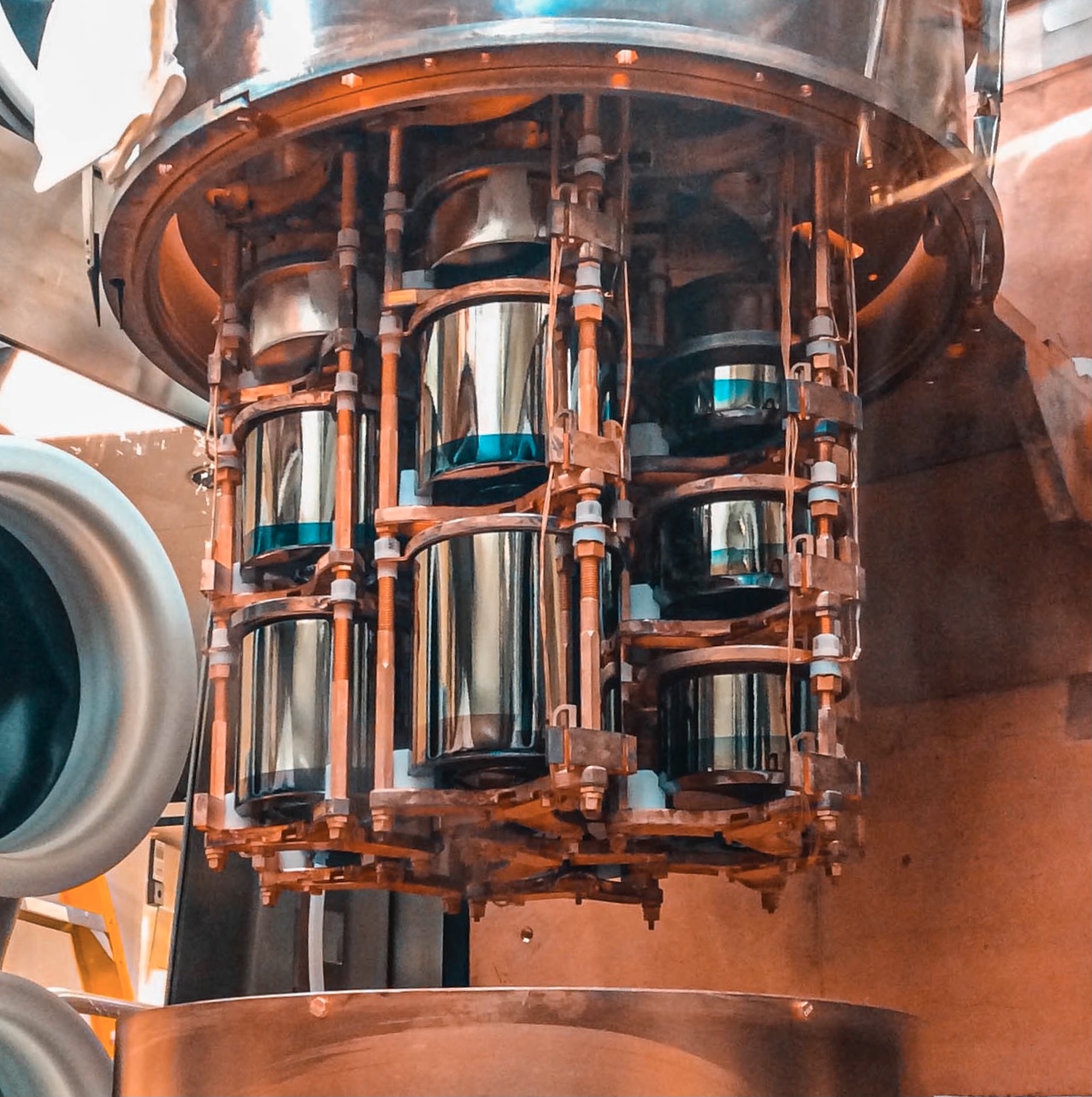}
  \includegraphics[height=0.35\columnwidth]{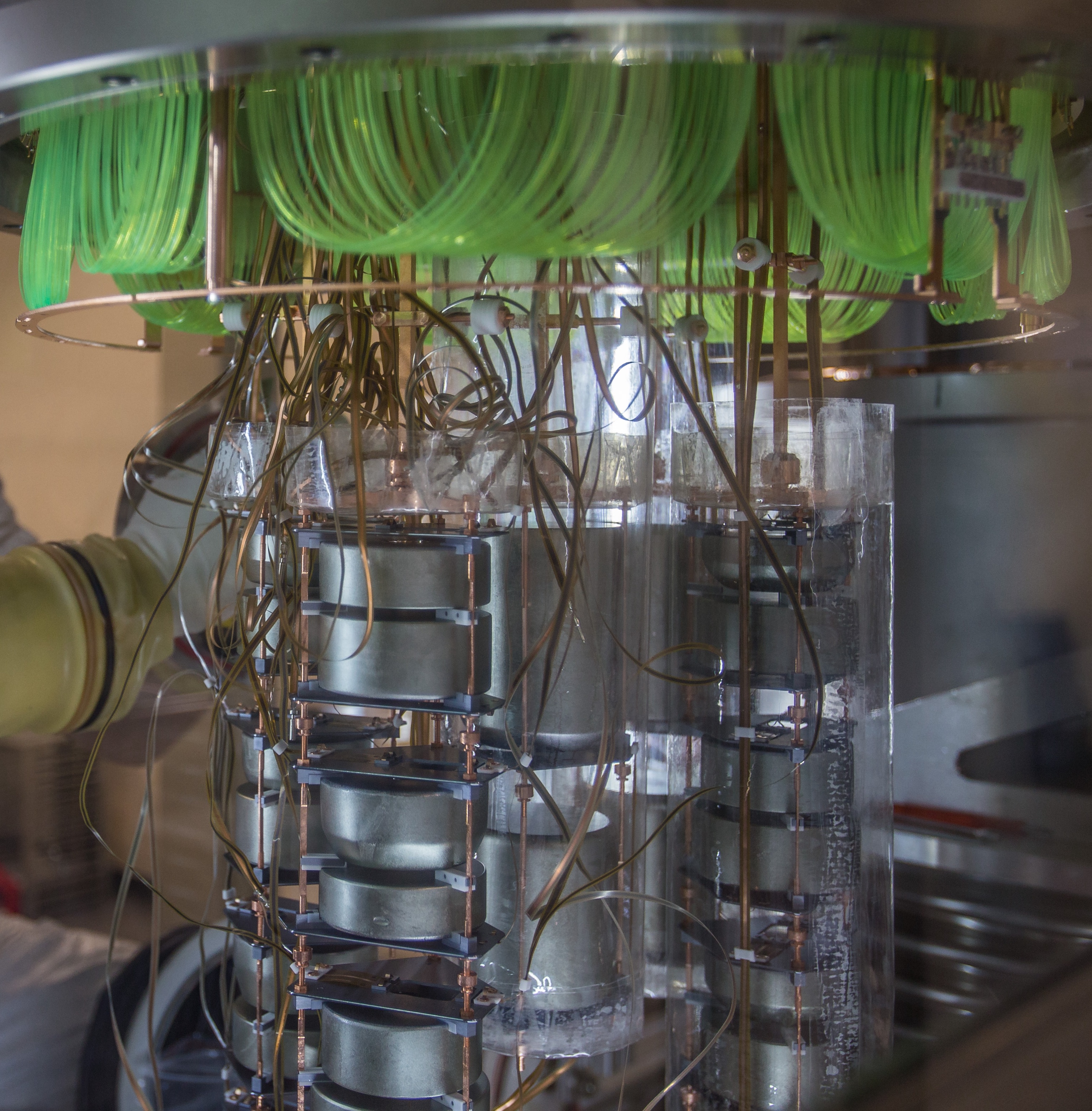}
  \caption{The \MJD\ (left) and GERDA (right) Ge detector arrays, with Ge
  detectors arranged in vertical strings. }
    \label{fig:strings}
  \end{figure}

The crystal-growth process for Ge material provides a significant purification
step resulting in unmeasurably low levels of internal \U- and \Th-chain contaminants.
The solid crystal bulk is not at risk of being contaminated during handling
and requires no further purification in situ.
Ge detectors are also mostly insensitive to surface activity, and the
cryogenic requirements for operating at \lnn\ temperatures are modest.
The \MJ\ and \Gerda\ collaborations pioneered the deployment of Ge detectors in
strings with minimal interstitial mass and shared shielding and cooling
infrastructure (see Fig.~\ref{fig:strings}), leading to orders-of-magnitude
background reduction with respect to previous-generation searches.

Germanium provides excellent discovery power, even for relatively small exposures compared to other isotopes.
Although the lower atomic number and \qval\ of \gesix\ give it a phase-space
factor \Gz\ that is relatively modest compared with other isotopes,
most of the precious target isotope is operated in a quasi-background-free\footnote{Quasi-background-free refers to background expectation values with high probability to observe less than one background count in the region of interest within the dataset. In this case, the sensitivity scales almost linearly with exposure.}
environment and is not used for self shielding as in other experiments. The
detection efficiency is hence very high, and nearly all the \BBz\ decay material
contributes to the sensitivity.

\begin{figure}[tb]
  \includegraphics[trim=10 30 10 20,clip,width=0.99\columnwidth]{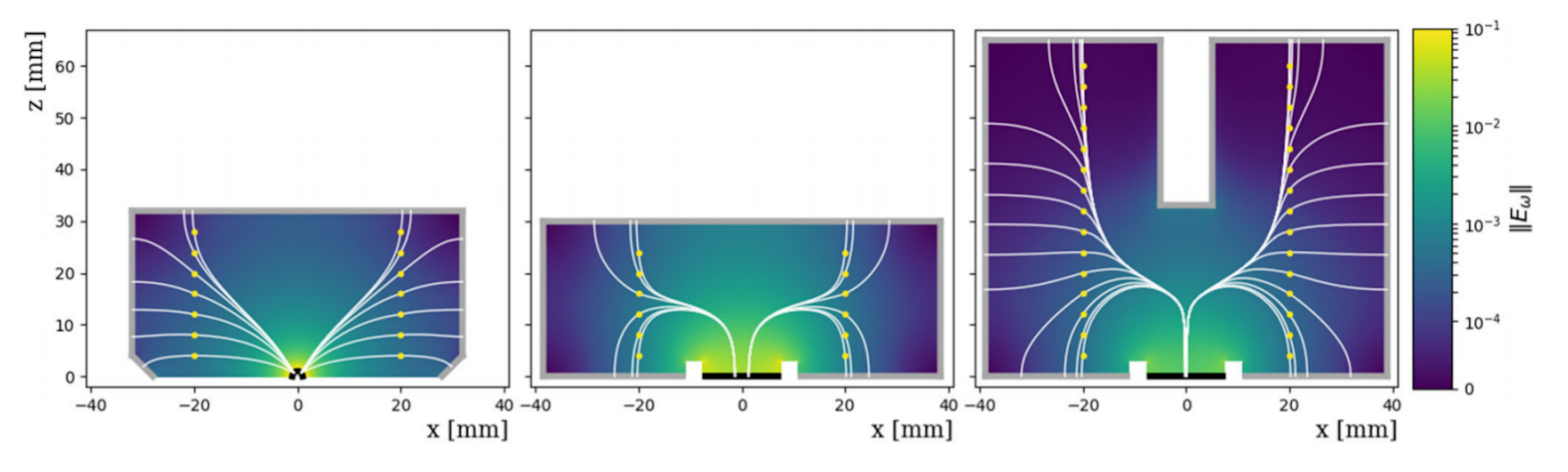}
  \caption{The three detector geometries used by the \MJ\ (PPC detector,
  left), \Gerda\ (BEGe detector, middle), and \LEG\ (\icpc\ detector, right)
collaborations. The mass of \icpc\ detectors are a factor of over four larger than that
of its predecessors. The plot gives the weighting field ($E_\omega$) within a cross section of each detector geometry. The thick black and gray lines are the \pPlus\ and \nPlus\ electrodes, respectively. The yellow points are locations of example energy depositions; the white trajectories connecting them to the \pPlus\
electrode are the drift paths of holes and those connecting them to the \nPlus\ electrode are the drift paths of electrons.
Figure from Ref.~\cite{Comellato:2020ljj}.}
    \label{fig:det-compare}
  \end{figure}

The \MJ\ and \Gerda\ collaborations have accumulated extensive experience in Ge crystal and detector usage. The two collaborations independently worked with vendors to utilize novel Ge-detector designs optimized for \BBz\ decay searches (see Fig.~\ref{fig:det-compare}).
\MJ\ used p-type, point-contact (\ppc) detectors, based on the original design of Ref.~\cite{Luke1989}, produced by ORTEC\footnote{AMETEK-ORTEC, Oak Ridge, TN, USA. \url{https://www.ortec-online.com/}}.
In collaboration with Mirion\footnote{Mirion Technologies, Olen, Belgium. \url{https://www.mirion.com/}},
\Gerda\ used a customized version\footnote{Commercial BEGe detectors have the top \nPlus\ surface ground off to allow better detection of low-energy $\gamma$ rays in a vacuum cryostat. The customized version retains the \nPlus\ contact wrapped around the whole detector.}
of broad-energy Ge (\bege) detectors~\cite{Agostini:2014hra},
which share the point-contact characteristics of the ORTEC \ppc\ detectors.
Both detector geometries exhibited outstanding energy resolution and advanced capabilities to reconstruct the event topology.
However, constrained by the electrode geometry, the maximum detector masses were limited to around 1~kg.

A major breakthrough for \LEG\ has been the invention of the p-type inverted-coaxial point-contact (\icpc) detectors~\cite{Cooper2011}.
These devices maintain the excellent energy and event reconstruction performance of \ppc\ and \bege\ detectors but are significantly larger, with typical masses around 2--3 kg.
The increased detector size allows the overall number of detectors operated in
\LEG\ to be reduced by a factor of three, also reducing by the same factor the number
of cables, read-out channels, and detector support materials that are all
sources of background. The reduction of the surface-to-volume ratio also helps
in mitigating the background due to radioactive decays on the detector surfaces.

The \icpc\ detectors were successfully deployed and operated in both \Gerda\ and the \MJD.
\Gerda's five enriched \icpc\ detectors were fabricated in 2018 by Mirion and ran
in \Gerda\ from July 2018 until November 2019.
With an average mass of 1.8~kg each, the \icpc\ detectors contributed to the
\Gerda\ \BBz-decay analysis with 8.5~\kgyr\ of high-quality data.
These detectors demonstrated excellent energy resolution,
lower surface contamination compared to the other \Gerda\ detectors,
and the lowest background level ever achieved in \Gerda: \cpowten{(4.9^{+7.3}_{-3.4})}{-4} \cpKkgy~\cite{Agostini:2021wzn}.
After analysis cuts, not a single high energy alpha-decay event was left in the
\icpc\ dataset while keeping the same cut efficiencies.
The four enriched \icpc\ detectors deployed in the \MJD\ were
fabricated in 2019--2020 by ORTEC and ran from August 2020 to March 2021. These detectors showed
an exquisite combined energy resolution of 2.4~keV FWHM at \Qbb. More details on
\icpc\ detector performance can be found in Sect.~\ref{subsec:energy-res}.

%% file: sec_ge/subsec_gerda.tex
\subsection{LAr Scintillation and \Gerda}
\label{subsec:gerda}

The  \Gerda\ experiment, located at \Lngs, pioneered the operation of bare enriched Ge detectors in LAr, serving simultaneously as radiation shield, coolant, and scintillation detector.
Exploiting the entire space-time signal topology in the novel Ge- and LAr-detector systems enabled the \Gerda\ Collaboration to distinguish background from \onbb\ decay signal candidate events with unprecedented efficiency.
This breakthrough experimental technology opened the path towards quasi-background-free searches for \onbb\ decay of \nuc{76}{Ge}.

\begin{figure}[]
 \centering
 \includegraphics[height=0.4\tw]{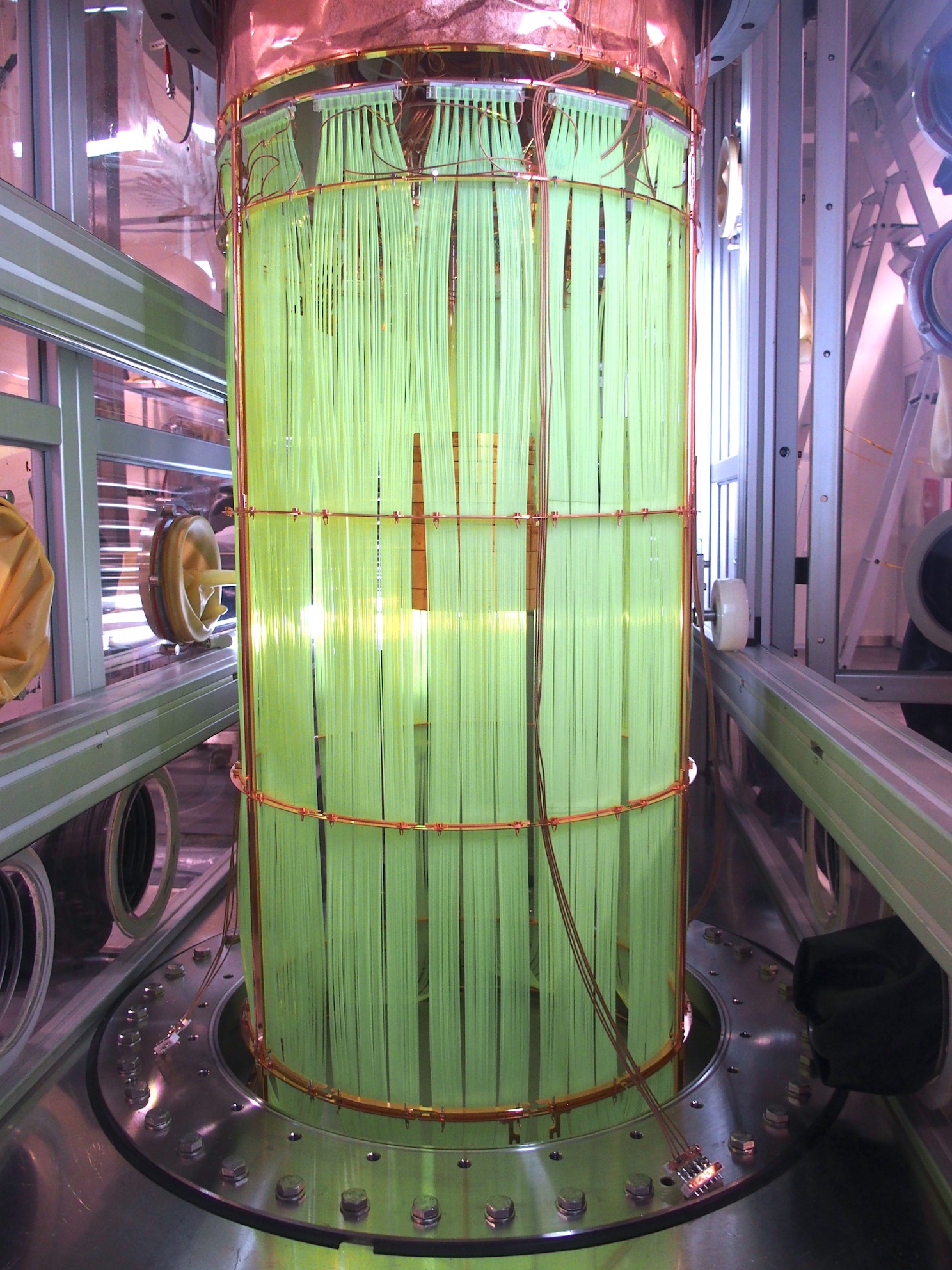}
 ~~~
  \includegraphics[trim=0 30 0 0,height=0.4\tw]{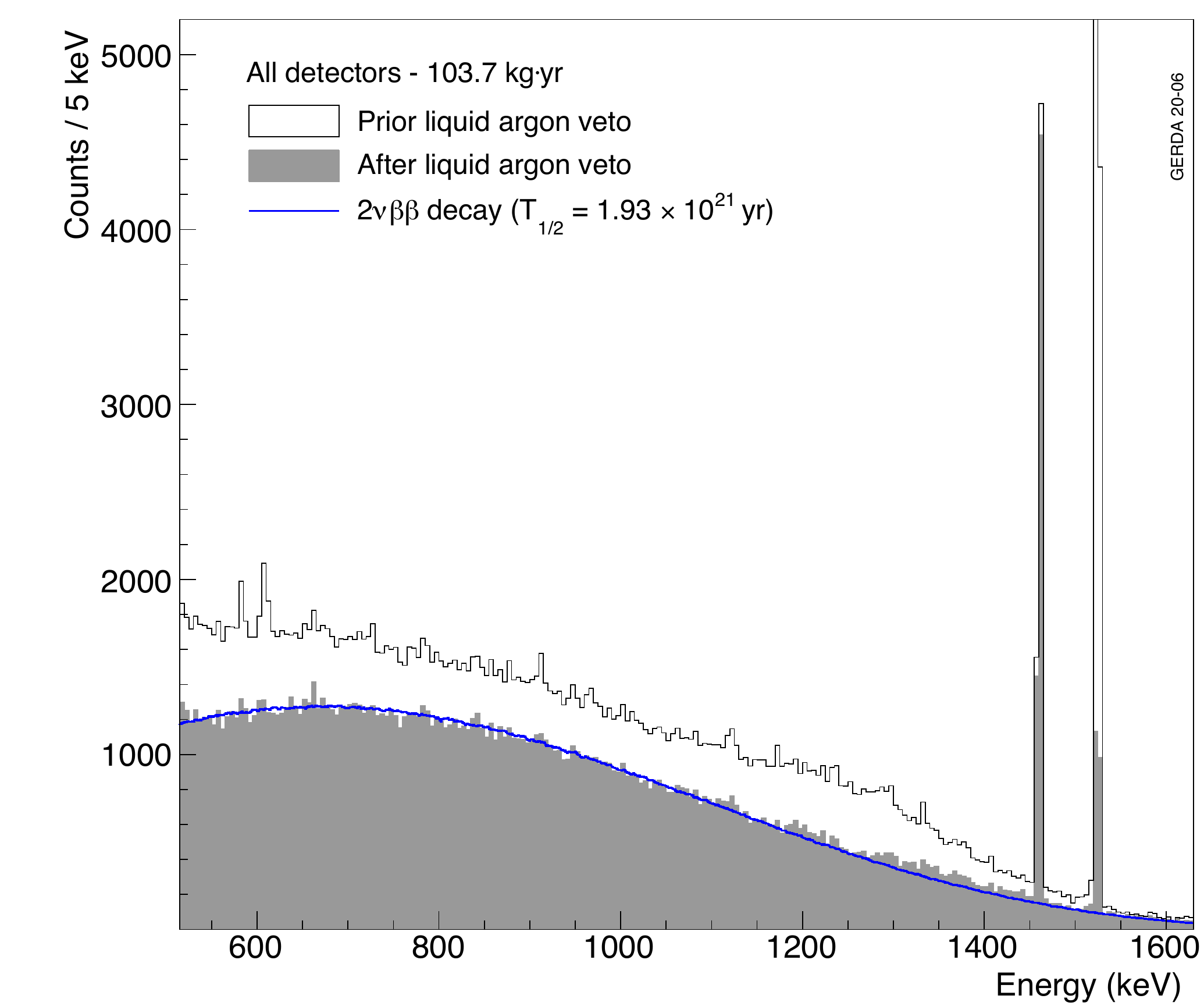}
 \caption{\label{fig:LArInstrumentation} Left:
\Gerda\ LAr scintillation light detection system enclosing the
Ge-detector array (not shown). Coated wavelength shifting fibers are read out by
low-background arrays of SiPMs. Right: Energy spectrum
measured by the Ge detectors in \ptwo\ (black), and the extremely pure sample
of \BBt\ events selected by rejecting background that creates LAr
scintillation light (grey). Note that the blue line is not a fit but rather the
expected contribution due to \BBt\ decays based on the measured half-life
value. The two visible peaks are the 1461~keV $\gamma$ ray from \nuc{40}{K} and the 1525\~keV $\gamma$ ray from \nuc{42}{K}.}
\end{figure}

The \Gerda\ experiment operated an array of 20~kg of enriched Mirion, custom-designed \bege\ detectors and 15.6~kg of enriched semi-coaxial detectors.
In the last year of data taking, the \Gerda\ collaboration deployed 9.6~kg of the novel \icpc\ detectors to verify their technical maturity for \LEG.
The LAr detector consists of a wavelength shifting fiber barrel that
surrounds the Ge-detector strings. The fibers convert Ar 128~nm scintillation
photons into green photons, which are guided to silicon photomultipliers (SiPMs)
attached at the ends of the fibers. Single photo-electron resolution was
observed, allowing the rejection of events with greater than $0.3$ photo-electrons in the LAr detector
within 5~$\mu$s of a Ge-detector signal. Figure~\ref{fig:LArInstrumentation}
shows the wavelength shifting fiber barrel coupled to low-background SiPM
arrays. The right panel demonstrates how the selection of events that are not
associated with LAr scintillation light creates an almost pure sample of
\BBt\ decay events which have the same topology as the sought-after \BBz\ decay signal.

\begin{figure}[]
 \centering
 \includegraphics[width=0.99\columnwidth]{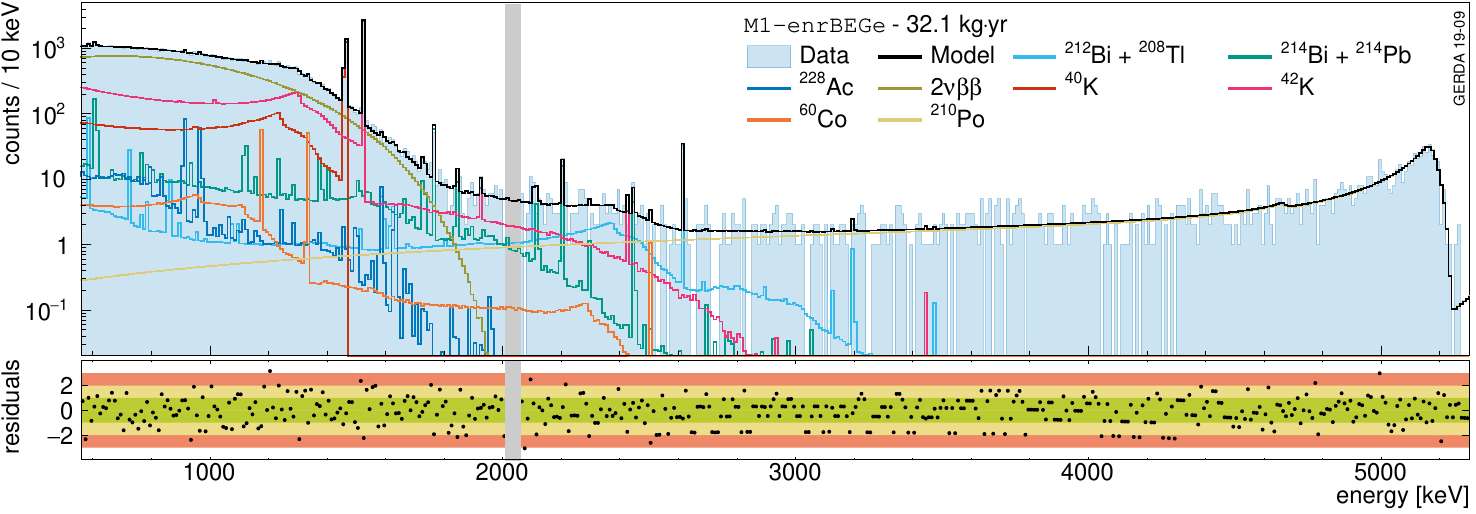}\\
 \caption{\label{fig:bkgModGerda}  Decomposition of the background observed before analysis cuts with the \bege\ detectors in \Gerda\ \ptwo~\cite{GERDA:2019cav}. The model combines screening measurements and a multi-variate fit.
 }
\end{figure}

The \Gerda\ background model before analysis cuts is shown in Fig.~\ref{fig:bkgModGerda}. The model is able to accurately describe all features  observed in the data set and is based on a comprehensive multivariate fit in which events are divided by detector and the number of detectors hit (multiplicity). The fit results are in good agreement with assay expectations, showing how Ge-based experiments can accurately make background projections.
The most prominent spectral features are the cluster of events at 5.2\,MeV tailing to lower energies. These are due to \nuc{210}{Po} decays on the surface of the Ge-detector \pPlus\ electrode. The $\gamma$ lines are due to radionuclides in the \nuc{232}{Th}- and \nuc{238}{U}-decay chains (e.g., \nuc{228}{Ac}, \nuc{212}{Bi}, \nuc{208}{Tl} and \nuc{214}{Bi}), and  \nuc{40}{K} and \nuc{42}{K} (a progeny of \nuc{42}{Ar}). At low energies the strong signal from \BBt\ decays is dominant. According to the background model, the dominant  background contributions originate from materials close to the detectors.

\begin{figure}[]
 \centering
 \includegraphics[width=0.99\columnwidth,]{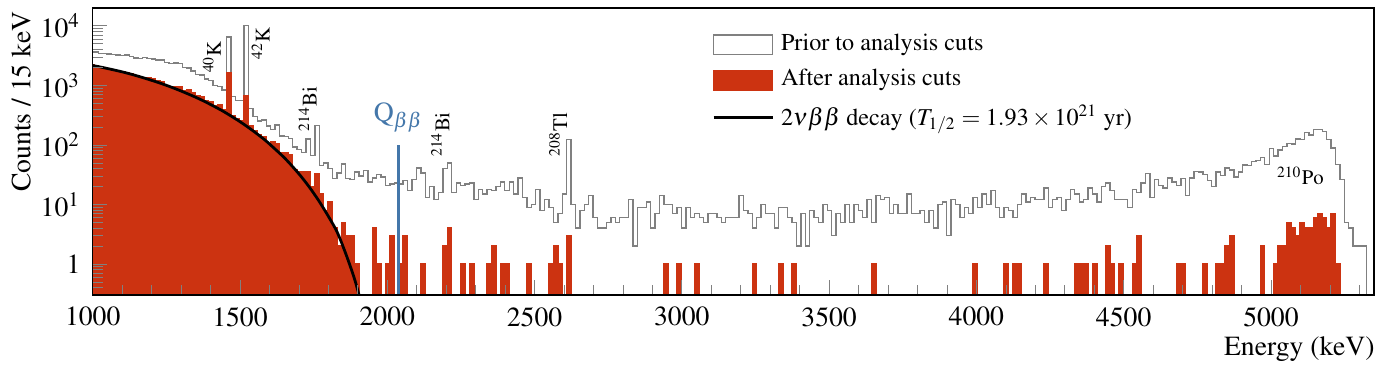}
 \caption{\label{fig:BackgroundSpectraGerda}
 Energy distribution of \Gerda\ \ptwo\ events before and after analysis
 cuts for the total exposure of 103.7~\kgyr~\cite{Agostini:2020xta}. The region near
 \qval\ is quasi-background-free. The analysis cuts
 include a LAr active veto and Ge pulse-shape discrimination, both of which are
 described in Sect.~\ref{subsec:topology}.
 }
\end{figure}

\Gerda\ \ptwo\ completed data taking in 2019. Its full data set is displayed in
Fig.~\ref{fig:BackgroundSpectraGerda}. Due to the complementarity of the
LAr scintillation light detection and the event topology reconstruction
of Ge detectors, \Gerda\ achieved the world-leading background index of
\cpowten{5.2^{+1.6}_{-1.3}}{-4}~\cpKkgy~\cite{Agostini:2020xta} with an
average FWHM energy resolution of 2.6~keV for the \bege\ detectors and 2.9~keV
for the \icpc\ detectors.
\Gerda\ was the first experiment to reach a half-life sensitivity above
$10^{26}$ years.
With only 127.2~\kgyr\ of exposure (Phase I + II), \Gerda\ set a lower limit at
\cpowten{1.8}{26}~yr ($90\%$ CL)~\cite{Agostini:2020xta}.

%% file: sec_ge/subsec_mjd.tex
\subsection{Materials, Electronics, and the \MJD}
\label{subsec:mjd}

\begin{figure}[t]
 \centering
 \includegraphics[height=0.34\columnwidth]{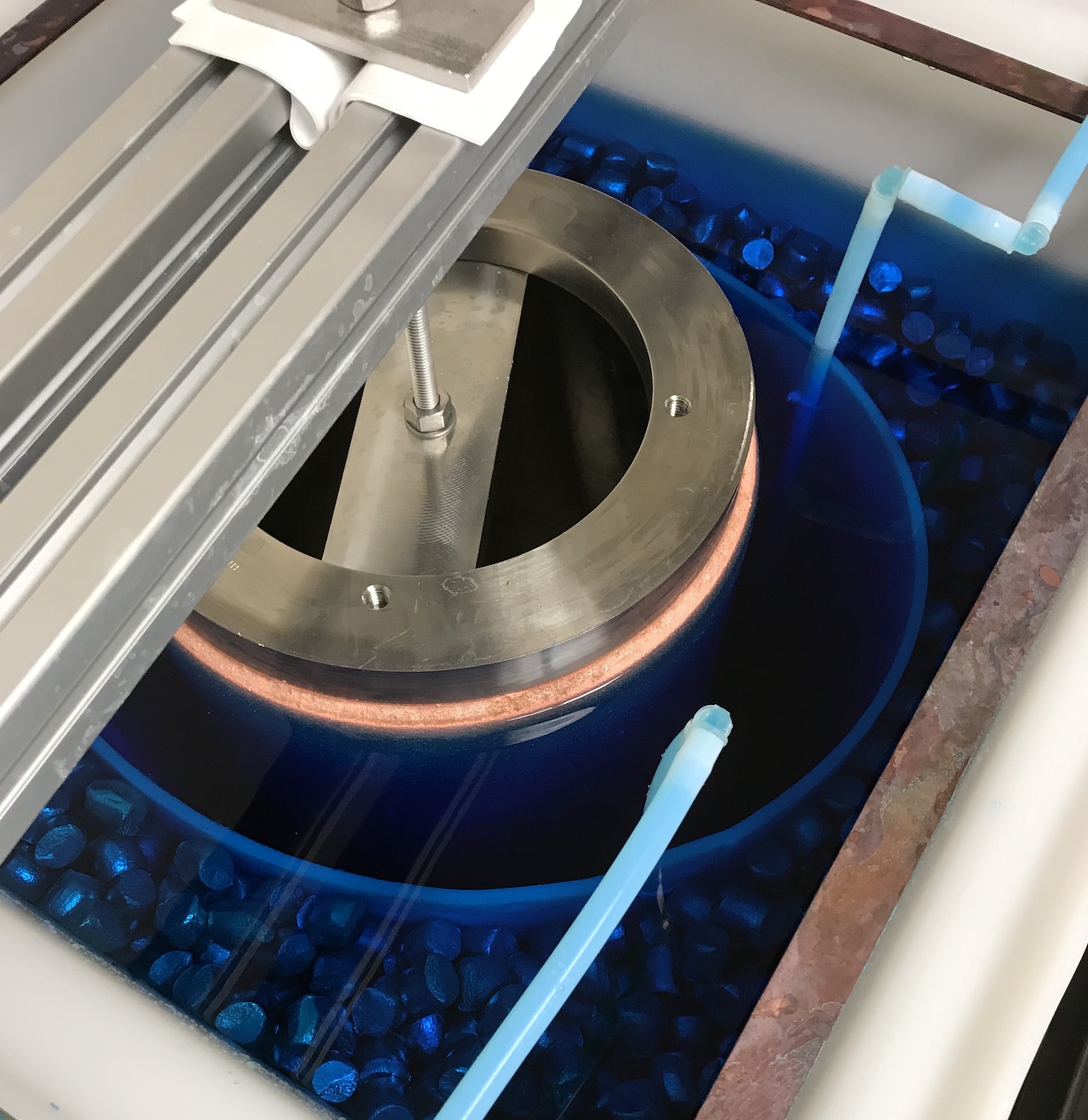}
 \includegraphics[height=0.34\columnwidth]{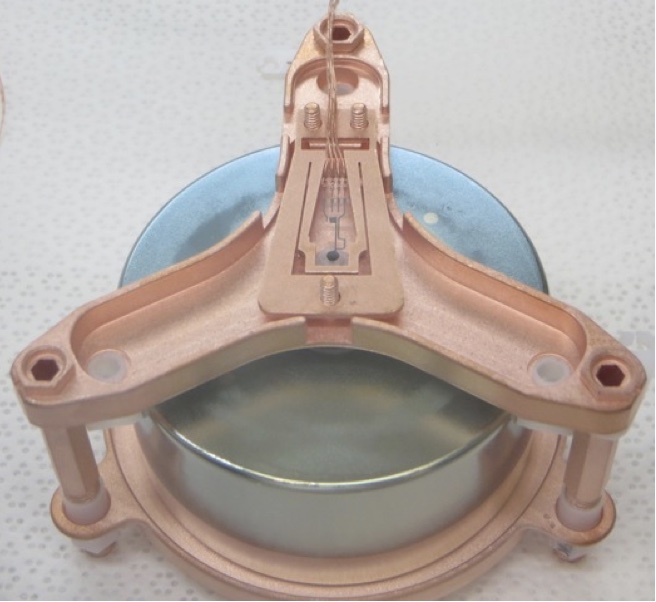}
 \caption{ \label{fig:mjdtech}
   Left: Cu electroforming bath running underground at SURF. Right: LMFE
   electronics board attached to four pico-coax cables, held in an
   EFCu spring clip affixed to a \MJD\ Ge-detector mount.
 }
\end{figure}

The \MJD\ experiment, located at SURF, operated an array of 30~kg of enriched ORTEC
\ppc\ detectors, split between two vacuum-insulated cryostats housed within a low-background shield.
The \MJD\ Collaboration developed low-noise electronics to improve energy resolution and developed clean materials to reduce internal component backgrounds.
One key innovation was the development of underground electroformed Cu (EFCu),
shown in Fig.~\ref{fig:mjdtech} and
described in Sect.~\ref{sec:materials}, in which one obtains the purest commercial
Cu available and then re-forms it underground. This eliminates cosmogenic activation
species like $^{60}$Co while rejecting bulk \U/\Th-chain natural radioactivity by a
factor of approximately 30, achieving radiopurities below 0.1~$\mu$Bq/kg~\cite{Abgrall:2016cct}.
This ultra-pure Cu served as the primary structural material inside the array.
It was also used to fabricate the cryostats and the innermost passive shield layer.
A number of other novel ultra-pure materials and
components were developed for the \MJD, including clean dielectrics, cables,
connectors, and coatings~\cite{Abgrall:2016cct}.

Another key development was the \MJD\ low-mass front-end (LMFE) electronics
board (see Fig.~\ref{fig:mjdtech}). In addition to being highly
radiopure, the LMFE achieved extremely low noise levels, contributing
a negligible 0.1~keV (FWHM) to the energy resolution.
This allowed the \MJD\ to achieve a record energy resolution for
large-scale \BBz\ decay experiments of $2.53 \pm 0.08$~keV ($0.124\%$) FWHM at \qval\
when combining all detectors~\cite{Alvis:2019sil}

\begin{figure}[t]
 \centering
 \includegraphics[height=0.26\columnwidth]{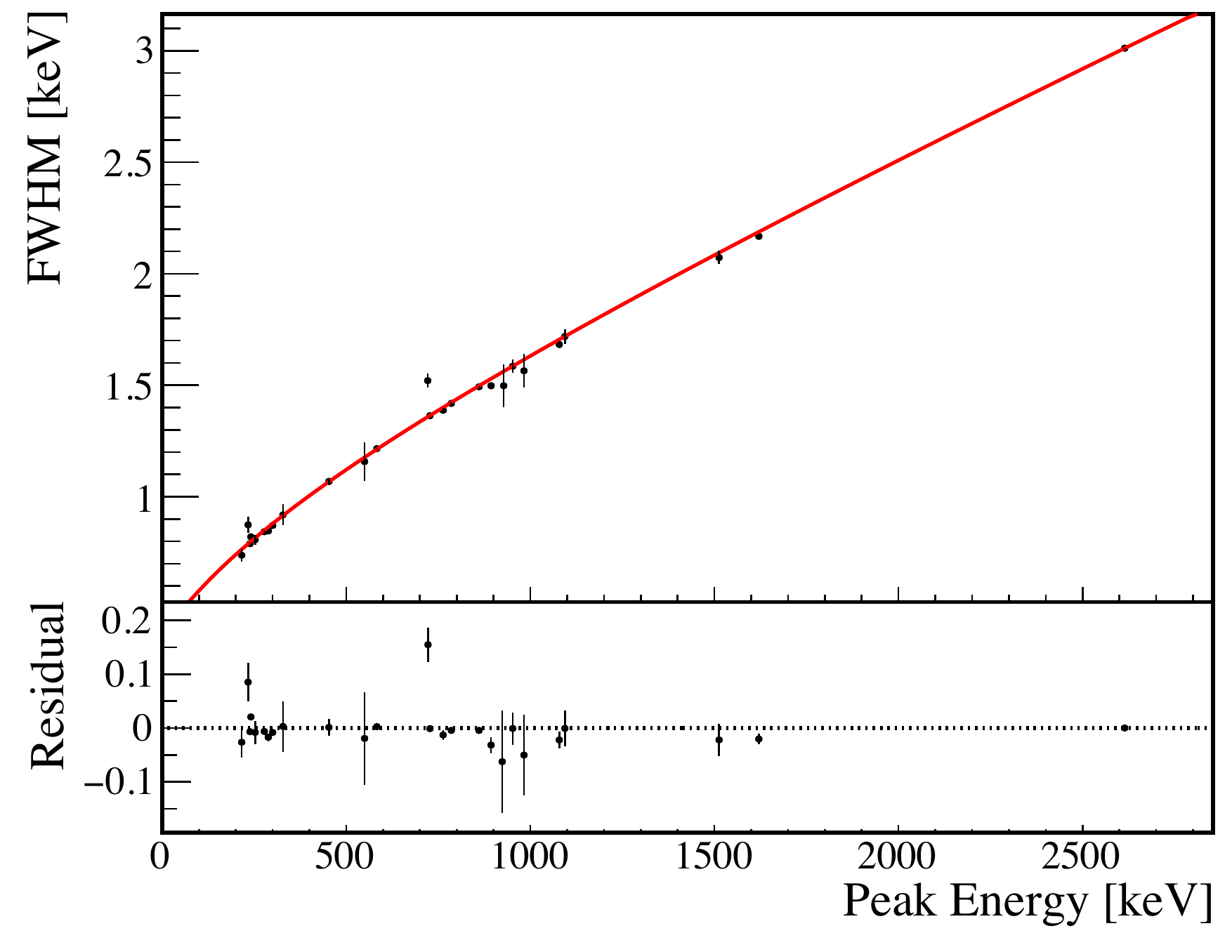}
 \includegraphics[height=0.265\columnwidth]{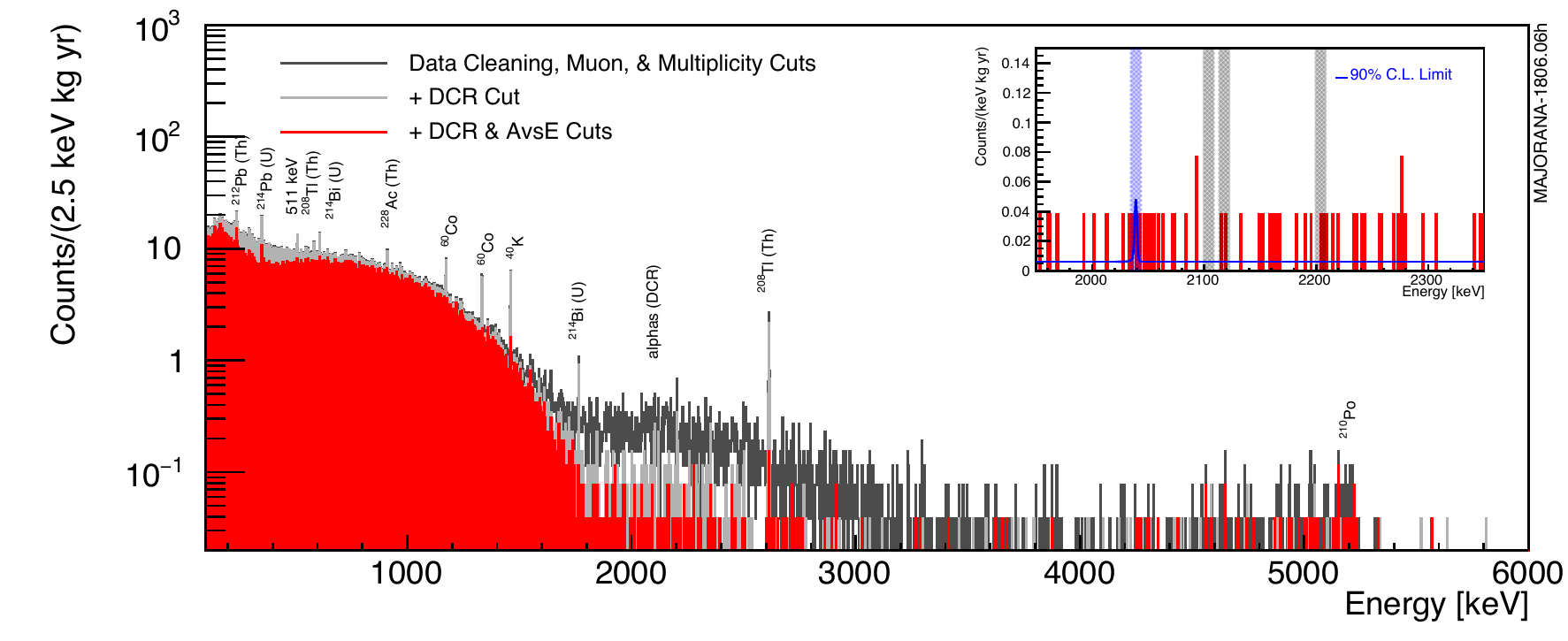}
 \caption{\label{fig:BackgroundSpectraMJD}
 Left: Measured energy resolution in \MJD\ calibration spectra vs energy, showing the record energy resolution of $2.53 \pm 0.08$~keV FWHM at \qval .
 Right: Energy spectrum before and after analysis cuts from the \MJD\ from an exposure
 of 26~\kgyr~\cite{Alvis:2019sil}. The main background contributors before cuts are noted on
 the spectrum. The delayed charge recovery (DCR) cut rejects degraded $\alpha$
 events, while the $AvsE$ cut removes multi-site $\gamma$ rays, both of which are
 described in Sect.~\ref{subsec:background-reduction}. The inset shows the background integration window around \Qbb. The $\pm$5-keV shaded bands are not considered in calculating the background index; the three gray bands cover known $\gamma$ rays from \nuc{208}{Tl} and \nuc{214}{Bi}, and the remaining blue band covers \Qbb.}
\end{figure}

Figure~\ref{fig:BackgroundSpectraMJD} shows the experiment's latest reported background spectrum.
The \MJD\ achieved a background rate of 
\cpowten{(4.7\pm0.8)}{-3}~\cpKkgy~\cite{Alvis:2019sil}
 in its low-background configuration.%
\footnote{Note that \MJ\ and \Gerda\ use different conventions to derive their respective background indices. \Gerda\ uses the full detector mass and an energy window of 1930--2190~keV whereas \MJ\ takes into account the active volume fraction and uses a 1950--2350~keV range. The \Lk\ background projections given in Sect.~\ref{subsec:background-budget}
use the full detector mass and an energy window of 1985--2095~keV centered around \Qbb.}
According to background model spectral fits, and supported by studies of relative peak intensities and coincident events, the major background contributor is \nuc{208}{Tl} outside the immediate vicinity of the detectors~\cite{Buuck2019, Gilliss2019}.

During a hardware upgrade in 2020, five of the \ppc\ detectors were removed for testing in LAr in advance of their use in \Ltwo.  At the same time, four of the ORTEC enriched \icpc\ detectors destined for \Ltwo\ were installed in the \MJD\ for characterization in a low-background vacuum environment. All \icpc\ detectors exhibited excellent PSD performance and energy resolution.
The experiment has completed data taking with enriched detectors and is
currently taking data with only natural detectors.

%% file: sec_ge/subsec_H-L200.tex

\begin{figure}[t]
 \centering
  \textbf{{\Ltwo}} \\
  \vspace{0.1in}
  \hfill
  \includegraphics[height=0.45\tw]{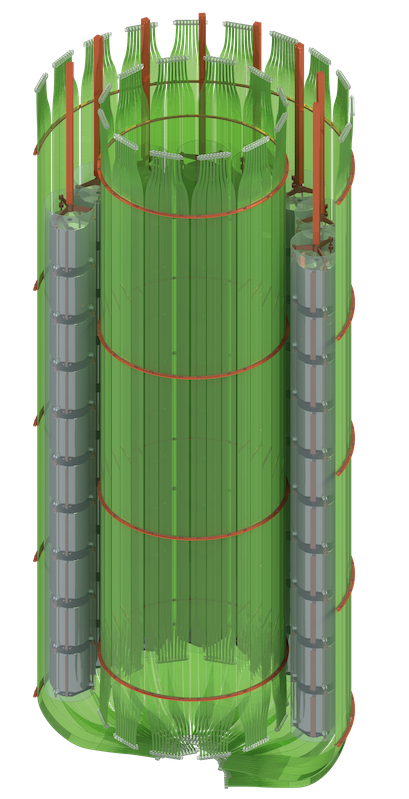}
  \hfill
  \includegraphics[height=0.45\tw]{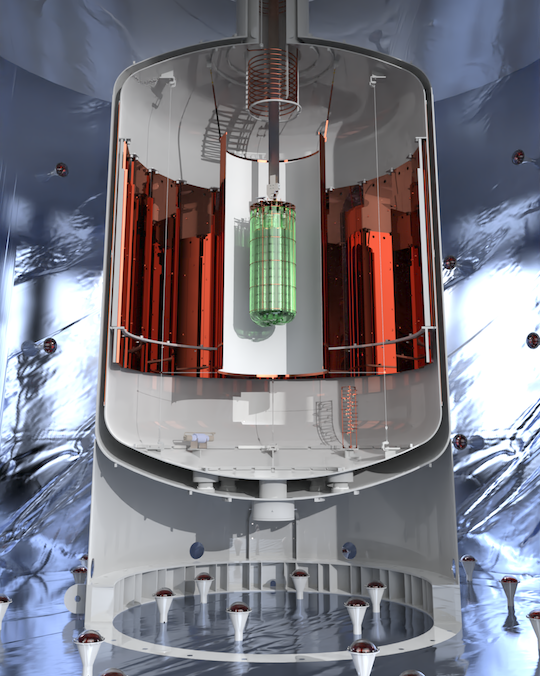}
  ~ ~ ~ ~ ~ ~ ~ ~ ~ ~ ~ ~
 \caption{\label{fig:L200} Left: \Ltwo\ Ge detectors mounted in strings and
 surrounded by optical fibers that are used to detect the LAr scintillation light. Right: Detector systems positioned in the center of a LAr cryostat equipped with wavelength-shifting reflectors. The cryostat is placed in a water tank instrumented with photomultipliers and used as a Cherenkov muon detector.
 }
\end{figure}

The \Ltwo\ experiment is planning to operate 200~kg of Ge detectors in a bath of LAr in an upgrade of the \Gerda\ infrastructure at \Lngs.
The \Ltwo\ design combines the best elements of \Gerda\ and the \MJD,
implementing many of the improvements envisioned for \Lk.
\Ltwo\ uses the existing 70~kg of enriched detectors from the \MJD\ and
\Gerda\ and an additional 130~kg of newly produced \icpc\ detectors.
\Ltwo\ is currently under construction and is expected to start commissioning in the fall of 2021.
The Ge- and LAr-detector systems and the setup are illustrated in
Fig.~\ref{fig:L200}.

\Ltwo\ is an integral part of the strategic, staged approach pursued by the \LEG\ Collaboration.
The design of \Ltwo\ integrates many of the key ideas that will remain in \Lk;
the ICPC Ge detectors, the LAr instrumentation, the holders, and many
other components and subsystems are very similar to what we propose for \Lk.
\Ltwo\ is a timely and concrete opportunity to integrate the best
elements of \Gerda\ and the \MJD\ and to test them before finalizing the \Lk\
baseline design proposed in this report.

\Ltwo\ will continue to be an asset during the design and
construction phases of \Lk.
The detector system of \Ltwo\, following the design of \Gerda, can be lifted
and lowered into the cryostat in just a few days. Technology improvements can hence be installed and tested in \Ltwo\ as soon as they become available.
\Ltwo\ thus acts
as a unique test-stand to refine our technological solutions in an ultra-low-background environment extremely similar to the final one.
The background data collected will also be used to further validate our
simulations, inform our design choices, and reduce risk for \Lk.

Until \Lk\ comes online, \Ltwo\ will be one of the leading experiments in the field, reaching a half-life sensitivity of $10^{27}$~yr after five years of operations.
To remain quasi-background-free for the design exposure of 1 \tyr, \Ltwo\ requires
the reduction of background by a modest factor of 2.5 with respect to
what has already been achieved by \Gerda.
This improvement is easily obtained due to:
\begin{myitemize}
\item The larger average detector mass, resulting in fewer nearby components,
  cables, and holder materials per kilogram of detector
\item The adoption of \MJD-style low-noise electronics, low-mass components, and
  clean materials with a lower level of radioimpurities (e.g., electroformed
  copper) as well as the incorporation of scintillating plastic components to
  minimize the inactive material around the detectors
\item  An improved design for the scintillation light readout and higher-purity
  LAr with better light transmission and light yield
\end{myitemize}
The additional 20-fold background reduction anticipated for \Lk\ with respect to \Ltwo\ is
obtained primarily by the usage of underground-sourced Ar, new less-radioactive
electronics and cables, and the presence of only ICPC detectors. These aspects
are discussed in depth in the following sections.

%% file: sec_legend-approach/sec_legend-approach.tex
\section{The \Lk\ Approach}\label{sec:legend}

\input{sec_legend-approach/subsec_exec}

\subsection{Conceptual Overview}

\subsubsection{Baseline Design}\label{subsec:baseline}
\input{sec_legend-approach/subsec_baseline}

\subsubsection{Energy Resolution}\label{subsec:energy-res}
\input{sec_legend-approach/subsec_energy-resolution}

\subsubsection{Multivariate Event Topology Discrimination}\label{subsec:topology}
\input{sec_legend-approach/subsec_topology}

\subsubsection{Projected Backgrounds}\label{subsec:projected_bg}
\input{sec_legend-approach/subsec_projected_bg}

\subsection{Discovery Potential}

\subsubsection{Discovery Sensitivity}\label{subsec:goals}
\input{sec_legend-approach/subsec_goals}

\subsubsection{Post-Discovery Validation}\label{subsec:postdiscovery}
\input{sec_legend-approach/subsec_post-discovery.tex}

\subsection{Key Experimental Parameters}\label{subsec:phys-params}
\input{sec_legend-approach/subsec_params}

%% file: sec_legend-approach/subsec_exec.tex
\begin{figure}[b]
\includegraphics[width=0.99\textwidth]{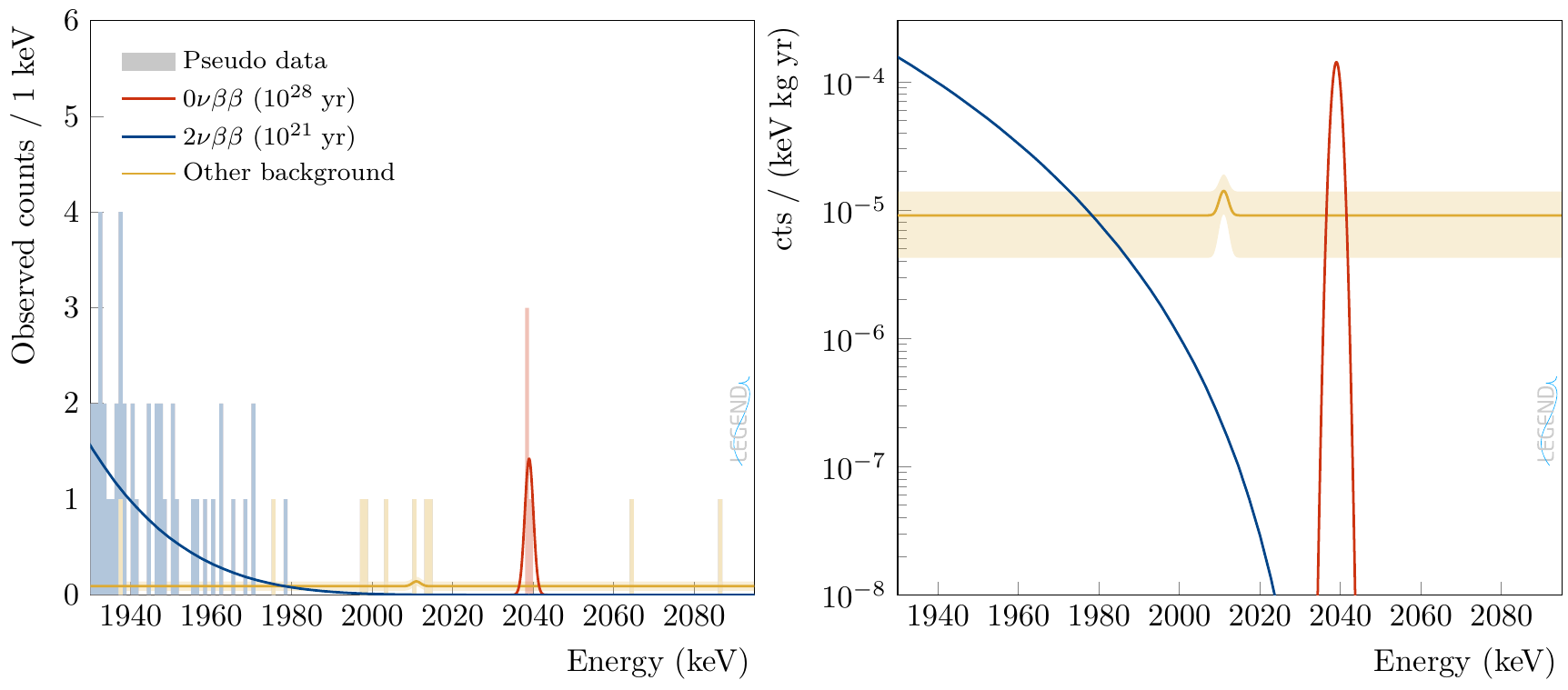}
\caption{
An illustrative Monte-Carlo pseudo-dataset of \Lk, generated for the full background model, 10~\tyr\ of exposure, and a \BBz\ decay half life of $10^{28}$ yr populated with the (left) representative observed counts and (right) normalized to exposure. The \BBt\ decays do not leak in the \BBz\ decay signal region, and their contribution is shown separately from the rest of the background sources. The yellow background model curve shows a small peak from \nuc{214}{Bi} decay as the nearest $\gamma$ line within the displayed energy range.  The uncertainty on the overall background model is covered by the yellow band.}
\label{fig:discovery_sim}
\end{figure}

The \Lk\ discovery power manifests as a signal visible
to the eye, appearing as an isolated peak at $Q_{\beta\beta}$ on a flat, featureless background continuum,
with no expected peaks nearby. The signal extraction
will not be affected by background modeling uncertainties, maximizing the
confidence that the signal is due to the sought-after \BBz\ decay.
Figure~\ref{fig:discovery_sim} illustrates the detection capability of \Lk\ through
an example Monte Carlo dataset generated assuming 10 years of data taking and a \BBz\ decay signal of $T^{0\nu}_{1/2} = 10^{28}$~yr. The \BBt\ decay spectrum
and all other expected backgrounds are derived from the comprehensive
\Lk\ background model, which has been tuned using data from \Gerda\ and the \MJD.
The expected number of background events within the 2.5-keV FWHM signal peak
with 10~\tyr\ of exposure is projected to be fewer than 0.25 counts and is
easily constrained from the sidebands.
This allows for even a simple analysis based
on counting statistics alone to yield unambiguous detection with just a handful
of counts.

In this section, we describe the baseline design concept of \Lk, the signal detection
capabilities of its \icpc\ detectors, the experiment's active background tagging
techniques, and their application to expected \Lk\ background sources, all of
which culminate in a 99.7\% CL discovery sensitivity\footnote{A 50\% chance of measuring a signal of at least 3$\sigma$ significance.} for \Tz\ of $1.3\times10^{28}$ years.

%% file: sec_legend-approach/subsec_baseline.tex
The \Lk\ baseline technical design is centered around the demonstrated
low background and excellent energy performance of p-type, point-contact,
high-purity Ge semiconductor detectors, enriched in \nuc{76}{Ge} and operated in
LAr. The innovative \icpc\ Ge-detector geometry was developed to
boost the experimental sensitivity and maximize synergy with the LAr scintillation light detection.
Combined with the \Ltwo\ \icpc\ detectors,
approximately 400 individual \icpc\ detectors with an average mass of 2.6~kg
are instrumented for a total detector mass of 1000~kg. The detectors are
mounted using underground electroformed Cu rods for physical support,
with electrical isolation provided via plastic insulators. Below each detector is a
baseplate supporting a wire-bonded signal cable and front-end application-specific integrated circuit (ASIC) board to
collect charges at the detector's \pPlus\ electrode. From there, flat flex
cables carry the amplified signals
out of the cryostat and water tank to a data-acquisition (DAQ) system for waveform
digitization and offline storage. A separate, single-conductor, flat flex cable
wire-bonded to the detector's \nPlus\ electrode provides a high-voltage bias to
deplete the Ge semiconductor.

The signal/background discrimination power of \icpc\ detectors begins with their
superior energy resolution, demonstrated to be 0.12\% FWHM (0.05\%
$\sigma$) at \Qbb,
making \nnbb\ decays a completely negligible background for \Lk.
Beyond the tight energy response, additional event-topology information
offers strong discrimination against background events with energy close to \Qbb.
Pulse shapes provide information on the interaction-site multiplicity, the
interaction-site location, and the presence of any delayed-charge collection,
allowing the prompt, single-site signature from bulk \BBz\ decay energy depositions
to be easily distinguished from multiply-interacting $\gamma$ rays, decays on the
surfaces of the detectors, and other backgrounds.
The granular nature of the Ge detector array and its immersion within an active
LAr scintillating volume offers a pixelation of the setup
that provides additional strong discrimination between
\BBz\ decay signal events, which are isolated to a single Ge detector, and
backgrounds generating coincident events in multiple regions.
\LEG's PSD and anti-coincidence techniques
and their background discrimination performance are described in greater detail in
Sect.~\ref{subsec:background-reduction}.

\begin{figure}
\includegraphics[width=0.95\textwidth]{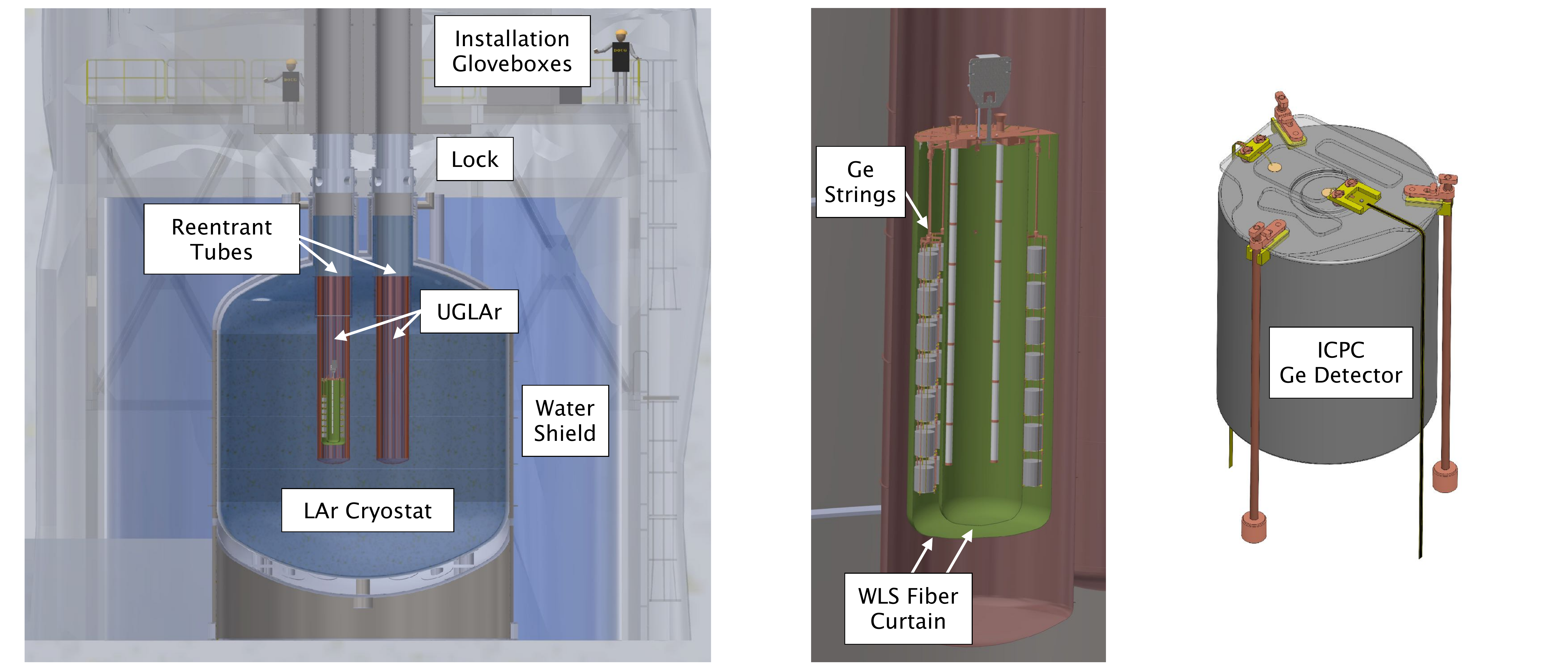}
\caption{
Left: A baseline conceptual design of the \Lk\ \nuc{76}{Ge} array that accommodates four 250-kg modules within a single vacuum-insulated LAr cryostat and water tank. Shown on top-right is a person for scale. Middle: A close-up of a single module, which contains UGLAr within the reentrant tubes, showing the strings of Ge detectors and LAr scintillation instrumentation. Right: An individual ICPC Ge detector unit that forms the strings in the array.}
\label{fig:baseline}
\end{figure}

The Ge detectors are distributed among four independently operating 250-kg
modules to allow commissioning of the array in stages (see
Fig.~\ref{fig:baseline}). In each module, the detector strings are immersed in
UGLAr sourced from radiopure underground Ar to provide direct reduction of the
primary background source observed in \Gerda. LAr scintillation light is read
out by a curtain of wavelength-shifting (WLS) polystyrene fibers.  The TPB
(tetraphenyl butadiene) coated fibers are coupled to SiPM
photodetectors, which provide single photo-electron detection capabilities.
Each of the four UGLAr modules is surrounded by
atmospheric-sourced LAr, with additional light collection, and supported within
a vacuum-insulated cryostat, itself inside a water tank providing infrastructure
and additional active and passive shielding.

Installation in the SNOLAB cryopit is assumed (see Sect.~\ref{sec:snolab}).
This reference cavity is used for cosmogenic background estimation, the cryostat conceptual design, and infrastructure needs.  The impact of the design and background contribution at shallower depths is described starting in
Sect.~\ref{sec:altsite}.
Further details of the \Lk\ technical design are given in
Sect.~\ref{sec:technical}. Descriptions of \Lk's unique materials are
given in Sect.~\ref{sec:materials}.

%% file: sec_legend-approach/subsec_energy-resolution.tex

The excellent energy
resolution of Ge detectors---the best of any ton-scale \BBz\ decay detection technology---provides
high discovery sensitivity with negligible background contribution from
irreducible \BBt\ decays and no $\gamma$ lines in the region of interest.
\MJ\ and \Gerda\ achieved an average energy resolution at \Qbb\ of
2.53~keV~\cite{Alvis:2019sil} and 2.6~keV~\cite{Agostini:2020xta} (FWHM),
respectively, with \ppc\ and \bege\ detectors operated over multi-year periods of
data taking.
Figure~\ref{fig:discovery_sim} (right)
shows how \BBt\ and \BBz\ decay events are fully
separated with this energy resolution. Fewer than $10^{-7}$ \BBt\
decay events are expected at the \Qbb\ region of interest (ROI) of the full \Lk\
dataset. This is the lowest rate expected by any \BBz\ decay experiment
and gives \Lk\ effectively double the sensitivity~\cite{Agostini:2017jim} with respect to experiments in
which a roughly exponentially falling tail of \BBt\ decay events leaks across the
\BBz\ decay ROI.

\begin{figure}[h]
	\includegraphics[width=0.99\textwidth]{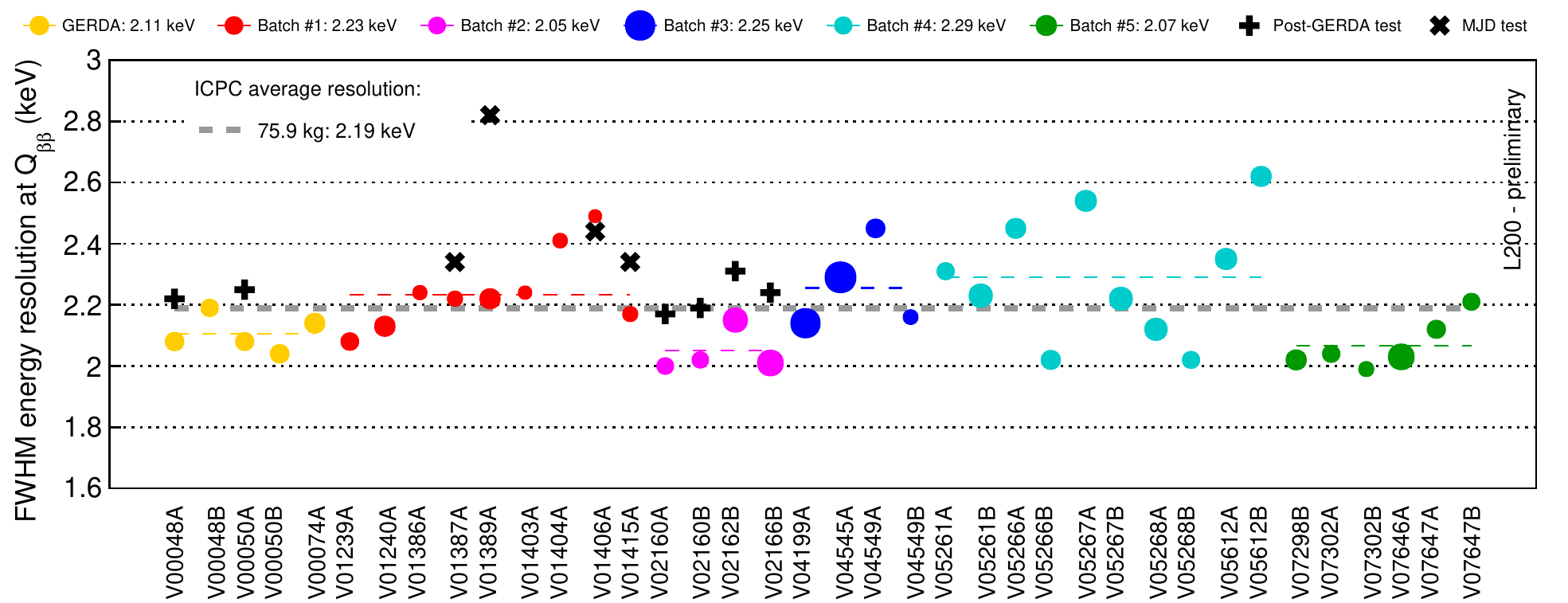}
	\caption{FWHM energy resolution of all \Ltwo\ \icpc\ detectors delivered to date, as measured
        in vendor vacuum cryostats (colored discs).  The dashed lines indicate
        the mass-weighted average per production batch (colored) and for all
        detectors combined (gray). Each data point's diameter scales with its
        detector mass; uncertainties are on the order of or smaller than the
        marker sizes. Also shown are the values measured during testing in the
        \Gerda\ (black plus) and \MJD\ (MJD, black cross) cryostats.}
	\label{fig:qbb-res}
\end{figure}

\begin{figure}[h]
        \hfill
	\includegraphics[height=0.32\columnwidth]{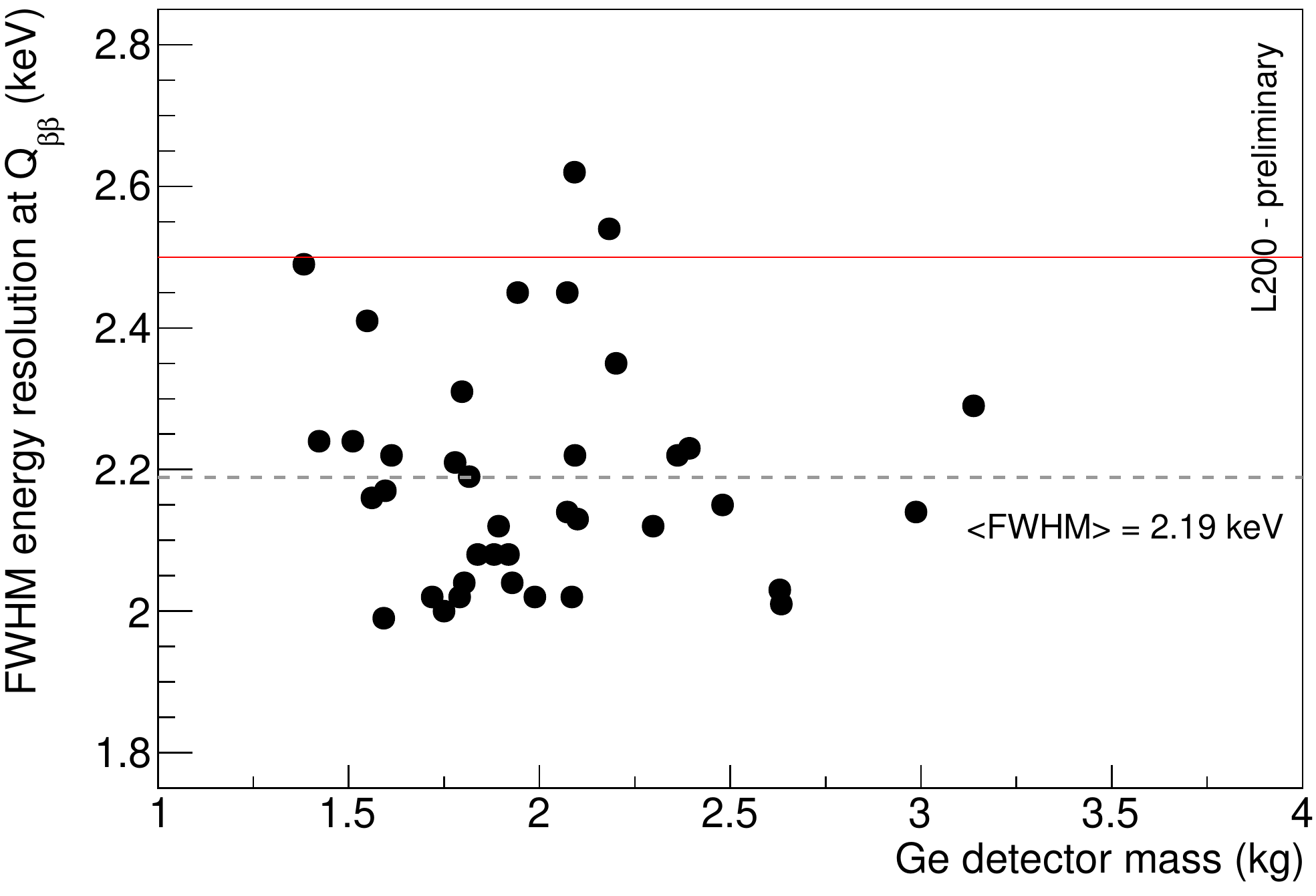}
        \hfill
	\includegraphics[height=0.32\columnwidth]{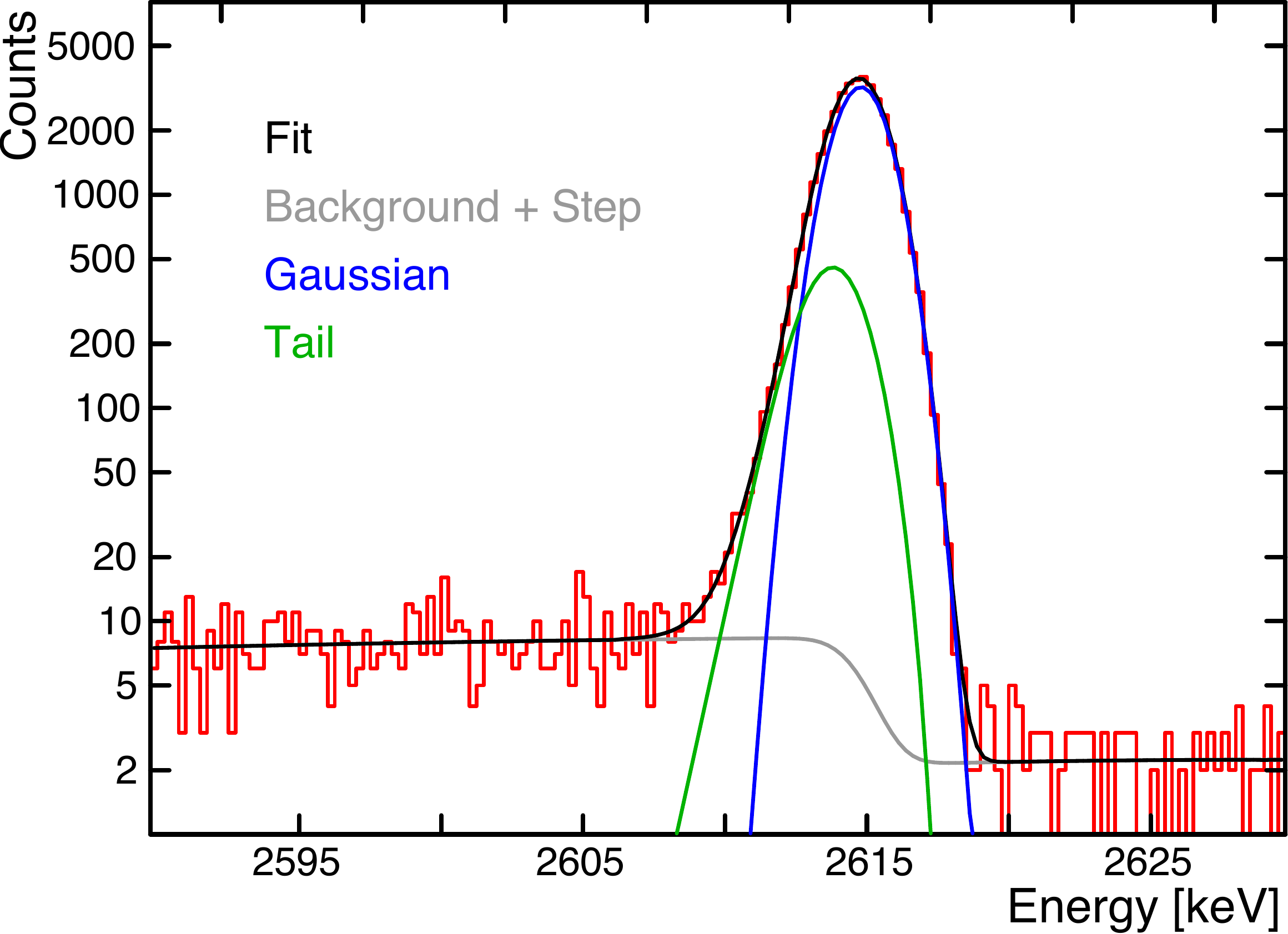}
        ~~~~
	\caption{Left: FWHM energy resolution of all \Ltwo\ ICPC detectors delivered to date, as measured
        in vendor vacuum cryostats, as a function of mass. Uncertainties
        are on the order of or smaller than the marker sizes.
	The dashed line indicates the overall weighted average and the solid red line the performance goal. Right: An example peak shape fit to the 2615-keV \nuc{208}{Tl} peak in a \LEG\ \icpc\ Ge detector from calibration data. The peak shape comprises a full-energy Gaussian, an exponentially modified Gaussian tail to account for incomplete charge collection, and a linear background with a step-like shift.}
	\label{fig:qbb-res-mass_pk-shape}
\end{figure}

The transition to the larger \icpc\ detector for both \Ltwo\ and \Lk\ preserves the excellent energy resolution of the \ppc\ and \bege\ detectors.
The performance of ICPC detectors has been demonstrated several ways.
Nine ICPC detectors from two different detector
vendors were deployed toward the end of the \Gerda\ and \MJD\ operational
periods.  In addition, we have characterized the delivered \Ltwo\ \icpc\
detectors in their vendor cryostats. As shown in Fig.~\ref{fig:qbb-res}, the first
75.9~kg of detectors characterized in their vendor cryostats has a
mass-weighted average energy resolution of 2.19~keV at 2039~keV
(0.11\% FWHM, 0.05\% $\sigma$). The in situ
tests in \Gerda\ and the \MJD\ revealed no significant degradation.
The increased detector mass does not affect the energy resolution as shown in Fig.~\ref{fig:qbb-res-mass_pk-shape} (left), leaving open the possibility to further increase the
average detector mass in the future.

In addition to the excellent energy resolution, a Ge detector's peak shape is well understood. The peak shape is modeled as the sum of a full-energy Gaussian component and an exponentially modified Gaussian tail to approximate the peak shape distortion due to incomplete charge collection. A linear step-like flat background is included on either side of the peak. An example fit to a 2615-keV \nuc{208}{Tl} peak in an ICPC Ge detector from calibration data is shown in Fig.~\ref{fig:qbb-res-mass_pk-shape} (right); additional examples of the consistent peak shape fits from the \MJD\ and \Gerda\ detectors can be found in Refs.~\cite{Alvis:2019sil} and \cite{Agostini:2021duc}, respectively.

%% file: sec_legend-approach/subsec_topology.tex
The \BBz\ decay events are inherently single-site, monoenergetic
events occurring within the bulk of a single Ge detector, whereas many background events have larger spatial distributions.
Extracting topological information about the spatial distribution of an event
allows additional background discrimination over a simple energy measurement and builds confidence in a possible observation (Sect.~\ref{subsec:postdiscovery}).
A number of such topological observables have been fully developed and
implemented successfully in both \Gerda\ and the \MJD. They include the following:

\begin{description}
\item[Pulse-Shape Information] ~

\paragraph{Interaction Site Multiplicity}
ICPC Ge detectors are in essence solid-state TPCs, with the
signal fidelity to identify the drift and arrival of distinct ionization clouds at
the point contact. A simple application of this capability is to count the
number of distinct ionization sites in an event. This allows for the discrimination of
single-site \BBz\ decay signals that deposit all their energy within approximately 1~mm from
multiple-site backgrounds like $\gamma$ rays, scattering sites for which can be 1~cm
apart.

\paragraph{Interaction Localization}
The shape of the signal induced during ionization drift carries additional
information that can help isolate the event location. A simple application of
this capability is discriminating the varying drift signals throughout the
detector bulk that would be exhibited by \BBz\ decay events from distinct drift
signals due to ionization from background events originating on the detector
surfaces.

\paragraph{Delayed Charge Collection}
\BBz\ decay events are generated throughout the detector bulk with effectively 100\% of the
ionization collected promptly. Events with significant charge collection at
delayed times, e.g., due to backgrounds generating ionization
on or near surfaces with significant charge trapping and/or low drift fields,
can be easily discriminated from signal events and removed.

\item[Event Coincidences] ~

\paragraph{Ge Multiplicity} The granularity of the Ge detector array configuration
allows the discrimination of single-hit \BBz\ decay events from background events, such
as scattered $\gamma$ rays with simultaneous energy deposition in more than one
Ge detector. It can also be used to identify coincident $\gamma$ rays indicating \BB\ decays to
excited states, serving effectively as a daughter
nucleus tag.

\paragraph{LAr Scintillation} Embedding the Ge-detector array in a LAr active medium
allows for the identification of simultaneous interactions occurring outside of the detectors, such as scattered $\gamma$ rays.
scintillation in \Lk\ over previous phases.

\paragraph{Muon Veto} The water tank surrounding the LAr is instrumented with photomultiplier tubes (PMTs) to act as a water Cherenkov muon veto.
This allows for additional muon-shower identification than that provided by the LAr shield to tag in situ radioisotope production.

\item[Timing Information]
In a \LEG\ ICPC detector, charge begins to drift
and generate a signal immediately after the decay. For a high-energy signal like
\BBz\ decay, this drift generates a measurable voltage change of the analog-to-digital converter (ADC) within just one or two
clock ticks, allowing for unambiguous and precise determination of
event times. This allows additional analysis handles:

\paragraph{Event-Time Distribution} A \BBz\ decay signal has a uniform time
distribution as opposed to, e.g., exponentially decaying backgrounds or other
sources with rates that vary during data taking.

\paragraph{Delayed Coincidences} \BBz\ decay has no progenitor event and is
unaccompanied by delayed de-excitations or daughter decays. Searches for events in
delayed coincidence with \BBz\ decay candidates (either as the prompt or delayed
event) can thus help discriminate between a \BBz\ decay signal and backgrounds, e.g.,
from cosmogenic $^{68}$Ge decay sequences, Bi-Po coincidences from Rn
progeny, and cosmic-ray-induced activation products.
\end{description}

\begin{figure}
  \includegraphics[width=0.99\textwidth]{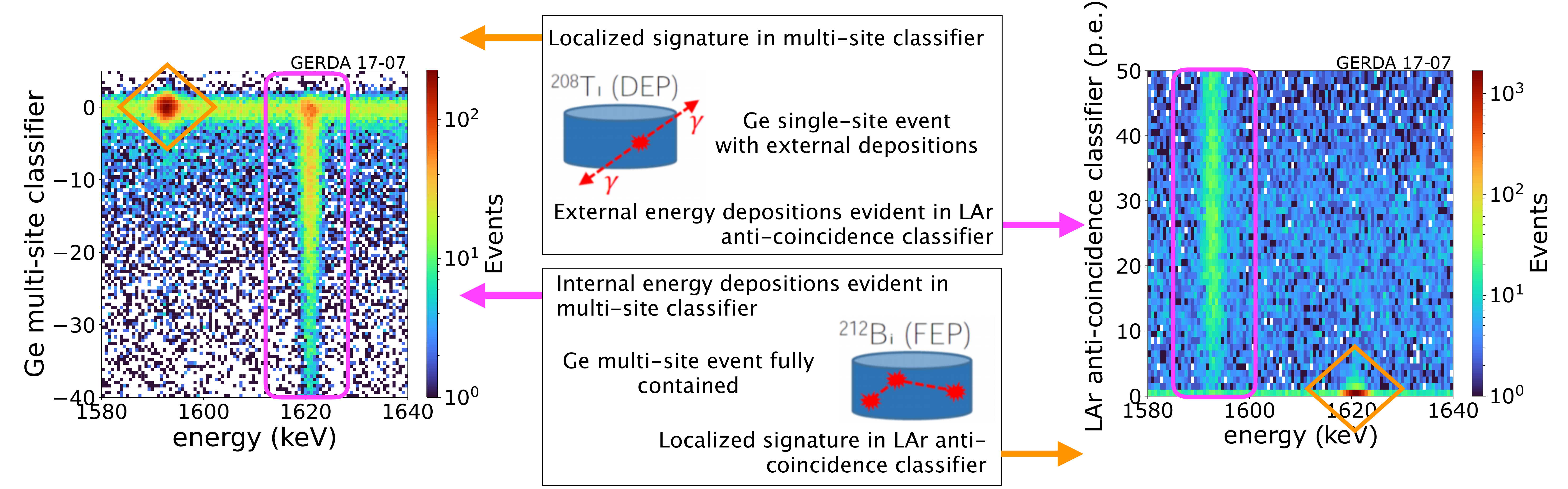}
\caption{An example of the clear separation of signal and background in the complementary PSD and anti-coincidence classifiers from \Gerda\ calibration data. Two types of events are considered: a \nuc{208}{Tl} double-escape peak (DEP) event and a \nuc{212}{Bi} full-energy peak (FEP) event. Both of these are characteristic of backgrounds that cause peaks in the Ge spectrum. The orange diamonds indicate the classification of events as signal-like in one of the two parameter spaces, the magenta rectangles highlight the distribution in each parameter for background like events. Unlike the two cases here, a \BBz\ decay event would be classified as a signal in both observable parameters.}
\label{fig:classifier-example}
\end{figure}

\Lk\ uses a multivariate analysis incorporating all of these observables
to identify and constrain not just signal events but also the nature and sources
of the backgrounds that are observed in the actual experiment.  Owing to the
clear separation of signal and background in the multivariate parameter space,
\Gerda\ and the \MJD\ have often applied discrete cuts on these observables (see Fig.~\ref{fig:classifier-example}). For
simplicity, we adopt this historical approach throughout the remainder of this
document and will refer to the following two cuts:
\begin{description}
\item[Anti-Coincidence (AC) Cut] Events producing energy depositions in multiple
Ge detectors (granularity cut), generating scintillation in the LAr volume (LAr veto cut), or creating Cherenkov light in the water tank (muon veto cut) are flagged as background and removed.
\item[Pulse Shape Discrimination (PSD) Cut] Events with waveforms indicative
either of multiple energy depositions within a single Ge detector or of
charge collection associated with surface events are flagged as background and removed.
\end{description}

%% file: sec_legend-approach/subsec_projected_bg.tex
We have estimated the expected background rate in the baseline design of \Lk\ through a comprehensive Monte Carlo simulation campaign, relying on the extensive radiopurity assays of selected materials, proven active background suppression through a LAr active veto, PSD capabilities of \icpc\ Ge detectors, opportunities afforded by such new capabilities as UGLAr, larger detectors, and new readout electronics development.
The anticipated background rate at \qval\ is \BGprojkev , consistent with
  our background goal of \bgkev.
Here we summarize the major sources of backgrounds, their impact, and background reduction improvements expected for \Lk:

\begin{description}
\item[\nuc{238}{U} and \nuc{232}{Th} Decay Chains]
Decays of long-lived radioisotopes \nuc{238}{U} and \nuc{232}{Th}---as well as their shorter-lived progeny in array components---have been simulated for the \Lk\ array.
The simulations assume secular equilibrium of the decay chains, however, more direct assay information is used to set the activities of the relevant sub-chains (see Sect.~\ref{sec:techmat})
  The decays of \nuc{214}{Bi} and \nuc{208}{Tl}, both emitting $\gamma$ rays with energies in excess of the 2039-keV $Q$ value for \onbb\ decay in \gesix, are of primary concern; however, the LAr veto is highly effective in rejecting Compton scatters from these $\gamma$ rays.  Backgrounds from the \nuc{238}{U}- and \nuc{232}{Th}-decay chains before analysis cuts are reduced compared with \Gerda\ by factors of 19 and 3.6, respectively, as a result of the following background reduction measures:
  \begin{myenumerate}
	\item \emph{Lower-background Kapton flex cables:} $\times$12 reduction in \nuc{238}{U} and $\times$13 reduction in \nuc{232}{Th} activity per channel compared with the \MJD\ cables; $\times$75 reduction in \nuc{238}{U}, and $\times$68 reduction in \nuc{232}{Th} activity per channel compared with the \Gerda\ cables \cite{GERDA:2019cav}, based on recent developments isolating the sources of contamination in the production process reported in Ref.~\cite{Arnquist:2019fkc}.
	\item \emph{Adoption of ASIC-based front-end electronics:} $\times$7 reduction in \U\ and  $\times$75 reduction in \Th\ activity per channel compared with the \MJD\ LMFE \cite{Abgrall:2016cct}, based on the ASIC chip activity reported in Ref.~\cite{Edzards:2020wfg} and direct mounting of ASIC chips on Kapton laminate signal cables (see Sect.~\ref{subsec:front-end}).  Additionally, this choice eliminates charge-sensitive amplifiers mounted near detectors.
	\item \emph{Reduced front-end substrate and connector mass:} $\times$7 reduction in the plastic mass at both the signal cable bonding point, enabled by the adoption of a ASIC-based front end chip, and the high-voltage cable bonding point, enabled by the adoption of Kapton flat-flex cable.
	\item \emph{Elimination of nylon mini-shrouds:} Use of UGLAr eliminates the need for nylon shrouds used in \Gerda\ to limit the drift of \nuc{42}{K} ions to detector surfaces (See item \emph{\nuc{42}{K} in LAr} below).
	\item \emph{Lower array packing density:} The circular string arrangement and increased spacing between detectors reduces the solid angle from $\gamma$-ray sources.
\end{myenumerate}

\item[\nuc{42}{K} in LAr]
Atmospheric Ar contains two long-lived radioisotopes, \nuc{39}{Ar} and \nuc{42}{Ar}, the latter of which is a background for \onbb\ decay searches through its decay to \nuc{42}{K}, which can drift in electric fields to detector surfaces and subsequently $\beta$ decay with a $Q$ value of 3525 keV.
The use of UGLAr near the arrays is expected to strongly reduce this background.  Because of the large uncertainty in the \nuc{42}{Ar} content of \UGLAr, we have estimated its activity conservatively.  Still, this approach represents a $\times12$ reduction in this source of background. See Sect.~\ref{subsubsec:ar42} for more details.

\item[Alpha Decays on Detector Surfaces]
Alpha particles originating in decays on the detector's thin \pPlus\ electrode and passivated insulating surfaces can lose sufficient energy to be a source of background at 2039~keV, while alphas impacting the thick \nPlus\ contacts are effectively stopped.  PSD is particularly effective at rejecting these surface events based on the fast (\pPlus) or delayed arrival of charge (passivated).
The areas of these sensitive surfaces are largely independent of detector mass, and the 2.6~kg \icpc\ detectors used in \Lk\ have a mass $4.3\times$ larger than that of the \bege\ detectors used in \Gerda; this directly translates to a $\times 4.3$ reduction in alpha backgrounds before cuts.

\item[Cosmogenics]
The cosmogenic isotopes \nuc{68}{Ge} and \nuc{60}{Co} are produced by exposure of Ge to cosmic-ray-produced neutrons, and they decay within enriched detectors with $Q$ values in excess of the \gesix\ \Qbb.
Backgrounds from \nuc{68}{Ge} ($T_{1/2} = 271$ days), averaged over a 10-year
operational lifetime, are projected to increase in \Lk\ as compared with
\Gerda; the reason is the two-year projected underground cool-down period for
the \LEG\ detectors prior to the start of data-taking as opposed to the roughly four-year cool down of the \Gerda\ \bege\ detectors.
The in situ production of muon-induced \geVII\ and \geVIIm\ backgrounds contributes to the cosmogenics backgrounds but to a lesser  extent at the SNOLAB reference depth.
\end{description}

\begin{figure}
\includegraphics[width=0.99\tw]{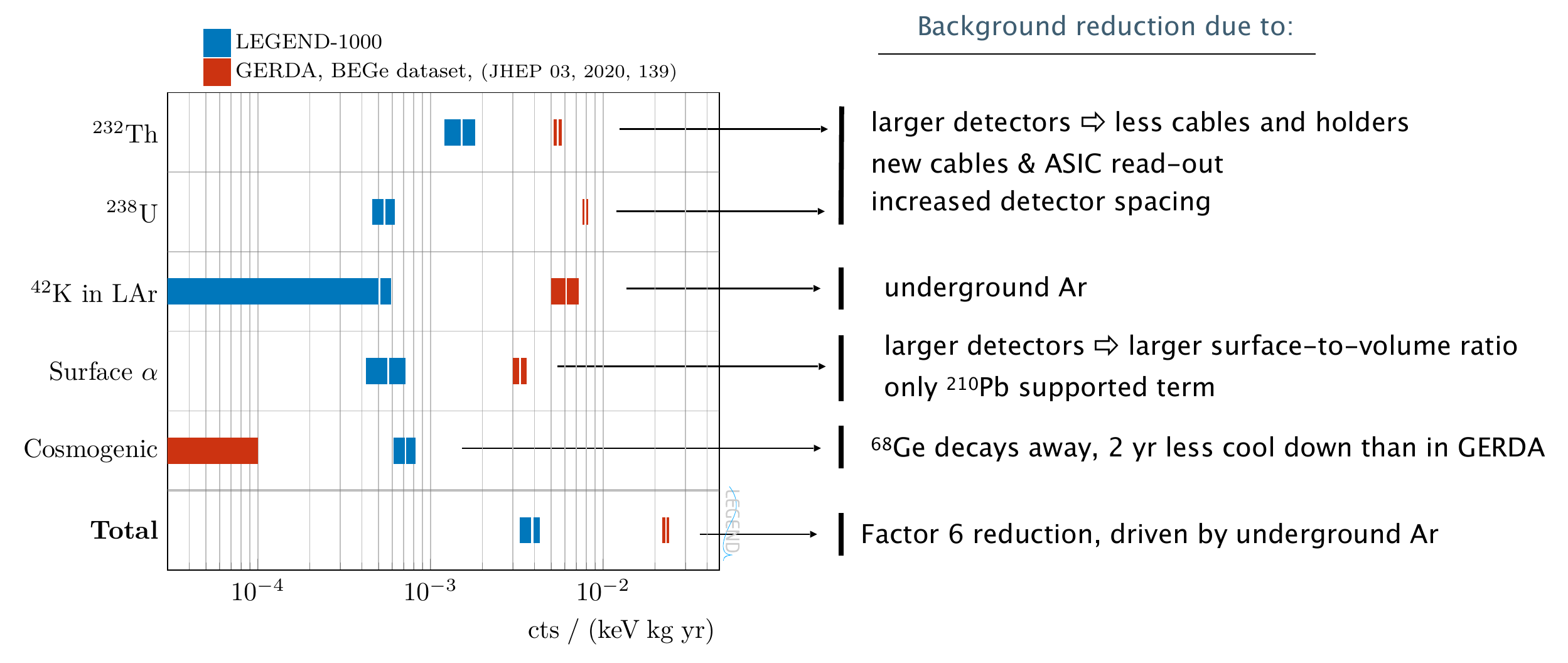}
\caption{The expected background index associated with each of the dominant sources, before applying analysis cuts, projected for LEGEND-1000 (blue) and measured in \Gerda\ (red).  Significant reductions in all categories, with the exception of internal cosmogenic backgrounds, are predicted based on the use of lower-background materials, Ar extracted from underground deposits, and the use of larger mass detectors.  For details of the calculation of these estimates, see Sect.~\ref{subsec:background-budget}.}
\label{fig:LGNDvsGerdaBIs}
\end{figure}

\begin{figure}
\includegraphics[width=0.99\tw]{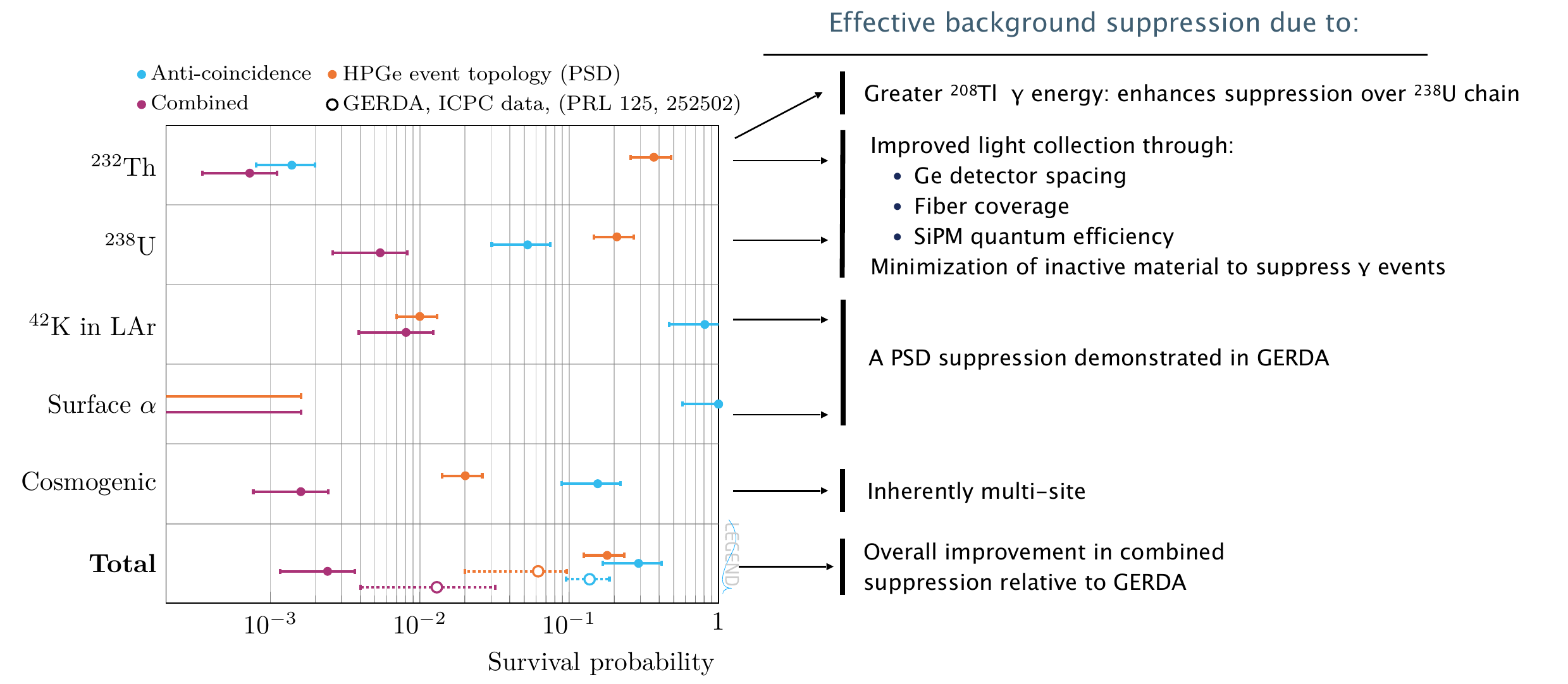}
\caption{A measure of the \Lk\ background suppression expressed as survival probabilities for the separate anti-coincidence, event topology (PSD), as well as their combined effect for each major source of backgrounds. A comparison with the total suppression achieved with the \Gerda\ \icpc\ data is included for comparison. The justification of the effectiveness of each background's suppression is noted. For details of the calculation of these estimates, see Sect.~\ref{subsec:background-budget}.}
\label{fig:BG-suppresion}
\end{figure}

The projected background index in the \Lk\ experiment prior to analysis cuts is
\cpowten{3.9^{+0.4}_{-0.6}}{-3}~\ctsper, and its components are compared to the
levels achieved in \Gerda\ in Fig.~\ref{fig:LGNDvsGerdaBIs}.
The effectiveness of the anti-coincidence and PSD cuts for each component is shown in
Fig.~\ref{fig:BG-suppresion} and compared with the values observed for the \Gerda\ ICPCs.
The background index prior to cuts corresponds to a factor of six reduction
over \Gerda, but the composition of the background is very different: the
before-cuts \Lk\ background is dominated by components that are strongly
rejected by one cut but not the other, while the \Gerda\ background index is
dominated by $^{238}$U whose efficient rejection requires the application of both cuts.
\Lk\ also benefits from improvements in light collection due to an increased
spacing between detectors, the reduction of inactive material near the
detectors, inner and outer fiber shrouds, and SiPMs with an improved quantum
efficiency, which enhances the AC cut performance.
The total rejection factor is thus expected to be more than a factor of five stronger
in \Lk\ over what was achieved in \Gerda\ ICPCs and a full factor of 10 over
the average suppression factor achieved for all \Gerda\ detectors.

\begin{figure}[ht]
  \includegraphics[height=0.4\tw]{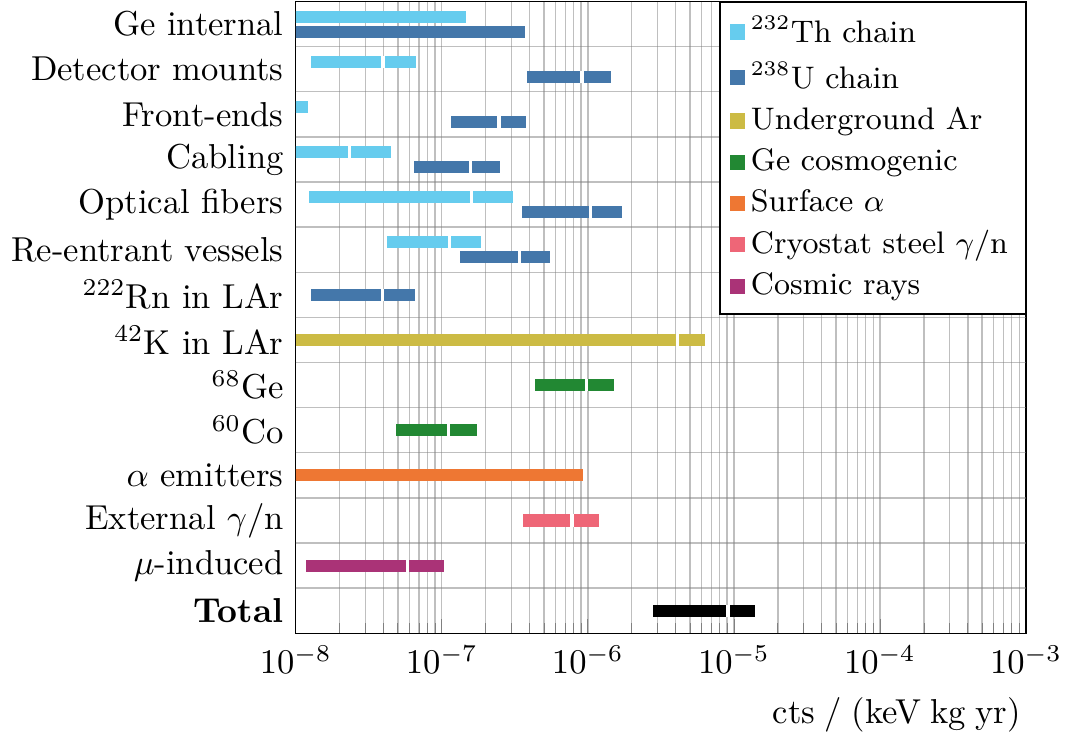}
 \caption{Estimated significant backgrounds for \Lk.  Bands indicate 1-$\sigma$ uncertainties (or 90\% CL upper limits) due to assay and Monte Carlo estimation of background-rejecting analysis cuts.  For Ge internal and surface alpha backgrounds, only upper limits are estimated.  For details of the calculation of these estimates, see Sect.~\ref{subsec:background-budget}.\label{fig:CutComparison}\label{fig:BackgroundBudget}}
\end{figure}

A summation of all non-negligible backgrounds expected in \Lk\ is shown in Fig.~\ref{fig:BackgroundBudget}; we estimate a final background index of \BGprojkev\ in \Lk.  At an energy resolution of 2.5~keV FWHM, an equivalent a background rate of \BGprojfwhm\ is consistent with the stated \Lk\ background goal of \bgfwhm.
A detailed treatment of the analysis methods is given in Sect.~\ref{subsec:data-analysis} and greater details on the background contributions in Sect.~\ref{subsec:background-budget}.

%% file: sec_legend-approach/subsec_goals.tex
\begin{figure}
\includegraphics[trim={10mm 0 10mm 0},clip=false,height=0.4\columnwidth]{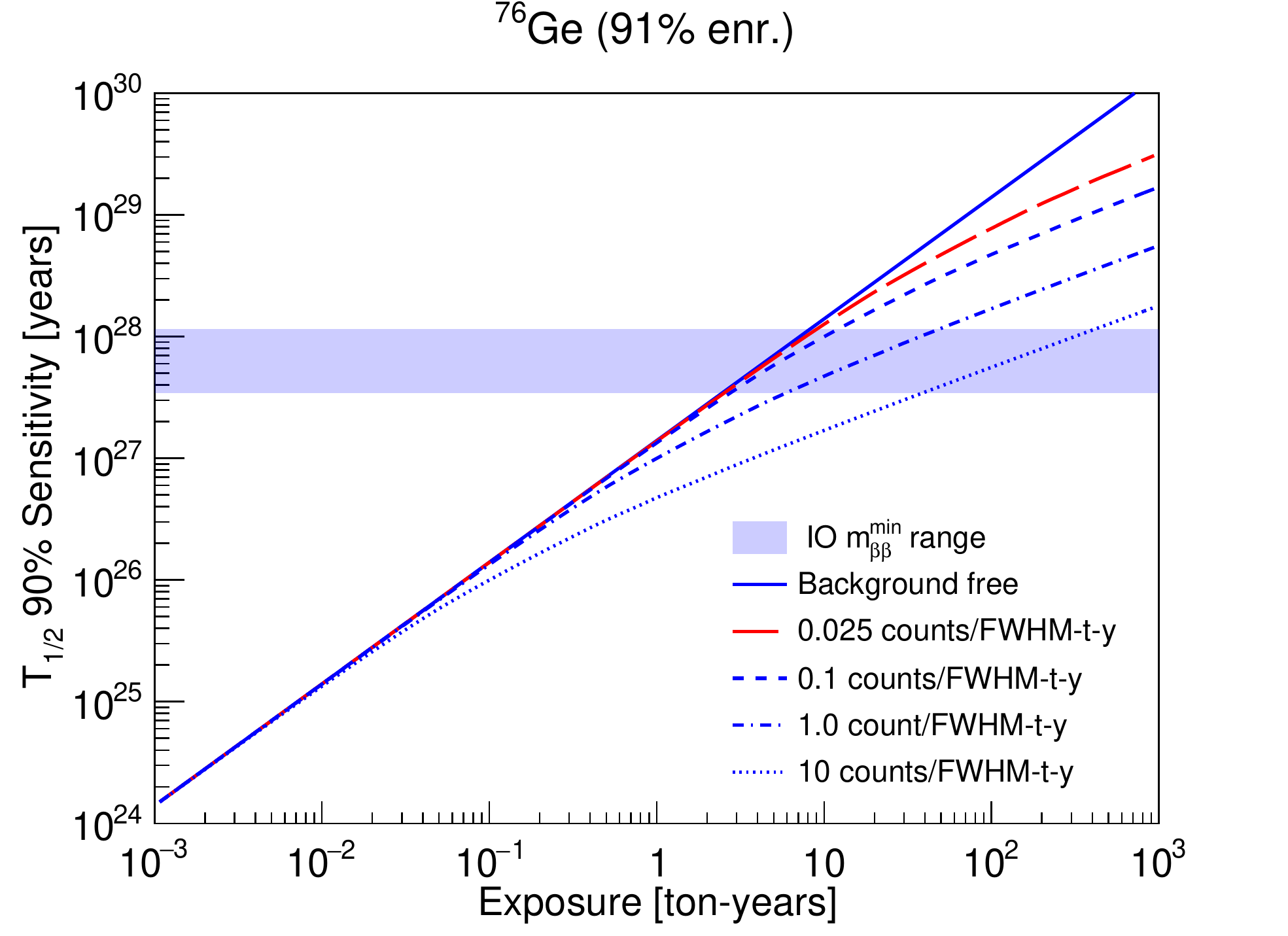}
\includegraphics[trim={5mm 0 15mm 0},clip=false,height=0.4\columnwidth]{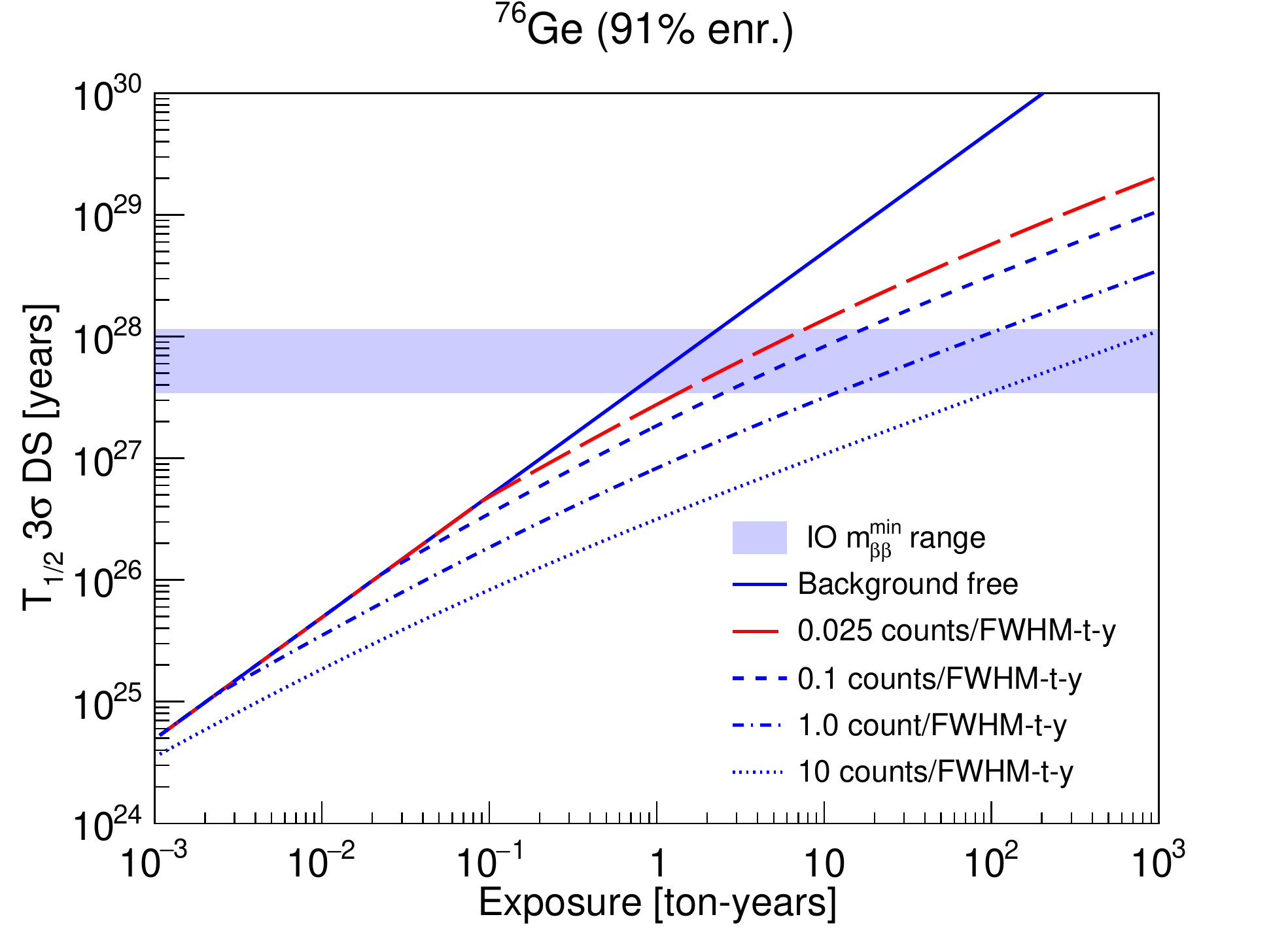}
\caption{\label{fig:ThreeSigDL} The sensitivity to a \onbb\ decay signal in \gesix\ as a function of exposure and background for a (left) 90\% CL exclusion
sensitivity and (right) 3$\sigma$ (99.7\% CL) discovery sensitivity (DS).
Note, the background rates are normalized to a 2.5 keV FWHM energy resolution.
}
\end{figure}

The sensitivity to a \onbb\ decay signal as a function of exposure and
background is shown in Fig.~\ref{fig:ThreeSigDL} separately for a 90\% CL exclusion
sensitivity and for a 99.7\% CL discovery sensitivity.
The exclusion sensitivity is defined as the median
half-life value that will be excluded assuming there is no signal, while the
discovery sensitivity gives the half-life value at which there is a 50\% chance of a 3$\sigma$ discovery.
The calculation assumes a total signal efficiency of 69\%, accounting for the
enrichment level, analysis cuts, active volume
fraction, and containment efficiency for \BBz\
decay events to have their full energy deposited within a crystal's active
volume.  If an experiment  background is {\em zero},
both the discovery sensitivity and the limit sensitivity scale linearly
with the exposure, whereas in the background-dominated
regime both sensitivities scale with the square root of exposure. The transition between
these two regimes is governed by Poisson statistics and is computed using the
approximation outlined in Ref.~\cite{Agostini:2017jim}. We neglect background
uncertainty under the assumption that it is well constrained from energy
side bands.  For signal discovery, a low background is especially important
because as the expected number of background counts increases, the signal level
required to obtain a 3$\sigma$ excess grows rapidly.

\LEG's staged approach provides a low-risk path to world-leading sensitivity.
The initial \Ltwo\ phase should easily achieve a modest
background improvement over \Gerda\ with a background index
of \cpowten{2}{-4}\,\cpKkgy\ or 0.5\,\cpFty\ at \Qbb. With this
background level, \Ltwo\ reaches a 3$\sigma$
discovery sensitivity of \powten{27}\,yr with an exposure of only 1\,\tyr\ within five years.
Using a nuclear matrix element range of 2.66 to 6.04 for
$^{76}$Ge (see Table~\ref{tab:MatrixElements}),
a phase space factor of
\cpowten{2.363}{-15}\,/yr~\cite{Kotila:2012zza} (consistent with \cpowten{2.37}{-15}\,/yr of Ref. ~\cite{Mirea:2015nsl}),
and a value of \gA=1.27,
the \Ltwo\ discovery sensitivity corresponds to an \mee\ upper limit in the range of $34-78$\,meV.

\begin{figure}
\includegraphics[height=0.45\columnwidth]{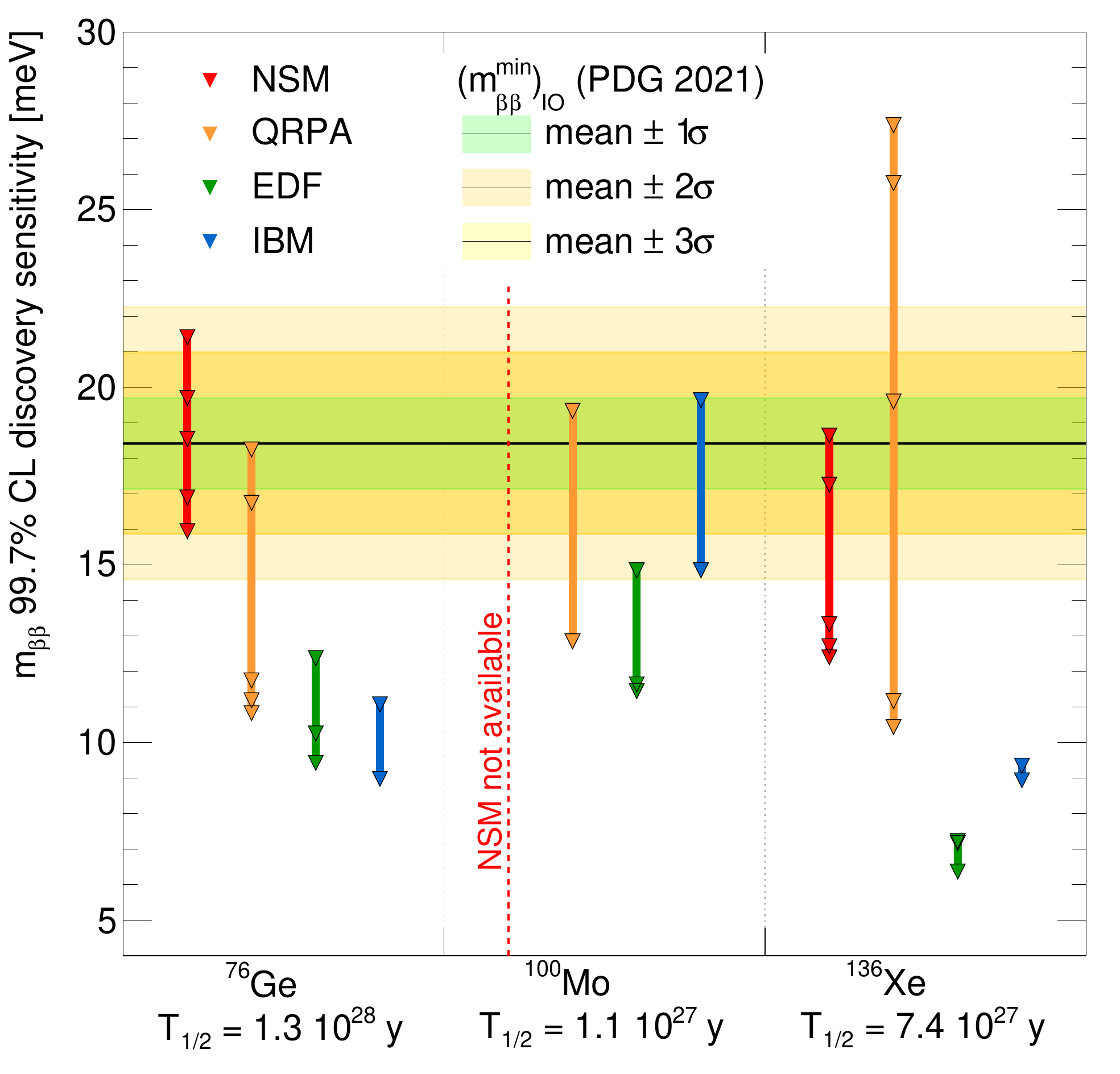}
\caption{\label{fig:mbb_compare} A comparison of \mee\ 99.7\% CL discovery
sensitivity for different isotopes at stated half life
sensitivities~\cite{CUPID:2019imh,nEXO:2021ujk} and for a range
of available nuclear matrix element calculations, grouped according to nuclear
framework~(using the references for Table~\ref{tab:MatrixElements} plus Ref.~\cite{Terasaki:2020ndc} and~\cite{Deppisch:2015qwa}). The horizontal bands show the probability distribution
for the minimally allowed \mee\ value assuming that neutrino masses follow the
inverted ordering. Uncertainties are due to the limited accuracy with which the
oscillation parameters have been measured.  Figure from Ref.~\cite{Agostini:2021kba}.
}
\end{figure}

\LEG's ultimate goal is to achieve 3$\sigma$ discovery sensitivity covering the full parameter space remaining for the inverted neutrino mass ordering, under the assumption of light left-handed neutrino exchange as the dominant mechanism.
As described previously, the \Lthou\ experiment should achieve a higher
signal/background discrimination than the present generation experiments.
The background goal for \Lthou\ is a background index of  \bgkev\ or \bgfwhm.
At this background level, \Lthou\ reaches a half-life discovery sensitivity of
\cpowten{1.3}{28}\,yr, corresponding to a \mee\ upper limit in the range of $9 -
21$\,meV in 10 yr of live time. As shown in Fig.~\ref{fig:mbb_compare}, the \Lk\
discovery sensitivity comes within a few
standard deviations of covering the inverted ordering parameter space for all
four of the primary many-body methods used in modern nuclear matrix element calculations.

%% file: sec_legend-approach/subsec_post-discovery.tex
In the event of a discovery, \Lk's rich data provide a number of additional
handles for signal verification and rejection of alternative hypotheses.
\Lk's exquisite energy resolution combined with the simple spectral shape of its
expected backgrounds near the ROI---a smooth continuum free of peaks within
an energy range equal to many multiples of the energy resolution
at \Qbb---makes \Lk\ capable of making a discovery even with
very few counts in the ROI.
 Figure~\ref{fig:discovery_sim} shows a Monte Carlo
example of what such a discovery might look like for a $^{76}$Ge \BBz\ decay
half life of 10$^{28}$ years, just within the \Lk\ discovery
sensitivity. With such a sparse background continuum after
applying all analysis cuts, a feature at \Qbb\ can be clearly discerned with just a handful of
counts.

This simple spectrum lends itself to a simple analysis; sensitivity is
near-optimal for even a rudimentary, Poisson-counting treatment based on
counting events within an ROI chosen based on the known detector resolution
function and then comparing that count with a background expectation derived from
interpolating count rates in sidebands straddling the ROI. In practice, we
maximize sensitivity from this experiment using a true spectral
analysis that accounts for the full detector resolution function and its
uncertainties, optimizes the separation of signal from background in the pulse
shape parameters, and accommodates potential variations in the energy dependence of
the background as derived from the background model. This can be achieved while incurring negligible
systematic uncertainty. We are also exploring more extensive
multivariate, as well as machine-driven, analyses using more of the available pulse shape
and coincidence event information, balancing improvements in statistics
with increases in systematic uncertainty.

In the event of a discovery, the additional
handles for signal verification and rejection of alternative hypotheses include the following:
\begin{myitemize}
\item The ROI-event energy distribution should be compatible with the detector
resolution function centered at \Qbb, with appropriate moments
(mean, standard deviation) or shape.
\item The ROI-event spatial information should be consistent with point-like
interactions uniformly distributed throughout the $^{\textrm{enr}}$Ge: no
clustering in specific detectors, no simultaneous energy deposition in adjacent
detectors or in the LAr, and pulse shapes consistent with single-site
energy depositions in the detector bulk.
\item The ROI events should occur randomly in time, with no correlations with
other detector events or delayed coincidences with the LAr veto.
\item The ROI events should be unique, with no other similar nearby peak structures
lacking explanation from known processes.
\end{myitemize}
These tests are informed by the observation that there exists no potential
detector signature that can mimic \BBz\ decay without also leaving
a telltale signature elsewhere in the data. For example:
\begin{myitemize}
\item As shown in Fig.~\ref{fig:classifier-example}, a faint $\gamma$ line appearing right at \Qbb\ and leaving just a handful of counts in the ROI after all cuts would also create approximately 10 times more events with a clear multi-site signature. Other $\gamma$-induced monoenergetic events in Ge such as double- and single-escape peaks would instead create a strong signature in the LAr scintillation light classifiers.
\item Other non-$\gamma$-induced monoenergetic transitions (e.g., electron capture, conversion
electrons, and $\alpha$ decay) are
likely to be accompanied by the emission of coincident $\gamma$-ray, x-ray, or
bremsstrahlung photons; by additional transitions appearing elsewhere in the energy
spectrum; or by time coincidences with progenitors or daughter decays.
\item Nearly all other conceivable backgrounds give a spectral
signature that either forms a continuum across the ROI or is significantly
broader than the detector energy-resolution function.
\end{myitemize}

%% file: sec_legend-approach/subsec_params.tex

The experimental parameters used in the \Lk\ discovery potential and background projections are listed in Table \ref{tab::params}.

\begin{table}[h]
\caption{Experimental parameters in the \Lk\ discovery potential and background projections.}
\label{tab::params}
\begin{tabularx}{0.75\textwidth}{ X X }
 \rowcolor{legendgrey}
 {\bf Parameter} &
{\bf Value}  \\
 \hline
 \rowcolor{legendblue}
 \multicolumn{2}{c}{Performance Parameters}\\
 \BBz\ decay isotope				&		\nuc{76}{Ge} \\
 \Qbb					&		2039 keV \\
 Total mass				&		1000 kg  \\
 Energy resolution at \Qbb		&		2.5 keV FWHM\\
 Overall signal acceptance\footnote{Includes an average 91\% \nuc{76}{Ge} enrichment, 92\%
 active volume, 92\% containment efficiency, and 90\% analysis cuts. An
 additional factor of 95\% is necessary for the fraction of events in a $\pm
 2\sigma$ optimal region of interest in a counting-based analysis. See
 Sect.~\ref{subsubsec:eff} for details.}
		&		0.69 \\
Live time	goal				& 		10 yr \\
Total exposure goal & 10 \tyr \\
Background goal			&		\mbgkev \\
	                 &		\mbgfwhm \\
 \Tz						&		\cpowten{1.3}{28}\,yr (99.7\% CL discovery) \\
 						& 		\cpowten{1.6}{28}\,yr (90\% CL sensitivity)\\
 \mee 					&		9.4--21.4\,meV (99.7\% CL discovery) \\
						&		8.5--19.4\,meV  (90\% CL sensitivity) \\
\rowcolor{legendblue}
\multicolumn{2}{c}{Physics Parameters}\\
 \Mz						& 		2.66--6.04
 \cite{Coraggio:2020hwx,Song:2017ktj}\\
 \Gz						& 		\cpowten{2.363}{-15}\,/yr \cite{Kotila:2012zza} \\
  \gA						&		1.2724
\end{tabularx}
\end{table}

%% file: sec_technical/sec_technical.tex
\section{The \Lk\ Design}
\label{sec:techmat}

\begin{figure}[h]
  \includegraphics[width=0.9\tw]{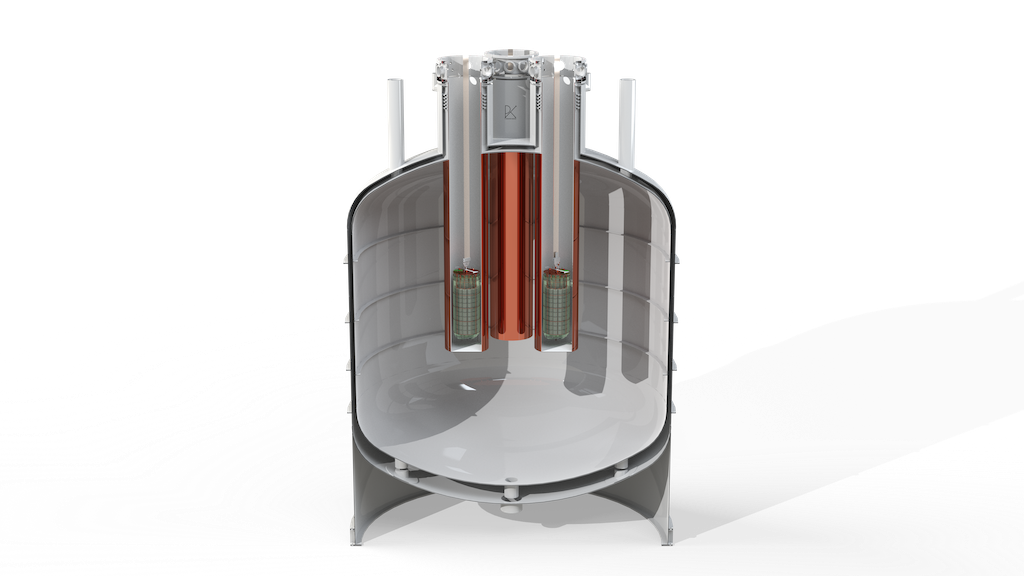}
\caption{
Conceptual design of the \Lk\ experiment.}
\label{fig:baseline-view}
\end{figure}

Figure \ref{fig:baseline-view} illustrates the conceptual design of the \Lk\ experiment. Large-mass \icpc\ detectors operate in a bath of LAr. The Ge detectors are distributed among four independently-operating, 250-kg modules to allow commissioning of the array in stages. Below, we provide greater detail on major components of the \Lk\ technical design. Next, the use of ultra-low-background materials, some of which have been developed by \LEG\ collaborators, that are certified for use in \Lk\ through material screening is described. Last, we specify the data-analysis techniques and conclude with a detailed treatment of the background projections for \Lk.

\subsection{Detailed Technical Design}\label{sec:technical}
\input{sec_technical/subsec_76ge}
\input{sec_technical/subsec_detectors}
\input{sec_technical/subsec_cryostat}
\input{sec_technical/subsec_uglar}
\input{sec_technical/subsec_active-shield}
\input{sec_technical/subsec_electronics}

\input{sec_technical/subsec_daq}

\input{sec_technical/subsec_calib}

\subsection{Materials and Assay}\label{sec:material-assay}
\input{sec_technical/subsec_materials}

\input{sec_technical/subsec_activities}

\subsection{Data Reduction and Analysis Techniques}\label{subsec:data-analysis}
\input{sec_technical/subsec_data-analysis}

\subsection{Background and Sensitivity Projections}\label{subsec:background-budget}
\input{sec_technical/subsec_background-budget}

%% file: sec_technical/subsec_76ge.tex
\subsubsection{Acquisition of \gesix}
\label{subsec:ge76-procurement}

The enriched \gesix\ \icpc\ detectors are fabricated by commercial vendors, as described next in Sect.~\ref{subsec:detector-fabrication}.
When dealing with enriched material, a series of additional steps are managed by the Collaboration to acquire and process the \gesix\ into the form usable by the detector manufacturers.

\paragraph{\nuc{76}{Ge} Procurement}

The \Ltwo\ project purchased the enriched $^{76}$Ge
isotope from Isotope JSC\footnote{Isotope JSC, Moscow, Russia; \url{http://www.isotop.ru/en/}}, which is enriched at the Electro Chemical Plant\footnote{Electrochemical Plant JSC, Zelenogorsk, Russia; \url{https://www.ecp.ru/eng/}} (ECP),
and Urenco\footnote{Urenco Nederland, Almelo, Netherlands; \url{https://www.urenco.com/}}.
Urenco material has now been successfully converted into working detectors for \Ltwo, while ECP delivered $^{76}$Ge
previously to the \Gerda\ and \MJD\ experiments.

In previous procurements, the \nuc{76}{Ge} isotopic enrichment fraction was specified to be at least 86\%, which was typically exceeded by 1--2\% in received shipments.
For \Ltwo, it became economical
to order material with at least 92\% enrichment.
If the detector enrichment is below 86\%, the fraction of \nuc{70}{Ge} becomes problematic due to the high cross section to produce \nuc{68}{Ge} while residing on the Earth’s surface. Higher enrichment fractions result in a more favorable signal-to-background ratio since most background
sources in the Ge material scale with the absolute detector mass, while the
signal strength scales with the enrichment fraction.  The higher
specification (92\%) boosts the sensitivity of the experiment
by about 6\% for the same amount of \nuc{76}{Ge}.
Based on considerations of  costs and benefits, we are planning for an enrichment fraction of 92\%.
Note that the experimental parameters listed in Sect.~\ref{subsec:phys-params} defining the \Lk\ sensitivity goals assume an enrichment fraction of 91\% since it represents the weighted average of the new material (at 92\%) and the addition of the \Ltwo\ detectors, which are described next.

\Lk\ plans to fabricate 870~kg of
new detectors and reuse 130~kg of existing \Ltwo\ detectors.
For a 75\% yield of detectors from the starting material, discussed below, and recycling
50~kg of existing small detectors, \Lk\ needs to procure
1100~kg of enriched $^{76}$Ge.
For \Ltwo, the two vendors produced 185~kg of enriched material, with a combined production rate of about 65~kg per year.
This production rate was not limited by vendor separation capabilities but rather by the availability of project funding.
From our \Ltwo\ experiences and based on discussions with vendors, we estimate that a conservative production rate is 220~kg per year.
This results in about five years to produce all of the material,
assuming a technically driven schedule.
The total amount of about a 12-ton quantity of
natural Ge input material to the enrichment process
spread over this period should not cause any disturbance of
the world market, considering that the current world production of natural Ge is about 130 tons per year.\footnote{2021 estimate from \url{https://www.usgs.gov/centers/nmic/germanium-statistics-and-information}}

\paragraph{\nuc{76}{Ge} Metal Production}

The enriched Ge is delivered in the form of GeO$_2$.
Established facilities, in the USA or Europe,
reduce the oxide to Ge metal and further purify the material to the level required by the detector manufacturers.
The Ge oxide has a purity of 4N (99.99\% Ge). For \Ltwo, the
conversion to metal and the initial purification by zone refinement to 5N to 6N
(99.999--99.9999\%) quality was performed by
PPM Pure Metals\footnote{\label{fn:ppm}PPM Pure Metals. Am Bahnhof 1, 38685 Langelsheim, Germany. Now Vital Pure Metal Solutions (VPMS)}. Additionally, the Leibniz-IKZ Institute in Berlin also has a similar capability and delivered 43~kg of enriched Ge metal to \Ltwo.
About 1--2\% of the material is lost in this processing step.
For \Lk, it is estimated this can be reduced to less than 1\% by collecting cutting
scraps (kerf) for recycling.
The enriched Ge is transported in a custom 16-ton steel housing mounted inside a standard 20~ft sea container~\cite{Agostini:2014hra,Abgrall:2017acl}.
The production rate of the radioactive isotopes $^{68}$Ge and $^{60}$Co, which contribute to experimental backgrounds, are reduced by a factor of 10 by the shield~\cite{Barabanov:2005cw}.

\paragraph{\nuc{76}{Ge} Losses and Recovery}

\begin{figure}
\includegraphics[height=0.4\textwidth]{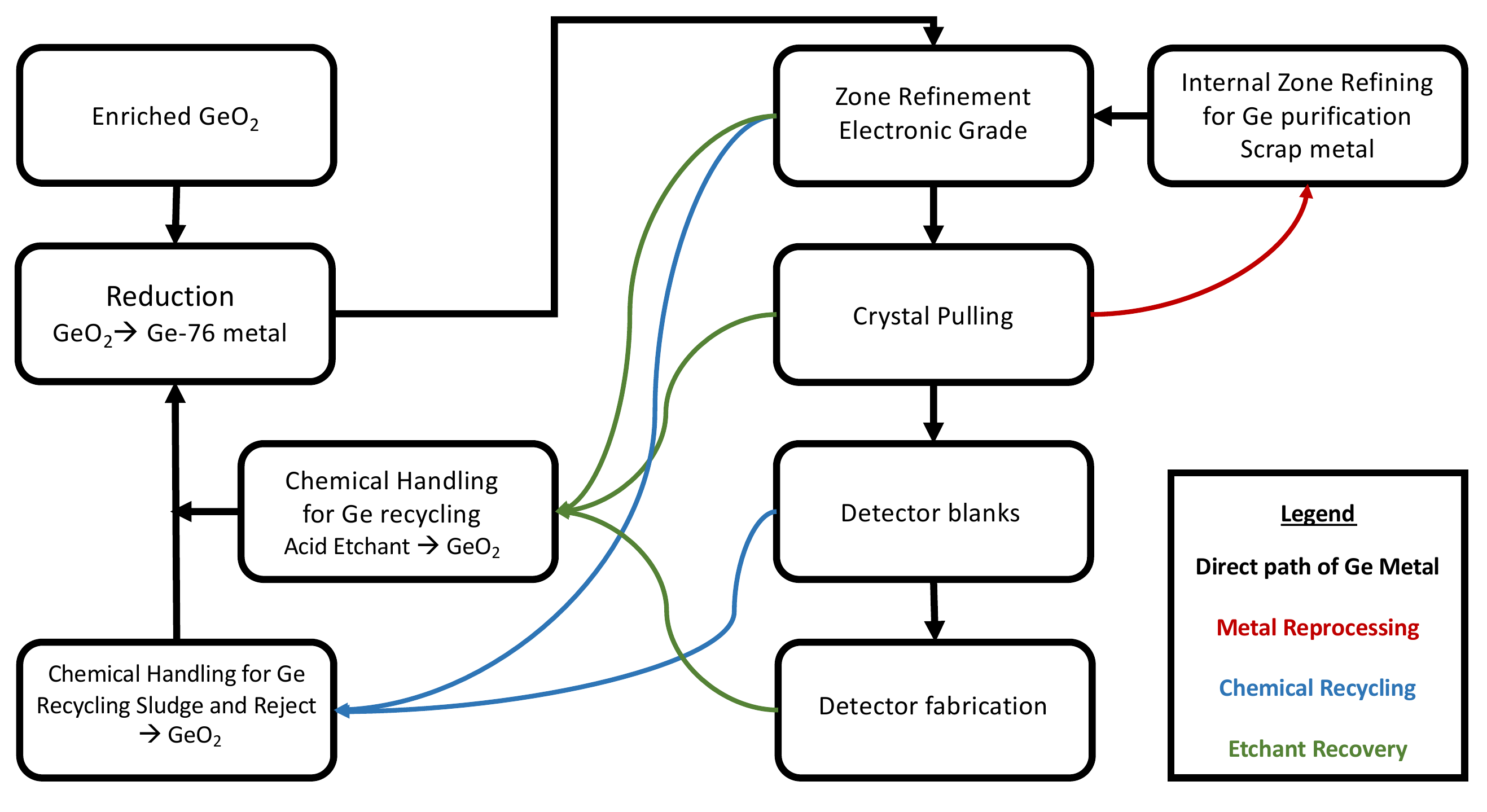}
\caption{A flow chart showing the movement of \nuc{76}{Ge} material through the processing, production, and recovery stages discussed in the text.}
\label{fig:GeFlowChart}
\end{figure}

During all processing steps starting from the reduction of GeO$_2$ to the final detector production, various
forms of Ge waste occur.
Due to the value of the enriched Ge, it is prudent to reduce losses and to collect and recycle waste material.
Figure~\ref{fig:GeFlowChart} shows schematically the production process and the material streams. The recycling steps are:

\begin{myenumerate}

\item Metal Reprocessing:  Cutting chips, crystal and bar tails that do not meet the
  required purity, can be reintroduced either directly into crystal pulling or through additional zone refinement during the
  detector manufacturing stages.
  While the fraction of this material stream depends on the viable size
  of the detector(s) cut from a crystal boule, the recovery yield from reprocessed metal is on the order of 96\%.

\item Chemical Recycling:  Cuttings and grindings with lubricant/water, lithiated contacts, and metal that has exceeded the
  contamination level for zone refining must be chemically recycled. Chemical recycling is also a normal industrial process, and experience exists within the collaboration. This step is discussed in more detail below.

\item Etchant Recovery:  Prior to zone refinement or crystal pulling, the metal bars are etched with a mixture of HF and HNO$_3$.
  Approximately 5\% of the input material per detector
  production run---a batch of typically 20--30~kg---is lost at this step. After the metal recycling step, this fraction increases.
  The extraction of the Ge from the etchant solution turns out to be difficult and is
  not part of the industry process. The collaboration is exploring if the etchant process can be adapted such that recycling becomes possible.

\end{myenumerate}

About 20\% of production run input material needs chemical reprocessing, with which
the collaboration has experience.
The \MJD\ chemical reprocessing~\cite{Abgrall:2017acl} of the cutting and grinding losses used the
following steps: burning in chlorine gas to produce GeCl$_4$, fractional distillation of GeCl$_4$, and
hydrolysis to convert to GeO$_2$, which was then fed back into the normal production cycle.
The same process was recently used by \Gerda\ for metal waste from the production of \bege\ detectors.

Another option for chemical reprocessing is to use a liquid process with HCl and H$_2$O$_2$ to produce GeCl$_4$. The advantage of this wet chemistry is that all forms of Ge and oxide can be recycled while not using Cl$_2$ gas, which poses other safety issues. This wet process was previously used by PPM Pure Metals on an industrial scale for recycling Ge. The company built a smaller unit
to reprocess the \Ltwo\ kerf.
This technology can be utilized near the detector manufacturing stages to maximize usable Ge and limit surface transportation of the material.

The \MJD\ demonstrated a yield of 70\% for the total conversion of input Ge to detector mass~\cite{Abgrall:2017acl}.
With larger amounts of input material and improvements in the process, we expect to achieve 75\% for \Ltwo.
If further improvements such as the recovery of etchant losses can be established, a yield above 80\% should
be possible.

%% file: sec_technical/subsec_detectors.tex
\subsubsection{Ge Detector Production}
\label{subsec:detector-fabrication}

The detector fabrication process starts when \LEG\ supplies enriched material to the detector fabrication vendor as electronic-grade material ($\textrm{R} \geq 47$~$\Omega$\,cm).
Prior to crystal growth, the vendor does a further zone refinement to detector-grade purity.
Crystals are then grown using the Czochralski process.
For \Ltwo\ detector fabrication, vendors have assigned at least one dedicated clean crystal puller while processing an entire batch of \LEG\ material
in order to prevent contamination of enriched Ge by natural material.
An analysis of the longitudinal net impurity concentration (IC) profile of the crystal ingots is used to decide where to cut it into slices for further conversion into working detectors.
\LEG\ collaborators are involved at this stage to ensure that the slices meet the impurity gradient specification, estimated case by case by means of electrostatic simulations using ADL3~\cite{Bruyneel:2016zih} and \Siggen~\footnote{Software for the field and signal generation in Ge detectors. \url{https://github.com/radforddc/icpc_siggen}} software.
Acceptable impurity values for the IC lie in the range of \cpowten{(0.5-2.5)}{10}\,$\textrm{cm}^{-3}$.
The conversion of the enriched Ge crystal ingots into an \icpc\ detector relies on four key steps:
\begin{myenumerate}
	\item Machining of the groove on the higher impurity concentration crystal side
	\item Machining of a well on the opposite side
	\item Diffusion of Li atoms into the outer crystal surface to produce the \nPlus\ contact
	\item Implanting B within the region enclosed by the groove for the \pPlus\ contact
\end{myenumerate}

\begin{figure}
\includegraphics[height=.22\textheight]{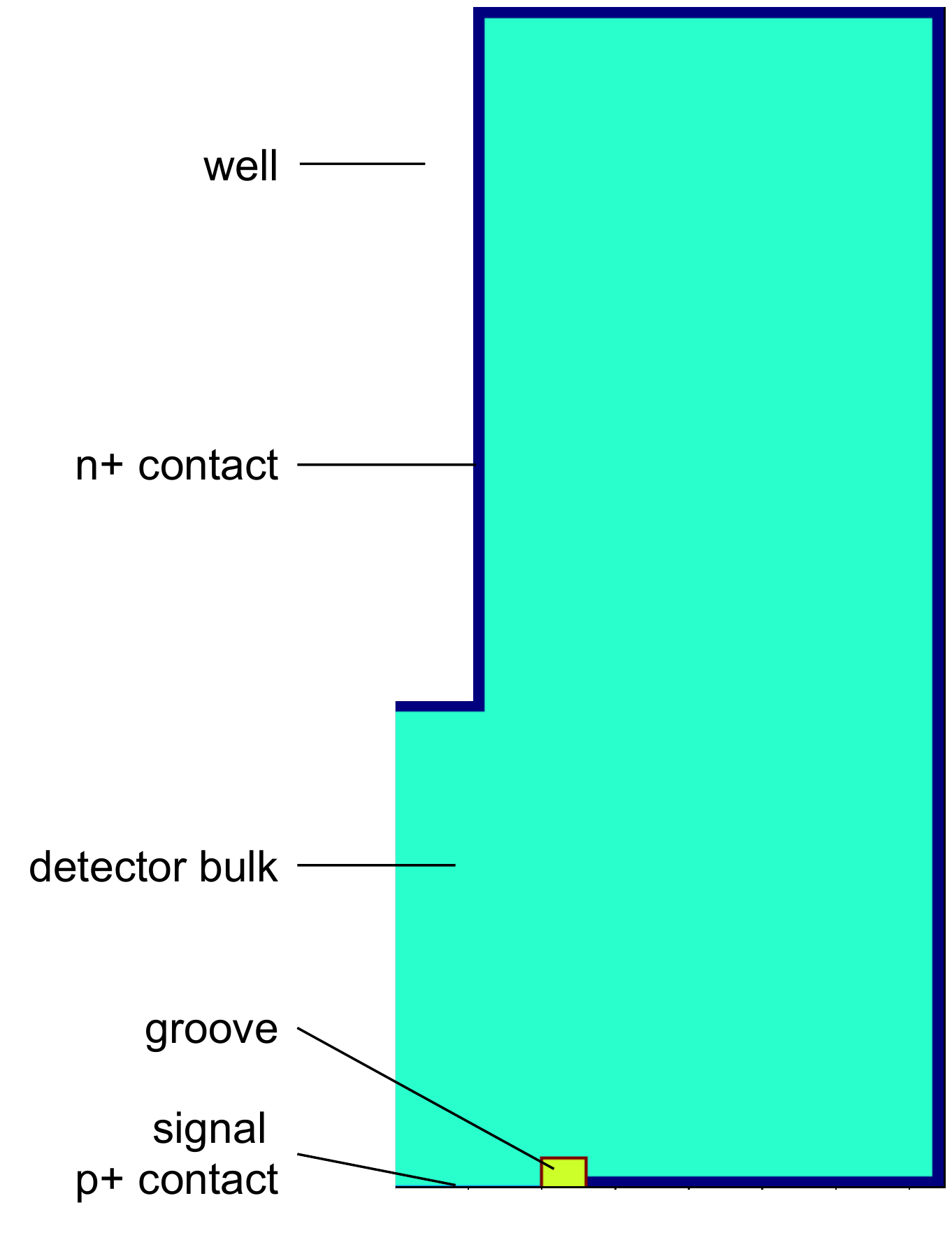}\hspace{2em}
\includegraphics[height=.22\textheight]{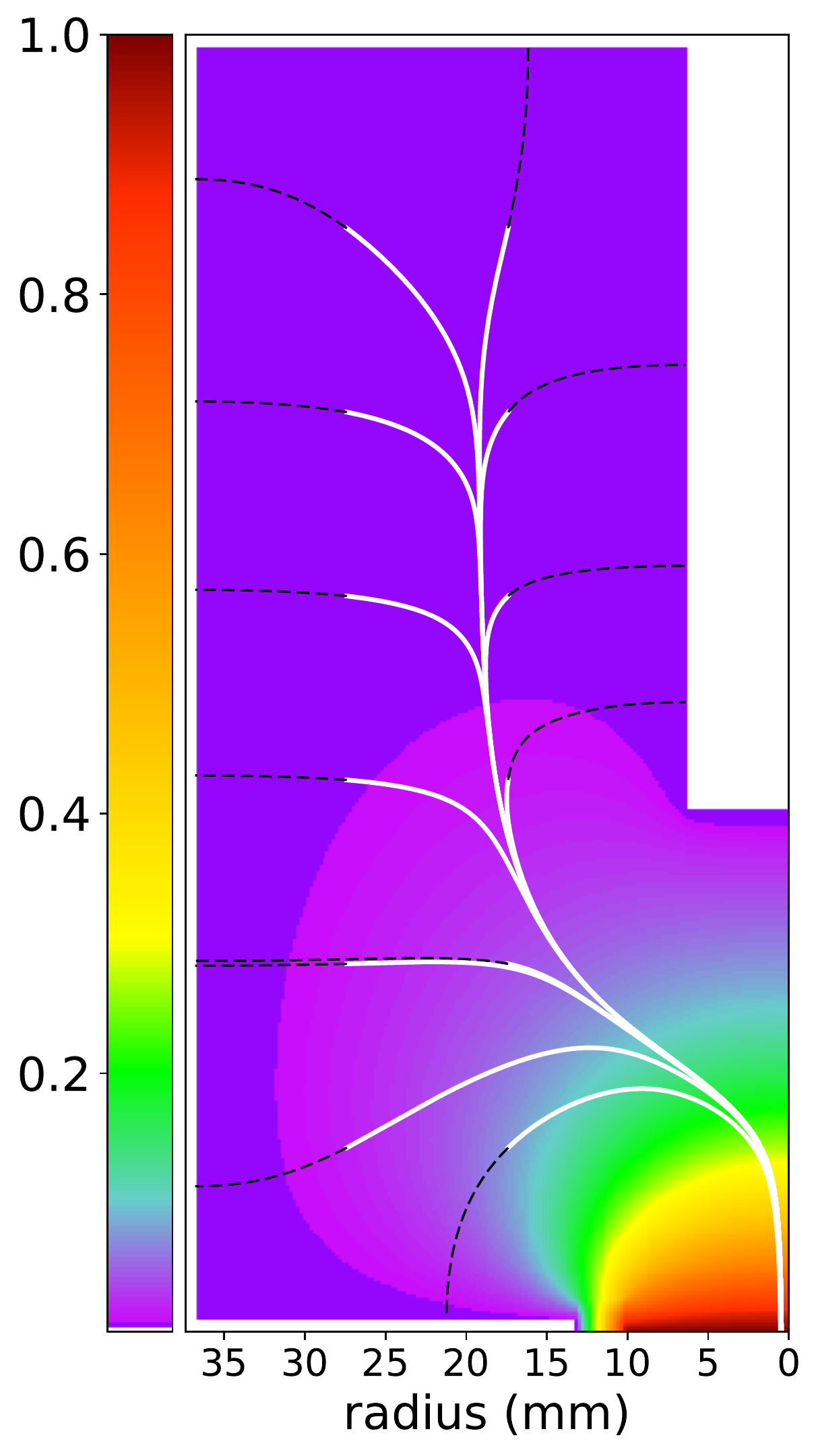}~
\includegraphics[height=.22\textheight]{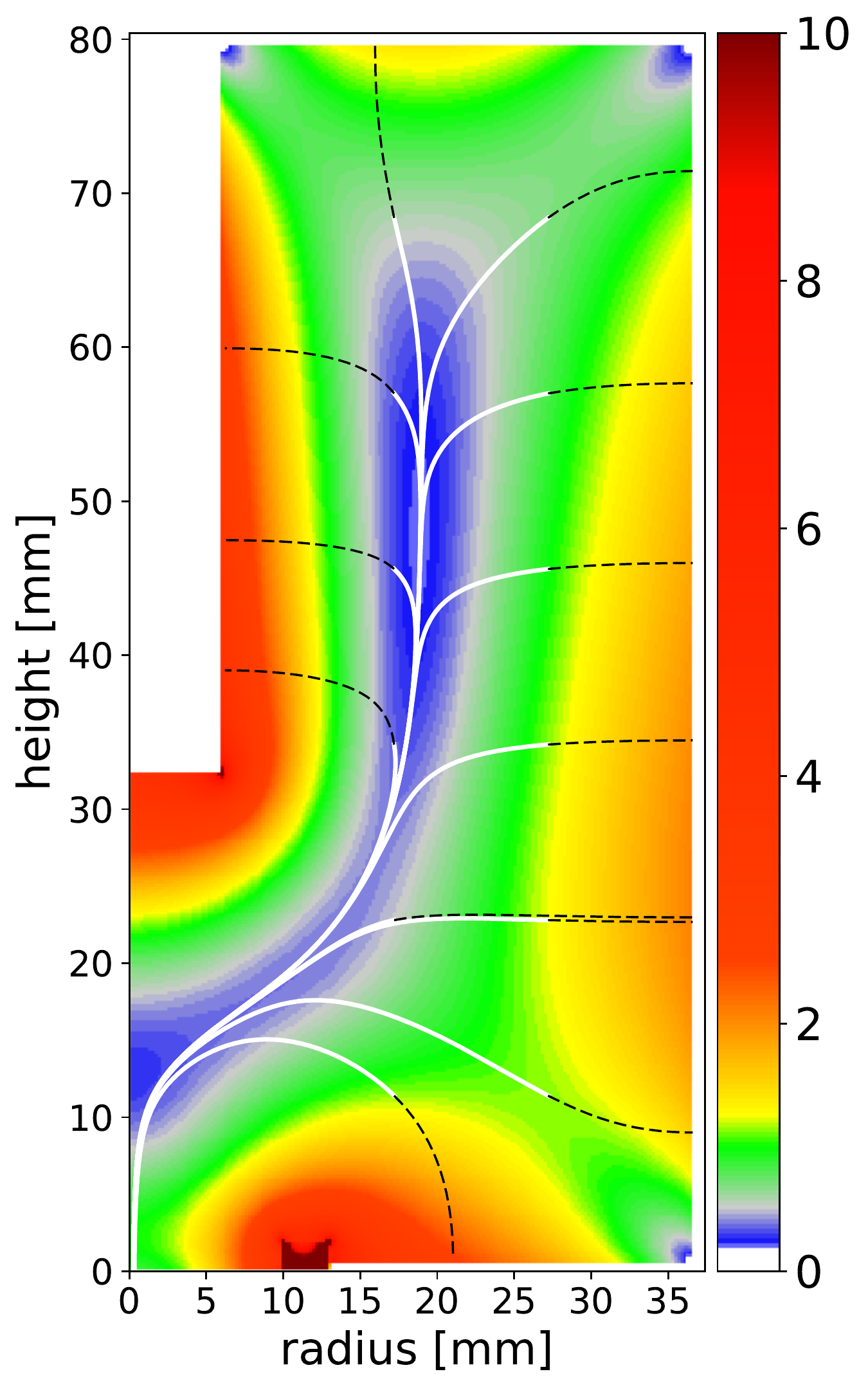}
\includegraphics[height=.22\textheight]{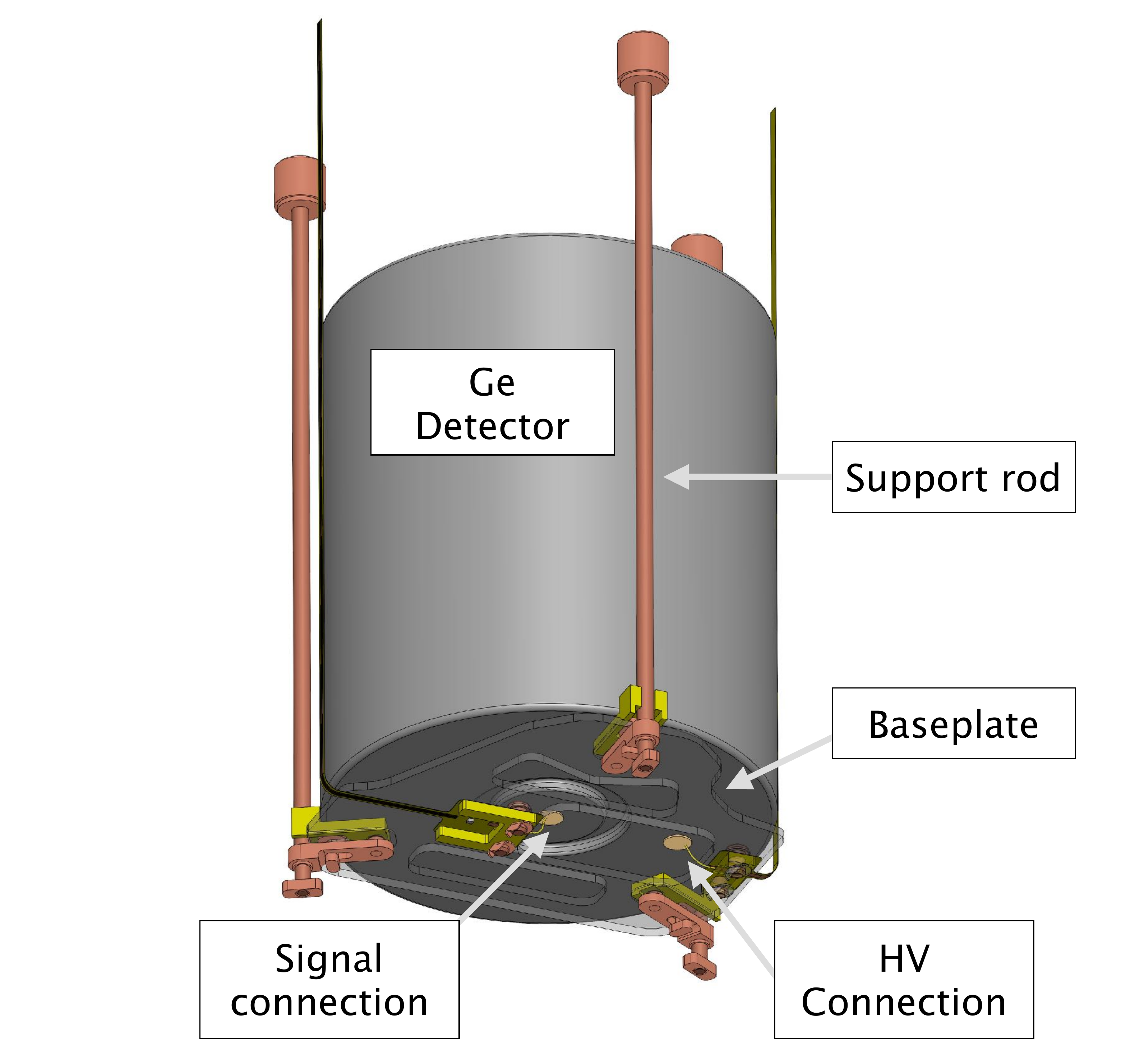}
\caption{Left: The main \icpc\ detector features. Middle-left: An ADL calculation of the weighting potential. Middle-right: The electric-field strength in kV/cm where the minimum allowed electric field is 200~V/cm (dark blue), and the black dashed lines show electron drift paths ending at the \nPlus\ contact while white solid lines are the hole drift paths reaching the signal contact from Ref.~\cite{Agostini:2021wzn}). Right: A conceptual drawing of a \Lk\ detector unit where the baseplate supports the high voltage (outer) and point-contact front-end (on center) connections and cabling.}
	\label{fig:ICPC_sketch}
\end{figure}

The specification for the \nPlus\ contact thickness---determined by the temperature and duration of the lithium diffusion process---results from a compromise between the loss of detector active volume and an efficient absorption of $\beta$ particles on the outer surface.
The accepted baseline thicknesses are within the range of 0.8~mm to 1.0~mm, but R\&D is ongoing to optimize this value for additional background reduction.
Other specifications are mainly driven by electrostatic simulations that take into account the measured impurity concentration.
The detector geometry is optimized by varying many parameters---overall dimensions, groove diameter and width, well diameter and depth, and point-contact size---to find values that maximize the detector mass while maintaining a minimal electric field in the detector bulk above 200~V/cm and a predicted depletion voltage below 4000~V.
These two parameters are of particular importance because of their influence on charge-collection efficiency and leakage current, respectively. Both contribute to the event-topology discrimination performance and energy resolution.

Mirion currently grows Ge crystals in Oak Ridge, TN, USA, and converts them into working diodes in Olen, Belgium.
They also have a detector fabrication facility in Meriden, CT, USA.
ORTEC processes Ge material and fabricates detectors entirely in Oak Ridge.
In all cases, care is taken to minimize the time above ground by making use of underground storage facilities and shielded shipping containers~\cite{Agostini:2014hra,Abgrall:2017acl}.

\paragraph{Detector Fabrication Baseline Design}
 The baseline design for the \Lk\ detector fabrication builds on the experience of \Ltwo.
 The realized mass increase of the \Ltwo\ \icpc\ detectors, which average 2.0~kg to date compared to the point-contact detectors from \Gerda\ (0.6~kg average)
 and the \MJD\ (0.9\,kg avg.), was a critical milestone in moving forward with \LEG.
 The \Lk\ baseline assumes a modestly larger average mass of 2.6~kg based on demonstrated
\icpc\ detectors with masses over 3~kg produced by the vendors.
 Larger detectors reduce the amount of surrounding material (e.g., front-end electronics and cables) and the surface-to-volume ratio.
 Although we continue to explore the fabrication of even larger detectors, they are not required for the \Lthou\ baseline design.
Critical aspects for \Lk\ detector fabrication are the production rate, material losses, and the cleanliness of the process.

\begin{figure}
\centering
\includegraphics[height=0.45\columnwidth]{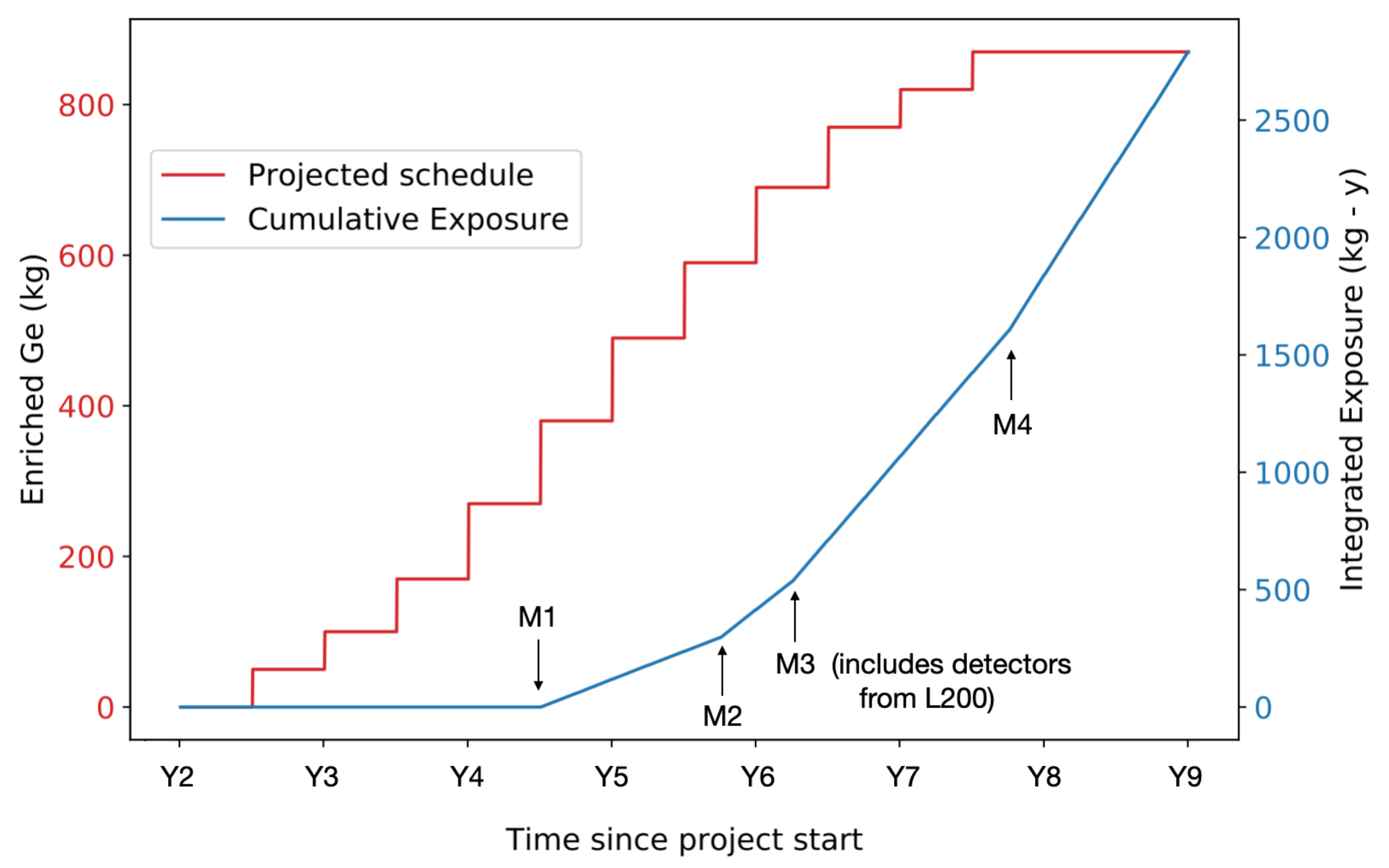}
\caption{\label{fig:l1000_det_sch} New detector production estimates (red, left axis) versus time since project start (Year 0) under an assumption that the rate gradually
increases from the start of production and decreases towards the end, as final detectors are fabricated from recycled material.
An overlay of the integrated \Lk\ exposure (blue, right axis) shows that data from the installed modules is collected as the detector production continues.
The projected start dates of the four modules are annotated in the figure where the exposure rate changes.
The \icpc\ detectors from \Ltwo\ are reused and installed into Module 3.
The integrated exposure accounts for array live time and the total detector mass.}
\end{figure}

\paragraph{Production Rate}
Based on \Ltwo\ detector deliveries and including an assumption that detector production is not limited by the availability of enriched \gesix\ material,
the estimated annual production rate for \Ltwo\ would be about 30 detectors per year, corresponding to 70~kg of detectors.
This projected throughput rate is constrained by the current vendor fabrication resources that were dedicated to working with the enriched material.
The \Lk\ estimates for detector production are based on input from vendors which include plans to have
dedicated production lines over much of the fabrication period.
Figure~\ref{fig:l1000_det_sch} shows the estimates for new detector production versus time
assuming an initial conservative production rate of 50 detectors per year that goes up to 110 detectors per year as fabrication becomes more routine.
This scenario also takes into account that the production rate decreases towards the end of the production cycle, accounting for delays associated with the
use of material that have been recycled from previous fabrication cycles.
At these projected rates, we estimate that it takes six years to produce all of the \Lk\ detectors.
With \Lk's staged approach, the detectors needed for the first module are ready in just over two years from the start of detector production.
The detector production rate is matched to the production of enriched isotope (see Sect.~\ref{subsec:ge76-procurement}).

\paragraph{Material Losses}  During the detector fabrication process one incurs material losses during the zone refining, crystal pulling,
crystal cutting and shaping, and final fabrication of the detectors.  However, as shown by \MJ,  by following careful materials management
it is possible to recover and recycle much of the material that would potentially be lost during etching, cutting, and grinding processes.
As discussed in detail in Sect.~\ref{subsec:ge76-procurement}, we estimate an overall yield of 70--80\% of detector mass per purchased enriched Ge.

\paragraph{Cleanliness of the Process} The process of zone-refinement and crystal pulling results in detectors which are intrinsically clean.
The current vendor standards are already sufficient in terms of bulk contaminants from the \nuc{238}{U} and \nuc{232}Th chain.
However, improved control of the air quality (low Rn concentration) and higher cleanliness requirements for detector handling (currently not under control of the collaboration) at the manufacturer sites could potentially reduce surface-associated $\alpha$ backgrounds.  For \Lk\ vendors may operate a dedicated fabrication line for the duration of the detector-production period.
Therefore, it is feasible for the \LEG\ collaboration to work with the vendors to institute improved cleanliness measures, further reducing the surface-associated $\alpha$ backgrounds.
However, the background-index projection described in Sect.~\ref{subsubsec:alpha} does not assume such an improvement.

\paragraph{Detector Characterization}\label{subsec:detector-char}

The detector fabrication follows a set of specifications on the mass, the maximum depletion voltage, the minimal electric field, the energy resolution, and the \nPlus-contact dead layer thickness of the detector.
In addition to these parameters, we are interested in the background rejection capabilities and the homogeneity of the detector response over its entire volume. For reference to recent characterization measurements made on \bege\ and \icpc\ detectors deployed in \Gerda, see Refs.~\cite{Agostini:2019mwn,Agostini:2021wzn}.

The working detectors received from the manufacturers undergo a set of specific characterization measurements that qualify and quantify the aforementioned parameters.
The comprehensive measurements are performed in a vendor-supplied vacuum cryostat with the best-achievable electronics noise and with high statistics, which cannot be reproduced once the detectors are installed in the final \Lk\ apparatus. The full length of a characterization campaign is typically up to a week per detector (using multiple setups in parallel), which makes use of available underground space near the manufacturer's facility in order to limit cosmic-ray activation of the detectors.

\new{The Ge detector's energy resolution at \Qbb\ is deduced through an interpolation of the measured $\gamma$ lines to a well-understood resolution curve as a function of energy.
The Ge detector's active volume is determined by measuring the amount of inactive material found at the detector surface.
The entire surface, except the groove and the \pPlus contact, undergo Li diffusion that makes the \nPlus\ contact and prevents $\alpha$ and most $\beta$ particles from penetrating into the active detector volume and depositing energy.
A precise estimate of its thickness can be achieved by comparing measurements of low-energy $\gamma$-ray absorption in the dead layer with Monte Carlo simulations, typically the $\gamma$-ray lines from $^{241}$Am (59\,keV and 99\,keV / 101\,keV) and $^{133}$Ba (79\,keV / 81\,keV and 356\,keV) to suppresses uncertainties due to the source activity. A precision on the dead layer thickness of better than 0.1\,mm has been achieved, leading to a systematic uncertainty on the active volume of a few percent.}

\new{The PSD performance of the detector is investigated in a test stand by means of a $^{228}$Th source---similar to the method by which it is measured in situ for \Lk. The source provides both single-site and multi-site events within a detector for which the PSD rejection and acceptance parameters can be determined.
The additional benefits of test-stand measurements are that significant statistics are collected within a few hours, instead of over a longer time through integrated \Lk\ calibration runs; the source position is changed from top to side in order to have a complete picture of the PSD as a function of position within the detector. In the case of a significant under-performance, the detector may be reprocessed and improved at the manufacturer site within a reasonable time frame.}

\paragraph{Detector Mounting}\label{subsec:detector-mounting}

Detector mounts provide physical support and electrical isolation for each detector. Additional requirements are that the design must accommodate the front-end electronics mounting, cable routing, and termination.
\new{The detector mounts provide a level of protection from damage during handling and installation. The mounts are made from ultra-pure materials to minimize their contribution to the background and have low volume in order to minimize any shadowing effects on the light collection for the active veto system.}
\Ltwo\ will serve as an early testbed for new designs. In addition, the 14-string array in \Lk\ allows some flexibility in detector mount design, and it is possible to deploy several design alternatives and iterations simultaneously.
Each detector has unique dimensions in order to maximize single pass yield from each crystal boule.

 The baseline \Lk\ detector holder design is an evolution of the \Gerda\ and \Ltwo\ detector mount design. Figure~\ref{fig:ICPC_sketch} shows the conceptual design of a \emph{detector unit} integrating the electrical and supporting connections to an individual Ge detector. The vertical support rods are machined from EFCu with laser welded features to provide a captured coupling nut at one end for mechanical attachment to an adjacent detector unit.
 Small blocks of \ultem\ provide electrical isolation from charged electrical surfaces as needed, primarily the outer diameter of the detectors that are at full bias voltage of up to 5\,kV.
 The baseplate is made out of polyethylene naphthalate (PEN), which can be produced sufficiently radiopure
 and has excellent mechanical properties. PEN is a scintillating material,
 described in more detail in Sect.~\ref{sec:materials}, which provides
 active shielding from nearby background sources (e.g., the detector's cables and front-end boards). Detector mounting parts produced through an additive machining process (also described in Sect.~\ref{sec:materials}) would add an on-demand production capability as well.

\begin{figure}
     \begin{center}
       \includegraphics[height=0.3\tw]{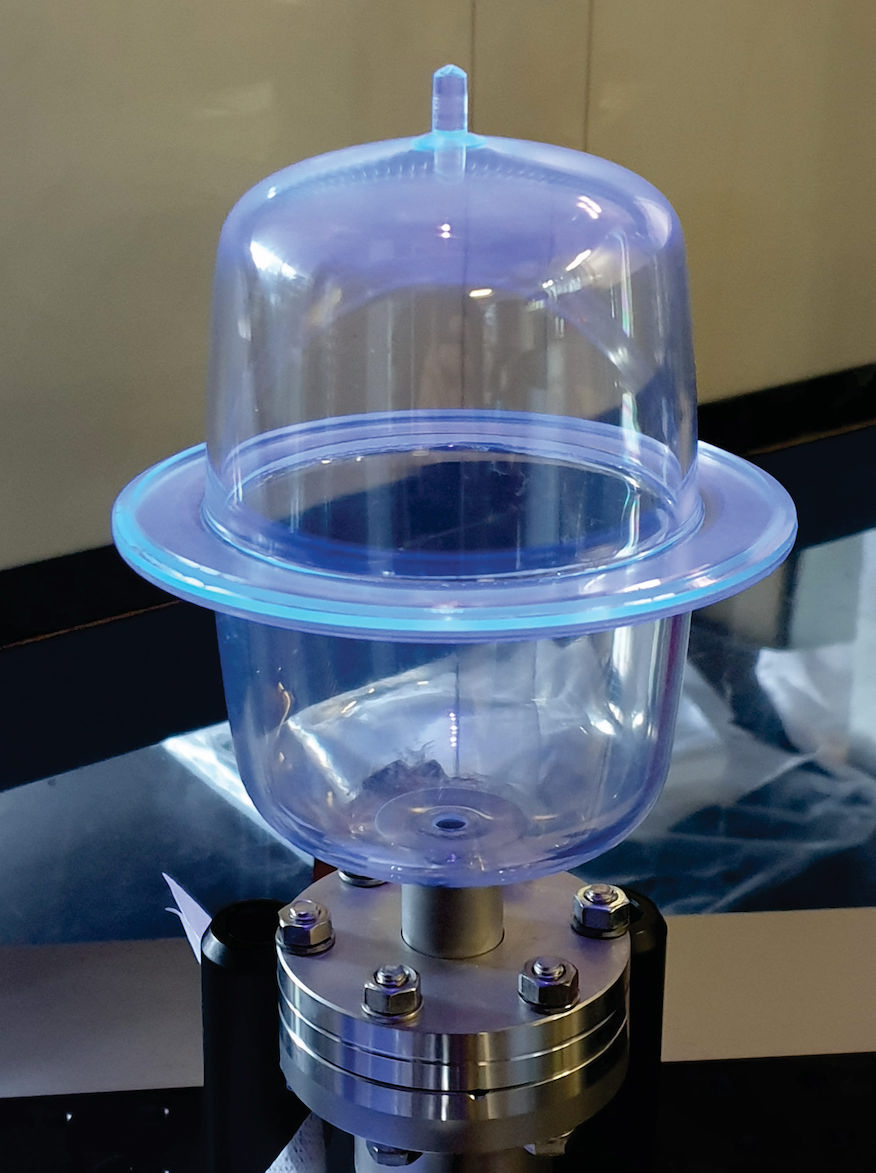}~
       ~
       \includegraphics[height=0.3\tw]{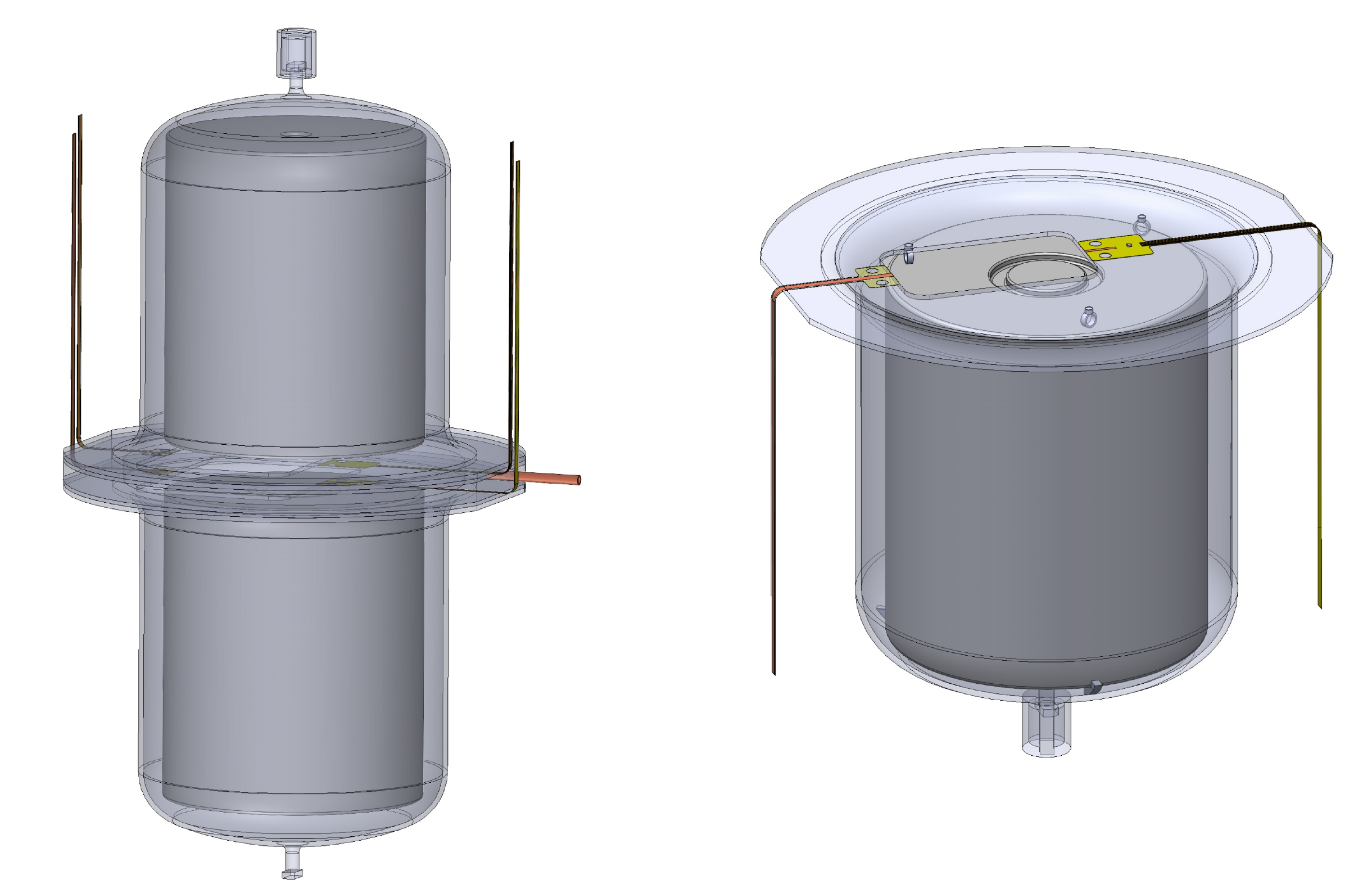}
       \caption{\new{An alternative design for the detector mounting concept. A prototype PEN capsule (left) assembled out of two capsules.
       A conceptual drawing showing how such a capsule could be used to form an  encapsulated detector unit (middle) with the same high voltage and point-contact front-end connections and cabling (right) used in the baseline design.}
       }
       \label{PEN:fig:capsule}
     \end{center}
 \end{figure}

 \new{\LEG\ is exploring alternative strategies to replace additional passive materials with active materials. The additional use of scintillating support structures would further reduce backgrounds by improving the geometric efficiency of the existing LAr veto and reduce the non-active mass.
 One option would be to replace the Cu vertical support rods in the detector mounts with similar support structures made from PEN to be integrated within the baseplate.
 It is also envisioned that an entirely new kind of detector mount could be made
 entirely out of PEN, which would surround a number of detectors and provide a closed
 volume to shield them from the environment.
 The mount would be a modular capsule with internal support structures with laminated Cu traces as cabling and
 serve as part of the active veto system.
 This detector encapsulation would be highly beneficial for a detector array
 directly submerged in LAr as it would limit the effect of the \nuc{42}{Ar} contamination. The produced \nuc{42}{K} ions would be unable to reach the detector surfaces, and any
 electrons reaching the surface of a PEN capsule would be absorbed, causing the event to be vetoed by the
 PEN scintillation response.
 Figure~\ref{PEN:fig:capsule} depicts a test capsule with a wall
 thickness of approximately 2\,mm assembled out of two
 capsules. As a proof of principle, such a capsule, filled with
 \nt\ gas, has been successfully tested in \lnn\ and LAr.
 Further testing with internal support structures, feed-throughs, and a
 Ge detector are underway.}

\paragraph{Detector R\&D}
\label{subsec:detector-rAndD}

\begin{figure}[htbp] \centering
  \includegraphics[width=0.55\tw]{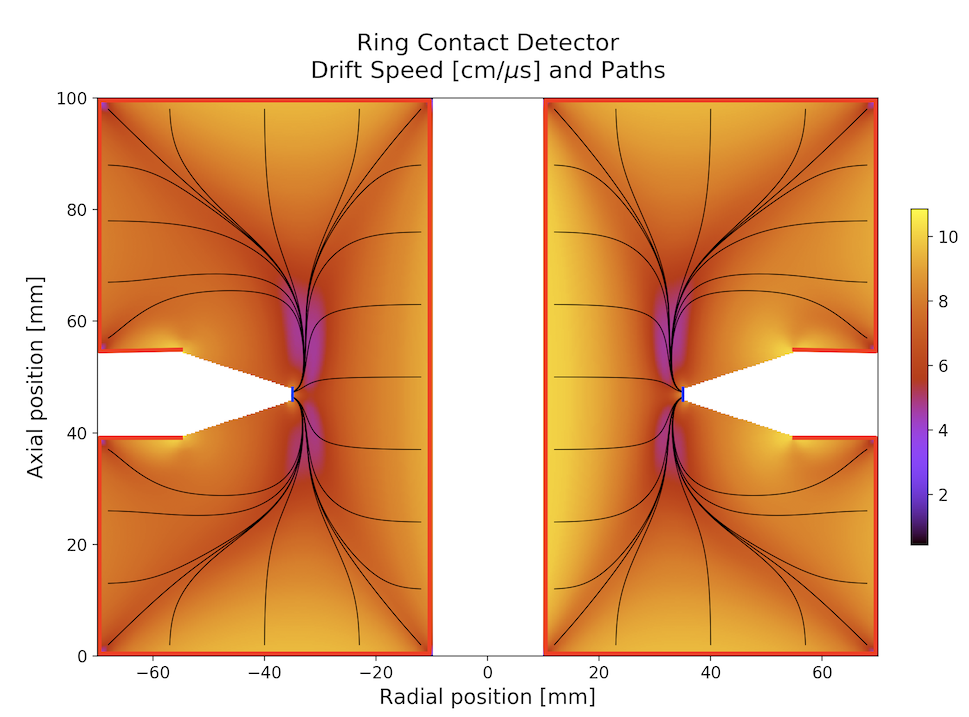}
  \caption{\new{A simulated electric field distribution of the Ge ring-contact detector design.}}
  \label{fig:f2}
\end{figure}

\new{The collaboration continues an R\&D program to further investigate and optimize the \icpc\ detector design for use in \Lk\ by characterizing available natural detectors and by testing larger prototype detectors of up to 6.0\,kg mass.
The effort includes detailed simulations of the depletion, field, and charge collection characteristics, and collaborative work with Mirion, ORTEC, and PHDS Co.\footnote{PHDS Co, Knoxville, TN, USA.} to produce and characterize large-scale prototype detectors.
Due to the long drift distance for charges produced in such large detectors and the relatively weak electric field, some charge trapping can be expected. Special digital-signal processing and analysis will be required to correct for this and to maintain the best energy resolution.
Alternative geometries to ICPC detectors are also being explored.
The collaboration is studying what is termed a Ge ring-contact detector.  The design detector has a true coaxial center hole, where the point contact is replaced by an outer contact ring that is centered near the mid-point of the detector (see Fig.~\ref{fig:f2}). Such detectors, if shown to be feasible, may allow for larger crystals than the ICPC design.}

\paragraph{Crystal R\&D}
\label{subsec:crystal-rAndD}

\new{While the collaboration will continue to partner with commercial vendors on R\&D studies,
several \LEG\ collaboration institutions also have the ability to grow crystals and fabricate detectors.
The `Institut f\"ur Kristallz\"uchtung' (IKZ), translated as Institute for Crystal Growth, is working to provide detector-grade high-purity Ge single crystals for detector fabrication to broaden the options in the supply chain.
Earlier crystal growth work for the \Gerda\ collaboration was supported by the experienced IKZ materials science team.
With extensive expertise in Ge crystal growth, IKZ is routinely growing high-purity Ge single crystals of 50\,mm in diameter, 15--20\,cm long, and 1.5\,kg in mass. It can now grow larger crystals of 75--80 mm diameter using a newly scaled-up facility.
For its participation in \LEG, IKZ has been enhancing its technique for Ge processing---including Ge reduction and purification---to establish a complete production process that starts with the reduction of the GeO$_2$ materials through the Ge crystal growth for detector fabrication.
The IKZ team has successfully established a high-yield
process of reducing both natural and isotopically enriched GeO$_2$ powder into Ge metal and zone-refining it to
semiconductor-grade Ge bars with an overall process yield of 99\% (94\% with 50~$\Omega$\,cm).
The University of South Dakota (USD) is also growing crystals and fabricating detectors as a member of the Partnerships for International Research and Education (PIRE) GErmanium Materials And Detectors Advancement Research Consortium (PIRE-GEMADARC). The global consortium was created to accelerate the Ge material platform used in R\&D for new generation dark matter and \BBz\ decay experiments while educating the next generation of scientists. PIRE-GEMADARC members USD and Texas A\&M provide in-house capabilities to grow crystals, develop detectors, and study detector performance.  Utilizing the in-house crystal growth ability, small Ge detectors can be fabricated on a weekly basis to study bulk and surface properties that are inconvenient or costly to study with large commercial detectors, such as contact technologies, passivation techniques, charge carrier mobility and trapping constants, or other properties. The group plans to fabricate a ring-contact detector (see Fig.~\ref{fig:f2}) from a USD-grown crystal with a mass of 2--3\,kg to characterize its energy resolution and pulse shape properties and determine its potential for use in \Lk.}

%% file: sec_technical/subsec_cryostat.tex
\subsubsection{Cryostat and Water Tank}
\label{subsec:cryogenics}

The \Lk\ cryostat and water tank infrastructure support the Ge-detector array and the means to provide shielding from backgrounds. \Lk\ backgrounds from the underground environment and laboratory infrastructure are very small components of the overall background budget.
Multiple modules of Ge detector arrays
allow commissioning of the detector system in stages, with operation of installed detectors remains undisturbed when additional modules are installed. The Ge is divided into four modules of about 250~kg each.

\begin{figure}
  \hfill
  \includegraphics[height=0.3\columnwidth]{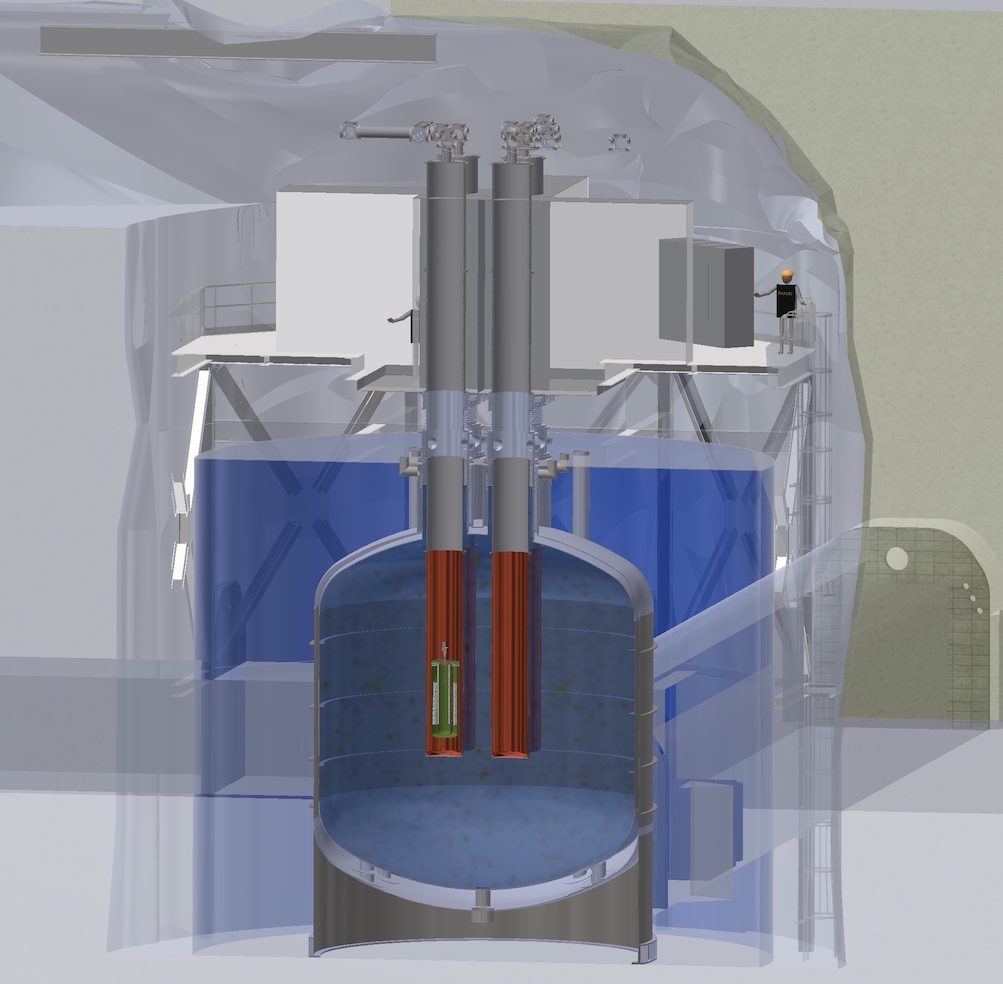}
  \hfill
  \includegraphics[height=0.35\columnwidth]{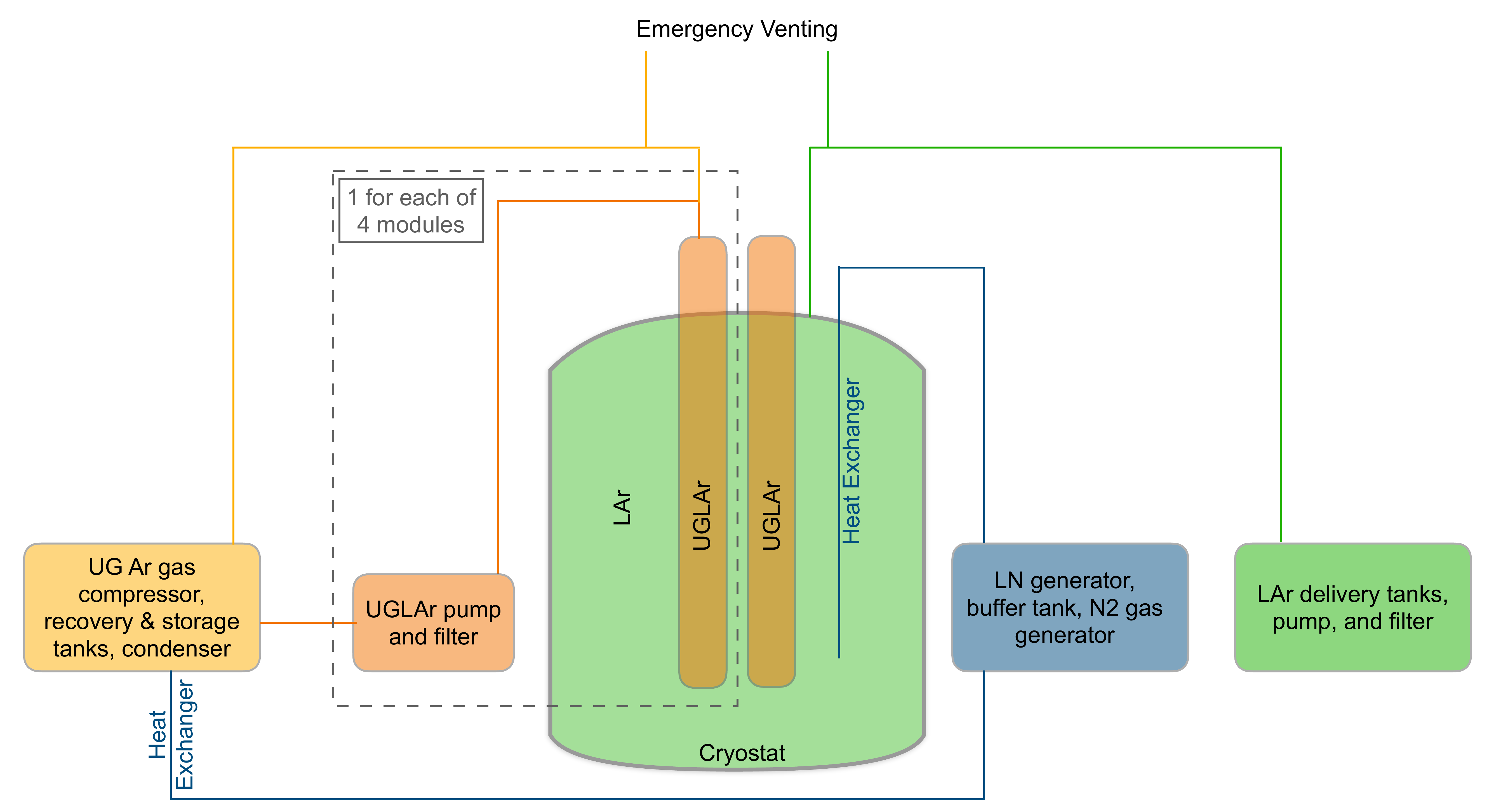}
\caption{
Left: Conceptual design of the of the \Lk\ baseline vacuum-insulated cryostat depicted in the SNOLAB Cryopit.
Right: A high-level illustration of the LAr cryostat auxiliary support systems and their notional interfaces.}
\label{fig::cryostat-vac}
\end{figure}

The reference design for \Lk\ is to
operate the Ge detector array in LAr within a single vacuum-superinsulated cryostat appropriately sized to fit, together with the enveloping water tank, in the Cryopit cavity at SNOLAB.
The Ge detectors comprise four modules within reentrant tubes containing UGLAr, which are submerged in a \lar\ cryostat and shielded from
site-related radioactivity by an external water tank.
The conceptual design of the \lar\ cryostat and
UGLAr-filled modules is shown in Fig.~\ref{fig::cryostat-vac}, along with its representation in the SNOLAB Cryopit.
As an alternative, we also consider this general cryostat design deployed at LNGS described in
Sect.~\ref{sec:altsite}.

The walls of the cryostat are made from welded stainless steel components
similar to 316L that are pre-screened for radioactivity. Due to the required mass for this cryostat, an entire ``heat" of 120\,t can be reserved and assayed, increasing the level of control the supplier has with input materials and consistency throughout the lot.
The walls contain multilayer insulation material within the vacuum gap.
The inner cryostat wall is 7\,m in diameter, and the outer diameter, without stiffening rings, extends to 7.2\,m. The cryostat holds 240\,m$^3$ or 332\,t of cryogenic LAr.
The cryostat has a 4-m high inner cylindrical portion height with 1.5-m tall domed shells on the top and bottom.
The reentrant tubes have a 950\,mm inner diameter and are
are arranged on a 2-m-diameter circle. The lower portion of the tubes that extend into the cryostat comprise multiple cylindrical rings grown from underground electroformed Cu (see Sect.~\ref{sec:materials}), rolled to a final thickness of 1.0\,mm, and electron-beam welded to form the full cylindrical height. Each tube holds 3.3\,m$^3$ or 4.6\,t of cryogenic UGLAr.
Each module's Ge detector array, of total mass 250\,kg,
 is inserted from the top into each of the reentrant tubes. A lock and shutter system provides a controlled interface between the reentrant tubes extending into the cryostat and the installation glovebox where the Ge-detector array is assembled. The lock systems also provide accommodations to facilitate the slack length of cabling before lowering or after raising the detector array. The design of the cryostat lock is based on the experience from \Gerda\ and \Ltwo. Openings accommodate a LAr purification loop, services for LAr filling, Ar exhaust gas, and \lnn\
 cooling,
as well as safety devices, cabling for LAr instrumentation,
 or other sensors.

 The physical dimensions and configuration of the cryostat are modeled through Monte Carlo simulations to ensure the backgrounds from the cryostat materials and outer environment are each at acceptable levels of 5--10\% of the experiment's background goal (see Sect.~\ref{subsubsec:external}).
Since the detectors in the array are arranged in rings offset from the cryostat center, the distance from a detector to the cylindrical cryostat wall varies between 2.1~m and 4.9~m. This thickness improves upon the $\gamma$-ray shielding achieved for the \Gerda\ cryostat \cite{Barabanov:2009zz}.
Therefore, a 7-m-diameter cryostat is expected to provide the required shielding, based on an assumed activity of 1~mBq/kg of $^{232}$Th for the stainless steel, compared to the 0.4~mBq/kg upper limit for the steel used in \Gerda, and conservative assumptions of the effectiveness of analysis cuts. The cryostat design allows the mounting copper plates on the inside of the inner wall---as implemented in the \Gerda\ cryostat---for additional shielding in case the radioactivity of the steel is higher than expected.

The cryostat is immersed in a water tank (visible in Fig.~\ref{fig::cryostat-vac}).
The combined shielding of LAr and water reduces the $\gamma$ ray and neutron (low-energy ($\alpha$,n) and fission reactions) backgrounds  from the laboratory environment to an acceptable level with a 2-m-thick water shield. Therefore, a water tank with an overall diameter of 12\,m is sufficient.
The water tank is instrumented with photon detectors along the walls of the tank to veto muons through the detection of Cherenkov light in water, as was done in \Gerda\ and planned for \Ltwo.
However, since the full LAr volume is separately instrumented (see Sect.~\ref{subsec:lar-veto}), muons can be sufficiently identified by the LAr
scintillation light. At the baseline SNOLAB depth, a water-based muon-veto instrumentation system is not a requirement to meet our background goal.

The cryostat design follows upon the experience of \Gerda, which operated its cryostat in a water
tank at LNGS for over a decade.
Deploying the cryostat at SNOLAB does introduce fabrication and installation constraints, which are factored into the infrastructure requirements, quality assurance and testing plans, and the overall schedule.
Safety is a prime concern for such a large cryostat, and
design studies are considering normal and accidental load cases. The cryostat must be designed according to a pressure vessel code to meet operational standards and ensure quality control of fabrication. The design must ensure that the cryostat can withstand not only normal operating conditions but also the conditions during leak checking and pressure tests.
Accident scenarios, including vacuum leaks that introduce air, water, or Ar and compromise the cryogenic insulation, are considered to inform the overall design and necessary precautions. The failure of the cryostat vessel within a large volume of water is a concern.
If the thermal insulation is compromised, the rate of Ar evaporation has to be limited to a tolerable level. Mitigation strategies may include using baffling to segregate the innermost volume of water, which can be drained quickly, or freezing the water entirely to minimize its heat load.
Additionally, dynamic loading from seismic events or mining activities must be considered to ensure adequate support of the cryostat walls. The inner vessel support tubes that exist between the cryostat walls must withstand the stress of any acceleration, in addition to supporting the full weight of the LAr and any supplemental shielding materials, while exhibiting low thermal conductivity and low radioactivity.

\paragraph{LAr Cryostat Auxiliary Systems}\label{subsec:lar-pruification}

Several auxiliary systems support the commissioning and operation of the LAr vacuum-insulated cryostat.
Figure~\ref{fig::cryostat-vac}
illustrates the major subsystems and their notional connections.
The UGLAr begins as a high-pressure gas in storage tanks prior to being condensed and fed to the UGLAr filtration system. A separate filtration system services the UGLAr contained within each of the reentrant tubes, both to purify the UGLAr upon initial filling and for re-purification during operation. Should the UGLAr need to be evacuated from the reentrant tubes, the evaporated gas is compressed and returned to the high-pressure storage tanks for later re-introduction to the cryostat. The handling of the atmospheric LAr begins with delivery in custom 2\,t transport skids prior to being fed to a filtration system during the filling of the cryostat. Should the atmospheric LAr need to be evacuated out of the cryostat, it is evaporated and relieved through a laboratory exhaust vent to the surface. A liquid nitrogen (\lnn) generating system and buffer tank provide a closed-loop heat exchanger to provide the cooling power for the LAr cryostat and the UGLAr condenser.  The \lnn\ system also generates N$_2$ gas for purging glove boxes and dry storage cabinets.

While most of the auxiliary systems support the cryogenic needs of the cryostat, the filtration systems are integral to the low-background operation of the experiment. The LAr active shield system (described separately in Sect.~\ref{subsec:lar-veto}) relies on the fact that ionizing radiation in LAr scintillates through decays of the excited singlet and triplet states. The lifetime of the triplet state reaches 1.6\,$\mu$s, which is significantly longer than the 6\,ns lifetime of the singlet state. The population ratio of these two states depends on the incident particle and enables the possibility of its identification, which is a key feature in LAr-based detectors searching for direct dark matter interactions. The presence of electronegative impurities in LAr can, however, lead to a strong suppression of the formation of excitons by capturing the free electrons needed for recombination. The light yield can be reduced by a depopulation of the excimer states caused by collisional destruction with impurities
like water, nitrogen, or oxygen. Taking into account different lifetimes of the two states, it is clear that this effect primarily influences the triplet state, where a longer lifetime leads to an increased probability to undergo a non-radiative decay. Measurements have shown that a small contamination (on the order of 1\,ppm of oxygen/water and/or 10\,ppm of nitrogen) leads to a strong reduction (50\%) of the light yield and a strong reduction of the triplet lifetime~\cite{Acciarri:2009xj, Acciarri:2008kx}. In \Gerda\ \ptwo\ the observed triplet lifetime did not change significantly from a value of 0.9\,$\mu$s\ during the course of the experiment. The attenuation length measured before the start of \ptwo\ was about 15\,cm, thus significantly lower than the theoretical Rayleigh scattering length of approximately 90\,cm. Both parameters indicate some airborne contamination. Either the \lar\ filled into the \Gerda\ cryostat was initially contaminated or the impurities were introduced during operation with the Ge detectors (the large volume of \lar\ prevents significant changes due to the dilution effect).

The \Lk\ cryostat has four reentrant tubes filled with UGLAr---the volume of each is about 3.3\,m$^3$---which must maintain sufficient purity to preserve its scintillation properties. After initial filling, the UGLAr slowly degenerates over time (i.e., due to internal degassing or operations on the detectors) as compared to the larger volume of LAr (65 m$^3$) in the \Gerda\ cryostat. A circulation loop  provides real-time filtration of the UGLAr in the liquid phase
to achieve purity at the sub-ppm level.
All filtration traps are constructed so that regeneration may be performed in situ, meaning that the traps are regenerated without removing them from the purification system (equipped with appropriate heaters, temperature sensors, and piping). The purification columns and the piping are vacuum insulated in order to minimize the heat load. The circulation speed is tuned to the needs (UGLAr degradation rate) or run periodically when the triplet lifetime reduces to an unacceptable level (below 1~$\mu$s).

%% file: sec_technical/subsec_uglar.tex
\subsubsection{Underground Liquid Argon}\label{subsec:uglar}
Beta decays from the isotope \nuc{42}{K}, progeny of \nuc{42}{Ar}, is a potential \Lk\ background intrinsic to the LAr shield.
The DarkSide collaboration has established that Ar sourced from a deep underground CO$_2$ well (underground argon, UGAr) has a reduced \nuc{39}{Ar} content 1400 times less than Ar sourced from the atmosphere~\cite{Agnes:2018fwg}.
Since both \nuc{42}{Ar} and \nuc{39}{Ar} are produced by cosmic-ray interactions, it is expected that \nuc{42}{Ar} is similarly or better reduced in UGAr.
This level of reduction would strongly reduce the background due to \nuc{42}{K}. Therefore, the baseline plan for \Lthou\ is to obtain the required quantity of underground-sourced LAr (\UGLAr) to immerse the detectors.

The Istituto Nazionale di Fisica Nucleare (INFN) is funding a plant to produce \UGLAr\ in connection to the DarkSide-20k project~\cite{Aalseth:2017fik,Wang2019}. The URANIA project plans to develop
95\,t/yr
of \UGLAr\ at the Kinder Morgan Doe Canyon Facility in Cortez, CO, USA. The plan is to produce 50\,t of UGAr for use in DarkSide-20k. It will be shipped to the ARIA plant in Sardinia to purify the Ar at a rate of up to 1\,t/d before it is transported to LNGS. The time scale for the start of operations of the URANIA plant is estimated to be 2022, and it requires about nine months to produce the \UGLAr\ required for DarkSide-20k. It is envisioned that \LEG\ receives \UGLAr\ from URANIA starting from 2023, which matches the needed schedule. It would then have to be further chemically purified at ARIA or a dedicated facility at SNOLAB so that the optical properties would be appropriate for the LAr scintillation detection.
The estimate of the required contained mass of \UGLAr\ for \Lthou\ is approximately 18\,t.
SNOLAB is making plans to store a few hundred tons of \UGLAr\ for a potential ARGO~\cite{Agnes:2020pbw}
project. Therefore, that facility is already preparing for transporting and storing large quantities of ultra-pure Ar.

The collaboration is investigating additional strategies to suppress the \nuc{42}{Ar} background. New tools could be used in combination with UGLAr to further lower the \Lk\ baseline background level and boost the design sensitivity.
The strategies that are currently being explored include: i) producing detectors with thicker \nPlus\ dead layers, ii) enclosing the strings with high-purity nylon mini-shrouds with reduced Ar volumes following the \Ltwo\ design, iii) enclosing the Ge detectors with wavelength shifting and scintillating high-purity PEN, iv) improving the pulse-shape discrimination of \nuc{42}{K} surface events, v) doping the LAr with a non-quenching impurity to neutralize the \nuc{42}{K} ions, and vi) moderately depleting the atmospheric Ar of \nuc{42}{Ar}.

%% file: sec_technical/subsec_active-shield.tex
\subsubsection{LAr Scintillation Detector: An Active Shield}
\label{subsec:lar-veto}

The LAr scintillation-light detector acts as an active shield. It shields the Ge detectors from any background source in the materials surrounding the array, and it suppresses background events that deposit energy in the LAr. In essence, its role is to detect radioactivity in the immediate vicinity of the Ge detectors.

The \Gerda\ collaboration operated a LAr detector system~\cite{Agostini:2015boa,Csathy:2016wdy,Agostini:2017hit} that mitigates such external background sources and, with improvements, is used in \Ltwo.
These background sources deposit energy in the LAr, which results in the emission of 128-nm vacuum ultraviolet (VUV) scintillation photons that can be efficiently detected.
As was the case in \Gerda, this can be accomplished by placing photomultiplier tubes in a position to collect the light or by collecting and guiding the scintillation light through WLS fibers coupled to SiPMs. The SiPMs require the shift in wavelength to better match their spectral response, which peaks at a wavelength around 450\,nm with a photon-detection efficiency greater than 35\%. The WLS fibers in \Ltwo\ are coated in a vacuum-evaporation process with a 1\,$\mu$m layer of tetraphenyl butadiene (TPB).
This layer absorbs the 128-nm VUV scintillation photons from the LAr scintillation and re-emits them in the blue spectrum. This light couples into the WLS fibers and is shifted to a spectrum above 500 nm (with details depending on the transport distance) guided via total internal reflection to the SiPMs.
The placement and coverage of the light detection and readout instrumentation are optimized to maximize light collection and reduce shadowing by the Ge array. For \Ltwo, this optimization led to the placement of two cylindrical fiber ``curtains" surrounding the detector strings arranged in a single ring. \Lk\ builds upon the LAr instrumentation system refined for \Ltwo\ to enhance background rejection through an improved geometrical fiber coverage, an increase of the photo-electron yield together with an improved LAr quality to maximize light collection, and improved front-end electronics to discard background events at the single photo-electron level.

\begin{figure}[h]
\centerline{
	\includegraphics[width=.7\textwidth]{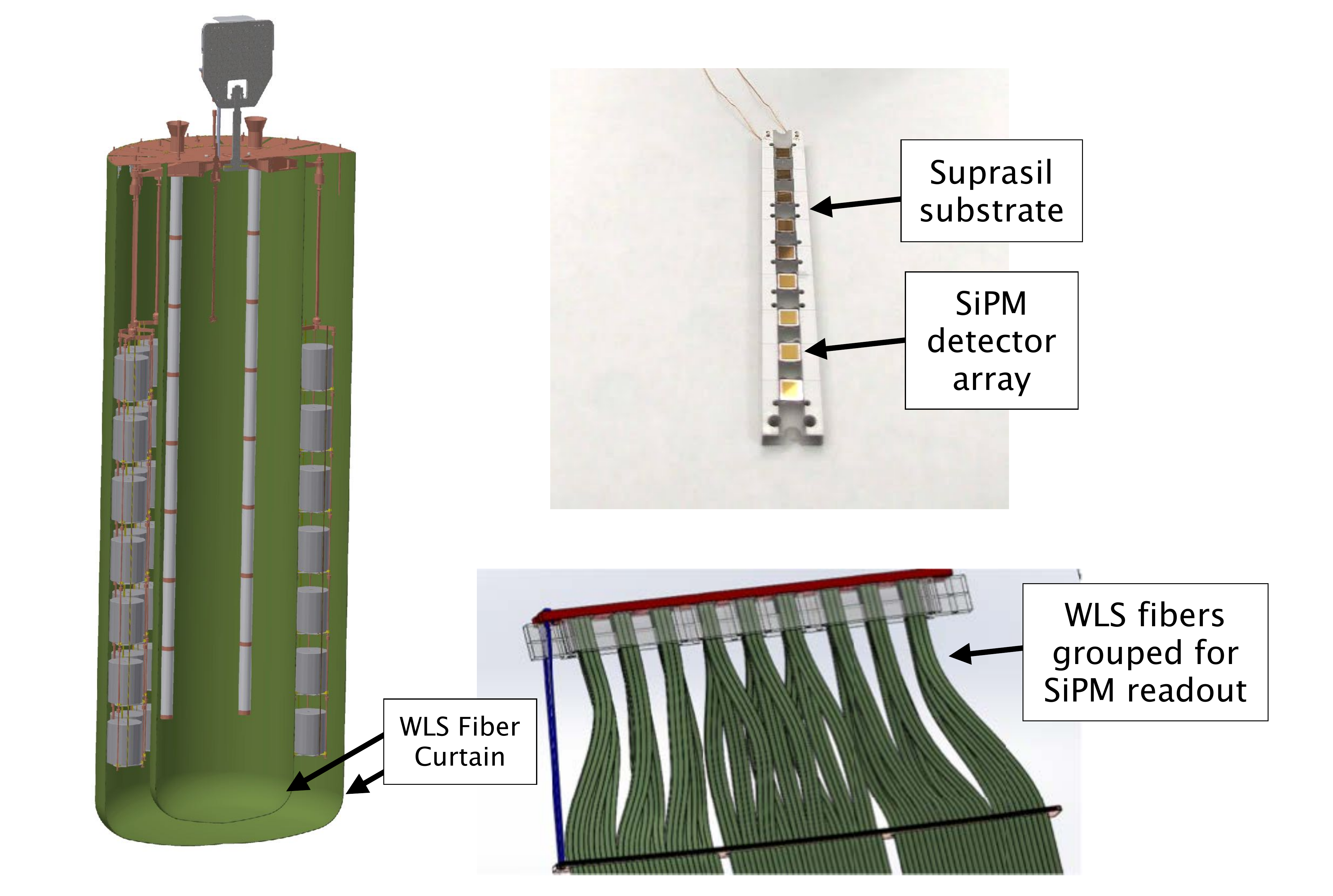}
}
\caption{The main components of the LAr veto system interior to the reentrant tubes for one of the four detector modules.  Left: The fiber curtains surround the Ge strings and are read out by a SiPM array (top right), which is coupled to the grouped fibers (bottom right).}
\label{fig:LAr-acive-veto-components}
\end{figure}

\paragraph{LAr Instrumentation Baseline Design}
Figure~\ref{fig:LAr-acive-veto-components} shows the principal components of a \Lk\ single-module LAr veto system within one of the four reentrant tubes of \Lk.
The detector strings are surrounded by a curtain of WLS polystyrene fibers.  The fibers are coated with TPB for shifting the LAr scintillation light from 128\,nm to into the blue-light range. The TPB emission, peaking around 430\,nm, is further shifted in the fiber doped with the K-27 green-emitting (around 500\,nm) compound. The shifted light is then transported in fibers that are grouped and coupled to an array of SiPMs mounted on a \suprasil\ substrate. Each of the Cu reentrant tubes can contain a wavelength-shifting and reflective surface for enhanced light collection.

The signal from each SiPM array is read out on a differential line by a receiver board placed outside of the cryostat to minimize the amount of radio-impurities in the LAr. Under the default plan, each board is equipped with a voltage operational amplifier at a high-gain value, optimized to ensure a sufficient resolution at a charge equivalent to a single photon above the white noise of the line. Other designs under consideration include an approach whereby the analog amplifier is placed closer to the SiPM to maximize the signal-to-noise ratio. This approach requires further R\&D of cryogenic performance and sufficient radio-purity of the components.

Outside of the Cu reentrant tubes, the LAr light collection requirements are less stringent. Two zones are defined in this region with different light collection requirements. First, wavelength-shifting material is mounted directly outside of the reentrant tubes and read out with a scheme similar to that for the inner region. This zone is sensitive to single photo-electron level of detection for inter-module backgrounds. An outer zone instrumentation scheme includes light read out at the outer radius of the LAr cryostat but with a photo-electron trigger raised to the level necessary for
detection of cosmic-ray muons and their secondary delayed neutrons that typically produce significant levels of light due to their energy loss. These zones and sub-zones therein are optically isolated to reduce pileup and dead time from the high \nuc{39}{Ar} rate in atmospheric LAr. While the exact design of this zone's instrumentation is under study to determine the needed photocathode coverage for the expected muon flux, likely readout options include 8-in-diameter PMTs or SiPM walls.

\paragraph{Alternatives and R\&D}
\new{While the baseline design is well tested and incremental improvements are expected, we intend to carry out further R\&D and refine the LAr active shield design where improvements are possible: lowering radio-impurities of its components and increasing the efficiency of detecting light from backgrounds in the proximity of the Ge-detector strings.}

\new{The current \Lk\ baseline background model assumes the use of the 1\,mm-diameter BCF-91A WLS fibers
used by \Gerda\ and \Ltwo, produced by Saint-Gobain Crystals\footnote{Saint-Gobain Crystals, Hiram, OH, USA; \url{https://www.crystals.saint-gobain.com/}}. While the backgrounds from the fibers are acceptable, a source of cleaner fibers is being pursued to improve upon the present \U/\Th-chain contamination at the level of 100\,\mubq/kg (see Table~\ref{tab:simulatedActivities}).
A collaborative effort with Saint-Gobain is helping identify the source or production step that introduces radio-impurities in WLS fibers.
This work includes sharing details on the production process and supplying raw materials available for measurements and screening.
\Lk\ is also collaborating with Kuraray\footnote{Kuraray Co., Ltd.; \url{https://www.kuraray.com/}} who has supplied high-quality fibers to several Fermilab experiments, including MINOS~\cite{Michael:2008bc,Adamson:2004mh} and NOvA~\cite{Ayres:2007tu,Ayres:2004js}.  Members of the \LEG\ collaboration have a long working experience with Kuraray~\cite{Pahlka:2019bxr} and plan to further engage with them on trial productions to evaluate the radiopurity of their fibers.}

\new{While the role of fibers is to efficiently detect scintillation light originating in LAr, other light collection geometries can be conceived of to accomplish the same goal. Instead of using fibers, one may use wavelength-shifting plates or cylinders, possibly made from PEN or doped plastic strips, to more fully enclose the Ge-detector strings.
Compared to fibers, the production of plates is much simpler and may result in a higher radiopurity product under the right conditions. With an established relationship with Eljen Technology\footnote{Eljen Technology, Sweetwater, TX, USA; \url{https://eljentechnology.com/}}, which is a manufacturer of such plates, the potential improvement in radiopurity of plates can be studied.
The base material of wavelength-shifting plates from Eljen is polyvinyl-toluene (PVT) with fluors added to give the plastic its desired properties. Eljen is interested in collaborating to achieve the cleanest ingredients as well as casting and polishing plates to achieve a high level of radiopurity.
Similarly, members of the collaboration have close contacts with the scintillator manufacturer Envinet\footnote{Envinet Nuvia Group, Prague, Czech Republic}, which have produced high-radiopurity scintillator blocks for the main calorimeter wall of the SuperNEMO Demonstrator~\cite{Barabash:2017sxf}.
Envinet's blocks for SuperNEMO are based on polystyrene for which both the monomer and polymerization are different than PVT. We can compare radiopurities of samples produced by Eljen and Envinet to guide our selection of a viable alternative to the WLS fibers.}

%% file: sec_technical/subsec_electronics.tex
\subsubsection{Front-End Electronics, Cables, and Connectors}
\label{subsec:front-end}

The \Lk\ electronics readout chain starts at the detector and requires local amplification to drive the 10-m-long cables that reach the lab environment from the center of the LAr shield.
The electronics design follows the general philosophy of amplifying the signal as close to the Ge detector as possible to maintain high signal fidelity, while minimizing the number of nearby components to improve radiopurity. Although the combination of the \Gerda\ and \MJD\ readout electronics designs is sufficient for \Ltwo, there are two additional design considerations for the ton-scale \Lk\ phase. First, the radiopurity requirements are more stringent, requiring even fewer (or more radiopure) components close to the enriched Ge detectors.  Second, the larger detector array necessitates longer cable paths from the internal detector to an exterior signal digitizer.

The \Ltwo\ design concept of low-mass front-end (LMFE) electronics and the closure of the gain loop in the charge-sensitive amplifier (CSA) about a meter away sets an upper limit on the bandwidth of signals and the rising edge---important for background event discrimination. A significantly increased gain loop length for \Lk\ would prohibitively slow down the rising edge of signals, necessitating an alternative charge-amplification topology.

\begin{figure}
\begin{center}
\includegraphics[height=0.33\tw]{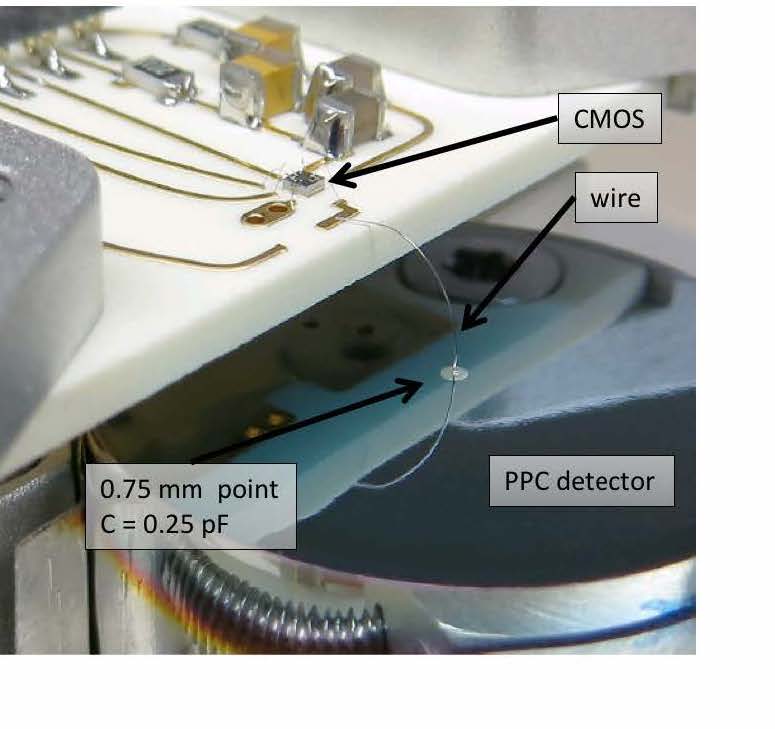}
\includegraphics[height=0.33\tw]{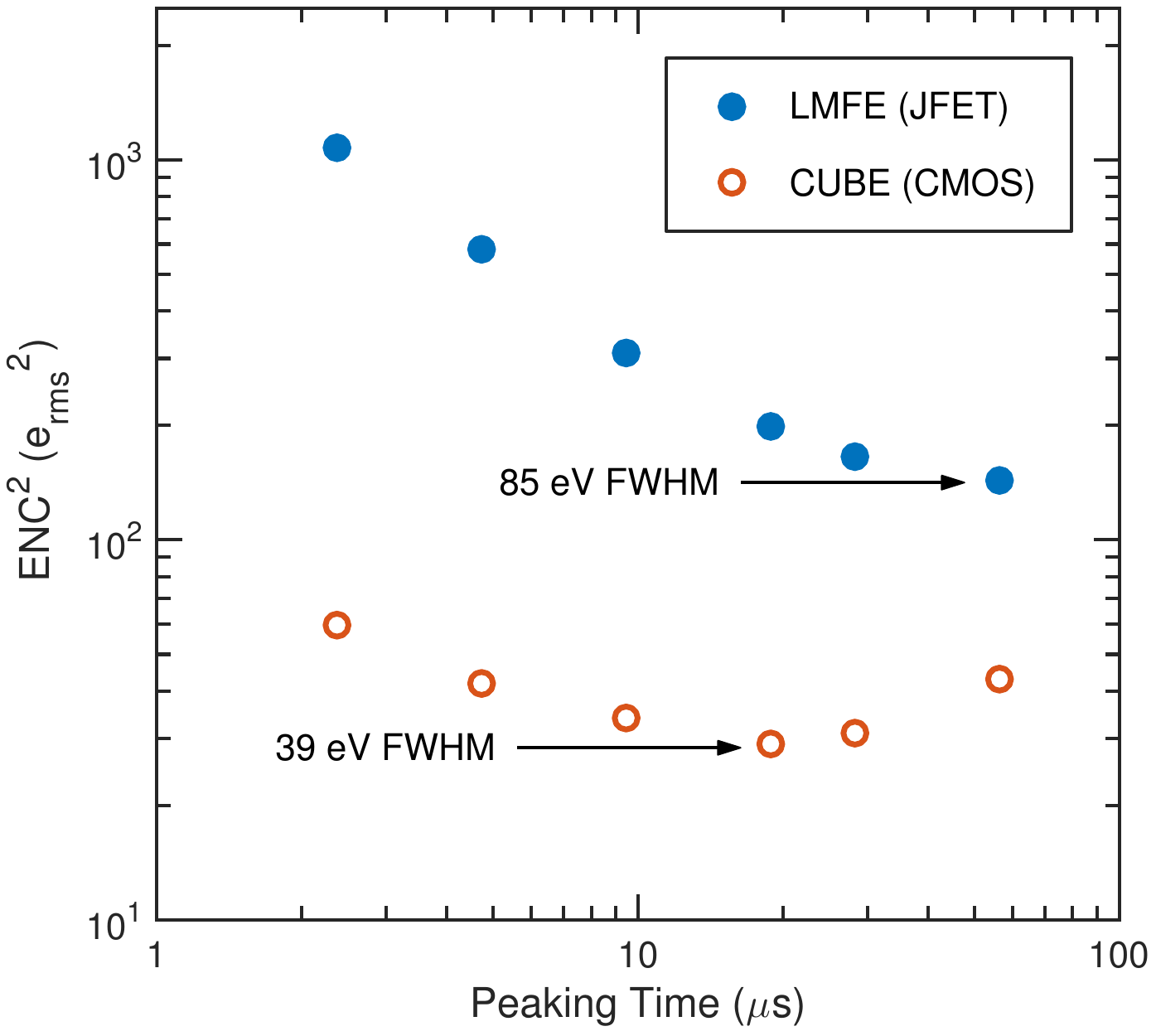}
\caption{Left: A commercial CUBE CMOS ASIC readout wire bonded to a \ppc\ detector. Right: Equivalent noise charge (ENC) versus peaking time of the \ppc\ with CUBE CMOS ASIC. The minimum resolution of 39~eV\,FWHM from
pulser peak widths is equivalent to a noise of 5.6~$e_\text{rms}$.
The higher-noise \MJD\ LMFE performance is included as a comparison to the CUBE ASIC readout. Figures adapted from Ref.~\cite{Barton:2015skb}.
}
\label{fig:lmfe_vs_cube}
\end{center}
\end{figure}

The \Lk\ design utilizes a CMOS\footnote{Complementary metal–oxide–semiconductor} ASIC charge sensitive preamplifier in close proximity to the detector.
This concept eliminates the radioactive background of the \Ltwo\ CSA while maintaining or improving the background near the detector.
The higher amplification level of the ASIC output compared to the LMFE also reduces the noise-performance specifications of the cables near the detector, allowing for cleaner cable fabrication methods and reduced cable backgrounds.
From the signal perspective, the \Lk\ ASIC preamplifier concept improves the bandwidth over the \Ltwo\ implementation while providing a differential output that can be received directly at the digitizer. With the low capacitance of the \Lk\ detectors, the MOSFET\footnote{Metal–oxide–semiconductor field-effect transistor} transistors in the CMOS ASIC process offer an electronic noise benefit over their JFET\footnote{Junction-gate field-effect transistor} counterparts. A recent demonstration of low electronic noise in a mechanically cooled low-capacitance Ge detector with a commercial CMOS ASIC preamp designed for silicon drift detectors~\cite{Bombelli:2011} can be found in Ref.~\cite{Barton:2015skb}.  Figure~\ref{fig:lmfe_vs_cube} shows a comparison of the noise level that was achieved between this CUBE\footnote{\url{https://www.xglab.it/products/cube/}} ASIC and the \MJD\ LMFE based on a low-noise JFET. Initial assays of this commercially available CUBE ASIC indicate that it meets radiopurity requirements~\cite{Edzards:2020wfg}.

\begin{table}
  \footnotesize
\caption{Specifications for a low-noise, low-capacitance readout ASIC for \Lk.}\label{tb:lbnl_asic_spec}
\begin{center}
  \begin{tabularx}{0.9\textwidth}{ Y Y}
  \rowcolor{legendgrey}
    {\bf Description} & {\bf Design Specifications} \\
    \hline
    Threshold & 1~keV \\
    Dynamic range & 10~MeV \\
    Bandwidth & 50~MHz \\
    Assumed detector capacitance & 5~pF \\
    Cabling & minimal (power, ground, pulser, diff.~out)\\
    External components & none \\
    Power supply & single \\
    Reset & internal \\
    Other & observable leakage current, testable warm or cold \\
  \end{tabularx}
\end{center}
\vspace{-5mm}
\end{table}

The specifications for the \Lk\ front-end electronics are listed in Table \ref{tb:lbnl_asic_spec}. The \Lk\ readout design team is pursuing two ASIC preamplifier implementations.  One approach involves broadening the dynamic range of the commercial CUBE ASIC. A recent revision of the chip has shown promising noise levels, bandwidth, and PSD capabilities when coupled with a 1\,kg Ge detector in vacuum~\cite{Edzards:2020wfg}. In order to remove the charge integrated on the feedback capacitor, the current prototype ASIC utilizes an external digital reset line. This extra signal cable is disadvantageous from the perspectives of cabling radiopurity and mixed signal noise immunity. Further development is ongoing to incorporate the external reset circuitry, reduce the number of power supplies, and add differential output.

\begin{figure}
  \begin{center}
        \includegraphics[height=0.2\tw]{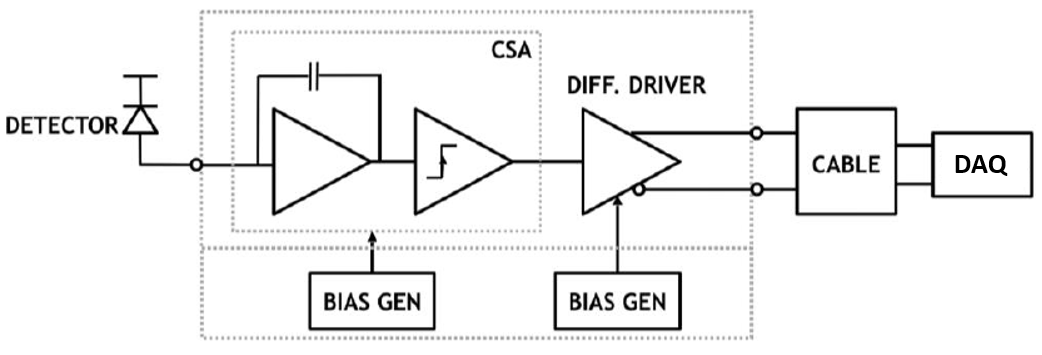}~~~
        \includegraphics[height=0.2\tw]{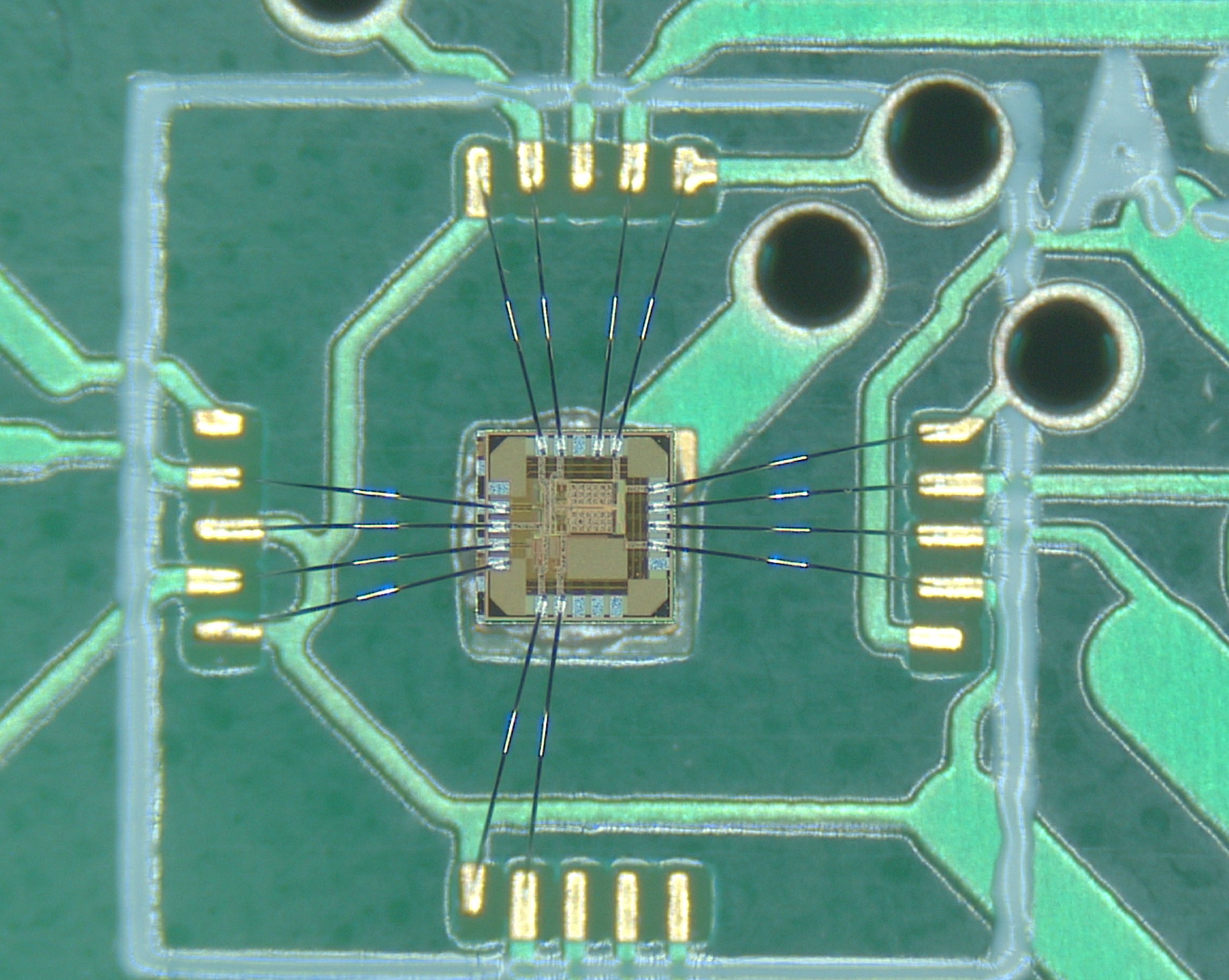}
		\caption{\label{fig:readout_block}  Left: Block diagram of the L1K charge-sensitive preamplifier ASIC, indicating internal continuous reset, voltage regulation, and differential driver.
    Right: The wire-bonded 1\,mm$^2$ L1K ASIC on its dedicated testboard.}
  \end{center}
\end{figure}

\paragraph{Signal readout baseline design}
The primary approach for the \Lk\ ASIC readout
is the custom design of a new ``L1K'' ASIC to match the needs and the operating environment of \Lk\ as set forth in Table~\ref{tb:lbnl_asic_spec}.
This ASIC consists of a CSA and a differential output driver. In contrast to the CUBE chip, L1K is designed with a continuous reset feedback circuit that does not require an external reset pulse. This prevents the need for transmitting potentially noise-inducing digital reset signals to each preamplifier. The L1K ASIC's differential outputs are transmitted to an external receiver at the DAQ over a meters-long cable, as illustrated in Fig.~\ref{fig:readout_block}.  This low-noise, low-capacitance ASIC was designed to be biased with a single power supply, which minimizes readout complexity, radioactive background from cabling, and the number of feedthroughs required for \Lk. The number of feedthroughs could further be reduced with an in-cryostat power and pulser distribution system, located far away from the detectors.

Simulations of the L1K performance for a 1~pF detector and leakage current of 5~pA estimate the equivalent noise charge (ENC) to be 70~eV\,FWHM.  For typical \icpc\ detectors with capacitance of 5~pF and leakage current of 20~pA, the simulated ENC is 130~eV\,FWHM.  Low electronic noise levels aid in achieving a low Ge-detector energy threshold.
The design was tested with the TSMC\footnote{Taiwan Semiconductor Manufacturing Company} 180-nm mixed signal/RF technology.
Alongside variants of the L1K ASIC,
test chips of the internal low-dropout voltage regulators intended for use in the next iteration of the design were also produced. These regulators stabilize the power supply voltage delivered from a distance and are undergoing characterization in LAr.
Figure~\ref{fig:readout_block}
also shows this first prototype ASIC attached to a test printed-circuit board.

Preliminary tests of the L1K ASIC prototype at room temperature and in LAr with cables of various lengths indicate good performance.
While most continuous-feedback preamplifiers discharge through a resistor, causing an exponential tail pulse, the internal reset method in the L1K ASIC leads to a more linear tail pulse and requires a non-linear digital pole-zero correction filter. The L1K ASIC was observed to have a bandwidth of $>$35~MHz and consume $<4$0~mW.  The chip design is currently being modified to include a low drop-out voltage regulator, which has been fabricated and characterized on a separate chip, and to adjust other internal bias currents and voltages for stable operation.

\paragraph{Cables}
The design for cables in \Lk\ is based on Cu on Kapton flat flex cables (FFCs).
The FFC design allows for integration of cable and front-end electronics circuits, further reducing parts count, mass, and installation complexity.
The FFC used in \Gerda\ has a background too high for use in \LEG, but recent development work at Pacific Northwest National Laboratory (PNNL)~\cite{Arnquist:2019fkc} has isolated and eliminated the main source of \U\ backgrounds in Kapton, hence making this convenient and low-mass solution appropriate for use in \Lk.
FFC cables can be sourced from pure Kapton with low friction additives removed from the process, as developed by PNNL for low-background use.

\begin{figure}
\centering
\includegraphics[height=.28\textwidth]{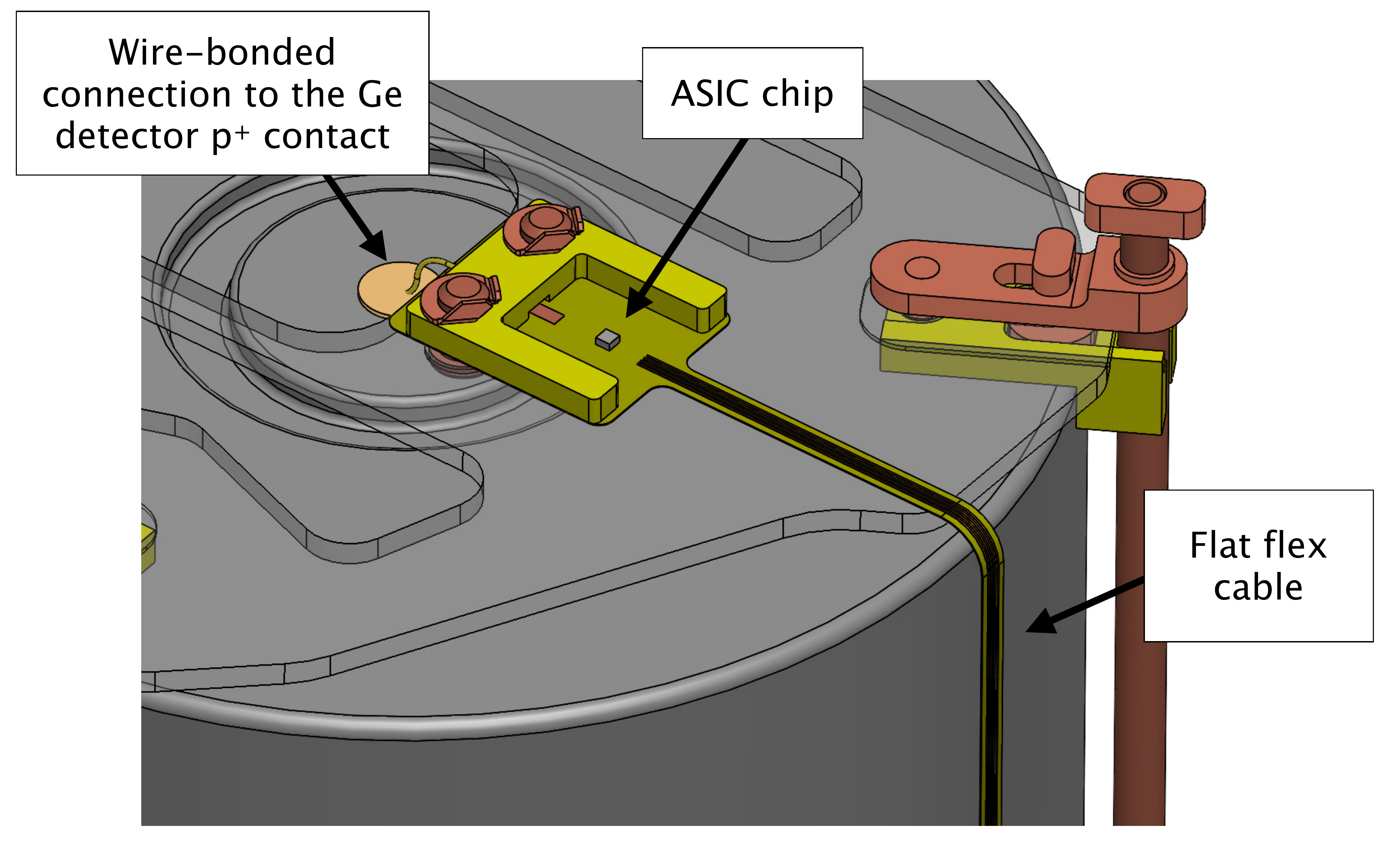}~
\includegraphics[height=.28\textwidth]{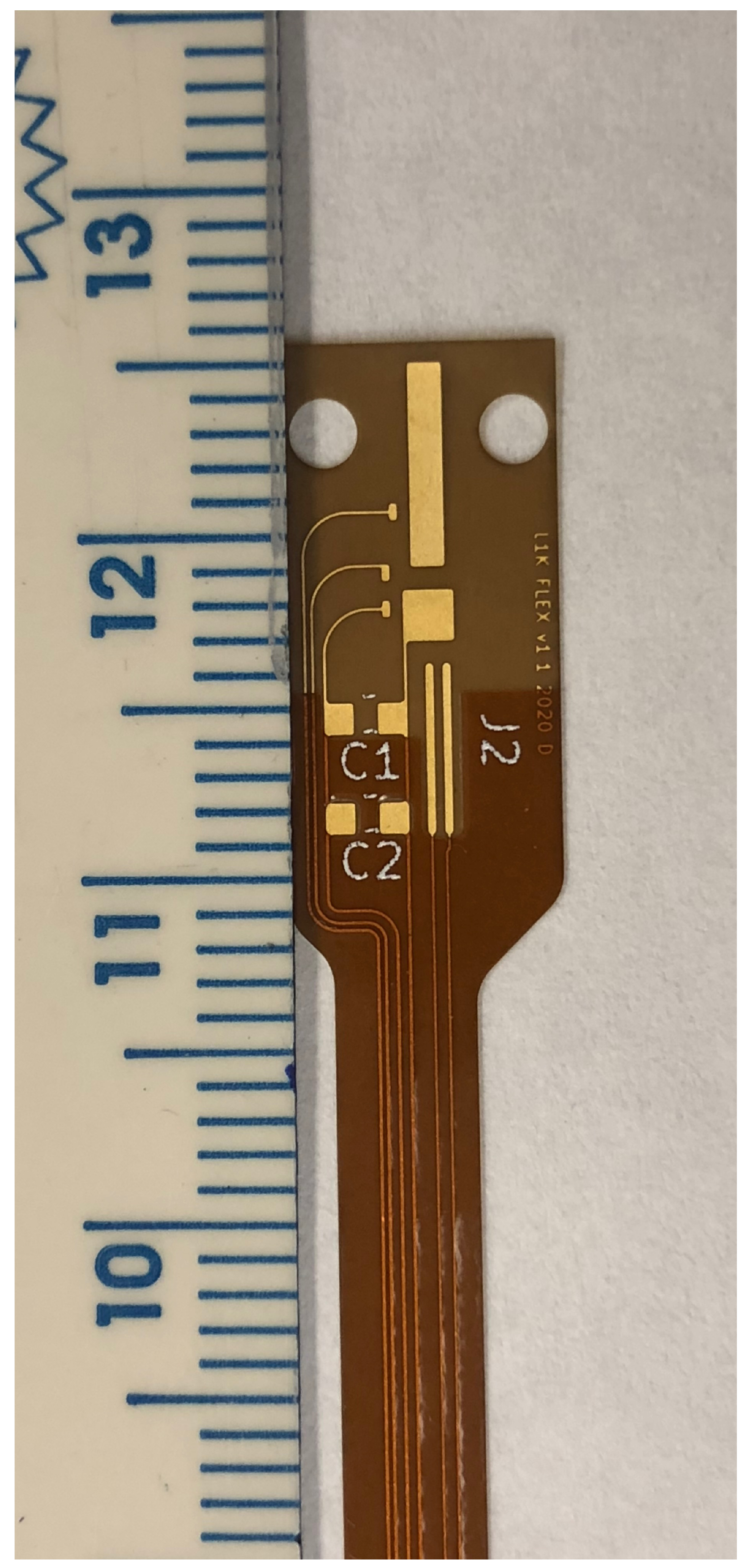}\;\;
\includegraphics[height=.28\textwidth]{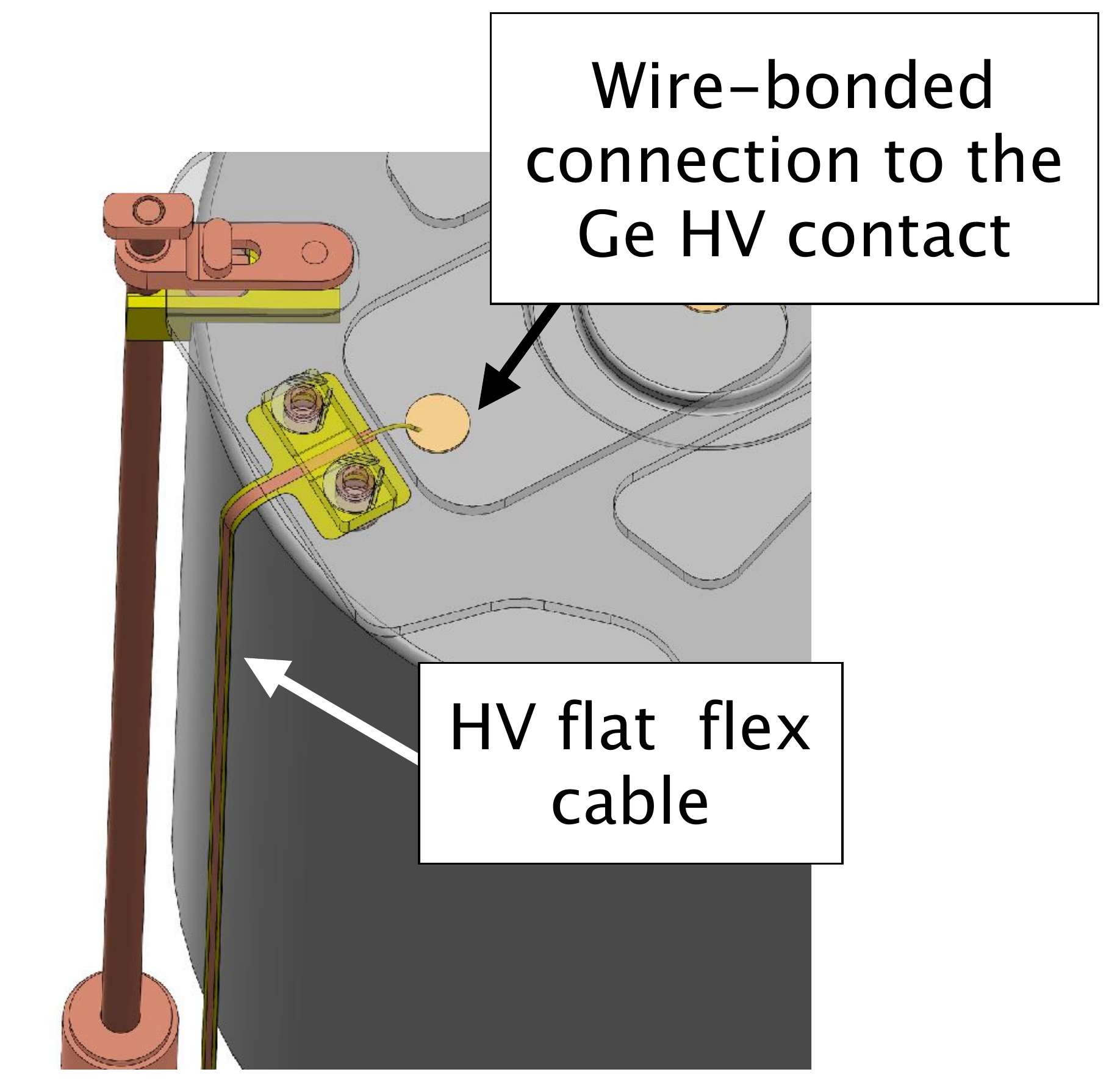}
\caption{Left: A conceptual drawing of the signal readout board and cable integrated into the detector unit. Middle: A prototype polyimide flex circuit comprising an L1K ASIC front-end board and cable. Right: The connection of the HV flat flex cable to the detector.
}
\label{fig:SignalDetectorUnit}
\end{figure}

\paragraph{Connections to the Detector Unit}
Figure~\ref{fig:SignalDetectorUnit} shows the conceptual design of how the ASIC is installed on the detector holder
and a
prototype flex circuit comprising a front-end board for the L1K ASIC and a 3\,m cable.
The detector readout electrode is wire-bonded to the same substrate trace as the L1K ASIC input.  This end of the cable is secured onto the PEN baseplate using an \ultem\ stiffener plate, EFCu retaining pins, and Ph-Br clips.
The high-voltage (HV) cable consists of a single conductor wire-bonded to the detector, as shown in Fig.~\ref{fig:SignalDetectorUnit}.
Wire bonding to the detector can be completed in a glovebox within an underground cleanroom environment in the same manner as for \Gerda\ and \Ltwo.  This technique is well established, reliable, and meets radiopurity requirements for \LEG.

\paragraph{Connections to the Detector Strings and Feedthrough Connectors}
In order to aid in the installation of the array, a cable connection point must be made inside the cryostat between the Ge-detector strings and the point where the feedthroughs exit the cryostat. Custom low-background zero insertion force (ZIF) connectors, based on previous designs used in \Gerda, are used above the array to connect the short, 1.5-m (on average) cables attached to the string of Ge detectors with the 10-m-long cable bundle that runs to the feedthrough flange at the top of the cryostat. For signal cables, the feedthrough flange can use vacuum industry standard ultra-high vacuum (UHV) feedthroughs.
For HV cables, a commercial HV feedthrough mates to an adapter board on either side of the flange.

%% file: sec_technical/subsec_daq.tex
\subsubsection{Data Acquisition and Slow Controls}\label{subsec:data-acquisition-slow-control}

The physics events collected by the data acquisition (DAQ) system extend over an energy range from
around 1~keV to around 10~MeV. Beside events in the energy region of the
$2\nu\beta\beta$ and $0\nu\beta\beta$ decays, it is necessary to collect events at higher energies to record
possible $\alpha$ contamination, while events at low energies enable
non-$0\nu\beta\beta$ physics searches.
The Ge detectors are subdivided into modules that contain their own dedicated LAr veto system.
The DAQ readout for each module operates as a stand-alone independent system.
All together there are four of these modular DAQ readout systems---a
separate DAQ system for the muon veto system and a LAr readout system external to the modules, and a slow control system for the modules and centralized systems.
Data from all systems include timestamps, allowing sequential event building.

The following describes the DAQ system for each module.
The signals from each Ge detector are read out through a digitizer, the characteristics of which
are defined by the large energy range to be covered and by the necessity to have excellent energy resolution
over the entire range. These requirements require a digitizer with a high number of bits (16-bit as adopted
in \Ltwo) or adopting a solution with lower number of bits but doubling the channel readout per detector
for low-gain and high-gain signals (as implemented in the \MJD). Due to the time development of a typical Ge signal and
the utilization of PSD methods, a sampling rate of at least 60~MHz is mandatory.
The baseline design of the SiPM-based LAr veto system calls for 60 channels of readout.
The DAQ of the LAr veto system has a structure similar to the Ge DAQ system and also requires a
digitizer for readout.
The baseline digitizer for all systems---Ge detectors, LAr SiPMs, and muon veto PMTs---is the FlashCam
digitizer being used in \Ltwo.  This digitizer is reconfigurable with an ADC resolution of 16/12~bits at sample rates of 62.5/250\,MHz.
Table~\ref{tbl:daq_mod_specs} summarizes the baseline DAQ specifications for a \Lk\ module.

\begin{table}[htbp]
  \centering
  \footnotesize
  \caption{\new{A summary of the DAQ specifications for each of the four \Lk\ modules. The last block of the table refers to the Ge detector high-voltage system.\label{tbl:daq_mod_specs}}}
  \scriptsize
  \begin{tabularx}{\textwidth}{L{0.38\tw} X }
    \rowcolor{legendgrey}
    {\bf Description} & {\bf Requirement} \\
    \hline
    Number of readout channels per Ge detector   & 1 \\
    Total number of channels Ge/SiPM    & Minimum 100--120/60--80 \\
    Clock system                                 & Synchronized clocks between subsystems, GPS timestamp \\
    \hline
    Typical background data rate         & 1 Hz/channel; $<$0.1~GB/s \\
    Dead time at typical data rate       & $<$0.01\% \\
    Typical calibration data rate        & 3--4 kHz; $<$1~GB/s \\
    Dead time at calibration data rate   & $<$0.1\% \\
    Maximum data rate                   & $>$3.5~GB/s \\
    Dead time at maximum data rate       & $<$1\% \\
    \hline
    Digitizer setting validation        & Ability to read back digitization settings at run start \\
    Digitizer initialization stability & Stable operation of the digitizer across initialization boundaries \\
    Channel identification              & Unique channel identifier recorded with each raw waveform \\
    DAQ data file naming                & Unique identifier in addition to run and cycle number \\
    DAQ data file size                  & Adjustable limit set to several GB during normal running \\
    DAQ log file storage                & Log files stored alongside data files for each run \\
    DAQ configuration repeatability     & Configuration file stored allowing system to return to any previous configuration \\
    DAQ software documentation          & Detailed description of the data format, all algorithms used (baseline, triggering, etc.), and readout control features \\
    DAQ software version control        & Git or svn with version identifier recoverable from data, configuration, and log files \\
    DAQ system real-time monitoring & Ability to view waveforms, baselines, rates, and energy histograms in real-time while acquiring data \\
    \hline
    Underground real-time data storage        & $>$50~TB \\
    Underground secondary RAID storage        & $>$250~TB\\
    Above ground storage                      & $>$500~TB \\
    Underground-to-surface network bandwidth  & $>$10~Gb/s \\
    \hline
    Digitizer ADC resolution (Ge)             & $\geq$16~bits \\
    Digitizer full-scale range (Ge)           & $\geq$10~MeV \\
    Digitizer intrinsic electronic noise (Ge) & $<$50~eV \\
    Digitization speed (Ge)                   & $\geq$62.5~MHz \\
    Digitization record length (Ge)           & Adjustable up to $\geq$48~$\mu$s \\
    \hline
    Digitizer ADC resolution (SiPM)           & $\geq$16~bits \\
    Digitization speed (SiPM)                 & $\geq$62.5~MHz \\
    Digitization record length (SiPM)         & Adjustable up to $\geq$48~$\mu$s \\
      \hline
    Triggering capability    & Single-channel readout with independent triggers OR global trigger \\
    Triggering functionality & Ability to run ALL trigger schemes for ALL types of physics data \\
    Software data reduction  & Optional mode for real-time data reduction on the DAQ machine(s) with simple, fully documented techniques \\
    \hline
    Number of HV channels (Ge) & $\geq$100--120 \\
    HV output range            & 0 to $\geq$5~kV \\
    Maximum current output     & $\geq10 \mu$A \\
    Voltage set precision      & $\leq$1~V \\
    Voltage ripple             & $\leq$10~mV \\
    Voltage monitor resolution & $\leq$0.1~V \\
    Current monitor resolution & $\leq$50~pA \\
    Minimum ramping speed      & $\leq$5~V/s \\

  \end{tabularx}
\end{table}

The DAQ system has flexible triggering options to maximize the physics capabilities.  The trigger capabilities include:
\begin{myitemize}
\item{Local trigger mode: Each readout channel has a threshold that can be set independently of other channels.  When an individual channel triggers, ideally only the waveform for that channel is read out.  However, if necessitated by the hardware/firmware, all channels on the respective card can be read out.}
\item{Global trigger mode: When a trigger is generated from one subsystem, all channels from one or several subsystems are read out.}
\item{Online data reduction: Upon readout due to one of the above trigger conditions, the DAQ system provides the ability to, in real time, examine the waveforms to determine which channel's waveforms should be written to disk.}
\item{Special trigger modes: Logic is implemented for specific items, i.e., for delayed coincidences.}
\end{myitemize}

Various mixtures of these triggering modes for the different subsystems are required depending on the type of data taking.  Through appropriate threshold adjustments and online data reduction, the bandwidth limitations allow for all types of data to be acquired with the same hardware triggering configuration.
The Ge and veto DAQ systems can receive a precise time signal from an external time generator so that all of the collected events have a timestamp.
Offline, the data streams coming from the Ge and veto DAQ systems can be unified on the basis of the timestamps.

A slow-control system monitors all the important parameters of the subsystems: cryostat, cleanroom,
high- and low-voltage power supplies, etc. It uses a database or similar tool to store information collected
from a distributed pool of clients that reads sensors within the the sub-systems.
The system can support alarm notifications
and has an accessible web interface for monitoring the status of all sub-component systems.
The DAQ system has access to the slow-control storage area to retrieve all the information useful for the
definition of the start running conditions.
All high-voltage distribution systems are under the supervision of the slow controls. There are four independent systems corresponding to individual bias supplies for each module's Ge detectors and a fifth system for the muon veto PMTs.
Commercial products, similar to what is used in \Gerda, the \MJD, and \Ltwo, provide autonomous computer-controlled systems with Ethernet interfaces.
The main specifications of the HV system are summarized in Table~\ref{tbl:daq_mod_specs}.

The DAQ systems monitor key parameters and issue alarms and notifications to operators and experts.  They are capable of protecting the
 hardware and placing modules or other systems into a safe state.
A web-based monitoring system provides near-time monitoring access to both the DAQ and slow controls systems.
Coupled to a database, it also allows access to historical data. This system is an extension of tools developed for \Gerda, the \MJD, and \Ltwo.

%% file: sec_technical/subsec_calib.tex
\subsubsection{\new{Calibration System}}\label{subsec:calibration}

Regular calibrations of the Ge detectors are necessary for \Lk\ to determine, monitor, and maintain a stable energy scale; achieve an optimal energy resolution; and provide event classes for the optimization and monitoring of the PSD parameters. By evaluating the position of $\gamma$ lines in the recorded energy spectra, the detectors can be calibrated, their resolution can be determined, and the analysis can be benchmarked to the spectra of well-known sources. Moreover, the calibration system is also needed to provide calibration data for the LAr veto system. Here, the data are used to compare experimental and simulated suppression factors for well-known locations of radioactivity inside the detector array. For this purpose, the Ge detectors and the LAr veto system should be regularly exposed to weak radioactive sources for a limited amount of time.

When not in use, radioactive calibration sources must be stored as far away as possible and shielded from the Ge detectors during physics data taking in order to not contribute to the background budget. During calibrations, a mechanical system lowers the sources into the cryostat and brings them close to the detectors. A weekly schedule of calibrations has proven to be a reasonable time frame to monitor Ge detector performance and account for any instabilities. By tuning the activity and distribution of the source material in the calibration assembly, an array of Ge detectors can be calibrated within several hours.

In general, such a system has to fulfill the following requirements:

\begin{myitemize}
\item {\it Precision:}
The deployed sources must be positioned precisely between tightly arranged detector strings. The mechanical system that controls the deployment must ensure full control over the absolute source position, not only to avoid contact with the detectors, but also to ensure sufficiently small source position uncertainty for the purpose of calibration source event modeling.

\item {\it Repeatability:}
The calibration system should position the sources relative to the detector array in a reproducible manner. Since the goal is to monitor the long-term stability of the
Ge detectors, the variability of the mechanical positioning should be negligible. To achieve this goal, the deployment system should be capable of repeatedly positioning the source within a one-millimeter range.

\item {\it Reliability:}
After commissioning, the calibration system should be operated over the duration of the project without the need to service components inside the cryostat.  Therefore, it must work reliably, needing almost no maintenance or inspection, and keep its full functionality and precision over the course of one decade.

\item {\it Background contribution:}
A source in the storage position should not contribute to the background budget of the experiment.
To guarantee this, storage positions must be located outside all major shielding parts,
posing a technical challenge of deploying the calibration sources over long distances.

\item {\it Safety:}
The handling of radioactive sources comes with the responsibility to guarantee the
safe performance of standard calibrations for all users, new or experienced. The calibration system shall be designed such that safe, remote operation is possible, even by users who are not familiar with the details of the system. The system should consist of several interlocks, mechanical and software, such that no contamination to users or the experiment can occur. Since the deployment mechanism is similar to that of \Gerda\ and \Ltwo, there is experience to mitigate possible failure modes.
\end{myitemize}

The LEGEND-1000 calibration system is designed to satisfy these requirements with encapsulated \nuc{228}{Th} sources.
The decay chain following a \nuc{228}{Th} decay combines the most beneficial characteristics for Ge-detector calibration.
The \MJD~\cite{Abgrall:2017gpr} and \Gerda~\cite{Baudis:2013kaa, Baudis:2015sba} both used $^{228}$Th sources
with activities between 10 and 40~kBq for standard calibrations. The $^{228}$Th isotope decays over a series of $\alpha$
and $\beta^-$ decays to \nuc{208}{Pb}. The progeny isotopes generate a wide range of intense
x- and $\gamma$ rays with energies ranging from below 100\,keV up to 2.6\,MeV. This range of energies can be
used for calibration of the energy scale and to explore systematic effects
such as ADC nonlinearity. The single- and multi-site events created by the 2615-keV $\gamma$-ray line can be used to tune event selection criteria at energies around \Qbb.

The activity of the source should allow calibrations in the shortest time
reasonable so that the loss of run time for physics data is minimal. Both pile-up rate and the readout electronics play a role in limiting the calibration event rate and source activity. While the DAQ system can be designed to handle higher count rates, a low-count rate, high-resolution calibration approach is preferred for the search of \BBz\ decay. Pileup events---events sitting on baselines that has not been completely restored---are not desirable in calibration since they rarely occur in the physics data. The readout scheme imposes a limitation due to the decay constant of the front-end electronics, which is on the order of 1~ms. To guarantee that most events start after the baseline has settled from a previous event, a rate of less than 300\,Hz per individual detector is favorable.
Simulations show that an activity of about 5\,kBq per source assembly meets these requirements. These low-activity radioactive sources are produced by electrodeposition of thorium on gold, which are then electron-beam welded in capsules and certified for use

\begin{figure}
  \centering
  \includegraphics[height=0.3\tw]{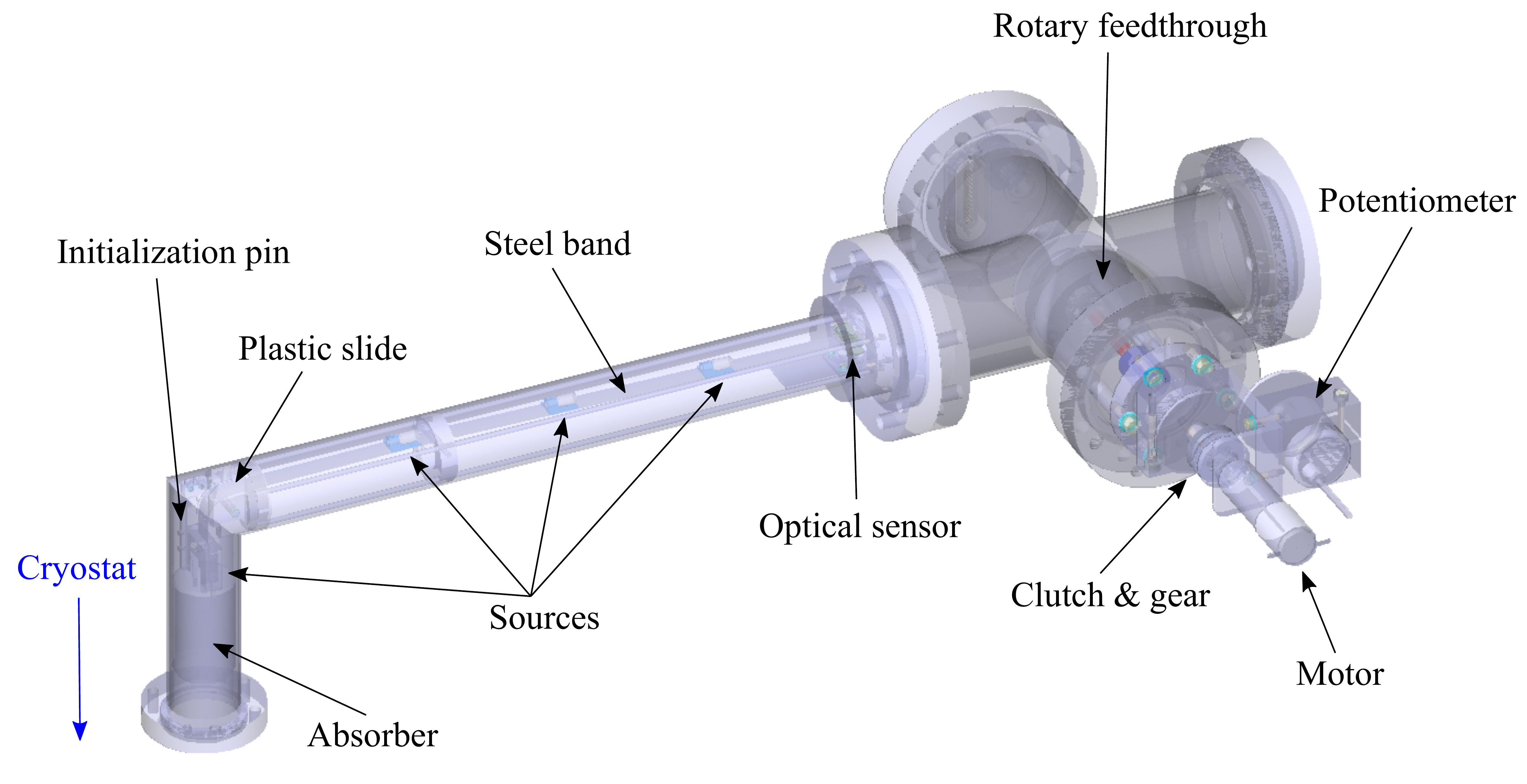}
  \hspace{0.7cm}
  \includegraphics[height=0.33\tw]{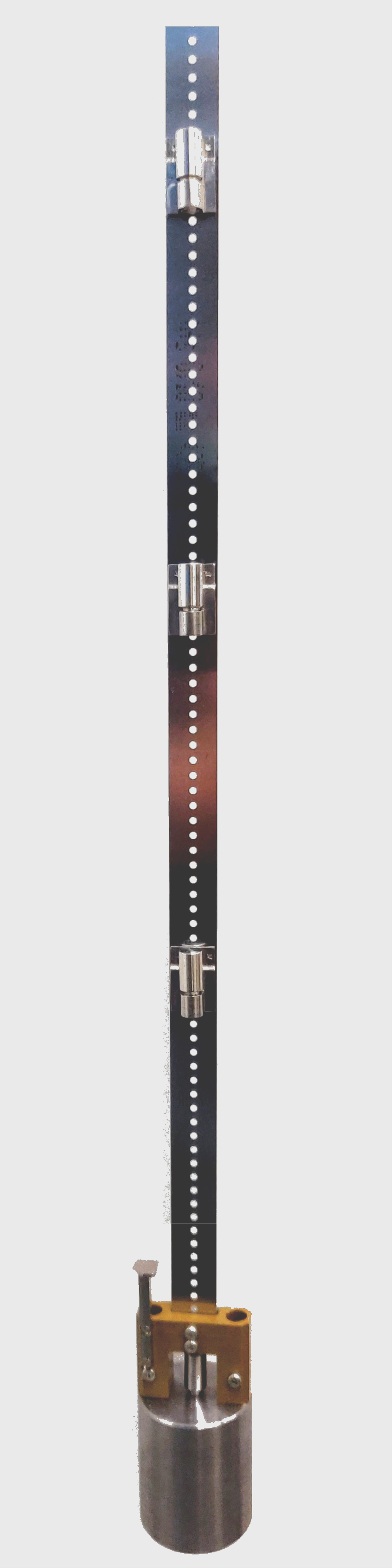}
  \caption{Left: The design of the calibration source insertion system for \Ltwo, which will be followed for \Lk. The sources are lowered down to the level of the detector array by a steel band when a calibration is performed. Right: Multiple source capsules are attached to the steel band with welded special adapters.}
  \label{fig:calibration_source_insertion_system}
\end{figure}

Access points along the top of the cryostat accommodate the installation of a mechanical system that can lower radioactive sources from above the detector array into the cryostat for detector calibration. The \Lk\ calibration system is modeled after those used in \Gerda\ and \Ltwo.
The source-insertion system for \Ltwo\ is shown in Fig.~\ref{fig:calibration_source_insertion_system}.
Individual source capsules are mounted onto a steel band that can be lowered into the detector array.
The advantage of deploying multiple low-activity sources spread across the array is that a number of detector units can be calibrated in parallel with good uniformity of statistics, without exceeding the desired individual rate. In addition, studies have shown that the performance of the PSD cuts in the \icpc\ detectors depend on the location of the $\gamma$-ray interaction inside the individual crystal. Therefore, it is desirable to irradiate a detector from various angles. A source assembly containing multiple sources would fulfill this requirement in one calibration deployment.

The relatively short \nuc{228}{Th} half-life of 1.9 years requires access to the calibrion sources during the runtime of \Lk\ for their replacement. Since the parking positions of sources are outside of the LAr and water shielding, we expect the replacement work to be convenient. The calibration sources of \Gerda, the \MJD, and \Ltwo\ all provided such access. Sources were replaced, or swapped in the case of special calibration runs, during the operation of these experiments without any problems.  The \Lk\ system will provide this capability. The source adapter design allows for the removal and rearrangement of sources as needed.

In addition to calibrations with \nuc{228}{Th} sources, the system has the capability to deploy alternative sources to address specific analysis questions.
For example, the isotope \nuc{56}{Co} has a short half-life of around 77 days and possesses several transitions around 2.5--3.5 MeV, which produce a number of single- and double-escape peaks surrounding the \Qbb\ region. These provide a useful cross-check for data-cleaning and background-rejection cuts developed in the standard \nuc{228}{Th} calibrations. The short half-life of \nuc{56}{Co} prohibits a long-term installation of this source, but over the 10-years expected run-time, two or three dedicated high-statistics deployments are planned. As another example, sources like \nuc{133}{Ba} deliver a number of low energy photons in the 30--400 keV range.  These photons
can be used to calibrate detectors down to low energies with high precision, which is of interest for physics searches beyond the Standard Model. Finally, in order to study and calibrate the light output of the LAr active veto system as a function of position and compare its performance to the complex Geant4 simulation, special sources can be prepared and installed during specific campaigns.

%% file: sec_technical/subsec_materials.tex
\subsubsection{Clean Materials}
\label{sec:materials}

Operating \BBz\ decay experiments deep underground mitigates backgrounds originating from cosmic rays. Of equal importance is constructing detector components using ultra-clean materials with the lowest levels of radio-contamination. Various metals and plastics have been found to be clean, after stringent production and handling, allowing their use as detector components.
Cu is one such metal; it also has excellent thermal and conductive properties, and can be easily electrolytically isolated, allowing further purification. New sources of scintillating plastics have been developed where structural integrity is needed to take the place of optically passive materials. Further, advances in additive manufacturing techniques would allow the convenience of on-demand printing of custom transparent, and potentially scintillating and wavelength-shifting, plastic parts that meet the stringent low-background specifications.

Advancements have been made in the understanding of post-machining contamination of both plastics and metals in order to maintain the materials as close to their bulk low-background purities as possible. All materials for \Lk\ utilize these cleaning techniques and handling procedures to reduce or eliminate the post-production contamination.

\paragraph{Copper Electroforming}\label{sec:clean-metals}

Due to its favorable thermal and electrical properties, ultra-pure Cu is an ideal material when levels of naturally occurring radioactive \nuc{238}{U} and \nuc{232}{Th}, or troublesome cosmogenically induced radioisotopes such as \nuc{60}{Co}, have been significantly reduced. Producing such Cu is possible under stringent cleanroom conditions within an underground electroforming facility.

The \MJ\ collaboration has extensive experience in setup, operations, and machining of EFCu to the size and thicknesses needed for \Lk\ components.
The \MJ\ Collaboration operated 16 electroforming baths split between its underground space at SURF and at PNNL's shallow underground laboratory to produce over 2830\,kg
of EFCu detector components and shield plates. The purity of the produced EFCu was below 0.3\,\mubq\ of \nuc{232}{Th} and \nuc{238}{U} per kg of Cu \cite{Abgrall:2016cct}. Although Cu may be viewed as a soft metal, the electroforming process enhances the strength; all \MJ-produced EFCu material exceeded the 10\,ksi tensile strength design criteria.
The electroforming facility at SURF remains in the \MJ\ laboratory clean space to supply the necessary clean EFCu detector components for \Ltwo\ as well as further study of low background metal production.

The \Lk\ design calls for the four reentrant tubes surrounding the Ge detectors to be fabricated from underground EFCu. This production of EFCu is a natural evolution of the collaboration's collective experience with Cu growth, welding of Cu components, and final post-production cleaning for low background use. Engineering calculations show that the EFCu used for the reentrant tubes can be grown in similar baths and on the same size cylindrical mandrels developed for \MJ, yet rolled to form larger-diameter cylindrical portions that can be welded to the necessary size. Testing of the EFCu rolling and welding processes are being performed along with engineering feasibility studies to validate the design. Since the reentrant tubes are constructed from portions of EFCu produced from a single electroforming bath, multiple baths can be operated in parallel to meet the desired production schedule. Additional baths operate to serve the needs for the production of small parts within the Ge-detector array.

Cu may not always be the ideal material of choice for certain parts within the experiment, so development of ultra-pure alloys can be utilized in locations that may need stronger or more durable metals.  One example is threaded Cu parts. Testing of Cu threads coated with Cr has shown good structural integrity and the elimination of plastic anti-galling coatings, which would make the components withstand  multiple mechanical and temperature cycles while not compromising cleanliness.  Further investigation of CuCr alloying could yield materials with the same or better structural capacity though requiring less mass.

\paragraph{Structural Scintillating Plastics: PEN}\label{sec:PEN}

Structural scintillating plastics are being developed as a replacement for some of the optically opaque and passive materials that support the Ge-detector array in previous experiments.
By replacing passive materials within the array with transparent scintillating materials, an improved background rejection can be achieved by enhancing
the light collection for the active veto and by creating a self-vetoing capability of the support structures.

PEN is an industrial polyester that scintillates in the blue region with a dominant decay constant
of 34.9~ns~\cite{Nakamura_2011, Majorovits:2017cqj, Bilki:2019lep}.
It also shifts the 128~nm VUV light emitted by LAr
into the blue region~\cite{Kuzniak:2018dcf,Boulay:2021njr,Garankin2018}, which can be
detected easily with standard photo-detection devices.
PEN was introduced in \Ltwo\ as the material of the detector baseplate that forms part of the
detector unit (see Sect.~\ref{subsec:detector-fabrication}).
In addition, PEN material is being considered for additional scintillating support structures within the \Lk\ design, including as encapsulation for an alternative detector mount design to separate the detector from any direct contact with the LAr.

\begin{table}[h]
    \footnotesize
\begin{center}
\caption{The physical properties of PEN.}
\label{tab:PEN:prop}
\begin{tabularx}{0.8\textwidth}{ L{0.25\tw} Z Z }
\rowcolor{legendgrey}
{\bf Property} &
\multicolumn{2}{c}{{\bf Value}} \\
 \hline
 Atomic composition & \multicolumn{2}{c}{[C$_{14}$H$_{10}$O$_{4}$]$_{n}$} \\
 Density: $\delta$  & \multicolumn{2}{c}{1.35\,g/cm$^{3}$} \\
 Melting point  & \multicolumn{2}{c}{270$^\circ$C} \\
 Peak emission $\lambda$&\multicolumn{2}{c}{$445 \pm 5$\,nm} \\
 Light yield & \multicolumn{2}{c}{$\approx$ 5000\,photons/MeV } \\
 Index of refraction & \multicolumn{2}{c}{$\approx$ 1.62 } \\
 Decay constant & \multicolumn{2}{c}{34.91 ns} \\
 Attenuation length & \multicolumn{2}{c}{$\approx$ 50\,mm at 450 nm } \\
 Young's modulus: E [GPa] &$1.855\pm0.011$ (296\,K)&$3.708\pm0.084$ (77\,K)\\
 Yield strength: $\sigma_{el}$ [MPa]&$108.6\pm2.6$ (296\,K)&$209.4\pm2.8$ (77\,K) \\
\end{tabularx}
\end{center}
\vspace{-5mm}
\end{table}

PEN, [C$_{14}$H$_{10}$O$_{4}$]$_n$, is primarily composed of only three elements with trace amounts of catalysts.
Its mechanical properties have been determined, both at room and
cryogenic temperatures~\cite{Efremenko:2019xbs}. The tensile strength of PEN is
higher than that of Cu and increases towards cryogenic temperatures (see Table~\ref{tab:PEN:prop}).
This qualifies PEN as a good structural material in any environment relevant
for \LEG.

\begin{figure}[h ]
    \begin{center}
        \includegraphics[width=5cm]{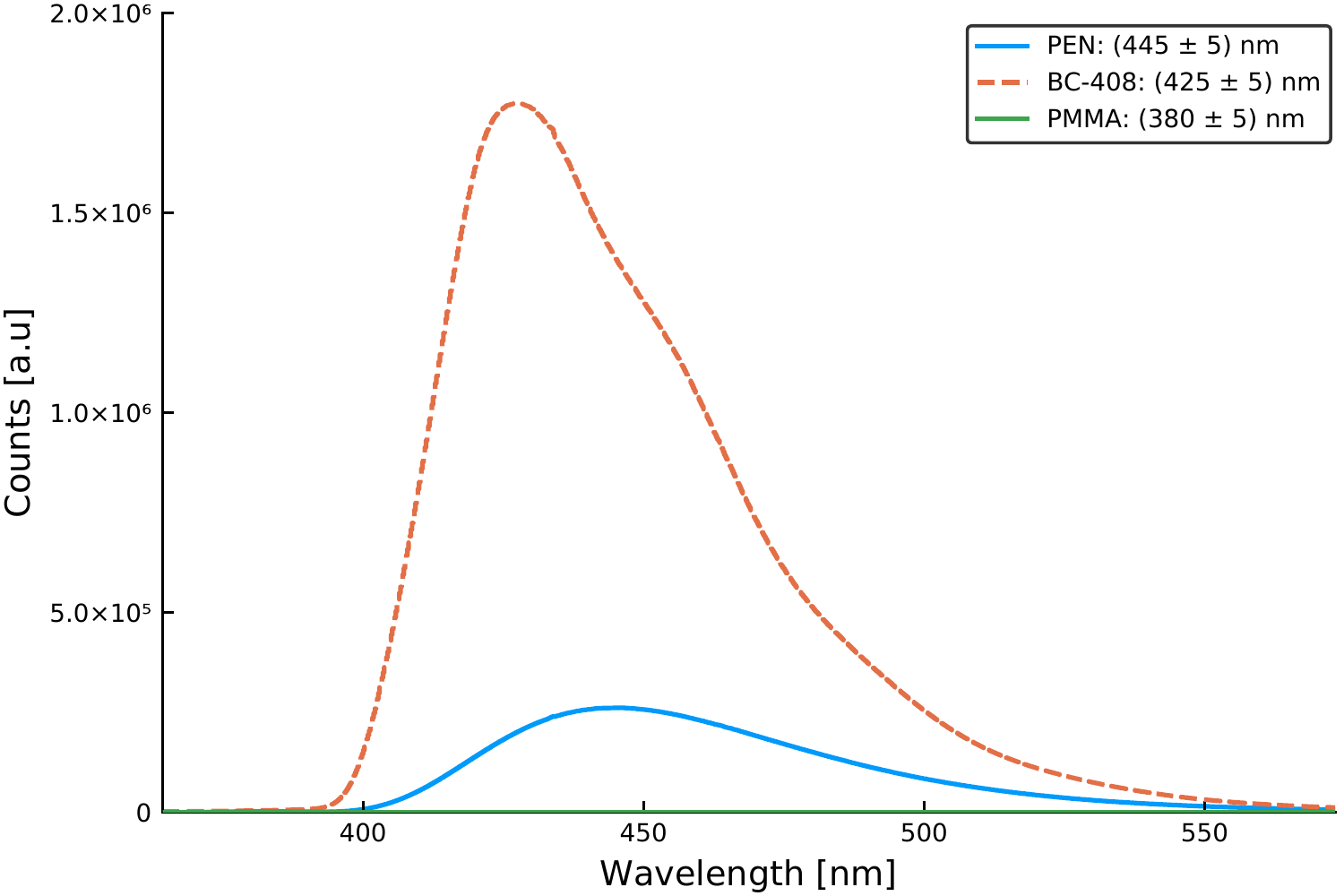}
        \hspace{1em}
        \includegraphics[width=5cm]{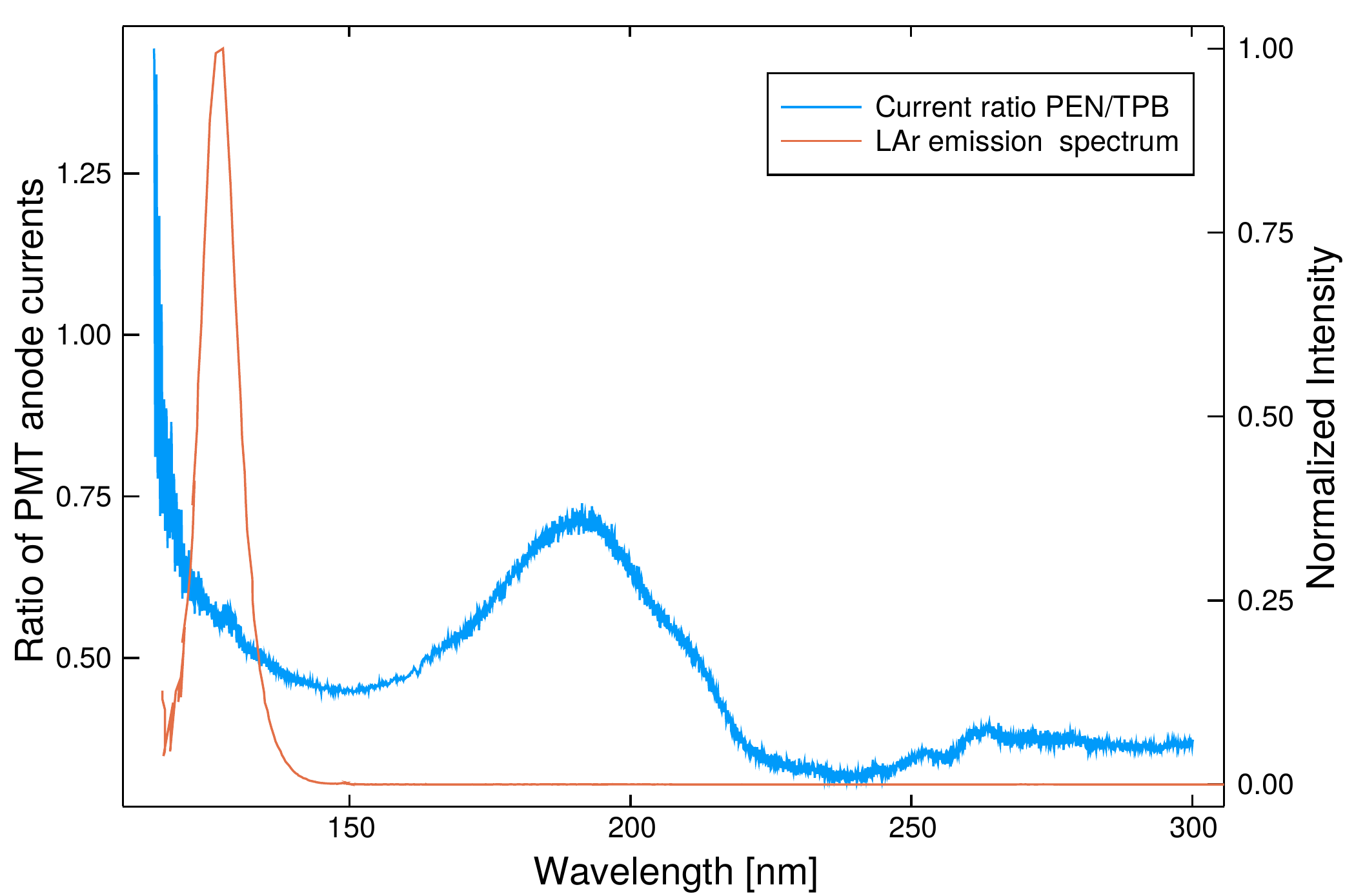}
        \hspace{1em}
       \includegraphics[width=4cm,angle=90]{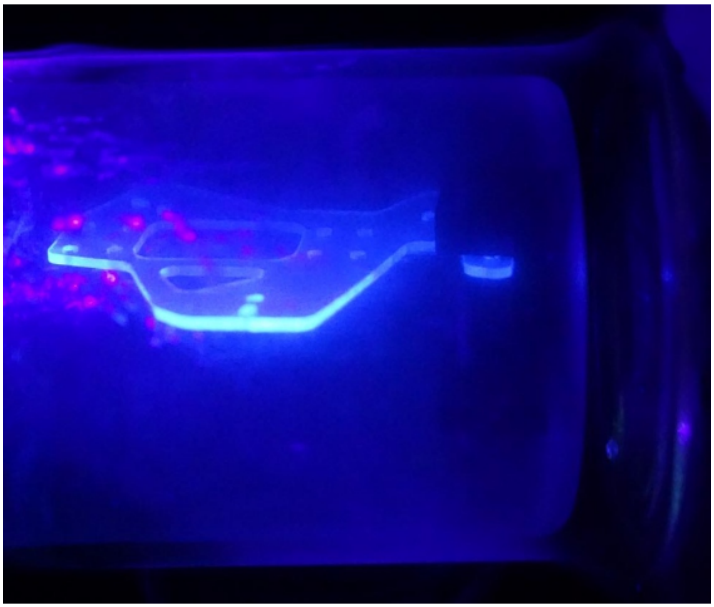}
        \caption{\new{
          Left: The emission spectrum of PEN compared to the common scintillator BC-408.
          Middle: The ratio of the wavelength-shifted light output of a PEN plate versus
                 a TPB-coated acrylic plate. The LAr
                 scintillation spectrum~\cite{Heindl:2010zz} is also shown.
        Right: Emission of blue light by a PEN \Ltwo\ baseplate in \lnn\ when excited with a UV lamp.}}
        \label{PEN:fig:emission1}
    \end{center}
\end{figure}

\new{The PEN scintillation spectra peaks in the blue region at 445\,nm, as shown in Fig.~\ref{PEN:fig:emission1}~(left panel).
For comparison, the emission spectrum of the standard plastic scintillator
BC-408 is also shown. Even though the light yield of PEN is significantly lower, it is sufficiently
high to make PEN a truly active material.
Figure~\ref{PEN:fig:emission1}~(center panel)
demonstrates the wavelength-shifting qualities of PEN~\cite{Efremenko:2019xbs}.
The wavelength-dependent efficiency to shift VUV radiation into
blue light was measured by comparing the light output of a PEN plate to the
light output of a TPB-coated acrylic plate. In the region of the LAr
emission spectrum, PEN reaches about 50\% of the light yield observed
for TPB.
Figure~\ref{PEN:fig:emission1}~(right panel) shows a prototype prototype \Ltwo\ PEN baseplate emitting blue light when excited with a UV lamp.
The PEN structure was submerged in \lnn\ for this test.
The scintillation light emerges predominantly from the edges of the plate, which is expected from simulations
and is favorable for light collection from a Ge-detector string.}

\new{PEN is commercially available as a granulate from Teijin-DuPont.
It has been demonstrated that after executing proper cleaning procedures, which mainly remove
surface contamination, it is sufficiently radiopure (see Tables~\ref{tab:simulatedActivities} and \ref{tab:bi})
to be used within the \Lk\ detector array.
Nevertheless, custom synthesis of PEN is being pursued to
reduce radioimpurities and
improve upon its properties such as
transparency, light yield, and processibility.
This has led to the development of an amorphous variation of PEN,
which is known as poly(ethylene-1,4-cyclohexanedimethylene-2,6-naphthalate)
(PECN).  The incorporation of 1,4-cyclohexanedimethylene units into
the polyester backbone significantly increases the crystallization
half-time and forms highly transparent amorphous material.
R\&D continues to purify components of PEN and PECN and screen them for
radioimpurities prior to use in a custom synthesis.}

\paragraph{Clean Additive Manufacturing}

Additive manufacturing (AM) can address outstanding R\&D challenges for fabrication of intricate scintillating and non-scintillating plastic components. One such AM technology which is well-suited for low-background experiments is laser-based stereolithographic (SLA) 3D printing. Unlike the commonly used fused filament fabrication method which uses a continuous filament of thermoplastic, SLA 3D printing is a contactless fabrication process where components are produced using just laser light. The SLA printing process starts with liquid resin. A laser or patterned light source shines UV light onto the resin, and the light interacts with photo-initiator molecules in the resin to generate free radicals. These free radicals solidify the resin into a polymer through free-radical polymerization.  This avoids the plastic coming in contact with cutting tools, extruder nozzles, or injection molds making it ideal for producing low-background materials.  The low cost and infrastructure overhead means that SLA 3D printers can be located at the experiment site, further reducing costs.

The aim is to use SLA 3D printing in the fabrication of intricate scintillating and non-scintillating plastic components for \Lk.  Components under consideration include electrical connectors, detector-unit support structures, and detector-encapsulation fixtures.  These materials serve not only as an active veto but also as wavelength shifters for LAr scintillation light. R\&D efforts focus on performing radio assay measurements of commercial and in-house produced SLA resins, optimization of the scintillation and wavelength-shifting properties of these materials, and detector-encapsulation schemes.

\paragraph{Post-Production Cleaning and Validation}
All materials within the experiment have construction procedures that specify the cleaning methods: controlled etching for metals or leaching for plastics.  Continuous validation of methods is imperative to ensure that contamination is mitigated during production steps leaving finished parts at the high purity of the original bulk material. As an example, this need is necessary for a material like EFCu, which is initially produced at ultra-low levels of radioactive contamination. The use of EFCU is only beneficial if it is demonstrated that its chemical cleaning, processing, post-production handling protocols, and protection against surface contamination show that the purity of a full part can be maintained. Such actions have been followed through high-sensitivity ICP-MS or other assay techniques through the production cycle of the material~\cite{Christofferson:2017nih}. This experience is applied to ensure that the custom clean materials described in this section can be produced and maintained at necessary cleanliness levels throughout the experiment's assembly stages.

Commercially sourced plastics, such as PTFE and \ultem, have been identified for multiple uses within \LEG.  While not produced in-house, these plastics can be utilized without compromising background.  These plastics are of low-mass elements that are essentially radiopure and simply must be selectively cleaned to remove surface imbedded impurities that occur in the machining and molding process.
Data from ICP-MS assays by the \MJ\ collaboration show that if the components are heavily leached in nitric acid after machining and stored in a controlled N$_2$ environment, the plastics remain as clean as the bulk material. Further assay and $\alpha$-counting data have shown that not only must plastics be cleaned prior to use for contamination reduction, but subsequent handling is of greater importance in reducing Rn plate out, leading to contamination with Rn progeny.
Plastics fabrication will follow detailed cleaning and handling procedures prior to their use in the experiment,
ensuring that additional contamination is not introduced to the clean starting materials.

\subsubsection{Material Screening}
\label{sec:radiopurity_assessment}

To ensure the radiopurity required by the background model of all relevant construction and active materials, an aggressive assay program will be followed.
Activities will include the selection of candidate materials, the development of material handling, surface cleaning, and clean assembly procedures, the
evaluation of impurities that may accumulate on surfaces during construction, and the purification of gases for use in clean environments.

Typically, the $^{238}$U and $^{232}$Th chains have several pathways of contaminating a material. A range of material screening methods and instruments are necessary to fully qualify a material for low-background use.
As an example, the $^{238}$U chain can be divided into three subchains. The first subchain contains the long-lived U and Th isotopes found in the bulk of terrestrial materials, which may be assayed with high precision down to a sub-ppt\footnote{ppt = parts per trillion, 10$^{-12}$ g/g} level by applying high-sensitivity ICP-MS techniques~\cite{LaFerriere:2014rva}. In the second subchain, one finds $^{226}$Ra, $^{222}$Rn, and its short-lived progeny, which are heavy metals that easily attach to aerosols and deposit on surfaces (the so-called Rn plate-out effect). If not supported by $^{222}$Rn, these isotopes decay away within about three hours. Activities in this part of the chain may be investigated either by application of high-sensitivity $\gamma$-ray spectrometry~\cite{NEDER:2000} or high-sensitivity $^{222}$Rn emanation techniques~\cite{Rau:2000, Zuzel:2009, Zuzel:2005hag}.
The final \U\ subchain contains the long-lived \nuc{222}{Rn} progeny, specifically $^{210}$Pb, $^{210}$Bi and $^{210}$Po, which may be responsible for serious surface activity. Their activities may be present over the duration of the experiment, due to the long half-life of $^{210}$Pb (22 yr), yet may not show up immediately due to the ingrowth rate of $^{210}$Po (138 d). The long-lived \nuc{222}{Rn} progeny (present in the bulk material and on the surface) may be assayed using large-surface, low-background $\alpha$ spectrometers~\cite{Zuzel:2017, Zuzel:2019, Zuzel:2018fzl} or applying appropriate chemical procedures to separate Po from the bulk material with subsequent $\alpha$-counting of its activity~\cite{Mroz:2019}.

Because U, Th, and Ra elements have different chemical properties, materials which undergo some production processes are usually enriched or depleted in Ra relative to the amount of elemental U contained. Therefore, it is not recommended to extract the specific activities of \nuc{226}{Ra} (and \nuc{222}{Rn}) from ICP-MS measurements. Additionally, due to the 22-yr half-life of \nuc{210}{Pb}, there can be disequilibrium between the short- and the long-lived \nuc{222}{Rn} progeny.  As a result, it is usually not possible to predict the activities in the \nuc{210}{Pb}--\nuc{210}{Po} sub-chain from only the high-sensitivity \nuc{222}{Rn} emanation measurements.
To predict the background rate in the experiment caused by a specific element of the detector, it is necessary to assay each part of the chain separately. For experiments like \Lk, the most important are the long-lived decay products from \nuc{226}{Ra} and \nuc{222}{Rn}. Therefore, strong emphasis has been placed on developing high-sensitivity $\gamma$-ray spectrometers in addition to Rn emanation techniques, as well as $\alpha$ spectroscopy capabilities.

The \LEG\ collaboration has significant expertise in ultra-low background experiments~\cite{Heusser:1995wd, Peurrung:1997wc, Amsbaugh:2007ke, Agostini:2013tek, Agostini:2015boa, Agostini:2016mof, Freund:2016fhz, Abgrall:2016cct, Lubashevskiy:2017lmf, Busch:2017kxq} improved on through past assay campaigns. However, each new phase of an ultra-low background project has new specific challenges. That experience will be relied upon to identify, address, and mitigate the material purity challenges anticipated throughout the fabrication and installation of the \Lk\ experiment.
The assays are organized into five main areas: mass spectrometry, direct $\gamma$-ray counting, neutron activation analysis, surface assay (contamination with long-lived Rn progeny), and Rn emanation (determination of Ra/Rn and the short-lived progeny).

\begin{table}
\centering
\caption{\label{tab:icpms_facil} \new{Mass spectrometry facilities available for use by the \LEG\ Collaboration.}}
\begin{tabularx}{0.7\textwidth}{ X X Z Z }
\rowcolor{legendgrey}
{\bf Institution (Location)} &
{\bf Instrument} &
{\bf $^{238}$U Sensitivity} [ppt] &
{\bf $^{232}$Th Sensitivity} [ppt] \\
\hline
PNNL (USA) & ICP-MS                              &  0.01 &  0.02   \\
LBNL (USA) & ICP-MS                              &   0.1 &  0.06   \\
ORNL (USA) & ICP-MS                              &  10   &  10     \\
VR Inc.~(USA)\footnote{VR Analytica, Bend, OR, USA; \url{http://www.vranalytical.com}} & ICP-MS      &  20   &  20     \\
LNGS (IT)/TUM & ICP-MS                   &  0.01--10   &  0.01--10     \\
UCL (UK)~\cite{Dobson:2017esw} & ICP-MS   &  0.1   &  0.1     \\
Comenius Univ.~(SK) & AMS                 &  0.1   &  0.1     \\
\end{tabularx}
\end{table}

\paragraph{Mass Spectroscopy Analysis}
\new{Glow Discharge Mass Spectroscopy (GD-MS), Inductively-Coupled Plasma Mass Spectrometry (ICP-MS), and Accelerator Mass Spectrometry (AMS) are sensitive methods where a small amount of material (of the order 1\,g) is processed. Application of these methods makes it possible to detect $^{238}$U and $^{232}$Th but not their decay products. Also, $^{40}$K cannot be detected directly, whereas concentrations of $^{39}$K are measured, and the $^{40}$K content is assumed based on its natural abundance of 0.0117\%.
Glow Discharge Mass Spectroscopy is useful for electrically conductive samples and requires very little sample preparation since the surface can be sputtered cleanly. Sensitivities down to the order of 1\,ppt can be achieved for high mass elements such as U and Th.
In general, samples for ICP-MS analysis need to be prepared as solutions. This requires preparations of solids through sample digestions. For the most sensitive measurements, chemical separation of matrix elements is required. For the utmost in accuracy and precision, isotope tracer dilution methods are preferred to account for sample preparation effects on the analyte during sample processing as well as plasma perturbations and instrument drift during analysis.
Further, samples must be prepared in validated containers and clean room conditions to avoid cross contamination.
In Accelerator Mass Spectroscopy, the ions are accelerated to high kinetic energies before mass analysis. This makes separation of masses very effective, and even rare isotopes may be decoupled from an abundant neighboring mass. Sensitivities down to 10$^{-16}$ g/g may be reached for determination of \U\ and \Th\ contents.
Noble Gas Mass Spectroscopy is used to measure residual (down to ppq in term of gas volume ratio~\cite{Lindemann:2013kna}) concentrations of noble gases in a carrier gas. 
Mass spectroscopy facilities available to \LEG, including the instrument sensitivites to the \U\ and \Th\ chains, are listed in Table~\ref{tab:icpms_facil}.}

\paragraph{$\gamma$-Ray Counting}
\new{Low-background $\gamma$-ray counting is a sensitive method to look for radioimpurities without the samples being destroyed. The process comprises a sample measured with low-background Ge detector, often operated within a multilayer shield with Rn exclusion and located deep underground.
A $\gamma$-ray assay is sensitive to the decay products of $^{238}$U and $^{232}$Th as well as $^{40}$K and $^{60}$Co.
Table~\ref{tab:gamma} lists the sensitivities given as a \U\ and \Th\ equivalent (assuming secular equilibrium in the chains).
The most sensitive instruments are operated at LNGS by the Max Planck Institute for Nuclear Physics in Heidelberg, reaching sensitivities of 10 \mubq/kg~\cite{NEDER:2000}.}

\begin{table}
\centering
\caption{\label{tab:gamma} \new{Example $\gamma$-ray spectrometers operated by institutions or facilities affiliated with the \LEG\ project.}}
\begin{tabularx}{0.86\textwidth}{ L{0.1\tw} L{0.12\tw} Y Z Z L{0.12\tw}}
\rowcolor{legendgrey}
{\bf Detector} &
{\bf Location} &
{\bf Crystal type} &
{\bf $^{238}$U Sensitivity} &
{\bf $^{232}$Th Sensitivity} &
{\bf Owner}  \\
\rowcolor{legendgrey}
& & & [mBq/kg] & [mBq/kg] & \\
\hline
Morgan   &  SURF (USA) 4300 m.w.e. & 2.1 kg p-type &  0.2 & 0.2 & Berkeley  \\
Maeve    & SURF (USA) 4300 m.w.e. & 1.7 kg p-type &  0.1 & 0.1 & Berkeley  \\
Mordred  &  SURF (USA) 4300 m.w.e. & 1.4 kg p-type &  0.7 & 0.7 & USD/Berkeley  \\
Various & LNGS (IT) 3500 m.w.e. & various &  0.01--0.1 & 0.01--0.1 & LNGS/MPIK-HD/UHZ    \\
Various & Bulby (UK) 2805 m.w.e. & various &   1 &    1 &          UCL        \\
Obelix   & LSM (FR) 4800 m.w.e.& various &  0.1     & 0.1 &     JINR, IEAP, CTU \\
Various & HADES (BE) 500 m.w.e. & various & 0.1 & 0.1 &                JRC Geel        \\
\end{tabularx}
\end{table}

\paragraph{Neutron Activation Analysis}
\new{For materials such as hydrocarbons with no long-lived neutron activation products, instrumental neutron activation analysis (NAA) can achieve substantially greater sensitivity than direct $\gamma$-ray counting. In this technique, samples are irradiated with neutrons from a nuclear reactor. When the neutrons are captured on $^{238}$U/$^{232}$Th, the $^{239}$Np/$^{233}$Pa isotopes are generated. After irradiation, the samples are counted by $\gamma$-ray spectrometers to search for characteristic lines at 106 keV and 312 keV from decays of $^{239}$Np and $^{233}$Pa, respectively. Using known or calibrated neutron capture probabilities, neutron flux, irradiation time, $\gamma$-ray detector efficiencies, and sample mass, the concentrations of \U\ and \Th\ in a sample can be calculated.
In Table~\ref{tab:ann}, facilities suitable for NAA and available to \LEG\ are listed.}

\begin{table}
\centering
\caption{\label{tab:ann} \new{NAA facilities available for \LEG.}}
\begin{tabularx}{0.9\textwidth}{ X Z Z Z Z}
\rowcolor{legendgrey}
{\bf Location} &
{\bf Name}   &
{\bf Neutron Flux}   &
{\bf $^{238}$U Sensitivity}  &
{\bf $^{232}$Th Sensitivity}  \\
\rowcolor{legendgrey}
& {\bf (Power)}   & n/cm$^2$/s    & [ppt]  & [ppt]  \\ \hline
ORNL (USA)  &  HFIR (85 MW)  & 4.0$\times$10$^{14}$ &  0.4 & 0.025       \\
UC-Davis (USA) &  MNRC  (2 MW) & 1.5$\times$10$^{13}$ &  0.4 & 0.09       \\
NC (USA)      &  PULSTAR (1 MW) & 6.0$\times$10$^{12}$ &  2.8 & 0.8       \\
TUM (DE)  &  FRM II (20 MW) & 1.0$\times$10$^{14}$ &  0.1 & 0.1       \\
Pavia (IT) &  TRIGA Mark II  & - &  0.1--1 & 0.1--1       \\
\end{tabularx}
\end{table}

\paragraph{Surface Assay}
\new{Exposure to environmental Rn during fabrication, assembly, and installation of a low-background system can lead to buildup of $^{210}$Pb~on surfaces. $^{210}$Pb, which has a 22-yr half-life, will act as an approximately constant source of radiation (from self decays and from decays of $^{210}$Bi and $^{210}$Po) throughout the full life of an experiment.
Sensitive surface and bulk assay of $^{210}$Po can be carried out within the \LEG\ collaboration using XIA\footnote{XIA LLC, Oakland, CA, USA; \url{https://xia.com/}} UltraLo-1800 large-surface low-background $\alpha$ spectrometers.
 Large-surface spectrometers allows for samples to be investigated with respect to their natural bulk and surface contamination with bulk $^{210}$Po sensitivities down to about 50~mBq/kg~\cite{Zuzel:2017} and surface contamination as low as 1~mBq/m$^2$.
The spectrometer also allows investigation of the efficiencies of various surface cleaning procedures for the removal of long-lived $^{222}$Rn progeny~\cite{Zuzel:2018fzl}, while measurements from higher specific activities ($^{222}$Rn-spiked samples) may be studied using low-background semiconductor $\alpha$ particle spectrometers~\cite{Christofferson:2019, Guiseppe:2017yah, Zuzel:2012a, Zuzel:2012b}.}

\paragraph{Radon Emanation}
\new{Rn emanation from detector components represents an additional pathway of backgrounds in ultra-sensitive experiments and one considered for \Lk.
One of the most sensitive methods of Rn detection is based on application of ultra-low-background proportional counters, which in combination with
Rn pre-concentration allow for detection of activities as low as 30\,\mubq~\cite{Rau:2000, Zuzel:2009, Zuzel:2005hag}.
Other systems dedicated to Rn emanation studies are based on electrostatic detectors with detection limits in the range of 1 mBq.}

%% file: sec_technical/subsec_activities.tex
\subsubsection{\Lk\ Component Activities}\label{sec:activities}

The candidate materials used to construct the \Lk\ experiment are selected based on their functional requirements and a certification of an acceptable purity levels. Material screening is performed by the collaboration using the methods and facilities described in Sect.~\ref{sec:radiopurity_assessment}. The reference materials initially certified for use in \Lk\ are listed in Table \ref{tab:simulatedActivities}, which also states the material screening method used to qualify the component. Many of the candidate materials were used by either the \MJD\ or \Gerda\ experiments, though a continued pursuit of cleaner materials remains. These measured activity levels and component masses are used as an input to the background projection effort discussed in Sect.~\ref{subsec:background-budget}.

\begin{table}[t]
\centering
\caption{
   Activities of the components and materials considered in the \Lk\ background projections.  Masses and activities listed correspond to a 4-module, 1000-kg \Lthou\ baseline design.
   Activities are based on ICP-MS (i), neutron activation (n), or direct gamma counting (g) measurements, as indicated, and the assumption of secular equilibrium where necessary.
}
\label{tab:simulatedActivities}
\footnotesize
\begin{tabularx}{1.0\tw}{L{1.4in} C{0.6in} U Y U Y U Y}
\rowcolor{legendgrey}
\thead[l]{Component}
& \thead{Isotope }
& \multicolumn{2}{c}{\textbf{Estimated Activity}}
& \multicolumn{2}{c}{\textbf{Quantity}}
& \multicolumn{2}{c}{\textbf{Total Activity}} \\\hline
\multirow{2}{*}{	Cabling - HV	} &	$^{238}$U	&$(	107	\pm	45	)$&	\mubq/kg (i)	&	\multirow{2}{*}{	0.2	} &	\multirow{2}{*}{	kg}	&$(	20	\pm	9	)$&	\mubq	\\
		&	$^{232}$Th	&$(	81	\pm	57	)$&	\mubq/kg (i)	&			&			&$(	15	\pm	11	)$&	\mubq	\\
\hline
\multirow{2}{*}{	Cabling -- Signal	} &	$^{238}$U	&$(	107	\pm	45	)$&	\mubq/kg (i)	&	\multirow{2}{*}{	0.2	} &	\multirow{2}{*}{	kg}	&$(	24	\pm	10	)$&	\mubq	\\
		&	$^{232}$Th	&$(	81	\pm	57	)$&	\mubq/kg (i)	&			&			&$(	18	\pm	13	)$&	\mubq	\\
\hline
\multirow{2}{*}{	Detector Mount (EFCu)	} &	$^{238}$U	&$(	0.97	\pm	0.48	)$&	\mubq/kg (i)	&	\multirow{2}{*}{	16	} &	\multirow{2}{*}{	kg}	&$(	16	\pm	8	)$&	\mubq	\\
		&	$^{232}$Th	&$(	0.70	\pm	0.32	)$&	\mubq/kg (i)	&			&			&$(	11	\pm	5	)$&	\mubq	\\
\hline
\multirow{2}{*}{	Detector Mount (\ultem)	} &	$^{238}$U	&$(	91	\pm	30	)$&	\mubq/kg (g)	&	\multirow{2}{*}{	0.4	} &	\multirow{2}{*}{	kg}	&$(	37	\pm	12	)$&	\mubq	\\
		&	$^{232}$Th	&$(	130	\pm	30	)$&	\mubq/kg (g)	&			&			&$(	53	\pm	12	)$&	\mubq	\\
\hline
\multirow{2}{*}{	Optical Fibers	} &	$^{238}$U	&$(	75	\pm	25	)$&	\mubq/kg (i)	&	\multirow{2}{*}{	15.0	} &	\multirow{2}{*}{	kg}	&$(	1.1	\pm	0.4	)$&	mBq	\\
		&	$^{232}$Th	&$(	93	\pm	28	)$&	\mubq/kg (i)	&			&			&$(	1.4	\pm	0.4	)$&	mBq	\\
\hline
\multirow{2}{*}{	Front End ASIC	} &	$^{238}$U	&$		<	12	$&	mBq/kg (g)	&	\multirow{2}{*}{	0.13	} &	\multirow{2}{*}{	g}	&$		<	1.61	$&	\mubq	\\
		&	$^{232}$Th	&$		<	8.1	$&	mBq/kg (g)	&			&			&$		<	1.05	$&	\mubq	\\
\hline
\multirow{2}{*}{	PEN Plates	} &	$^{238}$U	&$(	60	\pm	15	)$&	\mubq/kg (g)	&	\multirow{2}{*}{	2.0	} &	\multirow{2}{*}{	kg}	&$(	118	\pm	29	)$&	\mubq	\\
		&	$^{232}$Th	&$(	92	\pm	25.0	)$&	\mubq/kg (g)	&			&			&$(	181	\pm	49.2	)$&	\mubq	\\
\hline
\multirow{2}{*}{	HV Connector (\ultem)	} &	$^{238}$U	&$(	91	\pm	30	)$&	\mubq/kg (g)	&	\multirow{2}{*}{	0.04	} &	\multirow{2}{*}{	kg}	&$(	4	\pm	1	)$&	\mubq	\\
		&	$^{232}$Th	&$(	130	\pm	30	)$&	\mubq/kg (g)	&			&			&$(	6	\pm	1.3	)$&	\mubq	\\
\hline
\multirow{2}{*}{	HV Connector (Ph-Br)	} &	$^{238}$U	&$(	134	\pm	4	)$&	\mubq/kg (i)	&	\multirow{2}{*}{	0.10	} &	\multirow{2}{*}{	kg}	&$(	12.8	\pm	0.4	)$&	\mubq	\\
		&	$^{232}$Th	&$(	25	\pm	17	)$&	\mubq/kg (i)	&			&			&$(	2.3	\pm	1.6	)$&	\mubq	\\
\hline
\multirow{2}{*}{	FE Mount (\ultem)	} &	$^{238}$U	&$(	91	\pm	30	)$&	\mubq/kg (g)	&	\multirow{2}{*}{	0.08	} &	\multirow{2}{*}{	kg}	&$(	7	\pm	2	)$&	\mubq	\\
		&	$^{232}$Th	&$(	130	\pm	30	)$&	\mubq/kg (g)	&			&			&$(	11	\pm	2	)$&	\mubq	\\
\hline
\multirow{2}{*}{	FE Mount (Ph-Br)	} &	$^{238}$U	&$(	134	\pm	4	)$&	\mubq/kg (i)	&	\multirow{2}{*}{	0.10	} &	\multirow{2}{*}{	kg}	&$(	12.8	\pm	0.4	)$&	\mubq	\\
		&	$^{232}$Th	&$(	25	\pm	17	)$&	\mubq/kg (i)	&			&			&$(	2.3	\pm	1.6	)$&	\mubq	\\
\hline
\multirow{2}{*}{	CAPs\footnote{Cable Attachment Positions: regions above each detector array with cabling interconnects, SiPMs, and support structures.} (Various)	} &	$^{238}$U	&$(	420	\pm	105	)$&	\mubq~ea. (g)	&	\multirow{2}{*}{	4	} &	\multirow{2}{*}{	units}	&$(	1.7	\pm	0.4	)$&	mBq	\\
		&	$^{232}$Th	&$(	178	\pm	45	)$&	\mubq~ea. (g)	&			&			&$(	0.7	\pm	0.2	)$&	mBq	\\
\hline
\multirow{2}{*}{	Re-entrant Vessels (EFCu)	} &	$^{238}$U	&$(	0.97	\pm	0.48	)$&	\mubq/kg (i)	&	\multirow{2}{*}{	800	} &	\multirow{2}{*}{	kg}	&$(	0.77	\pm	0.38	)$&	mBq	\\
		&	$^{232}$Th	&$(	0.70	\pm	0.32	)$&	\mubq/kg (i)	&			&			&$(	0.56	\pm	0.25	)$&	mBq	\\
\hline
\multirow{2}{*}{	Reflectors (Tetratex/TPB)	} &	$^{238}$U	&$(	150	\pm	30	)$&	\mubq/m$^2$ (g)	&	\multirow{2}{*}{	10	} &	\multirow{2}{*}{	m$^2$}	&$(	1.46	\pm	0.29	)$&	mBq	\\
		&	$^{232}$Th	&$(	70	\pm	14	)$&	\mubq/m$^2$ (g)	&			&			&$(	0.68	\pm	0.14	)$&	mBq	\\
\end{tabularx}
\end{table}

%% file: sec_technical/subsec_data-analysis.tex
\subsubsection{Software Framework and Analysis Routines}
\label{subsubsec:analyframework}

The primary data-processing and analysis routines will follow the same procedures
implemented in the \Gerda\ and \MJD\ experiments, with some low-level
improvements implemented for \Ltwo. Detector signals recorded by the DAQ will be
converted into an on-disk format based on
HDF5~\footnote{HDF5$^\textrm{\textregistered}$ Group,
 \url{http://www.hdfgroup.org/solutions/hdf5}},
 a robust, portable, scientific data
format designed for supercomputing applications with large amounts of data. The
digital-signal processing (DSP) routines will be implemented using tools and
high-performance standard libraries written in
Python\footnote{Python Software Foundation, \url{https://www.python.org}},
NumPy\footnote{NumPy, \url{https://numpy.org}},
Numba\footnote{Numba, \url{http://numba.pydata.org/}},
and Julia\footnote{Julia Project, \url{https://julialang.org}}
and will incorporate parallel-processing capabilities.
The chief DSP routines will include:
\begin{myitemize}
  \item Digitizer ADC nonlinearity correction based on integral-nonlinearity
  lookup tables~\cite{Abgrall:2020jto}.
  \item Deconvolution of the high-pass components of the electronics response function with multi-pole-zero corrections.
  \item Offline re-triggering to optimally determine pulse onset times and pileup cases with trapezoidal filters~\cite{1994NIMPA.345..337J}.
  \item Optimal energy reconstruction algorithms, including fixed-time pickoff trapezoidal~\cite{Alvis:2019sil} and Zero-Area Cusp filters~\cite{gerda:2015:zac, Agostini:2015pta}.
  \item Optimized charge-trapping estimation~\cite{Alvis:2019sil},
  detection of delayed-charge collection~\cite{Arnquist:2020veq}, and slow-pulse identification~\cite{Abgrall:2016tnn} with additional fixed-time pickoff trapezoidal filters and optimized kernel convolutions.
  \item Determination of the interaction-site multiplicity with smooth
  derivatives or similar filters with optimized de-noising and upsampling techniques~\cite{Budjas:2009zu, Alvis:2019dzt}.
  \item Data cleaning based on characterization and monitoring of noise and baseline leakage current as well as discrimination of physics pulses from triggers on baseline fluctuations, transients, and other non-physics sources~\cite{Agostini:2011mh}.
\end{myitemize}

A strict blinded analysis policy will be implemented. All events
with energy falling within a wide energy window around \qval\ will be identified
through a very rough energy estimation performed by the DAQ hardware.
All evidence of the
existence of those events will be removed from the blinded data files, while the
original files including those events are stored in a protected area to which collaborators do not have
access prior to unblinding at the end of the analysis. Data will be stored and
synced among computing centers and processed in Docker\footnote{Docker.com, \url{https://www.docker.com/}}
virtual environments,
ensuring long-term data preservation and
redundancy.

The data will be cleaned of non-physical events and data quality assessed
through careful monitoring of multiple parameters.
Weekly calibration runs will be used to optimize the DSP routines and also to
tune parameters derived from the DSP and assess their uncertainties. The
associated metadata will be managed using light-weight database tools centered around
JSON\footnote{JSON.org, \url{https://www.json.org}}
file repositories managed using
Git.\footnote{GIT Project, \url{https://git-scm.com}}
The high-level data are built into event structures that can be used to identify
coincidences and perform multivariate spectral analyses. Advanced techniques
based on machine learning may be implemented at several steps in this process.

\subsubsection{Multivariate Analysis}\label{subsec:background-reduction}

As described in Sect.~\ref{subsec:topology}, in addition to the all-important energy
variable, the \Lk\ detector provides multiple extra analysis handles for separating
signal from background, based on pulse shape analysis, detector coincidences, and timing
information.  In this section we discuss in more detail how each of these
characteristics are used in the analysis.

\paragraph{Energy}
Energy is estimated for every Ge-detector signal using optimized filters
as described in Sect.~\ref{subsubsec:analyframework}. In detectors showing significant charge
trapping, a small correction is applied using the DSP parameters characterizing
the amplitude of the charge collection or the correlated presence of delayed
charge.  After this correction,
the energies have a variance that is dominated by the Fano contribution, and
the detector response function is very well characterized by the dominantly
Gaussian function described in Sect.~\ref{subsec:energy-res}.
The energy scale and resolution of each detector is extracted from weekly
calibration data and monitored regularly by injecting test pulses, permitting
spectral analyses in which each event is treated independently
and comes with specific systematic uncertainties~\cite{Agostini:2021duc}. Such
analyses are powerful for identifying a rare signal with a handful of events.
High-dimensionality spectral fits treating each detector independently also
provide a strong handle on identifying and characterizing backgrounds
with spatial variations and will be pursued in the \Lk\ background analysis.

\paragraph{Timing}
Event-timing information is obtained essentially without effort from the DAQ
system, since the DAQ records detector signals with sub-\mus\ triggering precision.
Within the recorded traces, pulse-shape analysis
can further extract signal start times with precision on the order of 10~ns.
This information allows us to tag time-correlated background events as well as
properly treat background contributions with a time-varying
rate.
Experience from \Gerda\ and the
\MJD\ indicate that some backgrounds, such as Bi-Po coincidences in the U/Th
decay chains or rapid decays of cosmogenic activation products generated
in-situ, can be identified and removed using simple cuts with high efficiency
and negligible live time impact~\cite{Csathy:2016wdy}.
The independent detection of neutron capture on \nuc{40}{Ar} allows an
improvement to the tagging efficiency for muon events with accompanying isotope
production, leading to reduced muon-induced background as discussed in
Sect.~\ref{subsubsec:muInduced} and based on Ref.~\cite{Wiesinger:2018qxt}.
An example of a background with a time-varying rate
is $^{68}$Ga events, the decay daughter of $^{68}$Ge which is generated by cosmogenic activation of the detectors during manufacturing and shipping.
 An analysis accounting for the exponentially decreasing rate of $^{68}$Ga events
following the half-life of $^{68}$Ge allows for the separation of this
background from a potential signal.
$^{68}$Ga background is also preceded within a few hours by an x-ray from the decay of
$^{68}$Ge. When such a candidate X-ray is identified, subsequent events in the
same detector can be vetoed for several half-lives of the $^{68}$Ga decay,
potentially achieving a factor of 5 to 10 reduction in this background source.
The live time loss due to such a veto can be optimized based on the rate of
$^{68}$Ge decay x-rays observed in the data.

\paragraph{Detector Coincidences}
Event coincidences are also obtained without effort from
the DAQ system, which records every signal above threshold occurring within a
short ($\sim$10~\mus) time window in the Ge detectors, the LAr SiPMs, and the PMTs in
the water tank. The \LEG\ design purposefully minimizes the use of opaque/passive materials, creating a local environment in which coincident depositions are almost always detected.  The tagging of $\gamma$ rays using
coincidences is significantly enhanced if additional energy from the parent
decay ($\alpha$ or $\beta$ component) is deposited in another detector
component. This is the case for decays inside the LAr, the optical fibers, the
PEN holders, or even a cosmogenic background in another Ge detector.  On the
other hand, surface contaminations of $\alpha$ and $\beta$ emitters generate little
energy deposition in the LAr and do so in a region that is well-shadowed,
thus exhibiting fewer detectable coincidences.

Since \BBz\ decay events are isolated to
single Ge crystals, it is almost always optimal to leverage coincidence information in
the analysis in the form of an anti-coincidence cut, since essentially 100\% of
the coincidences arise due to backgrounds.
The performance of the LAr veto cut has been successfully demonstrated in \Gerda\ following the same design concept (see Fig.~\ref{fig:BackgroundSpectraGerda} and \ref{fig:classifier-example}).
As an example, a more than three orders of magnitude suppression was achieved for a
\nuc{228}{Th} source in Ref.~\cite{Agostini:2015boa} using a simple cut-based
analysis.  However, amplitude and hit
pattern information in those coincidences can also be used in a full multivariate
analysis to help identify, isolate, and constrain those background
contributions, and this will be pursued in the full \Lk\ background
analysis.

\paragraph{Ge detector topology reconstruction}
\begin{figure}
\includegraphics[width=0.85\columnwidth]{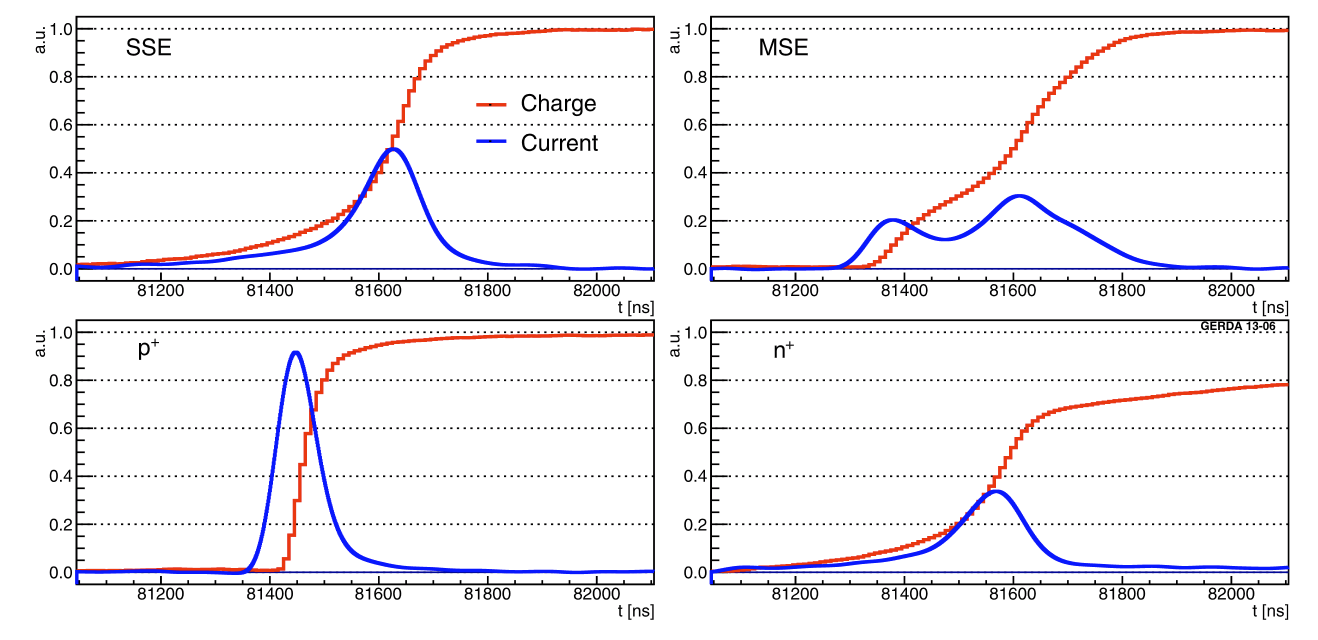}
\caption{
Charge (red) and current (blue) pulse shapes for different event topologies: $0\nu\beta\beta$-like single site event (top left), $\gamma$-like multi-site event (top right), $\alpha$-like \pPlus\ contact event (bottom left), and \nuc{42}{K}-like \nPlus\ contact event. Figure from Ref.~\cite{gerda:2013:psd}.}
\label{fig:pulseShapeTopologies}
\end{figure}

Ge detectors are effectively solid-state time projection chambers. Clusters of electrons and holes
are created when energy is deposited in the Ge material and then collected
through an electric field. The small size of the readout electrode of the ICPC
detectors ensures that a strong current signal is induced as the holes
approach it. Events with a single-site energy deposition such as those expected
by \BBz\ decays will generate a single-peak current signal. Background events
due to multi-site Compton-scattered $\gamma$ rays will instead exhibit a
multi-peak structure. These two typologies of signals are shown on the left and
right top panels of Fig.~\ref{fig:pulseShapeTopologies}, respectively. The
ns-scale time-resolution allows the resolution of current peaks generated by
clusters which are created only a millimeter apart.
A highly sensitive parameter to discriminate single- from multi-site events
is the maximum current of the pulse.
The maximum current $A$ is easily computed from a smoothed derivative of the
digitized waveforms. Then, either the $A$ variable can be normalized by
energy (as in the case of the $A/E$ parameter used by
\Gerda~\cite{Budjas:2009zu}) or its energy variation can be treated more
generically (with the $AvsE$ parameter used by \MJ~\cite{Alvis:2019dzt}).
Suppression of $A/E$ or $AvsE$ relative to the value typical for single-site
events thus indicates the presence and degree of
multiple interaction sites in an event.

\begin{figure}
\includegraphics[width=0.9\columnwidth]{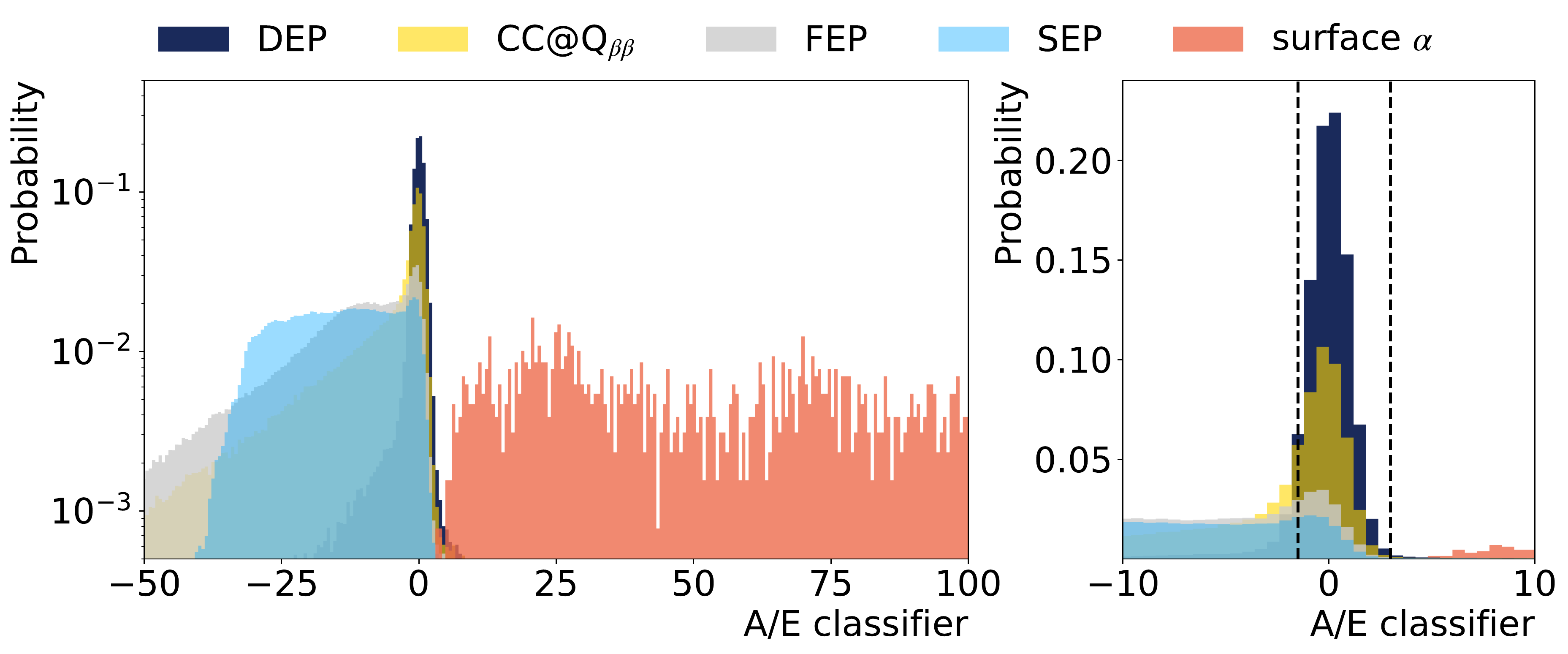}
\caption{$A/E$ distributions for a selection of event samples from BEGe-detector \nuc{228}{Th}-calibration and physics datasets of \Gerda\ Phase\,II. The samples are taken from the double-escape peak (DEP), Compton-continuum at 2 MeV (CC), full-energy peak (FEP) at 2.6~MeV, single-escape peak (SEP), and surface alpha decays with energy above 3.5~MeV.
About 90\% of \BBz\ decay events are expected to fall in the $A/E$ window between the dotted vertical lines.}
\label{fig:aepdfs}
\end{figure}

Figure~\ref{fig:aepdfs} shows the distributions of the $A/E$
parameter for a selection of event samples. Double-escape peak events are mostly
single-site and serve as a good proxy for \BBz\ decay.
Compton-continuum events include a significant component of multi-site events,
while full-energy peak events are mostly multi-site. These stark
differences in $A/E$ spectra can be leveraged in multi-variate analyses to
discriminate signal from background and strongly constrain possible background
sources. A simple cut in $A/E$, for example, can reject $\sim$90\% of the events
in full-energy gamma lines and $>$95\% of multi-site single-escape events while retaining $\sim$90\% of \BBz-decay-like events~\cite{Budjas:2009zu,Alvis:2019dzt}.

The maximum current can also be used to identify events resulting from energy
depositions on the detector surface. These events can be divided into
two groups. Energy depositions on the Li-diffused electrode that covers most of
the detector results in a cluster of holes that undergoes a \mus-scale
diffusion process in which only
a fraction of the holes produced reaches the active volume over an extended timescale.
The result is a mm-scale elongation of the hole cluster, and, in turn, a
broader current peak and hence lower value of $A$ compared to point-like
clusters in the Ge bulk.
Conversely, energy depositions on the readout electrode are characterized by a
signal driven by the cluster of electrons that moves twice as fast as the
holes. This creates a stronger current peak. Both topologies of signals are
shown in the bottom panels of Fig.~\ref{fig:pulseShapeTopologies}, and the
distribution for $\alpha$ events in \Gerda\ is shown in Fig.~\ref{fig:aepdfs}.
Pulse-shape discrimination of these surface events based on the $A/E$ method
is at the level of $\sim 99\%$ or better.

Although the $A/E$ method alone is sufficient to reach the \Lk\ background goals,
additional improvements can be made using information in the waveform beyond the
current peak.
Delayed-charge collection occurs at low levels in every pulse due to
the slow release of charges from trapping sites populated by the primary charge
cloud during its drift. In some detectors, this signal is strong enough to
provide an estimate of the total amount of trapped charge in the event, which
then can be used to correct event energies for charge-trapping losses, improving
resolution.
A stronger source of charge trapping is the detector passivated surface, where
trapping sites like charge impurities and crystal defects are abundant.
A ``delayed charge recovery'' (DCR) parameter able to identify the faint current
induced by these delayed charges was successfully used in
the \MJD\ to identify and remove $\alpha$-decay events on the surface of \ppc\ detectors
which have larger passivated surfaces~\cite{Arnquist:2020veq}. Such discrimination
could be applied in \Lk\ to achieve further background suppression beyond what is
assumed for our background model.

It is also possible to extract even more detailed interaction-location information
from the Ge pulses.  The leading edge of a pulse records the
history of how the drifting charges traversed the detector's weighting
potential. This information can be inverted to reconstruct the location of the
interaction site from which the drift started.
Initial studies
with \MJD\ \ppc\ detectors indicate that interaction sites can be localized with
sub-mm resolution in distance from the point contact and 1-mm resolution in
the transverse radial direction.
Although we do not require these methods to reach \Lk's
background goal, we plan to explore these options further to help gain further sensitivity out of the \Lk\ data.

\subsubsection{Efficiencies and Systematic Uncertainties}\label{subsubsec:eff}

As mentioned above, the \Lk\ detectors provide an abundance of event-topology information
beyond just the event energy.
However,
for ultra-faint signal levels at the very edge of the experiment's reach, the
signal will not have sufficient statistics to be constrained by any analysis
variable besides energy. All of the signal sensitivity then comes from the
relative count of events within a FWHM or so of \qval\ with respect to the rate of background
events in surrounding energies, and the primary effect of the other analysis
variables is the assignment of low or high signal likelihood to those events.
Thus, the ultimate sensitivity of \Lk's multivariate analysis can be well
approximated by a simple analysis based on the application of cuts in each
variable. We adopt this approach in planning purposes for \LEG, as it provides
a straightforward and efficient means of computing the background impact of
design changes and their ultimate effect on sensitivity. This in turn leads to
lower risk, as it decouples our sensitivity estimates from our ability to
maximize a high-dimensionality likelihood function with a very low statistics
dataset, which itself is a significant challenge~\cite{Buuck2019}.

For \Lk\ background and sensitivity estimates, we assume just two basic cuts, as
introduced in Sect.~\ref{subsec:topology}.  The first is an AC cut,
in which events which generate any coincident signal in the LAr, the water veto, or in
multiple Ge detectors are rejected as background.  The second is a
PSD cut, in which a cut on the $A/E$ parameter
is used to remove multi-site events and surface
events from $\alpha$ and $\beta$ decays.
Although time-varying backgrounds, time coincidence searches, delayed charge
identification, and ionization localization will
certainly be pursued in the full \Lk\ analysis to achieve the lowest
ultimate background and maximize sensitivity, in this document we
conservatively do not rely on any of these variables for removing background.
This makes our background estimates conservative with known room for improvement.

Of the two primary cuts, only the PSD cut has a significant impact on the
signal efficiency.  Double-escape peak (DEP) events from the \nuc{208}{Tl}
2614-keV $\gamma$
ray from calibration-source data can be used as signal proxies to extract the
peak in $A/E$ for single-site events. The cut is then tuned to keep a large
fraction of the DEP peak, with 90\% being very close to optimal, since this is
roughly the fraction of \BBz\ decay events that do not generate bremsstrahlung photons
that lead to a multi-site topology nor lie too close to the point contact so as
to generate a sharp surface-like signal. The dominant multi-site background
near \qval\ that is removed by this cut is from Compton scatters of
higher-energy $\gamma$ rays, which are reduced by a factor of 2--3.
Surface $\alpha$ decays are very strongly discriminated by this cut.
All \nuc{210}{Po} events observed in \Gerda\ were removed by the PSD cut, and an upper limit
for their survival probability of 0.16\% (90\% CL) was established. Slow pulses
from $^{42}$K-decay betas traversing the Li-diffused outer contact are also very
efficiently rejected by this cut with survival fractions at the 1\% level.

For the sensitivity estimates in this document, we assume a 90\% combined
efficiency for the AC and PSD cuts to retain \BBz\ decay signal events. For accuracy
of the simplified Poisson-counting analysis, we also have to account
for the 95\% efficiency of a $\pm$2$\sigma$ energy cut around \qval\ that
optimizes the width of the energy ROI for a quasi-background-free
experiment~\cite{Agostini:2017jim}. The energy cut also effectively removes signal events in
which not all of the energy is collected by the detector. This includes
approximately 8\% of events which originate in the outer 1~mm \nPlus\ contact
and thus suffer significant energy loss, giving an active
volume fraction of 92\%. The energy cut
also removes an additional 8\% of events which originate inside the
active volume but whose emitted $\beta$ particles at least partially traverse the dead
regions, where their energy depositions do not contribute to the signal, or
generate bremsstrahlung photons that leave the active region. This
loss is represented by a 92\% ``containment'' efficiency. With exposure defined
in terms of the total mass of Ge in the experiment, an additional efficiency
factor capturing the level of isotopic enrichment is included.

Systematic uncertainties in this simplified analysis picture arise only in the
estimates of the efficiencies and in the overall exposure of the experiment. The
analysis cut efficiency uncertainty dominates at the few percent level, as
pulse shape simulations must be used to translate the efficiencies observed
in DEP events from $^{208}$Tl to \BBz\ decay events. The latter have a more uniform
spatial distribution in the detector and lie at higher energy, with continuous
individual energy spectra for the emitted $\beta$ particles. Uncertainties in the active
volume fraction and the containment efficiency are at the percent level and are
dominated by uncertainty in measurements of the dead-layer thickness. The
containment efficiency also includes sub-dominant contributions from
imperfections in the Monte Carlo modeling of $\beta$ ranges and bremsstrahlung production.
Uncertainties in the enrichment fraction, and in the livetime estimates and total mass
measurements that go into the exposure calculation, are all well below the percent
level and are negligible. Uncertainties in energy are governed by the
uncertainties in the detector-resolution function shape and its energy and time
dependence as described above, which are typically on the order of 0.1~keV or
less. This translates to a negligible ROI cut efficiency uncertainty.
The total systematic uncertainty is $<5$\% and
is mostly irrelevant unless a strong signal is observed in \Ltwo, in which case \Lk\
would collect hundreds of signal counts. A full multivariate analysis would
result in a modest increase in systematic uncertainty that comes with the benefit of
a small boost in sensitivity.

%% file: sec_technical/subsec_background-budget.tex
\label{subsec:background-budget}

The comprehensive background projection for \Lk\
is derived from high-fidelity Monte Carlo simulations and detailed detector
response models.
A summary of the background projections is given in Sect.~\ref{subsec:projected_bg}, which includes a chart of the expected background contributors shown in Fig.~\ref{fig:BackgroundBudget}.
This section presents a more detailed description of the inputs, tools, analysis methods, and the major background components that lead to the total anticipated background index of \BGprojkev.

\subsubsection{Simulation Inputs and Methods}

Each material considered in constructing the \Lthou\ background model has been assayed for radioactive impurities
(see Sect.~\ref{sec:radiopurity_assessment} for a description of the material
screening techniques), and several of these materials are in use by either the \MJD\ or \Gerda\ experiment.
The lowest-background material appropriate for each purpose has been selected
from among those presently in use, and their activities are summarized in Table
\ref{tab:simulatedActivities} in Sect.~\ref{sec:activities}.

Extensive Monte Carlo simulations of the \Lthou\ design are performed using \MaGe~\cite{Boswell:2010mr}. \MaGe\ is a \GF-based~\cite{Agostinelli:2002hh} simulation framework jointly developed by the \Gerda\ and \MJ\ Collaborations, which has been successfully used in the modeling of the backgrounds in both experiments.
For this report, backgrounds are simulated for the full \Lk\ array based on the baseline technical design.

Radioactive decay backgrounds originating from the following sources were simulated in \MaGe\ based on their activities tabulated in Table~\ref{tab:simulatedActivities}:
\begin{myitemize}
 \item A total of 392 identical enriched Ge detectors with the average mass of 2.6\,kg.
 See Fig.~\ref{fig:baseline} for the arrangement of the Ge detectors in the array.
 \item EFCu and \ultem\ polyetherimide mounting hardware with PEN
 baseplates, and \ultem\ and phosphor bronze electronic and cable mounting hardware
 (see Fig.~\ref{fig:ICPC_sketch}).
 \item Low-mass ASIC-based front-end electronics mounted on a signal readout board (see Fig.~\ref{fig:SignalDetectorUnit}).
 \item Signal and high-voltage cabling spanning the distance between each individual detector and the interconnects at a region called the Cable Attachment Positions (CAPs) above each module's detector array.
 \item Inner and outer optical fiber curtains encompassing the ring of detector strings. See Fig.~\ref{fig:LAr-acive-veto-components} for the arrangement of the fiber curtains.
 \item The CAPs region above each detector array representing cabling interconnects, SiPMs, and support structures.
 \item The Cu reentrant tubes that encapsulate the UGLAr volume. See Fig.~\ref{fig:baseline} for the arrangement of the reentrant tubes.
\end{myitemize}

Radioactive decays are simulated uniformly throughout each of the components and, in some cases, uniformly across component surfaces.
Particle steps that result in energy deposition in active materials (Ge, LAr, PEN, optical fiber) are recorded.
Post-processing is applied to model the observables necessary for analysis
cuts. For this report, we only model the use of AC and
PSD cuts. Other analysis variables such as timing
information will lead to improved sensitivity but are not used here.

As described in Sect.~\ref{subsec:topology}, the AC cut comprises the following: the
granularity cut, which removes events triggering multiple Ge detectors; the LAr
veto cut, which removes events producing scintillation in active materials (LAr,
PEN, optic fibers) that trigger the SiPMs; and the external water Cherenkov muon
veto cut. Events which produce coincidences in the Ge detectors or the SiPMs
within a fixed time window (200~\mus\ in the model presented here) are removed by the AC
cut. The water Cherenkov muon veto is not currently modeled in the
post-processing; this only impacts muon-related events, which are treated
independently.

\begin{figure}[]
      \begin{center}
         \hfill
         \includegraphics[trim=0 150 0 100,clip,height=0.33\tw]{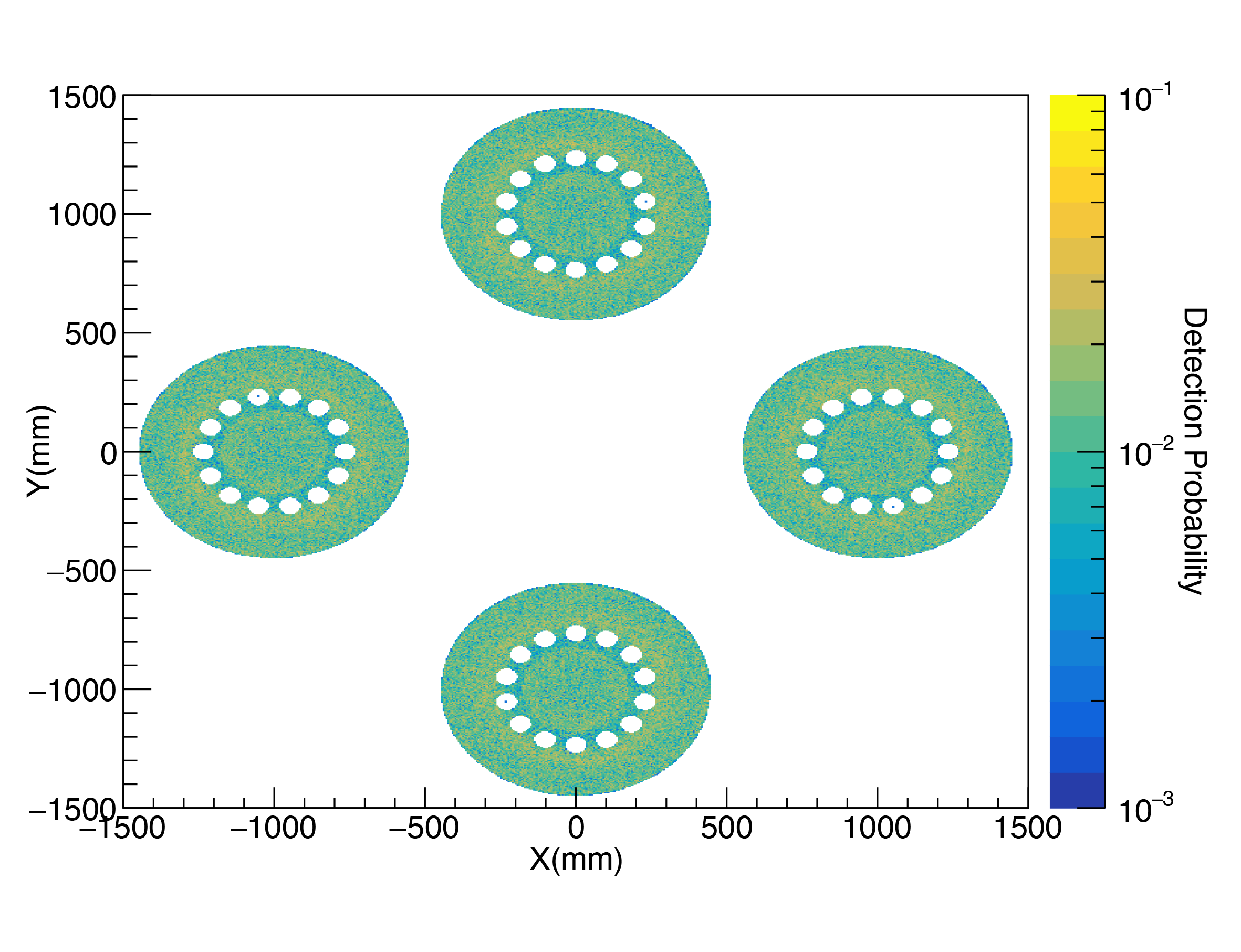}
         \hfill
         \includegraphics[trim=0 80 0 100,clip,height=0.33\tw]{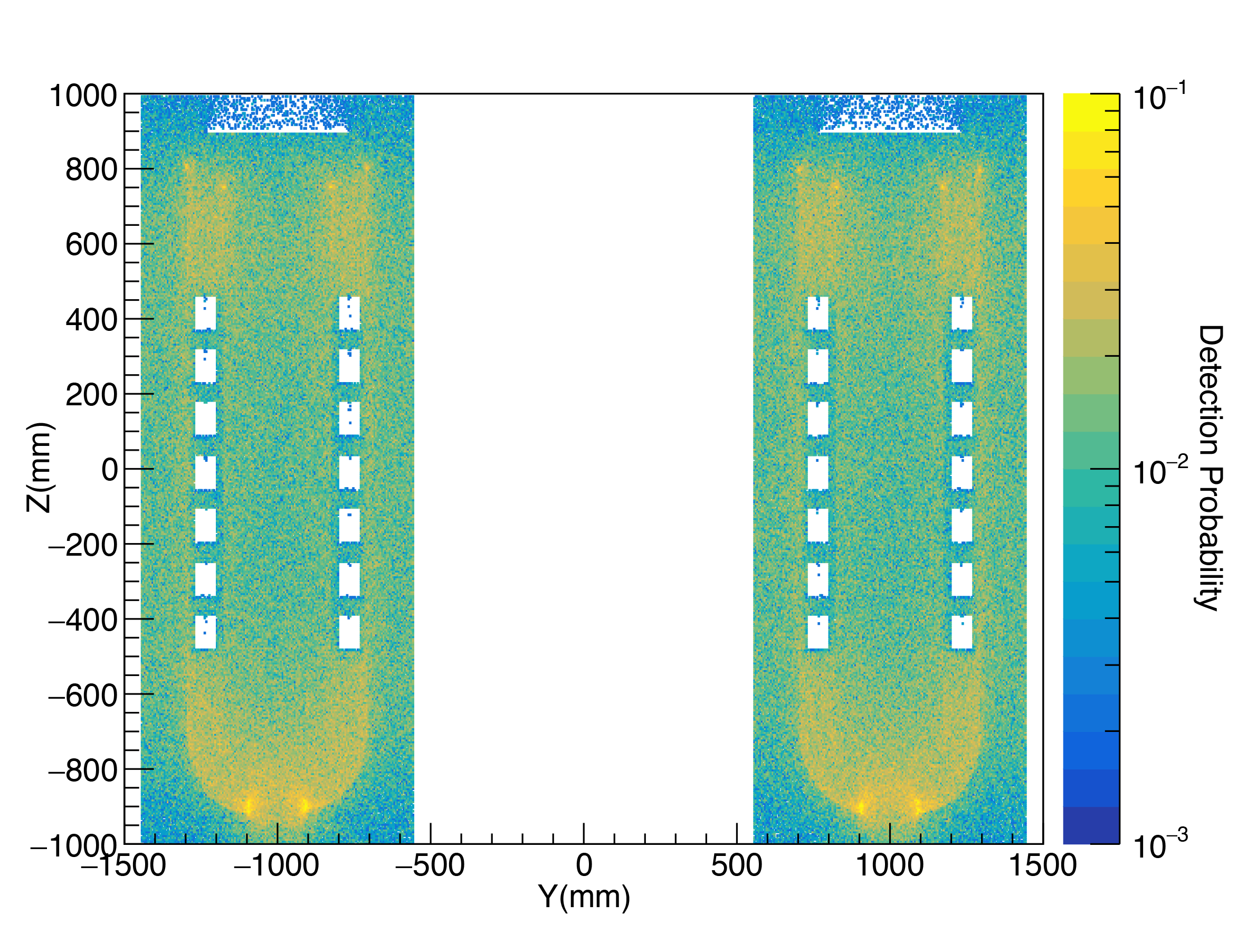}
         ~~~~
      \end{center}
 \caption{Monte Carlo detection-probability map for vacuum ultraviolet Ar scintillation photons generated within the \Lthou\ 14-string modules.  Slices of the 3D probability map in the (left) XY plane at $\text{Z}=0$ and the (right) YZ plane at $\text{X}=0$ indicate an expected detection probability for emitted scintillation photons of approximately 1\% within the arrays, not including the quantum efficiency of SiPM readout devices. Quantum efficiency and scintillation photon yield (a function of Ar purity) are left as parameters adjustable in post-processing. \label{fig:photon_prob_map}}
\end{figure}

The modeling of the granularity cut is achieved by requiring only one Ge
detector to have any non-zero energy deposition after the energy corrections
described below; the result is insensitive to the application of energy
thresholds up to 10~keV.
To accurately model the LAr veto, separate optical simulations have been
performed in \MaGe\ to calculate the conversion probability of a
VUV scintillation photon to the visible spectrum in
wavelength-shifter-coated materials. This is followed by the coupling of optical
photons into fibers and detection in the SiPMs.  For each set of scintillating
materials, a three-dimensional spatial map of photon detection probabilities is
generated (see Fig.~\ref{fig:photon_prob_map} for an example).
For each energy deposition in these materials, the expected number of
photo-electrons generated in the SiPMs is calculated from the scintillation yield of the material (LAr: 20~$\gamma$/keV, PEN: 5.0~$\gamma$/keV, fibers: 8.0~$\gamma$/keV) and the probability of photon detection from the location of the energy deposition in the map.
The expected number of detected photons is then summed for the whole event.
The use of three-dimensional maps to describe the LAr veto cut performance has been
successfully demonstrated in \Gerda\ which follows the same design concept for
the LAr veto and Monte Carlo post-processing~\cite{Wiesinger2021,Pertoldi2021}.

The PSD cut is simulated with a multi-step post-processing procedure.
First, energy depositions within the 1-mm partially-inactive transition region of the \nPlus\ high-voltage
electrode are corrected for reduced charge collection efficiency or removed,
depending on the deposition's distance from the surface.
The remaining energy depositions are summed within each detector, and an energy resolution function is applied to smear the energies.
Energy depositions within each detector are then accumulated into spatially localized clusters.
For each cluster, a charge drift time is calculated using a drift-time map
independently calculated using \Siggen.\footnote{Software for the field and
signal generation in Ge detectors.
\url{https://github.com/radforddc/icpc_siggen}} Clusters that are separated
in drift time by more than an energy-dependent threshold (approximately 25\,ns at
1\,MeV) are rejected as multi-site, simulating the effect of the PSD cut. This
pulse-shape emulation technique has been previously used in the \MJD\ with
reasonable accuracy at the \BBz\ decay energy region.
PSD post-processing of Ge surface interactions has not yet been implemented in
the full \Lk\ Monte Carlo simulations. Rejection efficiencies for $\alpha$ and
$\beta$ surface events are inserted manually based on observations in
characterization data and validations in stand-alone simulations.

\subsubsection{Internal \nuc{238}{U}, \nuc{232}{Th}, and \nuc{40}{K} Contaminations in the Array}\label{subsubsec:UAndTh}

Because of their long half-lives and natural abundances, \nuc{238}{U}, \nuc{232}{Th}, and \nuc{40}{K} of terrestrial origin are expected to be present in trace quantities in the materials used in the fabrication of all experimental components.
The \nuc{238}{U} and \nuc{232}{Th} chains decay through a series of unstable, shorter-lived progeny until arriving at \nuc{206}{Pb} and \nuc{208}{Pb}, respectively.
Of primary concern for \BBz\ decay searches are \nuc{214}{Bi} and \nuc{208}{Tl} decays, each of which produce $\gamma$ rays of sufficiently high energy to Compton scatter near the \Qbb\ region at 2039\,keV.
Although it is not a direct background for \BBz\ decay at 2039\,keV, the
1461-keV $\gamma$ ray from the decay of \nuc{40}{K} can impact the \BBt-decay
spectral analysis and can also hinder the performance of analysis cuts based on
signals at lower energies. This section considers only backgrounds from
\nuc{238}{U}, \nuc{232}{Th}, and \nuc{40}{K} decays occurring in the components
that define each of the detector arrays; external backgrounds from the cryostat walls
and beyond are covered separately in
Sect.~\ref{subsubsec:external}.

For each component group included in the \MaGe\ model, the full \nuc{238}{U}- and \nuc{232}{Th}-decay chains are simulated assuming secular equilibrium, with decays generated uniformly throughout the component volumes.
It is possible, however, that secular equilibrium in the \nuc{238}{U}- and \nuc{232}{Th}-decay chains may be broken. When direct assays of the bottom halves of the chains (Ra and below) are available, they are therefore preferred as more direct measurements of the activities of \nuc{214}{Bi} and \nuc{208}{Tl}, the leading contributors to the background index from these chains.
The efficiency at which a decay generates an event at \qval\ is calculated from the post-processed simulation output and scaled by the material activities and component masses listed in Table \ref{tab:simulatedActivities}.

 The 2.6-kg average detector mass considered in the \Lk\ baseline design reduces
 backgrounds from components that scale with channel count (cables, front-ends,
 plastic insulators).
 Additionally, the adoption of ASIC-based front-end readouts and low-background flex cables for signal and high voltage provide significant background reduction from these components.  Finally, the reduced background from \nuc{42}{Ar}/\nuc{42}{K} (see Sect.~\ref{subsubsec:ar42}) allows for the elimination of nylon shrouds used to limit K-ion drift in \Gerda\ and \Ltwo.

\begin{figure}[]
      \begin{center}
         \hfill
         \includegraphics[height=0.45\tw]{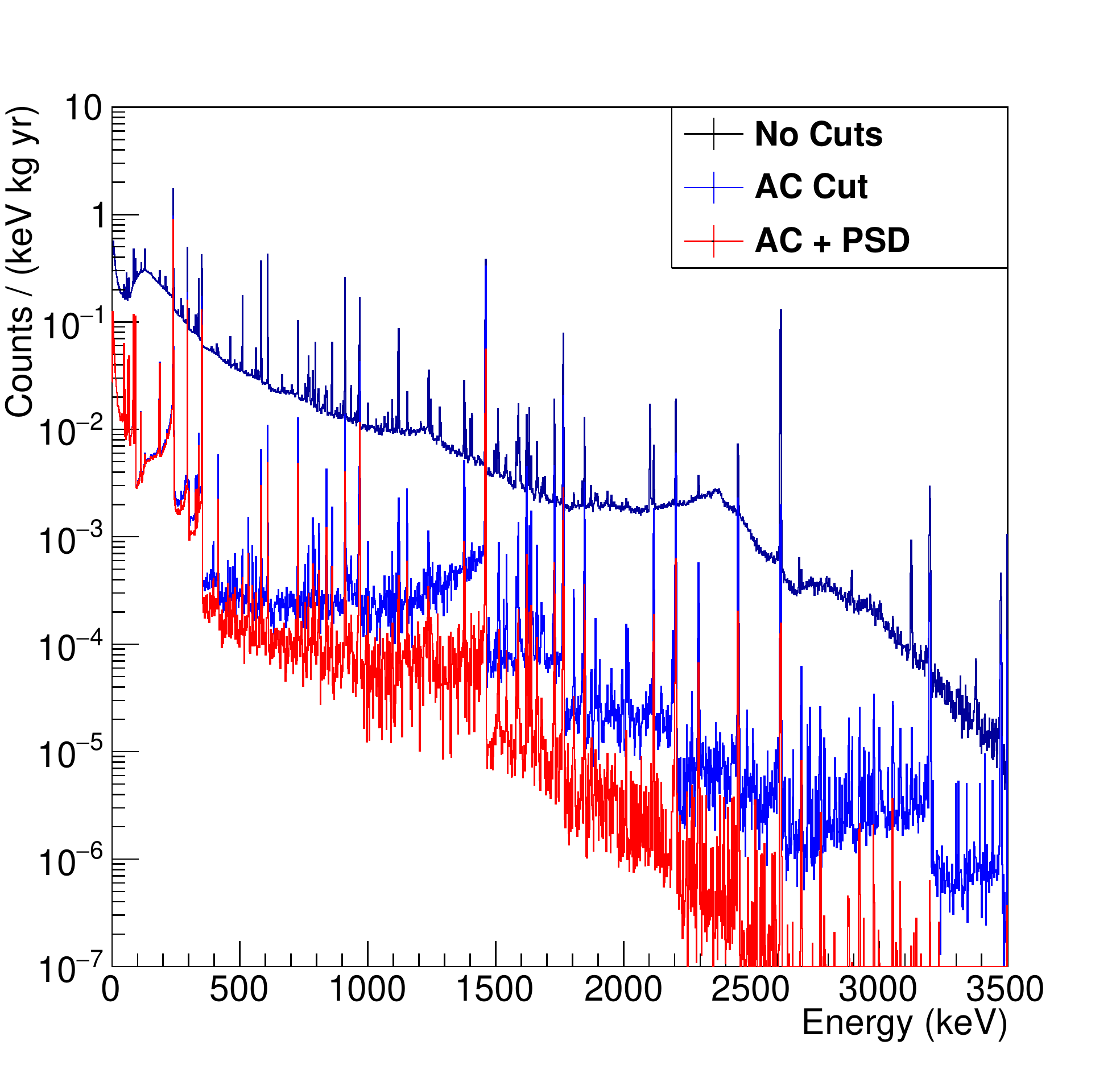}
         \hfill
         \includegraphics[height=0.45\tw]{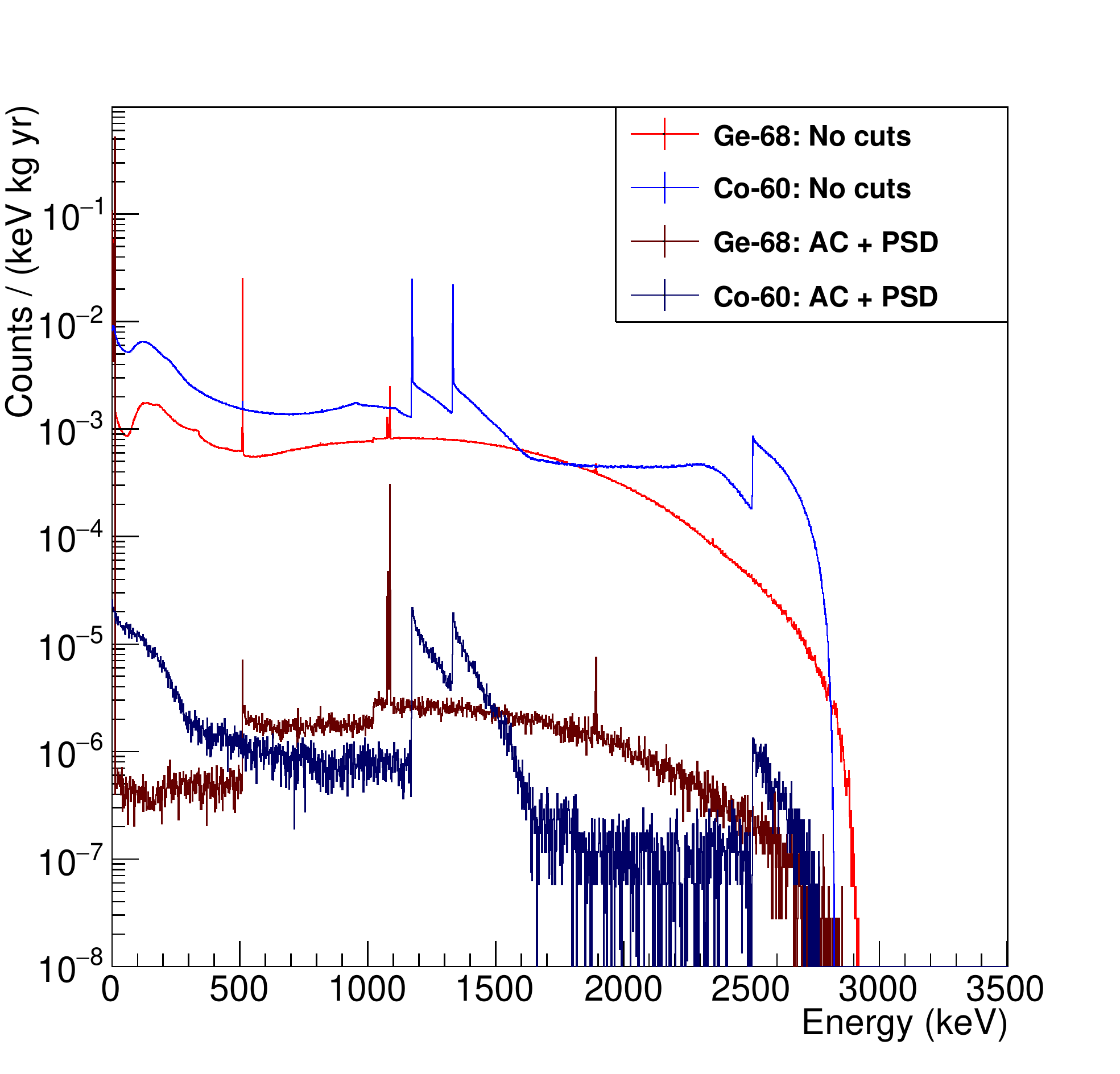}
         \hfill
      \end{center}
 \caption{Left: Expected total contribution of \nuc{238}{U} chain, \nuc{232}{Th}
 chain, and \nuc{40}{K} from the internal array components before cuts (black),
 after applying the AC cut (blue), and after apply the AC and PSD cuts (red).
 After all analysis cuts, the net
 background contribution is \cpowten{(3.0\pm1.7)}{-6}\,\cpKkgy\ in the \Qbb\
 background region.
 Right: Expected background spectra due to cosmogenic production of \nuc{68}{Ge}
 and \nuc{60}{Co} in the enriched detectors before analysis cuts and after
 applying AC and PSD cuts.}
\label{fig:UThK_Cosmo}
\end{figure}

Figure~\ref{fig:UThK_Cosmo} (left) shows the sums of expected
\nuc{238}{U} chain, \nuc{232}{Th} chain, and \nuc{40}{K} backgrounds from all
internal components of the detector array after each analysis cut stage:
Before cuts, after the AC cut, and after applying both the AC cut and the PSD cut.
The combination of analysis cuts provides very strong suppression of both \nuc{238}{U}- and \nuc{232}{Th}-chain events in the \BBz\ decay region.
With \nuc{208}{Tl} decays releasing 5\,MeV of decay energy and \nuc{214}{Bi} releasing 3.2\,MeV, approximately 2\,MeV must be deposited in a single detector with the remainder left in inactive (not Ge, LAr, PEN, or fiber) volumes for those decays to contribute to the background index at \Qbb.
The low-mass design of the array minimizes inactive volumes effectively, ensuring that energy depositions in active regions are highly likely.
The AC cut rejects greater than $99\%$ of \nuc{232}{Th}-chain backgrounds (\nuc{208}{Tl}) and greater than $75\%$~of \nuc{238}{U}-chain backgrounds (\nuc{214}{Bi}), depending on the source volume and location.

For those events that are not rejected by the AC cut, the
PSD cut is particularly effective.  These events have a high likelihood of
multiply scattering within a detector, as only $\gamma$-ray interactions
confined to a single Ge detector survive the AC cut.
The drift-time heuristic PSD emulation predicts a 5--25\%
survival probability for \U-chain backgrounds that pass the AC cut and a 2--20\%
survival probability for \Th-chain backgrounds.
The cuts are less effective in rejecting events from the \ultem\ detector
holders than the other components owing to their direct contact with the
detector surfaces.
The cut performance estimated with the simulations are consistent with the experimental values obtained in \Gerda\ and \MJ.
Not included in the figure, though included in
the overall background index tabulation, are the contributions from \nuc{238}{U}
and \nuc{232}{Th} contamination within the bulk of the Ge detectors. A
preliminary analysis of the full-exposure \Gerda\ dataset has shown no evidence
of \nuc{238}{U} or \nuc{232}{Th} contaminants in the Ge detectors, from which
upper limits of $<1$~nBq/kg in each isotope have been derived. These upper
limits on the \nuc{238}{U}- and \nuc{232}{Th}-chain activities are assumed to
hold for the \Lk\ purified Ge material.

Neutrons can also be generated in \nuc{238}{U} and \nuc{232}{Th} chains due to
($\alpha$,n) reactions and spontaneous fission. Due to the extra-low
radioactivities of all internal components, their contributions are estimated to
be negligible. Components that are more massive and less clean are the dominant
radiogenic neutron contributors in \Lk, but they are far away from the Ge
detectors; their contributions are discussed in Sect.~\ref{subsubsec:external}.

\subsubsection{Cosmogenic Isotopes in Ge}\label{subsubsec:cosmogenics}

Cosmic-ray exposure of the enriched Ge material and fabricated detectors leads to the production of \nuc{60}{Co} and \nuc{68}{Ge} in the detector volume.
Production of these isotopes is mitigated though control of cosmic-ray exposure throughout the production and transportation of the enriched material and detectors, as described in Sect.~\ref{subsec:detector-fabrication}.

At sea level, \nuc{60}{Co} is produced in enriched Ge material at a rate of 2.55\,atoms/(kg\,d)~\cite{Elliott:2009cw}.
The zone-refining and crystal-pulling processes effectively remove the
\nuc{60}{Co} produced in the feedstock, but once detector fabrication is
complete, subsequent cosmic-ray exposure leads to the accumulation of \nuc{60}{Co}.
The $\beta$ decay of \nuc{60}{Co} has a $Q$ value of 2.8\,MeV and \thalf\ of
5.3\,yr, and includes the emission of coincident 1173-keV and 1333-keV
$\gamma$ rays. The combined effect of the two $\gamma$ rays can contribute background at 2039\,keV in the case of one or more Compton scatters leaving partial energy deposition.
The $\beta$ particle with endpoint energy of 318\,keV can also contribute to the background signature, leading to a high-energy tail attached to each of the $\gamma$ peaks.
For background estimation purposes, we assume 20\,d of sea-level cosmic-ray
exposure following the detector fabrication process, followed by a cool-down period underground prior to data taking.
The cosmogenic activity of each detector varies as \Lthou\ ramps up its data taking in phases and operates over 10\,yr to reach
the total exposure goal. Following the baseline schedule for crystal fabrication
and deployment, we compute an array-wide average rate over the data taking
period that is 0.43 times the initial activity level.
The expected background from \nuc{60}{Co} activated in the detectors of \Lthou\
before and after cuts is shown in Fig.~\ref{fig:UThK_Cosmo} (right).
Because the pair of $\gamma$ rays emitted in
\nuc{60}{Co} decay give rise to summed energy depositions at \qval\ leaving
784\,keV of excess event energy, the events are highly multi-site and very
effectively cut by the AC and PSD cuts, with a survival probability of \cpowten{(1.6-3.7)}{-4}.

The \nuc{68}{Ge} isotope, with a half-life of 217\,d, is produced in the detector material at a rate of 2.12~atoms/(kg\,d)~\cite{Elliott:2009cw}. It is removed by the enrichment process but not by zone refinement or crystal pulling.
While its decay is not a background contributor, its shorter-lived \nuc{68}{Ga}
progeny decays through electron capture or $\beta^+$ emission with a 2.9-MeV $Q$
value.  To deposit energy at \qval, the $\beta^+$ particle and one or both of the
annihilation $\gamma$ rays must contribute, resulting in an inherently multi-site topology.
The expected activity assumed for \nuc{68}{Ge} is based on 40~d of sea-level
activation, and the detector production and deployment schedule gives an average
rate over the data taking period that is only 0.02 times the initial activity.
The expected background spectrum from \nuc{68}{Ge}--\nuc{68}{Ga} decays in the
detectors before and after analysis cuts is also shown in
Fig.~\ref{fig:UThK_Cosmo} (right). The analysis cuts are effective at removing
these events due to their multi-site nature.
Depending on the detector energy threshold and the background rate near 10 keV
and below, additional suppression of the \nuc{68}{Ga} background is possible
through time-coincidence tagging of the 10-keV and 1-keV Ga x-rays emitted in
the initial \nuc{68}{Ge} decay. No such time-coincidence cut is assumed here.

\subsubsection{\nuc{42}{Ar} in Liquid Argon}\label{subsubsec:ar42}

Cosmic-ray exposure of atmospheric Ar produces the long-lived radionuclide \nuc{42}{Ar}, which $\beta$ decays with a $Q$ value of 599\,keV and \thalf\ of 32.9\,yr.
While \nuc{42}{Ar} decays are not sufficiently energetic to be a background for \BBz\ decay,
its short-lived progeny \nuc{42}{K} can drift toward the detectors before
$\beta$ decaying ($Q = 3525$\,keV, $T_{1/2} = 12$\,h), in some cases with enough
energy to pose a background near \qval.

\Gerda\ mitigated this background risk by encapsulating detector strings in nylon shrouds~\cite{Lubashevskiy:2017lmf} (using nylon film produced for BOREXINO~\cite{Benziger:2007iv}) to limit the total volume of Ar from which \nuc{42}{K} ions can drift to detector surfaces. Whereas this technique is used in \Ltwo\ and is sufficient to meet its background goals, \Lthou\ surrounds the Ge detectors with UGLAr to meet its more aggressive background goals.
The DarkSide collaboration has found that underground Ar contains at least $1400
\pm 200$ times less \nuc{39}{Ar} than atmospheric Ar~\cite{DarkSide:2021mpp}.
Both \nuc{42}{Ar} and \nuc{39}{Ar} are produced from \nuc{40}{Ar} by cosmic-ray
and hadronic interactions, with \nuc{39}{Ar} having a wider variety of
production channels with lower reaction thresholds than \nuc{42}{Ar}.
We thus expect reduction factor for \nuc{42}{Ar} greater than has been measured for \nuc{39}{Ar}, though for this report, we conservatively assume only DarkSide's demonstrated
reduction factor of 1400.  See Sect.~\ref{subsec:uglar} for more details concerning the procurement plan for UGLAr.

The background contribution from \nuc{42}{Ar} depends not only on the activity of the LAr but also on ion drift in electric fields, which in turn depends on the Ar purity, the detector bias voltage, the location of grounded components, and the detector geometry.
Given the uncertainties associated with unfolding the effects of each of these factors, we instead choose to estimate the \nuc{42}{Ar} background index directly from measurements in the post-\Gerda\ test (PGT) prior to the start of \Ltwo\ assembly.
This approach relies on the fact that the geometry of the detector strings and mounting components in the \Lthou\ design is similar to that used in the PGT, including operating without \nuc{42}{Ar}-background-reducing mini-shrouds surrounding the detectors.
In the PGT, the background index contribution of \nuc{42}{Ar} $\beta$ decays before cuts is found to be 0.72\,\cpKkgy.
Applying the reduction factor for UGLAr,
we expect the background contribution from \nuc{42}{Ar} to fall to \cpowten{5.1}{-4}\,\cpKkgy\ before analysis cuts.

The AC cuts have limited effect on removing \nuc{42}{K} events due
to their decay location near detector surfaces, yielding only a 19\% suppression of the background index.
PSD, however, provides a further suppression by two orders of magnitude due to the slowness of the charge signal for events penetrating the transition layer at the \nPlus\ electrode. This effect has not yet been fully modeled in the post-processing of the \Lk\ simulations;
instead we assume an experimentally-derived survival probability of 1\%~\cite{Lubashevskiy:2017lmf}, which is also confirmed by dedicated stand-alone simulations~\cite{Lehnert:2016phd}.
The resulting background index from \nuc{42}{Ar} after cuts is found to be \cpowten{4.1^{+2.2}_{-4.1}}{-6}\,\cpKkgy\ and is the dominant singular background component in the \Lk\ model---albeit with conservative assumptions.

Validated pulse-shape simulations are being added to the simulation post
processing in order to refine this estimate. In addition, the detector response
to \nuc{42}{K} decays is being studied to optimize the detector's electrode
geometry, since the background index is strongly dependent on the thickness of
the \nPlus\ surface dead layer that absorbs part of the $\beta$ energy. While a
1-mm-thick dead layer is assumed in the current projections, this thickness can
be optimized to balance the reduction in the background rate from a thicker dead
layer with the greater pulse-shape information from a thinner dead layer.

\subsubsection{Surface Alpha Backgrounds}\label{subsubsec:alpha}

During detector and component fabrication, storage, and assembly, surface contaminants such as \nuc{210}{Po} ($\thalf=138$\,d) and \nuc{210}{Pb} ($\thalf=22$\,yr) can be introduced either as dust or through the deposition of ionized progeny of \nuc{222}{Rn} decays.
On the detector surfaces, these contaminants can emit high-energy $\alpha$
particles (e.g., 5.3\,MeV in the case of \nuc{210}{Po} decay) capable of partial
energy depositions in the active Ge detector volume.
Careful handling of the detectors in clean Rn-mitigated environments can limit the surface contamination introduced during handling and construction, and
point-contact Ge detectors have been shown to be highly effective in removing $\alpha$-induced surface events through a set of complementary pulse-shape cuts based on their fast rise-time
or delayed-charge collection 
characteristics.

It is difficult to model \emph{a priori} the detector response to surface
$\alpha$ events. However, efforts are under way to introduce \icpc\
$\alpha$ event modeling into the Monte Carlo
post-processing, based on the effective pulse-shape discriminators applied to
the \ppc\ and \bege\ detectors.
At present, we use the contamination level and PSD performance demonstrated in
\Gerda~\ptwo\ as the basis of our $\alpha$-background estimates.
The $\alpha$-induced event rate at energies of 3.5 to 5.3 MeV is 15
cts/(detector\,yr), corresponding to a rate of \cpowten{1.5}{-3}\,\cpKkgy\ at
\Qbb. Scaling to the 2.6\,kg average detector mass expected in \Lthou, this
results in an $\alpha$-induced background rate of \cpowten{5.7}{-4}\,\cpKkgy\
before the PSD is applied. From the total elimination of background events above
3.5\,MeV, where $\alpha$-induced events dominate the background spectrum, we can
extrapolate an upper limit on the $\alpha$-induced event survival probability of
0.16\% (90\% CL). Applying this survival probability to the expected $\alpha$-background rate, we find an expected background contribution of less than \cpowten{9.2}{-7}\,\cpKkgy\ in \Lk, which is a small component of the overall background composition.

\subsubsection{External $\gamma$-Ray and Neutron Backgrounds}\label{subsubsec:external}
In addition to backgrounds from components internal to the array modules
discussed in the previous sections, the stainless steel cryostat, water tank,
and the laboratory environment can potentially contribute to the background via
$\gamma$ rays and neutrons. A stand-alone \GF-based simulation
module~\cite{Barton:2020fiz} handles the computationally intensive simulations
of the of external sources through representative geometries of the water
shield, cryostat, the LAr volume, and the Ge detectors in a fashion similar
to Ref.~\cite{Barabanov:2009zz}.

For a stainless steel cryostat, the main $\gamma$-ray background contributor is the 2615-keV $\gamma$ ray from $^{208}$Tl decay in the $^{232}$Th chain.
For the steel used in \Gerda, batches with low concentrations of primordial radioisotopes in the range of 1\,mBq/kg and below have been measured~\cite{Maneschg:2008zz}.
We assume that steel with similar radiopurity can be sourced for \Lk.
The $\gamma$-ray background contribution is found to be \cpowten{1.1\pm0.2}{-5}\,\cpKkgy\ before analysis cuts.
A previous \Gerda\ analysis suggests a combined conservative suppression factor
of $20$ for the PSD and LAr veto cuts; we are refining this estimate for \Lthou\
with further Monte Carlo studies.
With the present inputs, the total expected background contribution from
cryostat material $\gamma$ rays is \cpowten{(5.3\pm1.0)}{-7}\,\cpKkgy.

Neutrons are another potential source of background originating from the cryostat material.
Alpha decays from the \nuc{238}{U} and \nuc{232}{Th} chains within the
apparatus can generate neutrons via ($\alpha$,n) reactions, adding to neutrons
from the spontaneous fission at a rate of \cpowten{1.1}{-6}\,n/(s\,Bq) for
$^{238}$U~\cite{Westerdale:2017kml}.
To estimate the neutron yield from ($\alpha$,n) reactions, the open-source application NeuCBOT~\cite{Westerdale:2017kml} is used.
Together, the total neutron flux from ($\alpha$,n) and fission is estimated to
be \cpowten{3.2}{-4}\,n/s.
These radiogenic neutrons are not associated with muon-induced triggers and difficult to tag.
They can be captured on \gesix\ nuclei, producing \geVII\ and \geVIIm\ states that can contribute to the background with probabilities and cut efficiencies reported in Ref.~\cite{Wiesinger:2018qxt}.
If the referenced delayed-coincidence is not applied, the radiogenic neutron contribution is found to be \cpowten{(2.0\pm0.5)}{-7}\,\cpKkgy.

The water tank enclosure and the laboratory environment can also contribute to the total background, which is dominated by the surrounding rock wall within the experimental hall. While neutrons can be very effectively moderated by the thick water shield surrounding the cryostat, the 2615-keV $\gamma$ ray from \nuc{208}{Tl} decay is considered.
Given its extremely small survival probability in traversing the water shield and LAr volume, its estimation is computationally intensive, so approximations were used. Similar to the approach in Ref.~\cite{Barabanov:2009zz}, the flux of 2615-keV $\gamma$ rays in the laboratory is assumed to be isotropic and to amount to up to $0.1\,\unit{cts/(cm^2\,s)}$, consistent with the expectation at SNOLAB~\cite{Smith:2012fq}. Similar to the cryostat $\gamma$ rays, a conservative suppression factor of $20$ for the PSD and LAr veto cuts is assumed for the laboratory environmental backgrounds. A conservative estimate of this contribution to the background index is approximately \cpowten{(5\pm2)}{-8}\,\cpKkgy and therefore negligible.

The total external gamma and neutron background in the \Lthou\ baseline
design is below $10\%$ of the background goal due to, in part, a
sufficient LAr shielding that is much larger than the LAr shielding of
\Gerda. With a smaller LAr shield, in the alternate case of siting \Lthou\ in Hall C at LNGS (discussed in Sect.~\ref{sec:altsite}), additional measures such
as adding Cu shielding can be considered.

\subsubsection{Muon-Induced Background}\label{subsubsec:muInduced}
Cosmic-ray muons penetrating the laboratory's rock overburden are able to induce prompt and delayed backgrounds.
Prompt events (in coincidence with the muon) are predominantly generated by accompanying secondaries in the muon-induced shower or by the muon itself, and they are efficiently reduced by the muon veto.
In \Gerda, a background contribution of \cpowten{2.9}{-5}\,\cpKkgy\ before the \lar\ veto and PSD cuts was estimated~\cite{Freund:2016fhz}.
The same study pointed out that the muon-induced background without a veto
condition would be \cpowten{3.2}{-3}\,\cpKkgy\ and indicated that the rejection
efficiency for coincident muon-Ge events is $99.2^{+0.3}_{-0.4}\%$. The \lar\
veto and PSD cuts are expected to reduce the prompt background below the level
of \powten{-6}\,\cpKkgy\ in \Ltwo\ with the \Lngs\ overburden. The factor of
100 reduction in the muon flux and associated prompt background at
\SL\ overburden is expected to render prompt cosmic-ray-induced backgrounds
negligible.

Delayed cosmic-ray-induced backgrounds are associated with the production of long-lived isotopes produced in Ge detectors or experimental components, primarily through photo-nuclear interactions or interactions with spallation neutrons.
Of these, production of isotopes in Ge detectors dominates, and backgrounds from other components are negligible.
Cosmogenic \geVII\ production in \Lk\ was first studied in Ref.~\cite{Wiesinger:2018qxt} for a \Gerda-like geometry at \Lngs. Analysis cuts and corresponding efficiencies for both ground (\geVII) and meta-stable (\geVIIm) states were carefully studied. A highly effective delayed coincidence cut to substantially suppress \geVIIm\ contributions was proposed. After all cuts, Ref.~\cite{Wiesinger:2018qxt} found the summed cosmogenic \geVII\ and \geVIIm\ background contributions in a \Gerda-like geometry at \Lngs\ is \cpowten{(2.7\pm0.3)}{-6}\,\cpKkgy.

The \Lthou\ geometry has \lar\ shielding that is much larger than the shielding  deployed in \Gerda. The effect of increased \lar\ thickness on cosmogenically-induced neutrons is studied in a stand-alone \Geant-based module~\cite{Barton:2020fiz},
which implements representative baseline and alternative geometries of \Lthou. The increased LAr thickness at \Lthou\ develops greater muon showers, increasing the rate of neutron production and thereby increasing the rate of cosmogenic \geVII\ production per unit mass of Ge. As a result, the final after-cut contributions to the background index by the \geVII\ and \geVIIm\ states in the \Lk\ baseline design is higher than that in a \Gerda-like geometry. Table~\ref{table:ge77} compares the after-cut contributions in the baseline design at the separate depths of three underground laboratories, which are discussed in Sect.~\ref{sec:labs}. The analysis cuts and corresponding efficiencies reported in Ref.~\cite{Wiesinger:2018qxt} are applied here. With the exception of the delayed coincidence cut, these efficiencies have been cross-checked with simulations of the \Lk\ baseline geometry. At the reference SNOLAB depth, the total muon-induced contribution to
the background index is \cpowten{(6\pm6)}{-8}\,\cpKkgy.

\begin{table}[h]
\centering
\caption{After-cut background indices (BI)
 of the cosmogenic \geVII\ and \geVIIm\ states at the depths of three
 underground laboratories. A delayed coincidence (DC) cut, as described
 in Ref.~\cite{Wiesinger:2018qxt}, is used to further reduce the
 \geVIIm\ background contribution.}
 \label{table:ge77}
 \begin{tabularx}{0.65\textwidth}{X C{0.15\tw} C{0.2\tw} C{0.2\tw}}
 \rowcolor{legendgrey}
 & & \textbf{BI before DC} & \textbf{BI after DC}  \\
 \rowcolor{legendgrey}
 \textbf{Source} & \textbf{Location}  &
 \textbf{$[$\cpKkgy$]$} & \textbf{$[$\cpKkgy$]$}  \\ \hline

 \geVII & SNOLAB            & \cpowten{3.2}{-8} & N/A \\

 \geVII & SURF              & \cpowten{5.0}{-7} & N/A \\

 \geVII & LNGS            & \cpowten{3.0}{-6} & N/A \\

 \hline \geVIIm & SNOLAB             & \cpowten{3.9}{-7} &
 \cpowten{2.6}{-8} \\

 \geVIIm & SURF              & \cpowten{6.0}{-6} & \cpowten{4.0}{-7} \\

 \geVIIm & LNGS              & \cpowten{3.6}{-5} & \cpowten{2.4}{-6} \\

\end{tabularx}

 \end{table}

\subsubsection{Total Background Projection}\label{subsec:background-projections}

The expected contributions to the background at \qval\ are summarized in Table~\ref{tab:bi}.
All values therein are normalized to the total detector mass and integrated over an analysis range of 1985--2095 keV around \Qbb.
For all background sources, with the exception of $\alpha$ emitters on the detector surfaces, Monte Carlo simulations were performed with the isotope or decay chain of interest uniformly distributed throughout component volumes or across surfaces.
Following post-processing, energy spectra are produced for each component group listed, scaling efficiency spectra by the expected activities listed in Table~\ref{tab:simulatedActivities}, at each stage of analysis cut application in order to examine the effect of each cut.

\begin{table}[htp]
\footnotesize
\caption{
Estimated background indices (BIs) before and after the application of analysis cuts.
The survival probabilities for the PSD
applied only to the events surviving the AC cut is enhanced by the complementarity of the two cuts.
Uncertainties correspond to $\pm$1$\sigma$ or limits at the 90\% CL.
}
\label{tab:bi}
\begin{tabularx}{1.0\tw}{L{0.7in} C{1.3in} C{1.2in} C{0.45in} C{0.45in} Z Z}
\rowcolor{legendgrey} & & \textbf{BI Before Cuts}    & \multicolumn{3}{c}{\textbf{Survival Probabilities $[\%]$}} & \textbf{BI After Cuts}  \\
\rowcolor{legendgrey} \textbf{Source} & \textbf{Location} & \textbf{$[$\cpKkgy$]$}& \textbf{AC} & \textbf{PSD} & \textbf{PSD After AC} & \textbf{$[$\cpKkgy$]$} \\
\hline
$^{238}$U chain	&	Cabling---HV	&\cpowten{(	1.2	\pm	0.5	)}{-	5	}&	11	&	19	&	5.0	&\cpowten{(	6.3	\pm	4.2	)}{-	8	}	\\
$^{238}$U chain	&	Cabling---Signal	&\cpowten{(	1.3	\pm	0.6	)}{-	5	}&	11	&	19	&	5.0	&\cpowten{(	7.3	\pm	4.9	)}{-	8	}	\\
$^{238}$U chain	&	Det.~Mount (EFCu)	&\cpowten{(	7.9	\pm	3.9	)}{-	6	}&	22	&	19	&	6.1	&\cpowten{(	1.0	\pm	0.8	)}{-	7	}	\\
$^{238}$U chain	&	Det.~Mount (\ultem)	&\cpowten{(	2.1	\pm	0.7	)}{-	5	}&	21	&	23	&	15	&\cpowten{(	6.8	\pm	4.2	)}{-	7	}	\\
$^{238}$U chain	&	Optical Fibers	&\cpowten{(	2.3	\pm	0.8	)}{-	4	}&	4.3	&	22	&	10	&\cpowten{(	1.0	\pm	0.7	)}{-	6	}	\\
$^{238}$U chain	&	Front End ASIC	&\cpowten{		<	1.0	}{-	6	}&	16	&	19	&	6.3	&\cpowten{		<	1.0	}{-	8	}	\\
$^{238}$U chain	&	PEN Plates	&\cpowten{(	7.3	\pm	1.8	)}{-	5	}&	2.7	&	20	&	6.5	&\cpowten{(	1.3	\pm	0.8	)}{-	7	}	\\
$^{238}$U chain	&	HV Conn.~(\ultem)	&\cpowten{(	2.0	\pm	0.6	)}{-	6	}&	17	&	21	&	7.4	&\cpowten{(	2.5	\pm	1.6	)}{-	8	}	\\
$^{238}$U chain	&	HV Conn.~(Ph-Br)	&\cpowten{(	6.3	\pm	0.2	)}{-	6	}&	17	&	21	&	7.4	&\cpowten{(	8.0	\pm	4.2	)}{-	8	}	\\
$^{238}$U chain	&	FE Mount (\ultem)	&\cpowten{(	4.4	\pm	1.5	)}{-	6	}&	19	&	18	&	6.0	&\cpowten{(	5.1	\pm	1.7	)}{-	8	}	\\
$^{238}$U chain	&	FE Mount (Ph-Br)	&\cpowten{(	7.7	\pm	0.2	)}{-	6	}&	19	&	18	&	6.0	&\cpowten{(	8.9	\pm	4.7	)}{-	8	}	\\
$^{238}$U chain	&	CAPs	&\cpowten{(	6.3	\pm	1.6	)}{-	6	}&	3.6	&	22	&	9.2	&\cpowten{(	2.1	^{+2.4}	_{-2.1}	)}{-	8	}	\\
$^{238}$U chain	&	Re-entrant Vessels	&\cpowten{(	8.3	\pm	4.1	)}{-	6	}&	13	&	22	&	11	&\cpowten{(	1.2	\pm	0.9	)}{-	7	}	\\
$^{238}$U chain	&	Tetratex \& TPB	&\cpowten{(	1.6	\pm	0.3	)}{-	5	}&	13	&	22	&	11	&\cpowten{(	2.2	\pm	1.4	)}{-	7	}	\\
\hline
$^{232}$Th chain	&	Cabling---HV	&\cpowten{(	2.3	\pm	1.6	)}{-	5	}&	0.068	&	35	&	4.6	&\cpowten{(	7.3	^{+7.7}	_{-7.3}	)}{-	10	}	\\
$^{232}$Th chain	&	Cabling---Signal	&\cpowten{(	2.7	\pm	1.9	)}{-	5	}&	0.068	&	35	&	4.6	&\cpowten{(	8.5	^{+9.0}	_{-8.5}	)}{-	10	}	\\
$^{232}$Th chain	&	Det.~Mount (EFCu)	&\cpowten{(	1.5	\pm	0.7	)}{-	5	}&	0.31	&	35	&	8.2	&\cpowten{(	3.9	\pm	2.8	)}{-	9	}	\\
$^{232}$Th chain	&	Det.~Mount (\ultem)	&\cpowten{(	7.8	\pm	1.8	)}{-	5	}&	0.32	&	36	&	6.8	&\cpowten{(	1.7	\pm	1.1	)}{-	8	}	\\
$^{232}$Th chain	&	Optical Fibers	&\cpowten{(	1.0	\pm	0.3	)}{-	3	}&	0.049	&	38	&	33	&\cpowten{(	1.6	\pm	1.5	)}{-	7	}	\\
$^{232}$Th chain	&	Front-End ASIC	&\cpowten{		<	1.6	}{-	6	}&	0.32	&	35	&	5.4	&\cpowten{		<	2.8	}{-	10	}	\\
$^{232}$Th chain	&	PEN Plates	&\cpowten{(	2.8	\pm	0.8	)}{-	4	}&	0.19	&	35	&	3.5	&\cpowten{(	1.9	\pm	2.8	)}{-	8	}	\\
$^{232}$Th chain	&	HV Conn.~(\ultem)	&\cpowten{(	7.9	\pm	1.8	)}{-	6	}&	0.18	&	37	&	9.3	&\cpowten{(	1.3	\pm	0.9	)}{-	9	}	\\
$^{232}$Th chain	&	HV Conn.~(Ph-Br)	&\cpowten{(	3.3	\pm	2.2	)}{-	6	}&	0.18	&	37	&	9.3	&\cpowten{(	5.5	\pm	5.2	)}{-	10	}	\\
$^{232}$Th chain	&	FE Mount (\ultem)	&\cpowten{(	1.6	\pm	0.4	)}{-	5	}&	0.35	&	35	&	8.4	&\cpowten{(	4.8	\pm	2.9	)}{-	9	}	\\
$^{232}$Th chain	&	FE Mount (Ph-Br)	&\cpowten{(	3.6	\pm	2.5	)}{-	6	}&	0.35	&	35	&	8.4	&\cpowten{(	1.1	\pm	1.0	)}{-	9	}	\\
$^{232}$Th chain	&	CAPs	&\cpowten{(	1.3	\pm	0.3	)}{-	5	}&	0.14	&	32	&	3.2	&\cpowten{(	6.1	^{+6.3}	_{-6.1}	)}{-	10	}	\\
$^{232}$Th chain	&	Re-entrant Vessels	&\cpowten{(	2.5	\pm	1.1	)}{-	5	}&	1.2	&	37	&	17	&\cpowten{(	5.1	\pm	4.0	)}{-	8	}	\\
$^{232}$Th chain	&	Tetratex \& TPB	&\cpowten{(	3.1	\pm	0.6	)}{-	5	}&	1.2	&	37	&	17	&\cpowten{(	6.2	\pm	4.2	)}{-	8	}	\\
\hline
$^{68}$Ge	&	Detector Material	&\cpowten{(	2.7	\pm	0.5	)}{-	4	}&	35	&	3.6	&	1.0	&\cpowten{(	1.0	\pm	0.5	)}{-	6	}	\\
$^{60}$Co	&	Detector Material	&\cpowten{(	4.5	\pm	0.9	)}{-	4	}&	3.7	&	1.1	&	0.67	&\cpowten{(	1.1	\pm	0.6	)}{-	7	}	\\
$^{238}$U chain	&	Detector Material	&\cpowten{		<	7.5	}{-	7	}&	65	&	53	&	77	&\cpowten{		<	3.7	}{-	7	}	\\
$^{232}$Th chain	&	Detector Material	&\cpowten{		<	4.3	}{-	7	}&	50	&	43	&	69	&\cpowten{		<	1.5	}{-	7	}	\\
\hline
$^{42}$Ar	&	Detector \nPlus\ Surf.	&\cpowten{(	5.1	^{+0.8}	_{-5.1}	)}{-	4	}&	81	&	1.0	&	1.0	&\cpowten{(	4.1	^{+2.2}	_{-4.1}	)}{-	6	}	\\
$^{222}$Rn	&	Underground Ar	&\cpowten{(	1.3	\pm	0.1	)}{-	4	}&	0.48	&	21	&	6.4	&\cpowten{(	3.9	\pm	2.1	)}{-	8	}	\\
\hline
Surface $\alpha$s	& 		&\cpowten{(	5.7	\pm	1.5	)}{-	4	}&	100	&	0.16	&	0.16	&\cpowten{		<	9.2	}{-	7	}	\\
\hline
$^{238}$U/$^{232}$Th	&	External ($\gamma$)	&\cpowten{(	1.1	\pm	0.2	)}{-	5	}&		&		&		&\cpowten{(	5.3	\pm	1.0	)}{-	7	}	\\
$^{238}$U/$^{232}$Th	&	External (n)	&						&		&		&		&\cpowten{(	2.0	\pm	0.5	)}{-	7	}	\\
\hline
$^{77}$Ge	& 	$\mu$-induced	&\cpowten{(	3.4	\pm	3.4	)}{-	7	}&		&		&		&\cpowten{(	3.2	\pm	3.2	)}{-	8	}	\\
$^{77m}$Ge	& 	$\mu$-induced	&\cpowten{(	5.4	\pm	5.4	)}{-	7	}&		&		&		&\cpowten{(	2.6	\pm	2.6	)}{-	8	}	\\
\hline
All Sources	& 		&\cpowten{(	3.9	^{+0.4}	_{-0.6}	)}{-	3	}&		&		&		&\cpowten{(	9.1	^{+4.9}	_{-6.3}	)}{-	6	}	\\
\hline
\end{tabularx}
\end{table}

\begin{figure}[]
      \begin{center}
         \includegraphics[width=.45\textwidth]{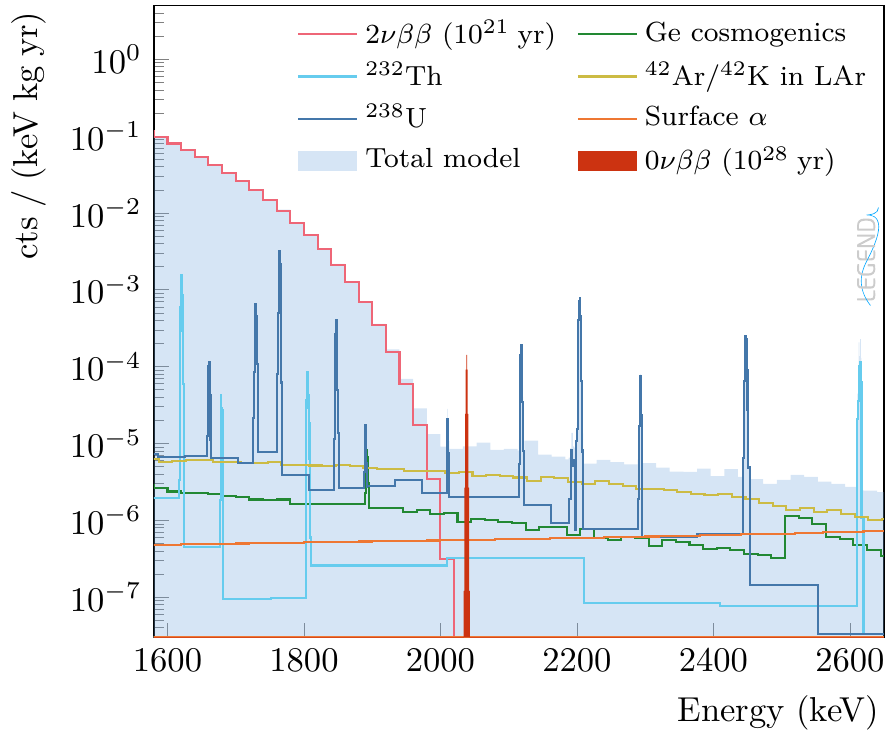}
         \includegraphics[width=.54\textwidth]{figs/lgnd1000-bkg-budget}
      \end{center}
 \caption{Left: Expected total spectrum from \BBt\ decay and from all background components listed in Table~\ref{tab:bi}, after analysis cuts.  The spectra are drawn with 1-keV-wide bins where peaks are presents and variable width binning elsewhere to maintain sufficient statistics. Right: Expected contributions to the background index of the sources of background comprising the \Lthou\ background model.  Colored bands represent $1\sigma$ uncertainties on best-estimate central values.  For internal Ge and surface $\alpha$ decay contributions, 90\% CL upper limits are shown.}
\label{fig:AllBackgrounds}
\end{figure}

The energy spectrum near \Qbb\ is shown after all analysis cuts in Fig.~\ref{fig:AllBackgrounds}.
Table~\ref{tab:bi} also shows the anticipated survival probabilities evaluated from simulations and measured values.
Monte Carlo simulations provide the AC and PSD cut survival probabilities and are in agreement with what has previously been observed experimentally.
The \nuc{232}{Th} chain backgrounds are generally strongly removed by the AC cut, owing to the large $Q$ value for \nuc{208}{Tl} decays.
The \nuc{238}{U} decays are less strongly rejected, with backgrounds from components farther from the detectors (e.g., CAPs) being more strongly rejected than those nearby (Detector Mount EFCu \& \ultem).

The AC cut pre-selects events which do not deposit significant
energy outside of a single detector, increasing the likelihood that the PSD cut
finds that an event is multi-site within that detector.
For this reason, we separately list the PSD survival probability and the PSD
survival probability after the AC cut.
As mentioned in the previous section, \nuc{42}{K} decays are expected to be strongly rejected by the PSD because of energy depositions in the transition layer giving rise to slow pulses.
As we have not yet implemented this expectation in our simulation post-processing, we insert this PSD survival probability by hand in Table~\ref{tab:bi}.
Similarly, $\alpha$- and $\beta$-induced backgrounds from contamination of the detector \pPlus\ surface are very efficiently removed by PSD, as described in Sect.~\ref{subsubsec:alpha}.
The active background suppression performance for these surface events are also taken from experimental data.

The last column of Table~\ref{tab:bi} shows the anticipated background indices near \qval\ integrated over the 1985--2095\,keV energy range after active background suppression based on the modeled survival probabilities; in total we expect a final background index of \BGprojkev\ after all cuts.
The dominant background component after all analysis cuts are applied is attributed to the \nuc{42}{Ar} content of the \UGLAr.
As mentioned in Sect.~\ref{subsubsec:ar42}, we expect the estimated activity for this background source to be quite conservative; there is a strong likelihood the actual activity, and resulting background index, will be significantly lower than what is projected here.
Backgrounds originating in optical fibers are the next-leading contributor.  The anti-coincidence suppression of backgrounds from this source is quite effective, due to the production of scintillation in the fibers from the primary $\alpha$/$\beta$ decay of the nucleus.  We are currently refining our estimates of this suppression with additional dedicated Monte Carlo studies.
Finally, cosmogenic activation of \nuc{68}{Ge} in the detectors is the remaining background component estimated to contribute at a level above $1\times10^{-6}$\cpKkgy.
With a 271-day half-life, most of these events are expected to occur in the
first year of data-taking, and additional operational lifetime time can
effectively reduce the contribution from this source. This background can also
potentially be vetoed by a significant factor due to its time coincidence with
x-rays from its progenitor's decay.

\subsubsection{Signal Extraction and Sensitivity}\label{subsec:background-projections}
With a background level of $1\times10^{-5}$~\cpKkgy\ and a resolution of 2.5\,keV, a total of 0.4 background events are expected in the region of interest ($\qval\pm2\sigma$) at the end of \Lk. Such a low background expectation makes \Lk\ able to identify a \BBz\ decay signal based on a handful of signal events. Using a simple Poisson counting, three events in the ROI would be sufficient to provide a $3\sigma$ statistical significance for a \BBz\ decay signal.

While an excess of counts can be observed by eye, \Gerda\ and \MJ\ have developed sophisticated statistical methods for the signal extraction. Unbinned likelihood analyses treating the full amount of information available for each event have been applied in both Frequentists and Bayesian frameworks~\cite{Agostini:2020xta,Alvis:2019sil}. Event-specific systematic uncertainties on the energy scale and detection efficiency are fully propagated with these techniques, which will incorporate additional multi-variate parameters in the final \Lk\ analysis.

\Lk's sensitivity is accurately estimated using a simplified analysis based on Poisson counting with known background.
A full multi-variate analysis will well-constrain the background using sidebands in energy and other observables.
An optimized energy window of $\qval\pm2\sigma$, which has 95\% probability of
containing the signal, is used for the inversion of Poisson probabilities
described in the appendix of Ref.~\cite{Agostini:2017jim}.
Using the key experimental parameters of Table~\ref{tab::params}, including an
overall \BBz\ decay detection efficiency of 69\%, we obtain a 90\% CL exclusion
sensitivity of $1.6\times10^{28}$~yr and a 99.7\% CL discovery sensitivity of
$1.3\times10^{28}$~yr.  The exclusion sensitivity is defined as the median
half-life value that will be excluded assuming there is no signal, while the
discovery sensitivity gives the half-life value at which \Lk\ has a 50\% chance of a 3$\sigma$ discovery. These half-life sensitivities can be converted into a
range of \mee\ values as discussed in Sect.~\ref{subsec:TheoryUnc}. Using a
range of matrix element values from 2.66 to 6.04 from the four primary many-body
methods (see Table~\ref{tab:MatrixElements}) and the quenching and phase space factors reported in Table~\ref{tab::params}, we obtain $\mee<[8.5-19.4]$~meV and $\mee<[9.4-21.4]$~meV for the 90\% CL exclusion and 99.7\% CL discovery sensitivity, respectively.

\begin{figure}
\includegraphics[width=0.8\columnwidth]{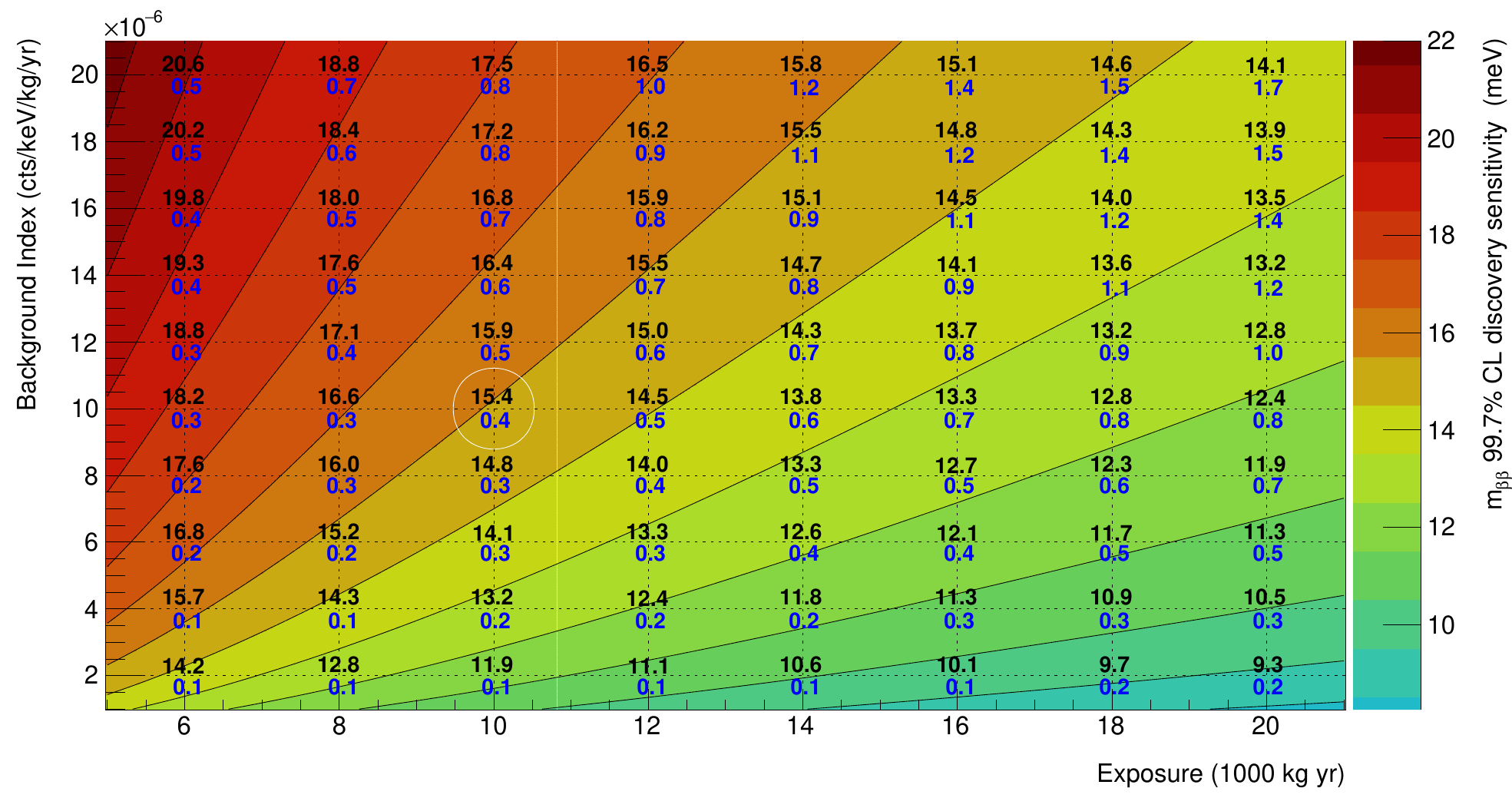}
\caption{Central value of the \mee\ discovery-sensitivity interval for \Lk\ as a
function of exposure and background level at \Qbb. The central value corresponds to an effective nuclear matrix element of 3.7. The black numbers show the \mee\ value at specific points on the grid, while the color bands show regions of the parameter space providing the same \mee\ sensitivity. The blue numbers show the expected number of background counts in the region of interest $\qval\pm2\sigma$.
The white circle marks the \Lk\ projected values.
}
\label{fig:legend_sens_2d}
\end{figure}

The central value of the \mee\ interval for the discovery sensitivity is shown in Fig.~\ref{fig:legend_sens_2d} as a function of exposure and background index. The \Lk\ background goal and baseline exposure corresponds to 15.4~meV, well below the bottom of the inverted ordering parameter space. The current uncertainties on the background model correspond to a modest $\sim6\%$ variation in terms of \mee\ sensitivity.
As new information becomes available and our conservative upper limits on some
background contributions decrease (e.g.~\nuc{42}{Ar} activity in \UGLAr), the
$m_{\beta\beta}$ discovery sensitivity may improve further.

%% file: sec_facilities/sec_facilities.tex

\section{\new{Underground Laboratory}}\label{sec:facilities}

\import{sec_facilities/}{subsec_requirements}

\import{sec_facilities/}{subsec_labs}

\import{sec_facilities/}{subsec_altsite}

%% file: sec_facilities/subsec_requirements.tex
The host laboratory provides a clean environment and necessary infrastructure located deep underground to assemble and operate the \Lk\ experiment. Due to the experiment's low-background requirements and strict material handling protocols, a clean-room environment with a series of dedicated clean work spaces is required for assembly of the detector. The size and scope of the experimental apparatus requires a range of facility space and process systems allowances. This section summarizes the facility requirements and elaborates on the candidate underground laboratories.

\subsection{\new{Facility Requirements}}

The principal facility requirement of \Lk\ is an underground detector cavity that can accommodate the full experimental apparatus. The Ge detectors will be housed within a LAr cryostat situated within a water tank. Figure~\ref{fig:baseline} shows the cryostat and water tank concept.
The water tank is approximately 12\,m in diameter and 11\,m in height. The nominal total height of the experimental hall including the water tank and an upper work deck area is 18\,m.
Part of the overhead space must allow substantial lifting devices to support the cryostat and water tank installation plan.
Some flexibility exists in the technical design to accommodate slight differences in an available cavity size. However, the module assembly and deployment above the work deck does impose strict working height requirements while the background considerations set requirements on the diameter of the shielding provided by the LAr cryostat and water tank.

The work deck space above the cryostat and water tank and adjacent areas need to operate as a clean space, either facility-wide or through dedicated clean spaces limited to \LEG's exclusive use.
Additionally, clean environments are established through the use of \nt-purged glove boxes and part storage cabinets. A series of glove boxes directly above the cryostat provide the means to perform final assembly of the Ge detector array and LAr instrumentation prior to deployment into one of the four reentrant tubes that complete a module. This space would include some hoisting capabilities to maneuver portions of the apparatus and the detector array during deployment.
The adjacent clean spaces support purged glove boxes and cabinets for part storage, detector mounting, and testing activities that take place in advance of detector installation. In addition, a separate clean space is used for final chemical cleaning of internal detector parts, which will require systems such as a fume hood, chemical storage and waste handling, and purified water. A unique feature of the \Lk\ design is the use of detector components fabricated from ultra-pure EFCu. It is preferable to operate the electroforming system underground to avoid cosmic activation of the Cu material, which requires a separate space with proper exhaust, chemical handling, and electrical services.
Beyond the dedicated clean spaces, work areas at general cleanliness levels are required. An area housing the electronics racks and data acquisition systems must be located close to the cryostat for detector readout while a nearby control room offers an additional interface to the experiment. A designated lay-down, storage, and assembly space supports construction work in the main detector hall.

The underground facility space must also support the detector process systems and services.
A water plant is necessary to supply the 980\,m$^3$ water tank  and supply purified water for chemical and cleaning activities.
An \lnn\ generator and/or buffer tanks will provide cooling of the LAr cryostat, support pre-cooling and heat exchangers for the LAr cryogenic systems, and \nt\ purge gas for the glove boxes and part storage cabinets.
The LAr cryogenic systems will surround the detector to support the cryostat initial fill and operations. Transport vessels that deliver the atmospheric and underground-sourced Ar will be staged near the detector hall. Several process loops will provide the needed Ar condensers, purifications systems, compressors, recovery, storage, and exhaust systems complete with system flow and monitoring controls. A volume of 240\,m$^3$ of LAr cryogen and 13.3\,m$^3$  of low-background UGLAr cryogen will be housed within the cryostat.

Specialized production and fabrication facilities are needed, though they may be located on the surface or fulfilled with existing facility capabilities. A standard machine shop may be used for fabrication of detector mounting parts, as well as housing specialized tools to support the needs of the \Lk\ EFCu production.
Surface clean space for production of plastic components and chemical cleaning activities, where cosmic activation is not a concern, may also be desirable.  Short- and long-term surface storage is required for staging materials and detector components in advance of transportation to the underground laboratory, while components needing protection from cosmic activation should be immediately stored underground upon arrival.
Computing and electronic storage infrastructure on the surface is necessary to serve as a buffer for the experiment's data stream from the underground and as a gateway to the collaboration's offsite computing resources. Modest surface space for office and administrative work is beneficial.

Uninterrupted electrical power is critical to maintain a smooth operation of the experiment, nominally demanding the use of uninterrupted power supplies and back-up power generators within the facility electrical infrastructure. Supplemental cooling and ventilation infrastructure may be needed for the separate clean spaces and electronics racks if not already covered by the facility's HVAC capacity. High-speed networking capability between the underground and the surface with a bandwidth of greater than 10~Gb/s is necessary to sustain the experiments's data stream and offsite detector monitoring.

%% file: sec_facilities/subsec_labs.tex
\subsection{\new{Existing Laboratories}}
\label{sec:labs}

While there is a wide range of worldwide underground laboratories that could potentially host \Lk, the top sites considered for \Lk\ are summarized in Table \ref{tab:LabSummary}. As the deepest laboratory, SNOLAB represents the preferred underground site with features that accommodate the \Lk\ technical design and provides the reference for the experimental details covered in Sect.~\ref{sec:techmat}. A brief description of the SNOLAB site is described in the next section.
A more thorough discussion of the LNGS alternative site follows, including its implications on the \Lk\ technical design and background model.

The two other underground laboratories  warrant consideration for \Lk, but do not have immediate space available.
The Sanford Underground Research Facility (SURF)~\cite{Heise:2017rpu} is
located in Lead, South Dakota in the former Homestake Gold mine and is the host laboratory for the \MJD.
While the DUNE project~\cite{DUNE:2020lwj} has begun excavation of several large cavities at SURF, the schedule for DUNE may not dove-tail well with the \Lk\ schedule. The construction of a new cavity for \Lk\ would take 2--3 years and could only start after DUNE has completed its construction (in 2025 or 2026). Nonetheless, the collaboration is holding ongoing discussions with the SURF management regarding these issues.
The Boulby Underground Laboratory is the United Kingdom’s (UK) deep underground science facility located in a working polyhalite and salt mine in North-East England. This underground laboratory offers a flat overburden of 1100\,m (2850 m.w.e.) and hosts several small experiments and an extensive radio-assay facility. The UK funding agency, STFC, recently funded a feasibility study for creating a new underground laboratory space with the aim to host a major international rare-event search experiment. The two main use-cases considered are a generation-three dark matter experiment with a noble liquid target and a \Lk-like \BBz\ decay detector.
Two sites for the new cavern are considered: one at the current depth of 1100\,m and a deeper site at 1400\,m.
The Boulby Feasibility Study report is expected to be published in July 2021.

\begin{table}
\caption{A summary of the potential underground sites. The column labeled {\em Occ.~Date} gives an estimate of the earliest beneficial occupancy date.}
\begin{center}
\begin{tabularx}{0.95\textwidth}{ L{0.18\tw} R{0.1\tw} C{0.3\tw} Z Z}
\rowcolor{legendgrey}
{\bf Site}		&
{\bf Depth}		&
{\bf Cavity}	&
{\bf Muon Flux} &
{\bf Occ.~Date}\\
\rowcolor{legendgrey}
		&	[m.w.e.]	&	&	[m$^{-2}$~s$^{-1}$]&	\\
\hline
Boulby (Upgrade-1) & 2850 & 25\,m \diameter\ x 25\,m & \cpowten{3.75}{-4} \cite{Reichhart:2013xkd} & 2028 \\
Boulby (Upgrade-2) & 3600 & 25\,m \diameter\ x 25\,m & \cpowten{1.13}{-4} \cite{BFSR:2021} & 2030 \\
\multirow{ 2}{*}{LNGS}	&	\multirow{ 2}{*}{3500}		& Hall A: 14.9\,m x 15.7\,m  & \multirow{ 2}{*}{\cpowten{3.5}{-4} \cite{GERDA:2016lhn}} & \multirow{ 2}{*}{2022}\\
	&			& Hall C:  17.7\,m x 19.0\,m & &\\
SURF	&	4300		& 50\,m x 20\,m x 20\,m 		&
 \cpowten{5.3}{-5} \cite{MAJORANA:2016ifg}&
2028/29\\
SNOLAB	&	6010		& Cryopit 15 m \diameter x 19.0 m	&
\cpowten{3}{-6} \cite{Smith:2012fq} &
Immediate\\
\end{tabularx}
\end{center}
\label{tab:LabSummary}
\end{table}%

\subsubsection{\new{Reference Site: SNOLAB}}
\label{sec:snolab}

SNOLAB~\cite{Duncan2010SNOLAB,Smith:2012fq} is located in the Canadian granite shield, close to
the town of Sudbury inside a working nickel mine. It is accessed
through the mine shaft.
It is a well established laboratory with all the required expertise
in running underground experiments with international collaborations requiring a clean environment.
The SNOLAB facility is at a depth of 6010\,m.w.e. and operates all experimental areas as Class 2000 clean rooms and accommodates additional cleanliness requirements as needed.

The access through the mine shaft would complicate some aspects of the \Lk\ construction.
The collaboration has been in close contact with SNOLAB management, and their engineering and project management staff to work through an experimental layout at the space referred to as the cryopit, a large cavity 15\,m in diameter, 19\,m in height designed to house a large liquid cryogen experiment; SNOLAB management has assigned the cryopit to a future \BBz\ decay experiment.
The placement of the \Lk\ cryostat and water tank within the cryopit is shown earlier in Fig.~\ref{fig::cryostat-vac}.  Since the available cryopit is compatible with our technical requirements and the muon flux at this depth introduces a negligible background contribution, the SNOLAB site sets the reference overburden depth and cavity size in the \Lk\ design. The \Lk\ technical design described in Sect.~\ref{sec:technical} and background model described in Sect.~\ref{subsec:background-budget} are based on the SNOLAB site location.

%% file: sec_facilities/subsec_altsite.tex
\subsubsection{\new{Alternative Site: LNGS}}\label{sec:altsite}

The alternative site considered for \Lthou\ is LNGS, the laboratory where \Ltwo\ is currently under construction.
The LNGS site is located under the Gran Sasso mountain in central Italy and allows for easy access through a freeway tunnel.
It is owned by the Italian state and operated by the Istituto Nazionale di Fisica Nucleare (INFN).
LNGS has about 3500~m.w.e.~overburden, and the muon flux is \cpowten{3.5}{-4}~m$^{-2}$~s$^{-1}$, about two orders of magnitude larger than at SNOLAB.
Consequently, the muon-induced background rate is greater, discussed below in Sec.~\ref{sec:lngsmuon}.

\begin{figure}[h]
\includegraphics[width=0.45\textwidth,angle=0]{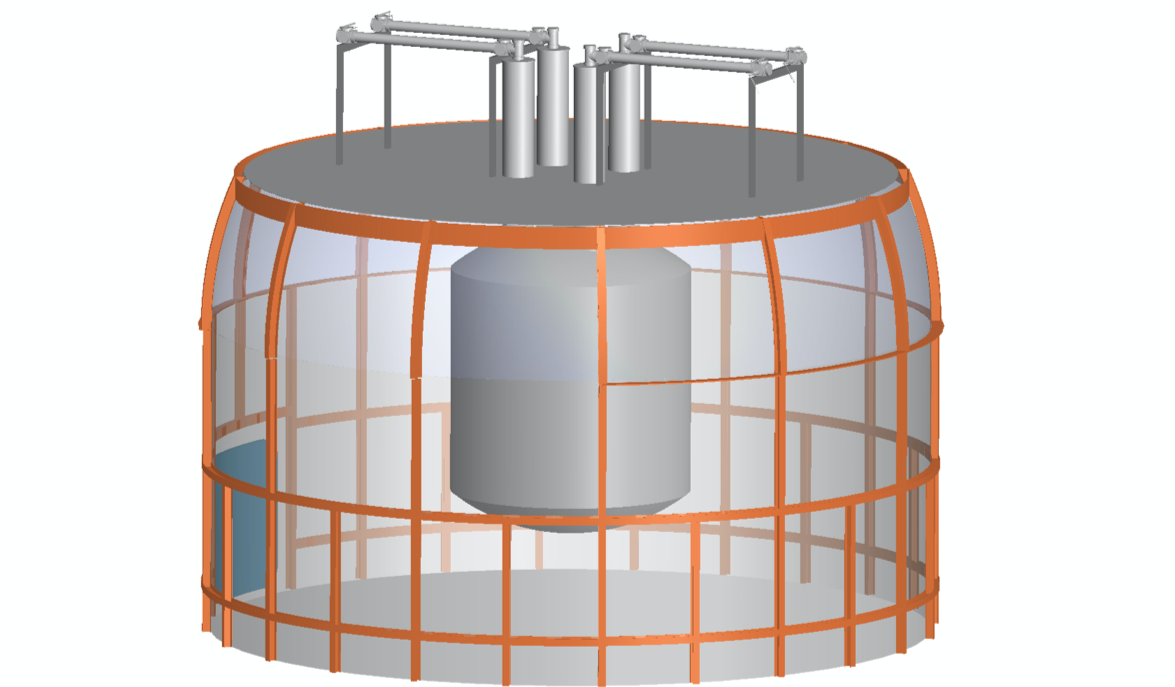}
\includegraphics[width=0.45\textwidth]{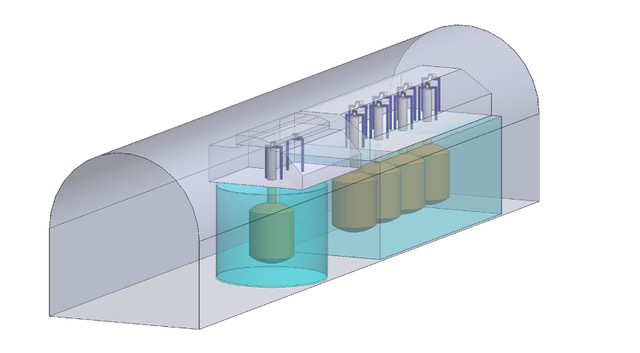}
\caption{
Alternative layouts of \Lthou\ sited at LNGS. Left:
\Lk\ vacuum-insulated cryostat within the Hall C Borexino tank.
Right: Four additional smaller cryostats in Hall A behind the existing \Ltwo\ infrastructure.
}\label{fig:cryo-LNGS}
\end{figure}

We consider two options for conducting the experiment at LNGS, the site of LVD~\cite{Aglietta:1992dy} in Hall A and Borexino~\cite{Borexino:2008gab} in Hall C (see Fig.~\ref{fig:cryo-LNGS}), which will both be decommissioned in 2022.
Hall C (height 19.0 m, width 17.7 m) is large enough to accommodate the baseline cryostat design, either in the existing Borexino water tank -- after reducing its height (see Fig.~\ref{fig:cryo-LNGS} left) -- or at a nearby site.
The latter requires the construction of a new water tank.
The load cases for the large cryostat are identical to that at SNOLAB, except for seismic acceleration due to earthquakes, which are smaller than those from the mining activities at SNOLAB.
The cryostat would be built according to the European pressure vessel code, which allows for larger maximum stress on the steel, and thus smaller wall thicknesses and 30\% less total steel mass.
The dimensions of Hall A (height 15.69 m, width 14.88 m) require smaller cryostats.
Up to four copies of the \Gerda-sized cryostat (4-m diameter) could be placed at this location, with each of them holding up to 250~kg of Ge detectors (Fig.~\ref{fig:cryo-LNGS} right).
Including \Ltwo, this arrangement would total a maximum of five cryostats.
The advantage here is that these cryostats would be built outside the laboratory, with no underground on-site welding for the cryostat construction.
The impact of a reduced LAr shielding on the external background from the cryostat steel, dominated by the 2.6-MeV \nuc{208}{Tl} $\gamma$ line, has been studied with Monte Carlo simulations.
This background can be efficiently suppressed by LAr instrumentation outside the reentrant tubes, and 10~cm of Cu inner shield attached to the wall of the cryostat, compared to 6~cm used in \Ltwo.

\subsubsection{\new{Background Considerations at Reduced Overburden}}
\label{sec:lngsmuon}

Siting the \Lk\ experiment at LNGS-depth results in a higher in-situ production of radioactive isotopes due to cosmic muon interactions, and requires special mitigation techniques.
As discussed extensively in Refs.~\cite{Pandola:2007hv, Wiesinger:2018qxt}, this depends critically on the experimental design, the available instrumentation, and the related analysis strategy.

\begin{figure}[h]
\includegraphics[width=1.0\textwidth]{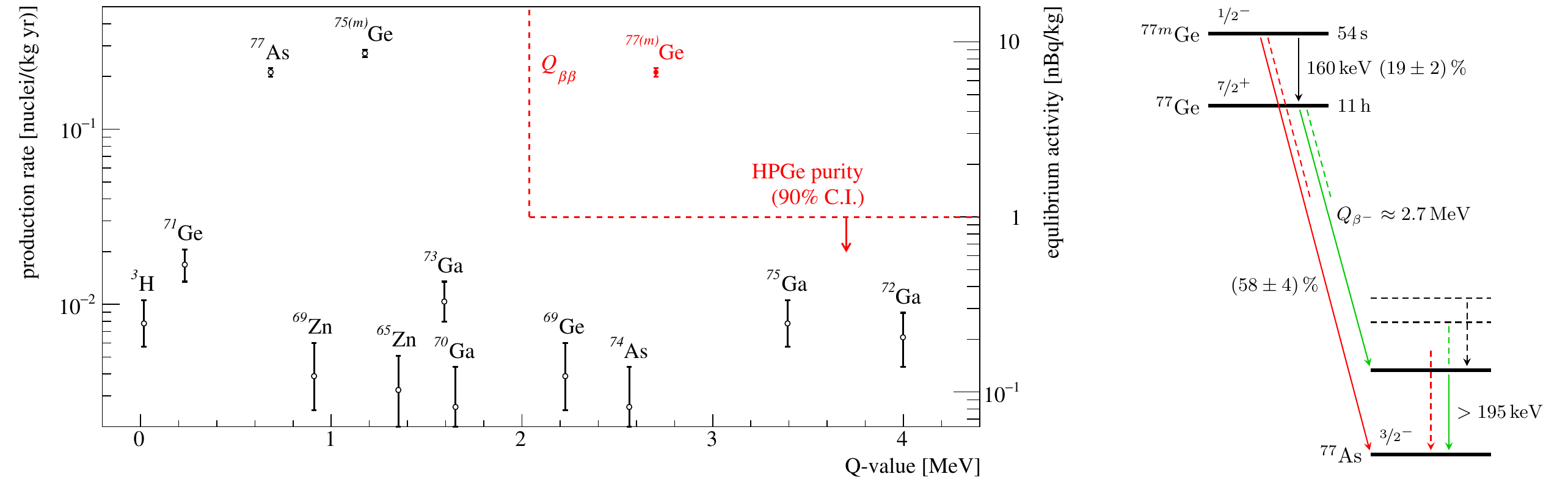}
\caption{Left: The production rate of isotopes from muon-induced reactions on enriched Ge plotted as a function of their decay $Q$ value. With a production rate of $0.21$ nuclei/(kg$\cdot$yr) in the simulations, \geVIIpm\ is the only in-situ produced radioactive isotope that exceeds both \Qbb\ and current limits on the Ge radiopurity.
Right: Due to $\gamma$ de-excitation after $\beta$ decay of ground state \nuc{77}{Ge}, it can be strongly suppressed by our standard topology cuts.
The comparably short lifetime (T$_{1/2}$ = 54 s) of meta-stable \geVIIm\ enables delayed coincidence rejection.}
\label{fig:cosmogenics}
\end{figure}

The underground production of \geVII\ and \geVIIm, together referred to as \geVIIpm, through capture of fast neutrons on \nuc{76}{Ge}, has been identified as the main in-situ cosmogenic background for \Lthou\ at LNGS.
Figure~\ref{fig:cosmogenics} shows the production rates obtained from Monte Carlo simulations for a \Gerda -like geometry.
This result has been reproduced in independent studies and a recent analysis of the \Gerda\ Phase-II data corroborates the accuracy of the predictions.
In 103.7~\kgyr\ exposure, the expected number of \geVIIm\ decays with energy deposition between 1900--2600~keV and within 3 life-times is $0.7\pm0.1$~counts.
The background derived from side-bands corresponds to $0.8\pm0.1$~counts.
Hence, the signal-plus-background expectation is 1.5~counts, compared to 3~counts observed in the \Gerda\ data set ($p$-value = 19\%).
Similarly, the \MJD\ at SURF compared predictions for $^{73\textrm{m}}$Ge and \geVIIm\ rates with experimental data.
Geant4 and Fluka simulations predicted 11 and 1 events for $^{73\textrm{m}}$Ge and \geVIIm\, compared to 7 and 0 observed events, respectively.
\Ltwo\ will allow a further test of the predictions with increased accuracy.

Applying event topology discrimination---but without delayed coincidence cuts--- \geVIIpm\ generates a background count rate at \Qbb\ of \cpowten{2}{-5}\,\ctsper\ for a 4-m-diameter cryostat at LNGS.
A delayed coincidence cut of 6~min subsequent to each muon-veto signal in coincidence with a Ge energy deposition reduces the \geVIIm\ background by almost a factor 20~\cite{Wiesinger:2018qxt}.
This results in a background contribution of \cpowten{2}{-6}\,\ctsper\ for the 4-m-diameter cryostats, with equal contributions from \geVII\ and \geVIIm\, at a one-percent-level dead time.
The results for the 7-m-diameter cryostat are about a factor two larger, since the increased shower depth in the LAr results in a higher neutron yield.
Additional techniques to reduce the in-situ cosmogenic background are under study, and summarized below.

First, 19\% of the \geVIIm\ decays proceed via isomeric transition (IT), i.e.,~160 keV $\gamma$ radiation or conversion electron emission from the excited metastable state to the \nuc{77}{Ge} ground state.
Identifying the Ge detector in which the neutron capture has occurred by detecting this IT within 6 min of the muon event, this detector can be removed from the analysis for several days (T$_{1/2}$ = 11 h).
This reduces the ground-state \nuc{77}{Ge} background contribution to $<$ \cpowten{1}{-6}\,\ctsper\ with negligible dead-time losses.

\begin{figure}[tb]
\centering
\includegraphics[width=0.7\textwidth]{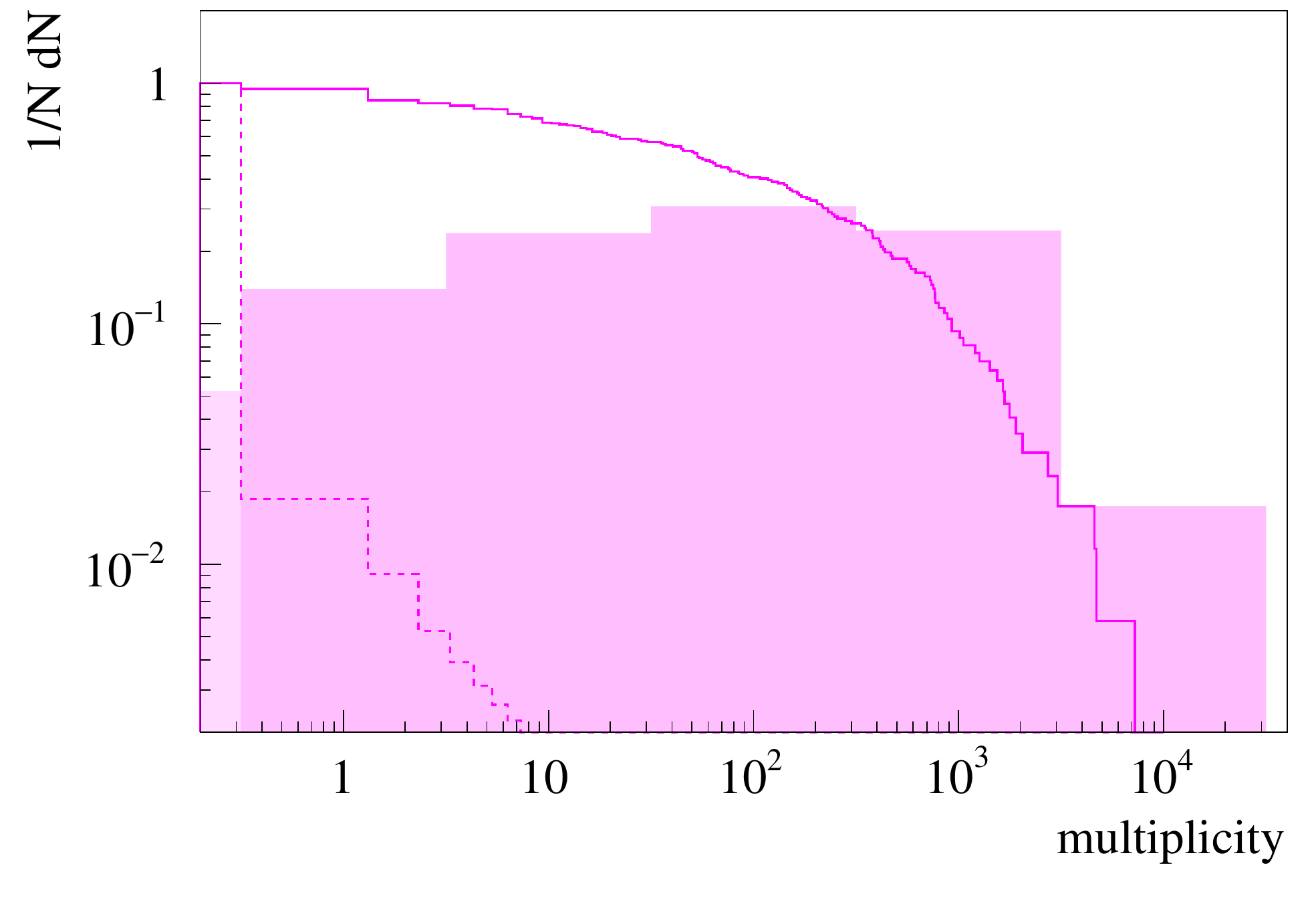}
\caption{Multiplicity of captured neutrons by Gd dissolved in the water shield.
Additionally, the dotted curve shows the inverse cumulative distribution of the neutron multiplicity for all muon events while the solid line corresponds to those events where \geVIIpm\ is produced.
The latter happens predominantly in so-called showering muon events, i.e., muon induced showers with a large neutron multiplicity.
In 70\% of all \geVIIpm\ production events, more than 10 captures appear on Gd, while for all muon-events this signature appears less than 0.2\% of the time.
}
\label{fig:NeutronMultiplicity}
\end{figure}

Second, exploiting the full sequence of muon-induced isotope production and their decays, based on the respective event topologies in the Ge, LAr, and water Cherenkov detector systems, the background contribution of \geVIIm\ can be reduced to $<$ \cpowten{1}{-6}\,\ctsper\ as well.
Crucial to this technique is the tagging of muon-induced showers with high neutron multiplicity, commonly referred to as showering muons.
Monte Carlo simulations predict that the median neutron multiplicity is approximately 75 in those showers in which \geVIIpm\ is produced.
Only a minor fraction of these sibling neutrons are captured in LAr while the larger fraction is captured by H in the water, or by Gd if such a compound is loaded in the instrumented water shield.
In about 70\% of muon-induced \geVIIpm\ productions, more than 10 neutrons are captured on Gd in the water shield (see Fig.~\ref{fig:NeutronMultiplicity}).
The high number of sibling captures on Gd or Ar gives an additional handle to identify potential \geVIIpm\ production and to improve the delayed coincidence cut efficiency.

Last, passive shielding options with low-Z materials installed within the LAr active shield moderate neutrons and reduce the \geVIIpm\ production~\cite{Barton:2020fiz}.
If made from clean---possibly scintillating---material, their radiogenic contribution can be marginal.

In summary, the \Lthou\ baseline design is compatible with siting at LNGS.
The in-situ cosmogenic contribution to the total background, however, can be up to 50\% for the 7-m-diameter cryostat and 20\% for the 4-m-diameter option without further measures.
Additional neutron moderators, improved analysis cuts (i.e., IT tagging) and muon shower identification will reduce this contribution further, potentially below 10\%, which is subdominant for the overall background budget.
The exposure loss due to dead-time from the delayed coincidence cuts may be compensated by faster underground construction and an earlier data taking start.

%% file: sec_appendix/sec_specs.tex
\section{\new{Design Specifications}}\label{appen:SpecTable}

\begin{table}[h]
\caption{A listing of the \Lk\ baseline design specifications.}
\begin{tabularx}{0.85\textwidth}{ | X X |}
  \hline
\rowcolor{legendgrey}
{\bf Parameter}                 & {\bf Specification}  \\
\hline \hline
\rowcolor{legendblue}
\multicolumn{2}{c}{Underground Site}\\
Depth                           &   6010 m.w.e.\\
\rowcolor{legendblue}
\multicolumn{2}{c}{Isotope Production}\\
Isotope                         & \nuc{76}{Ge}\\
Enrichment fraction			        &	92\% \nuc{76}{Ge}\\
Enrichment process					                & Gaseous centrifuge of GeF$_4$\\
Chemical form                   & GeO$_2$\\
Acquisition                     & 1100 kg $^{\textrm{enr}}$Ge\\
Overall detector yield          & 70--80\%\\
 \rowcolor{legendblue}
 \multicolumn{2}{c}{Detectors}\\
 Detector technology                 & \icpc: P-type inverted coaxial, point contact\\
 Total Ge detector mass 		    & 1000 kg\\
 ~~~~New detector mass       & ~~~~870 kg\\
 ~~~~Reused \Ltwo\ mass      & ~~~~130 kg \\
 Number of detectors            &	392 \\
 Avg.~detector mass			        &		2.6 kg \\
 Energy resolution at \Qbb		  &		2.5 keV FWHM\\
 \rowcolor{legendblue}
 \multicolumn{2}{c}{Detector Cooling}\\
 Technology            &  Vacuum-superinsulated cryostat\\
 Cryostat vessel material & Stainless steel\\
 Cryogen                        & Atmospheric liquid Ar \\
 Cryogen volume                 & 240 m$^3$\\
 Cryostat outer diameter        & 7.2 m\\
 Cryostat cylindrical height    & 4 m\\
 Cryostat overall height          & 10.7 m\\
 \rowcolor{legendblue}
 \multicolumn{2}{c}{Detector Deployment}\\
 Strategy                       & Modular\\
 Number of modules              & 4\\
 Detector mass per module      & 250 kg\\
 Module vessel material   & Electroformed Cu\\
 Module cryogen                 & Underground liquid Ar\\
 Module cryogen volume          & 3.31 m$^3$ \\
 \hline
 \end{tabularx}
\end{table}
\newpage
 \begin{table}[h]
 \caption{A listing of the \Lk\ baseline design specifications (continued).}
 \begin{tabularx}{0.85\textwidth}{ | X X |}
   \hline
 \rowcolor{legendgrey}
 {\bf Parameter}                 & {\bf Specification}  \\
 \hline \hline
 \rowcolor{legendblue}
 \multicolumn{2}{c}{Detector Mounting}\\
 Configuration                    & Strings of detectors\\
 String length                    & $\leq$1.2 m\\
 Detectors per string            &  $\leq$12     \\
 String detector mass           & $\leq$30\,kg \\
 Strings per module             & 14 \\
 Support rod material             & Underground electroformed Cu\\
 Electrical isolation material    & \ultem\\
 Baseplate material               & PEN \\
 \rowcolor{legendblue}
 \multicolumn{2}{c}{Detector Active Shield}\\
 Protection                     & Backgrounds that deposit energy in a medium surrounding the Ge detector array \\
 Scintillating medium           & Liquid Ar \\
 Scintilation light wavelength  & 128 nm\\
 Attenuation length (at 128 nm) & $>$60 cm \\
 LAr triplet lifetime           & 1.3 \mus \\

 Wavelength shifting material   & Tetraphenyl butadien (TPB) \\
 Light guide                    & TPB-coated polystyrene fibers\\
 Light collection and readout   & Silicon photo-multipliers (SiPM) \\
 \rowcolor{legendblue}
 \multicolumn{2}{c}{Detector Water Shield}\\
 Protection                     & Environmental $\gamma$ rays and low energy neutrons\\
 Medium                         & Liquid H$_2$O\\
 Water tank diameter           & 12.3 m\\
 Water tank height             & 10.6 m\\
 Water volume                  & 980 m$^3$\\
 \rowcolor{legendblue}
 \multicolumn{2}{c}{Cables and Connectors}\\
 Cable style                    & Kapton flat flex cables\\
 Connection to the detector     & Wire bond \\
 Inter-cable connection         & Custom low background ZIF connectors\\
 Signal feedthrough connection  & D-sub UHV feedthrough\\
 HV feedthrough connection      & Epoxy potted or commercial HV feedthrough\\
 \hline
 \end{tabularx}
 \end{table}
 \newpage
 \begin{table}[h]
 \caption{A listing of the \Lk\ baseline design specifications (continued).}
 \begin{tabularx}{0.85\textwidth}{ | X X |}
   \hline
 \rowcolor{legendgrey}
 {\bf Parameter}                 & {\bf Specification}  \\
 \hline \hline
 \rowcolor{legendblue}
 \multicolumn{2}{c}{Detector Front-End Electronics}\\
 Style                          & ASIC\\
 Noise/threshold                & ${<}$1~keV \\
 Dynamic range                  & 10~MeV \\
 Bandwidth                      & 50~MHz \\
 Minimum detector capacitance   & 1\,pF \\
 Cabling needs                  & 4: power, ground, pulser, diff.~output \\
 \rowcolor{legendblue}
 \multicolumn{2}{c}{Data Acquisition---Per Module}\\
 Number of readout channels per Ge detector   & 1 \\
 Total number of channels Ge/SiPM    & Minimum 100--120/60--80 \\
 Typical background data rate         & 1 Hz/channel; $<$0.1~GB/s \\
 Typical calibration data rate        & 3-4 kHz; $<$1~GB/s \\
 Maximum data rate of system          & $>$3.5~GB/s \\
 Dead time at maximum data rate       & $<$1\% \\
 Underground real-time data storage        & $>$50~TB \\
 Underground secondary RAID storage        & $>$250~TB\\
 Above ground storage                      & $>$500~TB \\
 Underground-to-surface network bandwidth  & $>$10~Gb/s \\
 Digitizer ADC resolution (Ge)             & $\geq$16~bits \\
 Digitizer full-scale range (Ge)           & $\geq$10~MeV \\
 Digitizer intrinsic electronic noise (Ge) & $<$50~eV \\
 Digitization speed (Ge)                   & $\geq$62.5~MHz \\
 Digitization record length (Ge)           & Adjustable up to $\geq$48~\mus \\
 Digitizer ADC resolution (SiPM)           & $\geq$16 \\
 Digitization speed (SiPM)                 & $\geq$62.5~MHz \\
 Digitization record length (SiPM)         & Adjustable up to $\geq$48~\mus \\
 Triggering capability    & Single channel readout with independent triggers OR global trigger \\
 \hline
\end{tabularx}
\end{table}
\newpage

%% file: sec_appendix/sec_glossary.tex
\section{\new{Glossary of Acronyms and Abbreviations}}


\begin{description}
  \item[\BBz] Neutrinoless double-beta decay
  \item[\BBt] Two-neutrino double-beta decay
\item[AC] Anticoincidence; an analysis cut
\item[ADC] Analog to digital converter; ``digitizer''
\item[AM] Additive manufacturing
\item[ASIC] Application specific integrated circuit
\item[\BB] A double-beta decay process, with or without neutrinos

\item[BEGe] Broad-energy Ge detector; a point contact detector manufactured by MIRION
\item[BSM] Beyond the standard model
\item[CL] Confidence level
\item[CSA] Charge sensitive amplifier
\item[DAQ] Data acquisition system; including detector readout and slow controls
\item[DCR] Delayed charge recovery; an analysis cut
\item[DEP] Double escape peak; refers to escaping annihilation $\gamma$ rays
\item[DSP] Digital signal processing
\item[EFCu] Electroformed copper
\item[ENC] Equivalent noise charge
\item[FEP] Full energy peak
\item[FFC] Flat flex cable
\item[FWHM] Full-width, half-maximum; a measure of a peak width
\item[GUT] Grand unified theory
\item[HV] High voltage
\item[IC] Impurity concentration
\item[ICPC] Inverted-coax point-contact Ge detector
\item[ICP-MS] Inductively-coupled plasma mass spectrometry
\item[IO] Inverted ordering; of neutrino masses
\item[LAr] Liquid argon
\item [\LEG] Large Enriched Germanium Experiment for Neutrinoless \BB\ Decay
\item[\Ltwo] The 200-kg phase of \LEG
\item[\Lk] The 1000-kg phase of \LEG
\item[LMFE] Low-mass front-end electronics
\item[\lnn] Liquid nitrogen
\item[LNGS] Laboratori Nazionali del Gran Sasso, Italy
\item[\mee] Effective Majorana neutrino mass in \BBz\ decay
\item[NO] Normal ordering; of neutrino masses
\item[PCB] Printed-circuit board
\item[PEN] Poly(ethylene napthalate)
\item[PGT] Post-\Gerda\ test
\item[PMT] Photomultiplier tube
\item[PPC] p-type, point-contact Ge detector, manufactured by ORTEC (AMETEK)
\item[PSD] Pulse shape discrimination; an analysis cut
\item[PTFE] Polytetrafluoroethylene; commercially known as Teflon\textsuperscript{\texttrademark}
\item[PVT] Polyvinyl-toluene
\item[\Qbb] $Q$-value of the double beta decay
\item[ROI] Region of interest
\item[SEP] Single escape peak; refers to escaping annihilation $\gamma$ rays
\item[SiPM] Silicon photomultiplier
\item[SLA] Stereolithographic printing
\item[SM] Standard Model of particle physics
\item[SNOLAB] A deep underground research laboratory in Sudbury, Canada
\item[SURF] Sanford Underground Research Facility, South Dakota
\item[TPB] Tetraphenyl butadiene
\item[TPC] Time projection chamber
\item[UGAr] Underground sourced Ar
\item[UGLAr] Underground sourced LAr
\item[UHV] Ultra-high vacuum
\item[VUV] Vacuum ultraviolet, 10 to 100 nm light
\item[WIMP] Weakly interacting massive particle
\item[WLS] wavelength shifting
\item[ZIF] Zero insertion force; a connector design
\end{description}

%% file: pcdr.bbl
\begin{thebibliography}{202}%
\makeatletter
\providecommand \@ifxundefined [1]{%
 \@ifx{#1\undefined}
}%
\providecommand \@ifnum [1]{%
 \ifnum #1\expandafter \@firstoftwo
 \else \expandafter \@secondoftwo
 \fi
}%
\providecommand \@ifx [1]{%
 \ifx #1\expandafter \@firstoftwo
 \else \expandafter \@secondoftwo
 \fi
}%
\providecommand \natexlab [1]{#1}%
\providecommand \enquote  [1]{``#1''}%
\providecommand \bibnamefont  [1]{#1}%
\providecommand \bibfnamefont [1]{#1}%
\providecommand \citenamefont [1]{#1}%
\providecommand \href@noop [0]{\@secondoftwo}%
\providecommand \href [0]{\begingroup \@sanitize@url \@href}%
\providecommand \@href[1]{\@@startlink{#1}\@@href}%
\providecommand \@@href[1]{\endgroup#1\@@endlink}%
\providecommand \@sanitize@url [0]{\catcode `\\12\catcode `\$12\catcode
  `\&12\catcode `\#12\catcode `\^12\catcode `\_12\catcode `\%12\relax}%
\providecommand \@@startlink[1]{}%
\providecommand \@@endlink[0]{}%
\providecommand \url  [0]{\begingroup\@sanitize@url \@url }%
\providecommand \@url [1]{\endgroup\@href {#1}{\urlprefix }}%
\providecommand \urlprefix  [0]{URL }%
\providecommand \Eprint [0]{\href }%
\providecommand \doibase [0]{http://dx.doi.org/}%
\providecommand \selectlanguage [0]{\@gobble}%
\providecommand \bibinfo  [0]{\@secondoftwo}%
\providecommand \bibfield  [0]{\@secondoftwo}%
\providecommand \translation [1]{[#1]}%
\providecommand \BibitemOpen [0]{}%
\providecommand \bibitemStop [0]{}%
\providecommand \bibitemNoStop [0]{.\EOS\space}%
\providecommand \EOS [0]{\spacefactor3000\relax}%
\providecommand \BibitemShut  [1]{\csname bibitem#1\endcsname}%
\let\auto@bib@innerbib\@empty
\bibitem [{\citenamefont {Weinberg}(2018)}]{Weinberg:2018apv}%
  \BibitemOpen
  \bibfield  {author} {\bibinfo {author} {\bibfnamefont {S.}~\bibnamefont
  {Weinberg}},\ }\href {\doibase 10.1103/PhysRevLett.121.220001} {\bibfield
  {journal} {\bibinfo  {journal} {Phys. Rev. Lett.}\ }\textbf {\bibinfo
  {volume} {121}},\ \bibinfo {pages} {220001} (\bibinfo {year}
  {2018})}\BibitemShut {NoStop}%
\bibitem [{\citenamefont {de~Gouvea}\ and\ \citenamefont
  {Vogel}(2013)}]{deGouvea:2013zba}%
  \BibitemOpen
  \bibfield  {author} {\bibinfo {author} {\bibfnamefont {A.}~\bibnamefont
  {de~Gouvea}}\ and\ \bibinfo {author} {\bibfnamefont {P.}~\bibnamefont
  {Vogel}},\ }\href {\doibase 10.1016/j.ppnp.2013.03.006} {\bibfield  {journal}
  {\bibinfo  {journal} {Prog. Part. Nucl. Phys.}\ }\textbf {\bibinfo {volume}
  {71}},\ \bibinfo {pages} {75} (\bibinfo {year} {2013})},\ \Eprint
  {http://arxiv.org/abs/1303.4097} {arXiv:1303.4097 [hep-ph]} \BibitemShut
  {NoStop}%
\bibitem [{\citenamefont {Rodejohann}(2011)}]{Rodejohann:2011mu}%
  \BibitemOpen
  \bibfield  {author} {\bibinfo {author} {\bibfnamefont {W.}~\bibnamefont
  {Rodejohann}},\ }\href {\doibase 10.1142/S0218301311020186} {\bibfield
  {journal} {\bibinfo  {journal} {Int. J. Mod. Phys. E}\ }\textbf {\bibinfo
  {volume} {20}},\ \bibinfo {pages} {1833} (\bibinfo {year} {2011})},\ \Eprint
  {http://arxiv.org/abs/1106.1334} {arXiv:1106.1334 [hep-ph]} \BibitemShut
  {NoStop}%
\bibitem [{\citenamefont {'t~Hooft}(1976)}]{tHooft:1976rip}%
  \BibitemOpen
  \bibfield  {author} {\bibinfo {author} {\bibfnamefont {G.}~\bibnamefont
  {'t~Hooft}},\ }\href {\doibase 10.1103/PhysRevLett.37.8} {\bibfield
  {journal} {\bibinfo  {journal} {Phys. Rev. Lett.}\ }\textbf {\bibinfo
  {volume} {37}},\ \bibinfo {pages} {8} (\bibinfo {year} {1976})}\BibitemShut
  {NoStop}%
\bibitem [{\citenamefont {Davidson}\ \emph {et~al.}(2008)\citenamefont
  {Davidson}, \citenamefont {Nardi},\ and\ \citenamefont
  {Nir}}]{Davidson:2008bu}%
  \BibitemOpen
  \bibfield  {author} {\bibinfo {author} {\bibfnamefont {S.}~\bibnamefont
  {Davidson}}, \bibinfo {author} {\bibfnamefont {E.}~\bibnamefont {Nardi}}, \
  and\ \bibinfo {author} {\bibfnamefont {Y.}~\bibnamefont {Nir}},\ }\href
  {\doibase 10.1016/j.physrep.2008.06.002} {\bibfield  {journal} {\bibinfo
  {journal} {Phys. Rept.}\ }\textbf {\bibinfo {volume} {466}},\ \bibinfo
  {pages} {105} (\bibinfo {year} {2008})},\ \Eprint
  {http://arxiv.org/abs/0802.2962} {arXiv:0802.2962 [hep-ph]} \BibitemShut
  {NoStop}%
\bibitem [{\citenamefont {Fukugita}\ and\ \citenamefont
  {Yanagida}(1986)}]{Fukugita:1986hr}%
  \BibitemOpen
  \bibfield  {author} {\bibinfo {author} {\bibfnamefont {M.}~\bibnamefont
  {Fukugita}}\ and\ \bibinfo {author} {\bibfnamefont {T.}~\bibnamefont
  {Yanagida}},\ }\href {\doibase 10.1016/0370-2693(86)91126-3} {\bibfield
  {journal} {\bibinfo  {journal} {Phys. Lett. B}\ }\textbf {\bibinfo {volume}
  {174}},\ \bibinfo {pages} {45} (\bibinfo {year} {1986})}\BibitemShut
  {NoStop}%
\bibitem [{\citenamefont {Haxton}\ and\ \citenamefont
  {Stephenson}(1984)}]{Haxton:1985am}%
  \BibitemOpen
  \bibfield  {author} {\bibinfo {author} {\bibfnamefont {W.~C.}\ \bibnamefont
  {Haxton}}\ and\ \bibinfo {author} {\bibfnamefont {G.~J.}\ \bibnamefont
  {Stephenson}},\ }\href {\doibase 10.1016/0146-6410(84)90006-1} {\bibfield
  {journal} {\bibinfo  {journal} {Prog. Part. Nucl. Phys.}\ }\textbf {\bibinfo
  {volume} {12}},\ \bibinfo {pages} {409} (\bibinfo {year} {1984})}\BibitemShut
  {NoStop}%
\bibitem [{\citenamefont {Barabash}(2020)}]{Barabash:2020nck}%
  \BibitemOpen
  \bibfield  {author} {\bibinfo {author} {\bibfnamefont {A.}~\bibnamefont
  {Barabash}},\ }\href {\doibase 10.3390/universe6100159} {\bibfield  {journal}
  {\bibinfo  {journal} {Universe}\ }\textbf {\bibinfo {volume} {6}},\ \bibinfo
  {pages} {159} (\bibinfo {year} {2020})},\ \Eprint
  {http://arxiv.org/abs/2009.14451} {arXiv:2009.14451 [nucl-ex]} \BibitemShut
  {NoStop}%
\bibitem [{\citenamefont {Schwingenheuer}(2013)}]{Schwingenheuer:2012zs}%
  \BibitemOpen
  \bibfield  {author} {\bibinfo {author} {\bibfnamefont {B.}~\bibnamefont
  {Schwingenheuer}},\ }\href {\doibase 10.1002/andp.201200222} {\bibfield
  {journal} {\bibinfo  {journal} {Annalen Phys.}\ }\textbf {\bibinfo {volume}
  {525}},\ \bibinfo {pages} {269} (\bibinfo {year} {2013})},\ \Eprint
  {http://arxiv.org/abs/1210.7432} {arXiv:1210.7432 [hep-ex]} \BibitemShut
  {NoStop}%
\bibitem [{\citenamefont {Cremonesi}(2013)}]{Cremonesi:2012av}%
  \BibitemOpen
  \bibfield  {author} {\bibinfo {author} {\bibfnamefont {O.}~\bibnamefont
  {Cremonesi}},\ }\bibfield  {booktitle} {\emph {\bibinfo {booktitle}
  {{Proceedings, Neutrino Oscillation Workshop (NOW 2012): Lecce, Italy,
  September 9-15, 2012}}},\ }\href {\doibase 10.1016/j.nuclphysbps.2013.04.045}
  {\bibfield  {journal} {\bibinfo  {journal} {Nucl. Phys. Proc. Suppl.}\
  }\textbf {\bibinfo {volume} {237-238}},\ \bibinfo {pages} {7} (\bibinfo
  {year} {2013})},\ \Eprint {http://arxiv.org/abs/1212.4885} {arXiv:1212.4885
  [nucl-ex]} \BibitemShut {NoStop}%
\bibitem [{\citenamefont {Elliott}\ and\ \citenamefont
  {Franz}(2015)}]{Elliott:2014iha}%
  \BibitemOpen
  \bibfield  {author} {\bibinfo {author} {\bibfnamefont {S.~R.}\ \bibnamefont
  {Elliott}}\ and\ \bibinfo {author} {\bibfnamefont {M.}~\bibnamefont
  {Franz}},\ }\href {\doibase 10.1103/RevModPhys.87.137} {\bibfield  {journal}
  {\bibinfo  {journal} {Rev. Mod. Phys.}\ }\textbf {\bibinfo {volume} {87}},\
  \bibinfo {pages} {137} (\bibinfo {year} {2015})},\ \Eprint
  {http://arxiv.org/abs/1403.4976} {arXiv:1403.4976 [cond-mat.supr-con]}
  \BibitemShut {NoStop}%
\bibitem [{\citenamefont {Engel}(2017)}]{Engel:2016xgb}%
  \BibitemOpen
  \bibfield  {author} {\bibinfo {author} {\bibfnamefont {J.}~\bibnamefont
  {Engel}, \bibfnamefont {J.~and~Men\'{e}ndez}},\ }\href {\doibase
  10.1088/1361-6633/aa5bc5} {\bibfield  {journal} {\bibinfo  {journal} {Rep.
  Prog. Phys.}\ }\textbf {\bibinfo {volume} {80}},\ \bibinfo {pages} {046301}
  (\bibinfo {year} {2017})},\ \Eprint {http://arxiv.org/abs/1610.06548}
  {arXiv:1610.06548 [nucl-th]} \BibitemShut {NoStop}%
\bibitem [{\citenamefont {Henning}(2016)}]{Henning:2016fad}%
  \BibitemOpen
  \bibfield  {author} {\bibinfo {author} {\bibfnamefont {R.}~\bibnamefont
  {Henning}},\ }\href {\doibase 10.1016/j.revip.2016.03.001} {\bibfield
  {journal} {\bibinfo  {journal} {Rev. Phys.}\ }\textbf {\bibinfo {volume}
  {1}},\ \bibinfo {pages} {29} (\bibinfo {year} {2016})}\BibitemShut {NoStop}%
\bibitem [{\citenamefont {Barabash}(2019)}]{Barabash:2019suz}%
  \BibitemOpen
  \bibfield  {author} {\bibinfo {author} {\bibfnamefont {A.~S.}\ \bibnamefont
  {Barabash}},\ }\href {\doibase 10.3389/fphy.2018.00160} {\bibfield  {journal}
  {\bibinfo  {journal} {Front. in Phys.}\ }\textbf {\bibinfo {volume} {6}},\
  \bibinfo {pages} {160} (\bibinfo {year} {2019})},\ \Eprint
  {http://arxiv.org/abs/1901.11342} {arXiv:1901.11342 [nucl-ex]} \BibitemShut
  {NoStop}%
\bibitem [{\citenamefont {Dolinski}\ \emph {et~al.}(2019)\citenamefont
  {Dolinski}, \citenamefont {Poon},\ and\ \citenamefont
  {Rodejohann}}]{Dolinski:2019nrj}%
  \BibitemOpen
  \bibfield  {author} {\bibinfo {author} {\bibfnamefont {M.~J.}\ \bibnamefont
  {Dolinski}}, \bibinfo {author} {\bibfnamefont {A.~W.}\ \bibnamefont {Poon}},
  \ and\ \bibinfo {author} {\bibfnamefont {W.}~\bibnamefont {Rodejohann}},\
  }\href {\doibase 10.1146/annurev-nucl-101918-023407} {\bibfield  {journal}
  {\bibinfo  {journal} {Ann. Rev. Nucl. Part. Sci.}\ }\textbf {\bibinfo
  {volume} {69}},\ \bibinfo {pages} {219} (\bibinfo {year} {2019})},\ \Eprint
  {http://arxiv.org/abs/1902.04097} {arXiv:1902.04097 [nucl-ex]} \BibitemShut
  {NoStop}%
\bibitem [{\citenamefont {Cirigliano}\ \emph
  {et~al.}(2018{\natexlab{a}})\citenamefont {Cirigliano}, \citenamefont
  {Dekens}, \citenamefont {de~Vries}, \citenamefont {Graesser},\ and\
  \citenamefont {Mereghetti}}]{Cirigliano:2018yza}%
  \BibitemOpen
  \bibfield  {author} {\bibinfo {author} {\bibfnamefont {V.}~\bibnamefont
  {Cirigliano}}, \bibinfo {author} {\bibfnamefont {W.}~\bibnamefont {Dekens}},
  \bibinfo {author} {\bibfnamefont {J.}~\bibnamefont {de~Vries}}, \bibinfo
  {author} {\bibfnamefont {M.~L.}\ \bibnamefont {Graesser}}, \ and\ \bibinfo
  {author} {\bibfnamefont {E.}~\bibnamefont {Mereghetti}},\ }\href {\doibase
  10.1007/JHEP12(2018)097} {\bibfield  {journal} {\bibinfo  {journal} {JHEP}\
  }\textbf {\bibinfo {volume} {12}},\ \bibinfo {pages} {097} (\bibinfo {year}
  {2018}{\natexlab{a}})},\ \Eprint {http://arxiv.org/abs/1806.02780}
  {arXiv:1806.02780 [hep-ph]} \BibitemShut {NoStop}%
\bibitem [{\citenamefont {Cirigliano}\ \emph
  {et~al.}(2018{\natexlab{b}})\citenamefont {Cirigliano}, \citenamefont
  {Dekens}, \citenamefont {De~Vries}, \citenamefont {Graesser}, \citenamefont
  {Mereghetti}, \citenamefont {Pastore},\ and\ \citenamefont
  {Van~Kolck}}]{Cirigliano:2018hja}%
  \BibitemOpen
  \bibfield  {author} {\bibinfo {author} {\bibfnamefont {V.}~\bibnamefont
  {Cirigliano}}, \bibinfo {author} {\bibfnamefont {W.}~\bibnamefont {Dekens}},
  \bibinfo {author} {\bibfnamefont {J.}~\bibnamefont {De~Vries}}, \bibinfo
  {author} {\bibfnamefont {M.~L.}\ \bibnamefont {Graesser}}, \bibinfo {author}
  {\bibfnamefont {E.}~\bibnamefont {Mereghetti}}, \bibinfo {author}
  {\bibfnamefont {S.}~\bibnamefont {Pastore}}, \ and\ \bibinfo {author}
  {\bibfnamefont {U.}~\bibnamefont {Van~Kolck}},\ }\href {\doibase
  10.1103/PhysRevLett.120.202001} {\bibfield  {journal} {\bibinfo  {journal}
  {Phys. Rev. Lett.}\ }\textbf {\bibinfo {volume} {120}},\ \bibinfo {pages}
  {202001} (\bibinfo {year} {2018}{\natexlab{b}})},\ \Eprint
  {http://arxiv.org/abs/1802.10097} {arXiv:1802.10097 [hep-ph]} \BibitemShut
  {NoStop}%
\bibitem [{\citenamefont {Zyla}\ \emph {et~al.}(2020)\citenamefont {Zyla} \emph
  {et~al.}}]{Zyla:2020zbs}%
  \BibitemOpen
  \bibfield  {author} {\bibinfo {author} {\bibfnamefont {P.}~\bibnamefont
  {Zyla}} \emph {et~al.} (\bibinfo {collaboration} {Particle Data Group}),\
  }\href {\doibase 10.1093/ptep/ptaa104} {\bibfield  {journal} {\bibinfo
  {journal} {Prog. Theor. Exp. Phys.}\ }\textbf {\bibinfo {volume} {2020}},\
  \bibinfo {pages} {083C01} (\bibinfo {year} {2020})}\BibitemShut {NoStop}%
\bibitem [{\citenamefont {Esteban}\ \emph {et~al.}(2017)\citenamefont
  {Esteban}, \citenamefont {Gonzalez-Garcia}, \citenamefont {Maltoni},
  \citenamefont {Martinez-Soler},\ and\ \citenamefont
  {Schwetz}}]{Esteban:2016qun}%
  \BibitemOpen
  \bibfield  {author} {\bibinfo {author} {\bibfnamefont {I.}~\bibnamefont
  {Esteban}}, \bibinfo {author} {\bibfnamefont {M.~C.}\ \bibnamefont
  {Gonzalez-Garcia}}, \bibinfo {author} {\bibfnamefont {M.}~\bibnamefont
  {Maltoni}}, \bibinfo {author} {\bibfnamefont {I.}~\bibnamefont
  {Martinez-Soler}}, \ and\ \bibinfo {author} {\bibfnamefont {T.}~\bibnamefont
  {Schwetz}},\ }\href {\doibase 10.1007/JHEP01(2017)087} {\bibfield  {journal}
  {\bibinfo  {journal} {J. High Energ. Phys.}\ }\textbf {\bibinfo {volume}
  {2017}},\ \bibinfo {pages} {087} (\bibinfo {year} {2017})},\ \Eprint
  {http://arxiv.org/abs/1611.01514} {arXiv:1611.01514 [hep-ph]} \BibitemShut
  {NoStop}%
\bibitem [{\citenamefont {Agostini}\ \emph
  {et~al.}(2017{\natexlab{a}})\citenamefont {Agostini}, \citenamefont
  {Benato},\ and\ \citenamefont {Detwiler}}]{Agostini:2017jim}%
  \BibitemOpen
  \bibfield  {author} {\bibinfo {author} {\bibfnamefont {M.}~\bibnamefont
  {Agostini}}, \bibinfo {author} {\bibfnamefont {G.}~\bibnamefont {Benato}}, \
  and\ \bibinfo {author} {\bibfnamefont {J.}~\bibnamefont {Detwiler}},\ }\href
  {\doibase 10.1103/PhysRevD.96.053001} {\bibfield  {journal} {\bibinfo
  {journal} {Phys. Rev. D}\ }\textbf {\bibinfo {volume} {96}},\ \bibinfo
  {pages} {053001} (\bibinfo {year} {2017}{\natexlab{a}})},\ \Eprint
  {http://arxiv.org/abs/1705.02996} {arXiv:1705.02996 [hep-ex]} \BibitemShut
  {NoStop}%
\bibitem [{\citenamefont {Agostini}\ \emph
  {et~al.}(2021{\natexlab{a}})\citenamefont {Agostini}, \citenamefont {Benato},
  \citenamefont {Detwiler}, \citenamefont {Men\'endez},\ and\ \citenamefont
  {Vissani}}]{Agostini:2021kba}%
  \BibitemOpen
  \bibfield  {author} {\bibinfo {author} {\bibfnamefont {M.}~\bibnamefont
  {Agostini}}, \bibinfo {author} {\bibfnamefont {G.}~\bibnamefont {Benato}},
  \bibinfo {author} {\bibfnamefont {J.~A.}\ \bibnamefont {Detwiler}}, \bibinfo
  {author} {\bibfnamefont {J.}~\bibnamefont {Men\'endez}}, \ and\ \bibinfo
  {author} {\bibfnamefont {F.}~\bibnamefont {Vissani}},\ }\href@noop {} {\
  (\bibinfo {year} {2021}{\natexlab{a}})},\ \Eprint
  {http://arxiv.org/abs/2107.09104} {arXiv:2107.09104 [hep-ph]} \BibitemShut
  {NoStop}%
\bibitem [{\citenamefont {Kotila}\ and\ \citenamefont
  {Iachello}(2012)}]{Kotila:2012zza}%
  \BibitemOpen
  \bibfield  {author} {\bibinfo {author} {\bibfnamefont {J.}~\bibnamefont
  {Kotila}}\ and\ \bibinfo {author} {\bibfnamefont {F.}~\bibnamefont
  {Iachello}},\ }\href {\doibase 10.1103/PhysRevC.85.034316} {\bibfield
  {journal} {\bibinfo  {journal} {Phys. Rev. C}\ }\textbf {\bibinfo {volume}
  {85}},\ \bibinfo {pages} {034316} (\bibinfo {year} {2012})},\ \Eprint
  {http://arxiv.org/abs/1209.5722} {arXiv:1209.5722 [nucl-th]} \BibitemShut
  {NoStop}%
\bibitem [{\citenamefont {Mirea}\ \emph {et~al.}(2015)\citenamefont {Mirea},
  \citenamefont {Pahomi},\ and\ \citenamefont {Stoica}}]{Mirea:2015nsl}%
  \BibitemOpen
  \bibfield  {author} {\bibinfo {author} {\bibfnamefont {M.}~\bibnamefont
  {Mirea}}, \bibinfo {author} {\bibfnamefont {T.}~\bibnamefont {Pahomi}}, \
  and\ \bibinfo {author} {\bibfnamefont {S.}~\bibnamefont {Stoica}},\
  }\href@noop {} {\bibfield  {journal} {\bibinfo  {journal} {Rom. Rep. Phys.}\
  }\textbf {\bibinfo {volume} {67}},\ \bibinfo {pages} {872} (\bibinfo {year}
  {2015})}\BibitemShut {NoStop}%
\bibitem [{\citenamefont {McKeown}\ \emph {et~al.}(2014)\citenamefont {McKeown}
  \emph {et~al.}}]{McKeown2014}%
  \BibitemOpen
  \bibfield  {author} {\bibinfo {author} {\bibfnamefont {R.}~\bibnamefont
  {McKeown}} \emph {et~al.},\ }\href
  {https://science.osti.gov/-/media/np/nsac/pdf/docs/2014/NLDBD_Report_2014_Final.pdf}
  {\enquote {\bibinfo {title} {{Neutrinoless Double Beta Decay: Report to the
  Nuclear Science Advisory Committee}},}\ } (\bibinfo {year} {2014}),\ \bibinfo
  {note} {available at
  \url{https://science.osti.gov/-/media/np/nsac/pdf/docs/2014/NLDBD_Report_2014_Final.pdf}}\BibitemShut
  {NoStop}%
\bibitem [{\citenamefont {Belley}\ \emph {et~al.}(2021)\citenamefont {Belley},
  \citenamefont {Payne}, \citenamefont {Stroberg}, \citenamefont {Miyagi},\
  and\ \citenamefont {Holt}}]{Belley:2020ejd}%
  \BibitemOpen
  \bibfield  {author} {\bibinfo {author} {\bibfnamefont {A.}~\bibnamefont
  {Belley}}, \bibinfo {author} {\bibfnamefont {C.~G.}\ \bibnamefont {Payne}},
  \bibinfo {author} {\bibfnamefont {S.~R.}\ \bibnamefont {Stroberg}}, \bibinfo
  {author} {\bibfnamefont {T.}~\bibnamefont {Miyagi}}, \ and\ \bibinfo {author}
  {\bibfnamefont {J.~D.}\ \bibnamefont {Holt}},\ }\href {\doibase
  10.1103/PhysRevLett.126.042502} {\bibfield  {journal} {\bibinfo  {journal}
  {Phys. Rev. Lett.}\ }\textbf {\bibinfo {volume} {126}},\ \bibinfo {pages}
  {042502} (\bibinfo {year} {2021})},\ \Eprint
  {http://arxiv.org/abs/2008.06588} {arXiv:2008.06588 [nucl-th]} \BibitemShut
  {NoStop}%
\bibitem [{\citenamefont {Horoi}\ and\ \citenamefont
  {Neacsu}(2016)}]{Horoi:2015tkc}%
  \BibitemOpen
  \bibfield  {author} {\bibinfo {author} {\bibfnamefont {M.}~\bibnamefont
  {Horoi}}\ and\ \bibinfo {author} {\bibfnamefont {A.}~\bibnamefont {Neacsu}},\
  }\href {\doibase 10.1103/PhysRevC.93.024308} {\bibfield  {journal} {\bibinfo
  {journal} {Phys. Rev. C}\ }\textbf {\bibinfo {volume} {93}},\ \bibinfo
  {pages} {024308} (\bibinfo {year} {2016})},\ \Eprint
  {http://arxiv.org/abs/1511.03711} {arXiv:1511.03711 [nucl-th]} \BibitemShut
  {NoStop}%
\bibitem [{\citenamefont {Men\'endez}(2018)}]{Menendez:2017fdf}%
  \BibitemOpen
  \bibfield  {author} {\bibinfo {author} {\bibfnamefont {J.}~\bibnamefont
  {Men\'endez}},\ }\href {\doibase 10.1088/1361-6471/aa9bd4} {\bibfield
  {journal} {\bibinfo  {journal} {J. Phys. G}\ }\textbf {\bibinfo {volume}
  {45}},\ \bibinfo {pages} {014003} (\bibinfo {year} {2018})},\ \Eprint
  {http://arxiv.org/abs/1804.02105} {arXiv:1804.02105 [nucl-th]} \BibitemShut
  {NoStop}%
\bibitem [{\citenamefont {Coraggio}\ \emph {et~al.}(2020)\citenamefont
  {Coraggio}, \citenamefont {Gargano}, \citenamefont {Itaco}, \citenamefont
  {Mancino},\ and\ \citenamefont {Nowacki}}]{Coraggio:2020hwx}%
  \BibitemOpen
  \bibfield  {author} {\bibinfo {author} {\bibfnamefont {L.}~\bibnamefont
  {Coraggio}}, \bibinfo {author} {\bibfnamefont {A.}~\bibnamefont {Gargano}},
  \bibinfo {author} {\bibfnamefont {N.}~\bibnamefont {Itaco}}, \bibinfo
  {author} {\bibfnamefont {R.}~\bibnamefont {Mancino}}, \ and\ \bibinfo
  {author} {\bibfnamefont {F.}~\bibnamefont {Nowacki}},\ }\href {\doibase
  10.1103/PhysRevC.101.044315} {\bibfield  {journal} {\bibinfo  {journal}
  {Phys. Rev. C}\ }\textbf {\bibinfo {volume} {101}},\ \bibinfo {pages}
  {044315} (\bibinfo {year} {2020})},\ \Eprint
  {http://arxiv.org/abs/2001.00890} {arXiv:2001.00890 [nucl-th]} \BibitemShut
  {NoStop}%
\bibitem [{\citenamefont {Mustonen}\ and\ \citenamefont
  {Engel}(2013)}]{Mustonen:2013zu}%
  \BibitemOpen
  \bibfield  {author} {\bibinfo {author} {\bibfnamefont {M.~T.}\ \bibnamefont
  {Mustonen}}\ and\ \bibinfo {author} {\bibfnamefont {J.}~\bibnamefont
  {Engel}},\ }\href {\doibase 10.1103/PhysRevC.87.064302} {\bibfield  {journal}
  {\bibinfo  {journal} {Phys. Rev. C}\ }\textbf {\bibinfo {volume} {87}},\
  \bibinfo {pages} {064302} (\bibinfo {year} {2013})},\ \Eprint
  {http://arxiv.org/abs/1301.6997} {arXiv:1301.6997 [nucl-th]} \BibitemShut
  {NoStop}%
\bibitem [{\citenamefont {Hyv\"arinen}\ and\ \citenamefont
  {Suhonen}(2015)}]{Hyvarinen:2015bda}%
  \BibitemOpen
  \bibfield  {author} {\bibinfo {author} {\bibfnamefont {J.}~\bibnamefont
  {Hyv\"arinen}}\ and\ \bibinfo {author} {\bibfnamefont {J.}~\bibnamefont
  {Suhonen}},\ }\href {\doibase 10.1103/PhysRevC.91.024613} {\bibfield
  {journal} {\bibinfo  {journal} {Phys. Rev. C}\ }\textbf {\bibinfo {volume}
  {91}},\ \bibinfo {pages} {024613} (\bibinfo {year} {2015})}\BibitemShut
  {NoStop}%
\bibitem [{\citenamefont {\v{S}imkovic}\ \emph {et~al.}(2018)\citenamefont
  {\v{S}imkovic}, \citenamefont {Smetana},\ and\ \citenamefont
  {Vogel}}]{Simkovic:2018hiq}%
  \BibitemOpen
  \bibfield  {author} {\bibinfo {author} {\bibfnamefont {F.}~\bibnamefont
  {\v{S}imkovic}}, \bibinfo {author} {\bibfnamefont {A.}~\bibnamefont
  {Smetana}}, \ and\ \bibinfo {author} {\bibfnamefont {P.}~\bibnamefont
  {Vogel}},\ }\href {\doibase 10.1103/PhysRevC.98.064325} {\bibfield  {journal}
  {\bibinfo  {journal} {Phys. Rev. C}\ }\textbf {\bibinfo {volume} {98}},\
  \bibinfo {pages} {064325} (\bibinfo {year} {2018})},\ \Eprint
  {http://arxiv.org/abs/1808.05016} {arXiv:1808.05016 [nucl-th]} \BibitemShut
  {NoStop}%
\bibitem [{\citenamefont {Fang}\ \emph {et~al.}(2018)\citenamefont {Fang},
  \citenamefont {Faessler},\ and\ \citenamefont {Simkovic}}]{Fang:2018tui}%
  \BibitemOpen
  \bibfield  {author} {\bibinfo {author} {\bibfnamefont {D.-L.}\ \bibnamefont
  {Fang}}, \bibinfo {author} {\bibfnamefont {A.}~\bibnamefont {Faessler}}, \
  and\ \bibinfo {author} {\bibfnamefont {F.}~\bibnamefont {Simkovic}},\ }\href
  {\doibase 10.1103/PhysRevC.97.045503} {\bibfield  {journal} {\bibinfo
  {journal} {Phys. Rev. C}\ }\textbf {\bibinfo {volume} {97}},\ \bibinfo
  {pages} {045503} (\bibinfo {year} {2018})},\ \Eprint
  {http://arxiv.org/abs/1803.09195} {arXiv:1803.09195 [nucl-th]} \BibitemShut
  {NoStop}%
\bibitem [{\citenamefont {Barea}\ \emph {et~al.}(2015)\citenamefont {Barea},
  \citenamefont {Kotila},\ and\ \citenamefont {Iachello}}]{Barea:2015kwa}%
  \BibitemOpen
  \bibfield  {author} {\bibinfo {author} {\bibfnamefont {J.}~\bibnamefont
  {Barea}}, \bibinfo {author} {\bibfnamefont {J.}~\bibnamefont {Kotila}}, \
  and\ \bibinfo {author} {\bibfnamefont {F.}~\bibnamefont {Iachello}},\ }\href
  {\doibase 10.1103/PhysRevC.91.034304} {\bibfield  {journal} {\bibinfo
  {journal} {Phys. Rev. C}\ }\textbf {\bibinfo {volume} {91}},\ \bibinfo
  {pages} {034304} (\bibinfo {year} {2015})},\ \Eprint
  {http://arxiv.org/abs/1506.08530} {arXiv:1506.08530 [nucl-th]} \BibitemShut
  {NoStop}%
\bibitem [{\citenamefont {Rodriguez}\ and\ \citenamefont
  {Martinez-Pinedo}(2010)}]{Rodriguez:2010mn}%
  \BibitemOpen
  \bibfield  {author} {\bibinfo {author} {\bibfnamefont {T.~R.}\ \bibnamefont
  {Rodriguez}}\ and\ \bibinfo {author} {\bibfnamefont {G.}~\bibnamefont
  {Martinez-Pinedo}},\ }\href {\doibase 10.1103/PhysRevLett.105.252503}
  {\bibfield  {journal} {\bibinfo  {journal} {Phys. Rev. Lett.}\ }\textbf
  {\bibinfo {volume} {105}},\ \bibinfo {pages} {252503} (\bibinfo {year}
  {2010})},\ \Eprint {http://arxiv.org/abs/1008.5260} {arXiv:1008.5260
  [nucl-th]} \BibitemShut {NoStop}%
\bibitem [{\citenamefont {Rodriguez}\ and\ \citenamefont
  {Martinez-Pinedo}(2013)}]{Rodriguez:2012rv}%
  \BibitemOpen
  \bibfield  {author} {\bibinfo {author} {\bibfnamefont {T.~R.}\ \bibnamefont
  {Rodriguez}}\ and\ \bibinfo {author} {\bibfnamefont {G.}~\bibnamefont
  {Martinez-Pinedo}},\ }\href {\doibase 10.1016/j.physletb.2012.12.063}
  {\bibfield  {journal} {\bibinfo  {journal} {Phys. Lett. B}\ }\textbf
  {\bibinfo {volume} {719}},\ \bibinfo {pages} {174} (\bibinfo {year}
  {2013})},\ \Eprint {http://arxiv.org/abs/1210.3225} {arXiv:1210.3225
  [nucl-th]} \BibitemShut {NoStop}%
\bibitem [{\citenamefont {L\'{o}pez~Vaquero}\ \emph {et~al.}(2013)\citenamefont
  {L\'{o}pez~Vaquero}, \citenamefont {Rodr\'{i}guez},\ and\ \citenamefont
  {Egido}}]{Vaquero:2014dna}%
  \BibitemOpen
  \bibfield  {author} {\bibinfo {author} {\bibfnamefont {N.}~\bibnamefont
  {L\'{o}pez~Vaquero}}, \bibinfo {author} {\bibfnamefont {T.~R.}\ \bibnamefont
  {Rodr\'{i}guez}}, \ and\ \bibinfo {author} {\bibfnamefont {J.~L.}\
  \bibnamefont {Egido}},\ }\href {\doibase 10.1103/PhysRevLett.111.142501}
  {\bibfield  {journal} {\bibinfo  {journal} {Phys. Rev. Lett.}\ }\textbf
  {\bibinfo {volume} {111}},\ \bibinfo {pages} {142501} (\bibinfo {year}
  {2013})},\ \Eprint {http://arxiv.org/abs/1401.0650} {arXiv:1401.0650
  [nucl-th]} \BibitemShut {NoStop}%
\bibitem [{\citenamefont {Song}\ \emph {et~al.}(2017)\citenamefont {Song},
  \citenamefont {Yao}, \citenamefont {Ring},\ and\ \citenamefont
  {Meng}}]{Song:2017ktj}%
  \BibitemOpen
  \bibfield  {author} {\bibinfo {author} {\bibfnamefont {L.}~\bibnamefont
  {Song}}, \bibinfo {author} {\bibfnamefont {J.}~\bibnamefont {Yao}}, \bibinfo
  {author} {\bibfnamefont {P.}~\bibnamefont {Ring}}, \ and\ \bibinfo {author}
  {\bibfnamefont {J.}~\bibnamefont {Meng}},\ }\href {\doibase
  10.1103/PhysRevC.95.024305} {\bibfield  {journal} {\bibinfo  {journal} {Phys.
  Rev. C}\ }\textbf {\bibinfo {volume} {95}},\ \bibinfo {pages} {024305}
  (\bibinfo {year} {2017})},\ \Eprint {http://arxiv.org/abs/1702.02448}
  {arXiv:1702.02448 [nucl-th]} \BibitemShut {NoStop}%
\bibitem [{\citenamefont {Towner}(1987)}]{Towner:1987zz}%
  \BibitemOpen
  \bibfield  {author} {\bibinfo {author} {\bibfnamefont {I.~S.}\ \bibnamefont
  {Towner}},\ }\href {\doibase 10.1016/0370-1573(87)90138-4} {\bibfield
  {journal} {\bibinfo  {journal} {Phys. Rept.}\ }\textbf {\bibinfo {volume}
  {155}},\ \bibinfo {pages} {263} (\bibinfo {year} {1987})}\BibitemShut
  {NoStop}%
\bibitem [{\citenamefont {\v{S}imkovic}\ \emph {et~al.}(2013)\citenamefont
  {\v{S}imkovic}, \citenamefont {Rodin}, \citenamefont {Faessler},\ and\
  \citenamefont {Vogel}}]{Simkovic:2013qiy}%
  \BibitemOpen
  \bibfield  {author} {\bibinfo {author} {\bibfnamefont {F.}~\bibnamefont
  {\v{S}imkovic}}, \bibinfo {author} {\bibfnamefont {V.}~\bibnamefont {Rodin}},
  \bibinfo {author} {\bibfnamefont {A.}~\bibnamefont {Faessler}}, \ and\
  \bibinfo {author} {\bibfnamefont {P.}~\bibnamefont {Vogel}},\ }\href
  {\doibase 10.1103/PhysRevC.87.045501} {\bibfield  {journal} {\bibinfo
  {journal} {Phys. Rev. C}\ }\textbf {\bibinfo {volume} {87}},\ \bibinfo
  {pages} {045501} (\bibinfo {year} {2013})},\ \Eprint
  {http://arxiv.org/abs/1302.1509} {arXiv:1302.1509 [nucl-th]} \BibitemShut
  {NoStop}%
\bibitem [{\citenamefont {Ejiri}\ and\ \citenamefont
  {Suhonen}(2015)}]{Ejiri:2015wna}%
  \BibitemOpen
  \bibfield  {author} {\bibinfo {author} {\bibfnamefont {H.}~\bibnamefont
  {Ejiri}}\ and\ \bibinfo {author} {\bibfnamefont {J.}~\bibnamefont
  {Suhonen}},\ }\href {\doibase 10.1088/0954-3899/42/5/055201} {\bibfield
  {journal} {\bibinfo  {journal} {J. Phys. G}\ }\textbf {\bibinfo {volume}
  {42}},\ \bibinfo {pages} {055201} (\bibinfo {year} {2015})}\BibitemShut
  {NoStop}%
\bibitem [{\citenamefont {Suzuki}\ \emph {et~al.}(2018)\citenamefont {Suzuki},
  \citenamefont {Chiba}, \citenamefont {Yoshida}, \citenamefont {Takahashi},\
  and\ \citenamefont {Umeda}}]{Suzuki:2018aey}%
  \BibitemOpen
  \bibfield  {author} {\bibinfo {author} {\bibfnamefont {T.}~\bibnamefont
  {Suzuki}}, \bibinfo {author} {\bibfnamefont {S.}~\bibnamefont {Chiba}},
  \bibinfo {author} {\bibfnamefont {T.}~\bibnamefont {Yoshida}}, \bibinfo
  {author} {\bibfnamefont {K.}~\bibnamefont {Takahashi}}, \ and\ \bibinfo
  {author} {\bibfnamefont {H.}~\bibnamefont {Umeda}},\ }\href {\doibase
  10.1103/PhysRevC.98.034613} {\bibfield  {journal} {\bibinfo  {journal} {Phys.
  Rev. C}\ }\textbf {\bibinfo {volume} {98}},\ \bibinfo {pages} {034613}
  (\bibinfo {year} {2018})},\ \Eprint {http://arxiv.org/abs/1807.02367}
  {arXiv:1807.02367 [nucl-th]} \BibitemShut {NoStop}%
\bibitem [{\citenamefont {Lovato}\ \emph {et~al.}(2020)\citenamefont {Lovato},
  \citenamefont {Carlson}, \citenamefont {Gandolfi}, \citenamefont {Rocco},\
  and\ \citenamefont {Schiavilla}}]{Lovato:2020kba}%
  \BibitemOpen
  \bibfield  {author} {\bibinfo {author} {\bibfnamefont {A.}~\bibnamefont
  {Lovato}}, \bibinfo {author} {\bibfnamefont {J.}~\bibnamefont {Carlson}},
  \bibinfo {author} {\bibfnamefont {S.}~\bibnamefont {Gandolfi}}, \bibinfo
  {author} {\bibfnamefont {N.}~\bibnamefont {Rocco}}, \ and\ \bibinfo {author}
  {\bibfnamefont {R.}~\bibnamefont {Schiavilla}},\ }\href {\doibase
  10.1103/PhysRevX.10.031068} {\bibfield  {journal} {\bibinfo  {journal} {Phys.
  Rev. X}\ }\textbf {\bibinfo {volume} {10}},\ \bibinfo {pages} {031068}
  (\bibinfo {year} {2020})},\ \Eprint {http://arxiv.org/abs/2003.07710}
  {arXiv:2003.07710 [nucl-th]} \BibitemShut {NoStop}%
\bibitem [{\citenamefont {Pastore}\ \emph {et~al.}(2018)\citenamefont
  {Pastore}, \citenamefont {Baroni}, \citenamefont {Carlson}, \citenamefont
  {Gandolfi}, \citenamefont {Pieper}, \citenamefont {Schiavilla},\ and\
  \citenamefont {Wiringa}}]{Pastore:2017uwc}%
  \BibitemOpen
  \bibfield  {author} {\bibinfo {author} {\bibfnamefont {S.}~\bibnamefont
  {Pastore}}, \bibinfo {author} {\bibfnamefont {A.}~\bibnamefont {Baroni}},
  \bibinfo {author} {\bibfnamefont {J.}~\bibnamefont {Carlson}}, \bibinfo
  {author} {\bibfnamefont {S.}~\bibnamefont {Gandolfi}}, \bibinfo {author}
  {\bibfnamefont {S.~C.}\ \bibnamefont {Pieper}}, \bibinfo {author}
  {\bibfnamefont {R.}~\bibnamefont {Schiavilla}}, \ and\ \bibinfo {author}
  {\bibfnamefont {R.~B.}\ \bibnamefont {Wiringa}},\ }\href {\doibase
  10.1103/PhysRevC.97.022501} {\bibfield  {journal} {\bibinfo  {journal} {Phys.
  Rev. C}\ }\textbf {\bibinfo {volume} {97}},\ \bibinfo {pages} {022501}
  (\bibinfo {year} {2018})},\ \Eprint {http://arxiv.org/abs/1709.03592}
  {arXiv:1709.03592 [nucl-th]} \BibitemShut {NoStop}%
\bibitem [{\citenamefont {Gysbers}\ \emph {et~al.}(2019)\citenamefont {Gysbers}
  \emph {et~al.}}]{Gysbers:2019uyb}%
  \BibitemOpen
  \bibfield  {author} {\bibinfo {author} {\bibfnamefont {P.}~\bibnamefont
  {Gysbers}} \emph {et~al.},\ }\href {\doibase 10.1038/s41567-019-0450-7}
  {\bibfield  {journal} {\bibinfo  {journal} {Nature Phys.}\ }\textbf {\bibinfo
  {volume} {15}},\ \bibinfo {pages} {428} (\bibinfo {year} {2019})},\ \Eprint
  {http://arxiv.org/abs/1903.00047} {arXiv:1903.00047 [nucl-th]} \BibitemShut
  {NoStop}%
\bibitem [{\citenamefont {Cirigliano}\ \emph {et~al.}(2019)\citenamefont
  {Cirigliano}, \citenamefont {Dekens}, \citenamefont {De~Vries}, \citenamefont
  {Graesser}, \citenamefont {Mereghetti}, \citenamefont {Pastore},
  \citenamefont {Piarulli}, \citenamefont {Van~Kolck},\ and\ \citenamefont
  {Wiringa}}]{Cirigliano:2019vdj}%
  \BibitemOpen
  \bibfield  {author} {\bibinfo {author} {\bibfnamefont {V.}~\bibnamefont
  {Cirigliano}}, \bibinfo {author} {\bibfnamefont {W.}~\bibnamefont {Dekens}},
  \bibinfo {author} {\bibfnamefont {J.}~\bibnamefont {De~Vries}}, \bibinfo
  {author} {\bibfnamefont {M.~L.}\ \bibnamefont {Graesser}}, \bibinfo {author}
  {\bibfnamefont {E.}~\bibnamefont {Mereghetti}}, \bibinfo {author}
  {\bibfnamefont {S.}~\bibnamefont {Pastore}}, \bibinfo {author} {\bibfnamefont
  {M.}~\bibnamefont {Piarulli}}, \bibinfo {author} {\bibfnamefont
  {U.}~\bibnamefont {Van~Kolck}}, \ and\ \bibinfo {author} {\bibfnamefont
  {R.~B.}\ \bibnamefont {Wiringa}},\ }\href {\doibase
  10.1103/PhysRevC.100.055504} {\bibfield  {journal} {\bibinfo  {journal}
  {Phys. Rev. C}\ }\textbf {\bibinfo {volume} {100}},\ \bibinfo {pages}
  {055504} (\bibinfo {year} {2019})},\ \Eprint
  {http://arxiv.org/abs/1907.11254} {arXiv:1907.11254 [nucl-th]} \BibitemShut
  {NoStop}%
\bibitem [{\citenamefont {Wirth}\ \emph {et~al.}(2021)\citenamefont {Wirth},
  \citenamefont {Yao},\ and\ \citenamefont {Hergert}}]{Wirth:2021pij}%
  \BibitemOpen
  \bibfield  {author} {\bibinfo {author} {\bibfnamefont {R.}~\bibnamefont
  {Wirth}}, \bibinfo {author} {\bibfnamefont {J.~M.}\ \bibnamefont {Yao}}, \
  and\ \bibinfo {author} {\bibfnamefont {H.}~\bibnamefont {Hergert}},\
  }\href@noop {} {\  (\bibinfo {year} {2021})},\ \Eprint
  {http://arxiv.org/abs/2105.05415} {arXiv:2105.05415 [nucl-th]} \BibitemShut
  {NoStop}%
\bibitem [{\citenamefont {Davoudi}\ \emph {et~al.}(2021)\citenamefont
  {Davoudi}, \citenamefont {Detmold}, \citenamefont {Orginos}, \citenamefont
  {Parre\~no}, \citenamefont {Savage}, \citenamefont {Shanahan},\ and\
  \citenamefont {Wagman}}]{Davoudi:2020ngi}%
  \BibitemOpen
  \bibfield  {author} {\bibinfo {author} {\bibfnamefont {Z.}~\bibnamefont
  {Davoudi}}, \bibinfo {author} {\bibfnamefont {W.}~\bibnamefont {Detmold}},
  \bibinfo {author} {\bibfnamefont {K.}~\bibnamefont {Orginos}}, \bibinfo
  {author} {\bibfnamefont {A.}~\bibnamefont {Parre\~no}}, \bibinfo {author}
  {\bibfnamefont {M.~J.}\ \bibnamefont {Savage}}, \bibinfo {author}
  {\bibfnamefont {P.}~\bibnamefont {Shanahan}}, \ and\ \bibinfo {author}
  {\bibfnamefont {M.~L.}\ \bibnamefont {Wagman}},\ }\href {\doibase
  10.1016/j.physrep.2020.10.004} {\bibfield  {journal} {\bibinfo  {journal}
  {Phys. Rep.}\ }\textbf {\bibinfo {volume} {900}},\ \bibinfo {pages} {1}
  (\bibinfo {year} {2021})},\ \Eprint {http://arxiv.org/abs/2008.11160}
  {arXiv:2008.11160 [hep-lat]} \BibitemShut {NoStop}%
\bibitem [{\citenamefont {Acero}\ \emph {et~al.}(2019)\citenamefont {Acero}
  \emph {et~al.}}]{Acero:2019ksn}%
  \BibitemOpen
  \bibfield  {author} {\bibinfo {author} {\bibfnamefont {M.~A.}\ \bibnamefont
  {Acero}} \emph {et~al.} (\bibinfo {collaboration} {NOvA}),\ }\href {\doibase
  10.1103/PhysRevLett.123.151803} {\bibfield  {journal} {\bibinfo  {journal}
  {Phys. Rev. Lett.}\ }\textbf {\bibinfo {volume} {123}},\ \bibinfo {pages}
  {151803} (\bibinfo {year} {2019})},\ \Eprint
  {http://arxiv.org/abs/1906.04907} {arXiv:1906.04907 [hep-ex]} \BibitemShut
  {NoStop}%
\bibitem [{\citenamefont {Abe}\ \emph {et~al.}(2020)\citenamefont {Abe} \emph
  {et~al.}}]{Abe:2019vii}%
  \BibitemOpen
  \bibfield  {author} {\bibinfo {author} {\bibfnamefont {K.}~\bibnamefont
  {Abe}} \emph {et~al.} (\bibinfo {collaboration} {T2K}),\ }\href {\doibase
  10.1038/s41586-020-2177-0} {\bibfield  {journal} {\bibinfo  {journal}
  {Nature}\ }\textbf {\bibinfo {volume} {580}},\ \bibinfo {pages} {339}
  (\bibinfo {year} {2020})},\ \bibinfo {note} {[Erratum: Nature 583, E16
  (2020)]},\ \Eprint {http://arxiv.org/abs/1910.03887} {arXiv:1910.03887
  [hep-ex]} \BibitemShut {NoStop}%
\bibitem [{\citenamefont {Esteban}\ \emph {et~al.}(2020)\citenamefont
  {Esteban}, \citenamefont {Gonzalez-Garcia}, \citenamefont {Maltoni},
  \citenamefont {Schwetz},\ and\ \citenamefont {Zhou}}]{Esteban:2020cvm}%
  \BibitemOpen
  \bibfield  {author} {\bibinfo {author} {\bibfnamefont {I.}~\bibnamefont
  {Esteban}}, \bibinfo {author} {\bibfnamefont {M.~C.}\ \bibnamefont
  {Gonzalez-Garcia}}, \bibinfo {author} {\bibfnamefont {M.}~\bibnamefont
  {Maltoni}}, \bibinfo {author} {\bibfnamefont {T.}~\bibnamefont {Schwetz}}, \
  and\ \bibinfo {author} {\bibfnamefont {A.}~\bibnamefont {Zhou}},\ }\href
  {\doibase 10.1007/JHEP09(2020)178} {\bibfield  {journal} {\bibinfo  {journal}
  {JHEP}\ }\textbf {\bibinfo {volume} {09}},\ \bibinfo {pages} {178} (\bibinfo
  {year} {2020})},\ \Eprint {http://arxiv.org/abs/2007.14792} {arXiv:2007.14792
  [hep-ph]} \BibitemShut {NoStop}%
\bibitem [{\citenamefont {Kelly}\ \emph {et~al.}(2021)\citenamefont {Kelly},
  \citenamefont {Machado}, \citenamefont {Parke}, \citenamefont
  {Perez-Gonzalez},\ and\ \citenamefont {Funchal}}]{Kelly:2020fkv}%
  \BibitemOpen
  \bibfield  {author} {\bibinfo {author} {\bibfnamefont {K.~J.}\ \bibnamefont
  {Kelly}}, \bibinfo {author} {\bibfnamefont {P.~A.~N.}\ \bibnamefont
  {Machado}}, \bibinfo {author} {\bibfnamefont {S.~J.}\ \bibnamefont {Parke}},
  \bibinfo {author} {\bibfnamefont {Y.~F.}\ \bibnamefont {Perez-Gonzalez}}, \
  and\ \bibinfo {author} {\bibfnamefont {R.~Z.}\ \bibnamefont {Funchal}},\
  }\href {\doibase 10.1103/PhysRevD.103.013004} {\bibfield  {journal} {\bibinfo
   {journal} {Phys. Rev. D}\ }\textbf {\bibinfo {volume} {103}},\ \bibinfo
  {pages} {013004} (\bibinfo {year} {2021})},\ \Eprint
  {http://arxiv.org/abs/2007.08526} {arXiv:2007.08526 [hep-ph]} \BibitemShut
  {NoStop}%
\bibitem [{\citenamefont {Capozzi}\ \emph {et~al.}(2020)\citenamefont
  {Capozzi}, \citenamefont {Di~Valentino}, \citenamefont {Lisi}, \citenamefont
  {Marrone}, \citenamefont {Melchiorri},\ and\ \citenamefont
  {Palazzo}}]{Capozzi:2020add}%
  \BibitemOpen
  \bibfield  {author} {\bibinfo {author} {\bibfnamefont {F.}~\bibnamefont
  {Capozzi}}, \bibinfo {author} {\bibfnamefont {E.}~\bibnamefont
  {Di~Valentino}}, \bibinfo {author} {\bibfnamefont {E.}~\bibnamefont {Lisi}},
  \bibinfo {author} {\bibfnamefont {A.}~\bibnamefont {Marrone}}, \bibinfo
  {author} {\bibfnamefont {A.}~\bibnamefont {Melchiorri}}, \ and\ \bibinfo
  {author} {\bibfnamefont {A.}~\bibnamefont {Palazzo}},\ }\href {\doibase
  10.1103/PhysRevD.101.116013} {\bibfield  {journal} {\bibinfo  {journal}
  {Phys. Rev. D}\ }\textbf {\bibinfo {volume} {101}},\ \bibinfo {pages}
  {116013} (\bibinfo {year} {2020})}\BibitemShut {NoStop}%
\bibitem [{\citenamefont {de~Salas}\ \emph {et~al.}(2021)\citenamefont
  {de~Salas}, \citenamefont {Forero}, \citenamefont {Gariazzo}, \citenamefont
  {Mart\'\i{}nez-Mirav\'e}, \citenamefont {Mena}, \citenamefont {Ternes},
  \citenamefont {T\'ortola},\ and\ \citenamefont {Valle}}]{deSalas:2020pgw}%
  \BibitemOpen
  \bibfield  {author} {\bibinfo {author} {\bibfnamefont {P.~F.}\ \bibnamefont
  {de~Salas}}, \bibinfo {author} {\bibfnamefont {D.~V.}\ \bibnamefont
  {Forero}}, \bibinfo {author} {\bibfnamefont {S.}~\bibnamefont {Gariazzo}},
  \bibinfo {author} {\bibfnamefont {P.}~\bibnamefont {Mart\'\i{}nez-Mirav\'e}},
  \bibinfo {author} {\bibfnamefont {O.}~\bibnamefont {Mena}}, \bibinfo {author}
  {\bibfnamefont {C.~A.}\ \bibnamefont {Ternes}}, \bibinfo {author}
  {\bibfnamefont {M.}~\bibnamefont {T\'ortola}}, \ and\ \bibinfo {author}
  {\bibfnamefont {J.~W.~F.}\ \bibnamefont {Valle}},\ }\href {\doibase
  10.1007/JHEP02(2021)071} {\bibfield  {journal} {\bibinfo  {journal} {JHEP}\
  }\textbf {\bibinfo {volume} {02}},\ \bibinfo {pages} {071} (\bibinfo {year}
  {2021})},\ \Eprint {http://arxiv.org/abs/2006.11237} {arXiv:2006.11237
  [hep-ph]} \BibitemShut {NoStop}%
\bibitem [{\citenamefont {Lu}(2021)}]{Lu:2021lfe}%
  \BibitemOpen
  \bibfield  {author} {\bibinfo {author} {\bibfnamefont {H.}~\bibnamefont {Lu}}
  (\bibinfo {collaboration} {JUNO}),\ }\href {\doibase
  10.1088/1402-4896/ac0a29} {\bibfield  {journal} {\bibinfo  {journal} {Phys.
  Scripta}\ }\textbf {\bibinfo {volume} {96}},\ \bibinfo {pages} {094013}
  (\bibinfo {year} {2021})}\BibitemShut {NoStop}%
\bibitem [{\citenamefont {Aiello}\ \emph {et~al.}(2021)\citenamefont {Aiello}
  \emph {et~al.}}]{KM3NeT:2021ozk}%
  \BibitemOpen
  \bibfield  {author} {\bibinfo {author} {\bibfnamefont {S.}~\bibnamefont
  {Aiello}} \emph {et~al.} (\bibinfo {collaboration} {KM3NeT}),\ }\href@noop {}
  {\  (\bibinfo {year} {2021})},\ \Eprint {http://arxiv.org/abs/2103.09885}
  {arXiv:2103.09885 [hep-ex]} \BibitemShut {NoStop}%
\bibitem [{\citenamefont {Abi}\ \emph {et~al.}(2020)\citenamefont {Abi} \emph
  {et~al.}}]{DUNE:2020lwj}%
  \BibitemOpen
  \bibfield  {author} {\bibinfo {author} {\bibfnamefont {B.}~\bibnamefont
  {Abi}} \emph {et~al.} (\bibinfo {collaboration} {DUNE}),\ }\href {\doibase
  10.1088/1748-0221/15/08/T08008} {\bibfield  {journal} {\bibinfo  {journal}
  {JINST}\ }\textbf {\bibinfo {volume} {15}},\ \bibinfo {pages} {T08008}
  (\bibinfo {year} {2020})},\ \Eprint {http://arxiv.org/abs/2002.02967}
  {arXiv:2002.02967 [physics.ins-det]} \BibitemShut {NoStop}%
\bibitem [{\citenamefont {Abe}\ \emph {et~al.}(2018)\citenamefont {Abe} \emph
  {et~al.}}]{Abe:2018uyc}%
  \BibitemOpen
  \bibfield  {author} {\bibinfo {author} {\bibfnamefont {K.}~\bibnamefont
  {Abe}} \emph {et~al.} (\bibinfo {collaboration} {Hyper-Kamiokande}),\
  }\href@noop {} {\  (\bibinfo {year} {2018})},\ \Eprint
  {http://arxiv.org/abs/1805.04163} {arXiv:1805.04163 [physics.ins-det]}
  \BibitemShut {NoStop}%
\bibitem [{\citenamefont {Caldwell}\ \emph {et~al.}(2017)\citenamefont
  {Caldwell}, \citenamefont {Merle}, \citenamefont {Schulz},\ and\
  \citenamefont {Totzauer}}]{Caldwell:2017mqu}%
  \BibitemOpen
  \bibfield  {author} {\bibinfo {author} {\bibfnamefont {A.}~\bibnamefont
  {Caldwell}}, \bibinfo {author} {\bibfnamefont {A.}~\bibnamefont {Merle}},
  \bibinfo {author} {\bibfnamefont {O.}~\bibnamefont {Schulz}}, \ and\ \bibinfo
  {author} {\bibfnamefont {M.}~\bibnamefont {Totzauer}},\ }\href {\doibase
  10.1103/PhysRevD.96.073001} {\bibfield  {journal} {\bibinfo  {journal} {Phys.
  Rev. D}\ }\textbf {\bibinfo {volume} {96}},\ \bibinfo {pages} {073001}
  (\bibinfo {year} {2017})},\ \Eprint {http://arxiv.org/abs/1705.01945}
  {arXiv:1705.01945 [hep-ph]} \BibitemShut {NoStop}%
\bibitem [{\citenamefont {Agostini}\ \emph
  {et~al.}(2021{\natexlab{b}})\citenamefont {Agostini}, \citenamefont {Benato},
  \citenamefont {Dell'Oro}, \citenamefont {Pirro},\ and\ \citenamefont
  {Vissani}}]{Agostini:2020oiv}%
  \BibitemOpen
  \bibfield  {author} {\bibinfo {author} {\bibfnamefont {M.}~\bibnamefont
  {Agostini}}, \bibinfo {author} {\bibfnamefont {G.}~\bibnamefont {Benato}},
  \bibinfo {author} {\bibfnamefont {S.}~\bibnamefont {Dell'Oro}}, \bibinfo
  {author} {\bibfnamefont {S.}~\bibnamefont {Pirro}}, \ and\ \bibinfo {author}
  {\bibfnamefont {F.}~\bibnamefont {Vissani}},\ }\href {\doibase
  10.1103/PhysRevD.103.033008} {\bibfield  {journal} {\bibinfo  {journal}
  {Phys. Rev. D}\ }\textbf {\bibinfo {volume} {103}},\ \bibinfo {pages}
  {033008} (\bibinfo {year} {2021}{\natexlab{b}})},\ \Eprint
  {http://arxiv.org/abs/2012.13938} {arXiv:2012.13938 [hep-ph]} \BibitemShut
  {NoStop}%
\bibitem [{\citenamefont {Lahav}\ and\ \citenamefont
  {Liddle}(2019)}]{Lahav:2019bbc}%
  \BibitemOpen
  \bibfield  {author} {\bibinfo {author} {\bibfnamefont {O.}~\bibnamefont
  {Lahav}}\ and\ \bibinfo {author} {\bibfnamefont {A.~R.}\ \bibnamefont
  {Liddle}},\ }\href@noop {} {\  (\bibinfo {year} {2019})},\ \Eprint
  {http://arxiv.org/abs/1912.03687} {arXiv:1912.03687 [astro-ph.CO]}
  \BibitemShut {NoStop}%
\bibitem [{\citenamefont {Aker}\ \emph {et~al.}(2021)\citenamefont {Aker} \emph
  {et~al.}}]{Aker:2021exx}%
  \BibitemOpen
  \bibfield  {author} {\bibinfo {author} {\bibfnamefont {M.}~\bibnamefont
  {Aker}} \emph {et~al.} (\bibinfo {collaboration} {KATRIN}),\ }\href@noop {}
  {\  (\bibinfo {year} {2021})},\ \Eprint {http://arxiv.org/abs/2103.04755}
  {arXiv:2103.04755 [physics.ins-det]} \BibitemShut {NoStop}%
\bibitem [{\citenamefont {Ashtari~Esfahani}\ \emph {et~al.}(2017)\citenamefont
  {Ashtari~Esfahani} \emph {et~al.}}]{Project8:2017nal}%
  \BibitemOpen
  \bibfield  {author} {\bibinfo {author} {\bibfnamefont {A.}~\bibnamefont
  {Ashtari~Esfahani}} \emph {et~al.} (\bibinfo {collaboration} {Project 8}),\
  }\href {\doibase 10.1088/1361-6471/aa5b4f} {\bibfield  {journal} {\bibinfo
  {journal} {J. Phys. G}\ }\textbf {\bibinfo {volume} {44}},\ \bibinfo {pages}
  {054004} (\bibinfo {year} {2017})},\ \Eprint
  {http://arxiv.org/abs/1703.02037} {arXiv:1703.02037 [physics.ins-det]}
  \BibitemShut {NoStop}%
\bibitem [{\citenamefont {Dvorkin}\ \emph {et~al.}(2019)\citenamefont {Dvorkin}
  \emph {et~al.}}]{Dvorkin:2019jgs}%
  \BibitemOpen
  \bibfield  {author} {\bibinfo {author} {\bibfnamefont {C.}~\bibnamefont
  {Dvorkin}} \emph {et~al.},\ }\href@noop {} {\  (\bibinfo {year} {2019})},\
  \Eprint {http://arxiv.org/abs/1903.03689} {arXiv:1903.03689 [astro-ph.CO]}
  \BibitemShut {NoStop}%
\bibitem [{\citenamefont {Peng}\ \emph {et~al.}(2016)\citenamefont {Peng},
  \citenamefont {Ramsey-Musolf},\ and\ \citenamefont {Winslow}}]{Peng:2015haa}%
  \BibitemOpen
  \bibfield  {author} {\bibinfo {author} {\bibfnamefont {T.}~\bibnamefont
  {Peng}}, \bibinfo {author} {\bibfnamefont {M.~J.}\ \bibnamefont
  {Ramsey-Musolf}}, \ and\ \bibinfo {author} {\bibfnamefont {P.}~\bibnamefont
  {Winslow}},\ }\href {\doibase 10.1103/PhysRevD.93.093002} {\bibfield
  {journal} {\bibinfo  {journal} {Phys. Rev. D}\ }\textbf {\bibinfo {volume}
  {93}},\ \bibinfo {pages} {093002} (\bibinfo {year} {2016})},\ \Eprint
  {http://arxiv.org/abs/1508.04444} {arXiv:1508.04444 [hep-ph]} \BibitemShut
  {NoStop}%
\bibitem [{\citenamefont {Abgrall}\ \emph
  {et~al.}(2017{\natexlab{a}})\citenamefont {Abgrall} \emph
  {et~al.}}]{Abgrall:2016tnn}%
  \BibitemOpen
  \bibfield  {author} {\bibinfo {author} {\bibfnamefont {N.}~\bibnamefont
  {Abgrall}} \emph {et~al.} (\bibinfo {collaboration} {{\sc Majorana}}),\
  }\href {\doibase 10.1103/PhysRevLett.118.161801} {\bibfield  {journal}
  {\bibinfo  {journal} {Phys. Rev. Lett.}\ }\textbf {\bibinfo {volume} {118}},\
  \bibinfo {pages} {161801} (\bibinfo {year} {2017}{\natexlab{a}})},\ \Eprint
  {http://arxiv.org/abs/1612.00886} {arXiv:1612.00886 [nucl-ex]} \BibitemShut
  {NoStop}%
\bibitem [{\citenamefont {Agostini}\ \emph
  {et~al.}(2020{\natexlab{a}})\citenamefont {Agostini} \emph
  {et~al.}}]{GERDA:2020emj}%
  \BibitemOpen
  \bibfield  {author} {\bibinfo {author} {\bibfnamefont {M.}~\bibnamefont
  {Agostini}} \emph {et~al.} (\bibinfo {collaboration} {{\sc Gerda}}),\ }\href
  {\doibase 10.1103/PhysRevLett.125.011801} {\bibfield  {journal} {\bibinfo
  {journal} {Phys. Rev. Lett.}\ }\textbf {\bibinfo {volume} {125}},\ \bibinfo
  {pages} {011801} (\bibinfo {year} {2020}{\natexlab{a}})},\ \Eprint
  {http://arxiv.org/abs/2005.14184} {arXiv:2005.14184 [hep-ex]} \BibitemShut
  {NoStop}%
\bibitem [{\citenamefont {Xu}\ and\ \citenamefont
  {Elliott}(2017)}]{Xu:2016tap}%
  \BibitemOpen
  \bibfield  {author} {\bibinfo {author} {\bibfnamefont {W.}~\bibnamefont
  {Xu}}\ and\ \bibinfo {author} {\bibfnamefont {S.~R.}\ \bibnamefont
  {Elliott}},\ }\href {\doibase 10.1016/j.astropartphys.2017.01.008} {\bibfield
   {journal} {\bibinfo  {journal} {Astropart. Phys.}\ }\textbf {\bibinfo
  {volume} {89}},\ \bibinfo {pages} {39} (\bibinfo {year} {2017})},\ \Eprint
  {http://arxiv.org/abs/1610.03886} {arXiv:1610.03886 [hep-ex]} \BibitemShut
  {NoStop}%
\bibitem [{\citenamefont {Agostini}\ \emph
  {et~al.}(2015{\natexlab{a}})\citenamefont {Agostini} \emph
  {et~al.}}]{Agostini:2015nwa}%
  \BibitemOpen
  \bibfield  {author} {\bibinfo {author} {\bibfnamefont {M.}~\bibnamefont
  {Agostini}} \emph {et~al.},\ }\href {\doibase 10.1140/epjc/s10052-015-3627-y}
  {\bibfield  {journal} {\bibinfo  {journal} {Eur. Phys. J. C}\ }\textbf
  {\bibinfo {volume} {75}},\ \bibinfo {pages} {416} (\bibinfo {year}
  {2015}{\natexlab{a}})},\ \Eprint {http://arxiv.org/abs/1501.02345}
  {arXiv:1501.02345 [nucl-ex]} \BibitemShut {NoStop}%
\bibitem [{\citenamefont {Bolton}\ \emph {et~al.}(2021)\citenamefont {Bolton},
  \citenamefont {Deppisch}, \citenamefont {Gr\'af},\ and\ \citenamefont
  {\v{S}imkovic}}]{Bolton:2020ncv}%
  \BibitemOpen
  \bibfield  {author} {\bibinfo {author} {\bibfnamefont {P.~D.}\ \bibnamefont
  {Bolton}}, \bibinfo {author} {\bibfnamefont {F.~F.}\ \bibnamefont
  {Deppisch}}, \bibinfo {author} {\bibfnamefont {L.}~\bibnamefont {Gr\'af}}, \
  and\ \bibinfo {author} {\bibfnamefont {F.}~\bibnamefont {\v{S}imkovic}},\
  }\href {\doibase 10.1103/PhysRevD.103.055019} {\bibfield  {journal} {\bibinfo
   {journal} {Phys. Rev. D}\ }\textbf {\bibinfo {volume} {103}},\ \bibinfo
  {pages} {055019} (\bibinfo {year} {2021})},\ \Eprint
  {http://arxiv.org/abs/2011.13387} {arXiv:2011.13387 [hep-ph]} \BibitemShut
  {NoStop}%
\bibitem [{\citenamefont {Agostini}\ \emph
  {et~al.}(2021{\natexlab{c}})\citenamefont {Agostini}, \citenamefont {Bossio},
  \citenamefont {Ibarra},\ and\ \citenamefont {Marcano}}]{Agostini:2020cpz}%
  \BibitemOpen
  \bibfield  {author} {\bibinfo {author} {\bibfnamefont {M.}~\bibnamefont
  {Agostini}}, \bibinfo {author} {\bibfnamefont {E.}~\bibnamefont {Bossio}},
  \bibinfo {author} {\bibfnamefont {A.}~\bibnamefont {Ibarra}}, \ and\ \bibinfo
  {author} {\bibfnamefont {X.}~\bibnamefont {Marcano}},\ }\href {\doibase
  10.1016/j.physletb.2021.136127} {\bibfield  {journal} {\bibinfo  {journal}
  {Phys. Lett. B}\ }\textbf {\bibinfo {volume} {815}},\ \bibinfo {pages}
  {136127} (\bibinfo {year} {2021}{\natexlab{c}})},\ \Eprint
  {http://arxiv.org/abs/2012.09281} {arXiv:2012.09281 [hep-ph]} \BibitemShut
  {NoStop}%
\bibitem [{\citenamefont {D\'\i{}az}\ \emph {et~al.}(2013)\citenamefont
  {D\'\i{}az}, \citenamefont {Kosteleck\'y},\ and\ \citenamefont
  {Lehnert}}]{Diaz:2013saa}%
  \BibitemOpen
  \bibfield  {author} {\bibinfo {author} {\bibfnamefont {J.~S.}\ \bibnamefont
  {D\'\i{}az}}, \bibinfo {author} {\bibfnamefont {A.}~\bibnamefont
  {Kosteleck\'y}}, \ and\ \bibinfo {author} {\bibfnamefont {R.}~\bibnamefont
  {Lehnert}},\ }\href {\doibase 10.1103/PhysRevD.88.071902} {\bibfield
  {journal} {\bibinfo  {journal} {Phys. Rev. D}\ }\textbf {\bibinfo {volume}
  {88}},\ \bibinfo {pages} {071902} (\bibinfo {year} {2013})},\ \Eprint
  {http://arxiv.org/abs/1305.4636} {arXiv:1305.4636 [hep-ph]} \BibitemShut
  {NoStop}%
\bibitem [{\citenamefont {D\'\i{}az}(2014)}]{Diaz:2013ywa}%
  \BibitemOpen
  \bibfield  {author} {\bibinfo {author} {\bibfnamefont {J.~S.}\ \bibnamefont
  {D\'\i{}az}},\ }\href {\doibase 10.1103/PhysRevD.89.036002} {\bibfield
  {journal} {\bibinfo  {journal} {Phys. Rev. D}\ }\textbf {\bibinfo {volume}
  {89}},\ \bibinfo {pages} {036002} (\bibinfo {year} {2014})},\ \Eprint
  {http://arxiv.org/abs/1311.0930} {arXiv:1311.0930 [hep-ph]} \BibitemShut
  {NoStop}%
\bibitem [{\citenamefont {Azzolini}\ \emph {et~al.}(2019)\citenamefont
  {Azzolini} \emph {et~al.}}]{CUPID:2019kto}%
  \BibitemOpen
  \bibfield  {author} {\bibinfo {author} {\bibfnamefont {O.}~\bibnamefont
  {Azzolini}} \emph {et~al.} (\bibinfo {collaboration} {CUPID}),\ }\href
  {\doibase 10.1103/PhysRevD.100.092002} {\bibfield  {journal} {\bibinfo
  {journal} {Phys. Rev. D}\ }\textbf {\bibinfo {volume} {100}},\ \bibinfo
  {pages} {092002} (\bibinfo {year} {2019})},\ \Eprint
  {http://arxiv.org/abs/1911.02446} {arXiv:1911.02446 [nucl-ex]} \BibitemShut
  {NoStop}%
\bibitem [{\citenamefont {Deppisch}\ \emph {et~al.}(2020)\citenamefont
  {Deppisch}, \citenamefont {Graf},\ and\ \citenamefont
  {\v{S}imkovic}}]{Deppisch:2020mxv}%
  \BibitemOpen
  \bibfield  {author} {\bibinfo {author} {\bibfnamefont {F.~F.}\ \bibnamefont
  {Deppisch}}, \bibinfo {author} {\bibfnamefont {L.}~\bibnamefont {Graf}}, \
  and\ \bibinfo {author} {\bibfnamefont {F.}~\bibnamefont {\v{S}imkovic}},\
  }\href {\doibase 10.1103/PhysRevLett.125.171801} {\bibfield  {journal}
  {\bibinfo  {journal} {Phys. Rev. Lett.}\ }\textbf {\bibinfo {volume} {125}},\
  \bibinfo {pages} {171801} (\bibinfo {year} {2020})},\ \Eprint
  {http://arxiv.org/abs/2003.11836} {arXiv:2003.11836 [hep-ph]} \BibitemShut
  {NoStop}%
\bibitem [{\citenamefont {Arnold}\ \emph {et~al.}(2020)\citenamefont {Arnold}
  \emph {et~al.}}]{NEMO-3:2020mcq}%
  \BibitemOpen
  \bibfield  {author} {\bibinfo {author} {\bibfnamefont {R.}~\bibnamefont
  {Arnold}} \emph {et~al.} (\bibinfo {collaboration} {NEMO-3}),\ }\href@noop {}
  {\  (\bibinfo {year} {2020})},\ \Eprint {http://arxiv.org/abs/2011.07657}
  {arXiv:2011.07657 [nucl-ex]} \BibitemShut {NoStop}%
\bibitem [{\citenamefont {Liu}\ \emph {et~al.}(2019)\citenamefont {Liu} \emph
  {et~al.}}]{Liu:2019kzq}%
  \BibitemOpen
  \bibfield  {author} {\bibinfo {author} {\bibfnamefont {Z.~Z.}\ \bibnamefont
  {Liu}} \emph {et~al.} (\bibinfo {collaboration} {CDEX}),\ }\href {\doibase
  10.1103/PhysRevLett.123.161301} {\bibfield  {journal} {\bibinfo  {journal}
  {Phys. Rev. Lett.}\ }\textbf {\bibinfo {volume} {123}},\ \bibinfo {pages}
  {161301} (\bibinfo {year} {2019})},\ \Eprint
  {http://arxiv.org/abs/1905.00354} {arXiv:1905.00354 [hep-ex]} \BibitemShut
  {NoStop}%
\bibitem [{\citenamefont {Alvis}\ \emph
  {et~al.}(2019{\natexlab{a}})\citenamefont {Alvis} \emph
  {et~al.}}]{Alvis:2018pne}%
  \BibitemOpen
  \bibfield  {author} {\bibinfo {author} {\bibfnamefont {S.}~\bibnamefont
  {Alvis}} \emph {et~al.} (\bibinfo {collaboration} {{\sc Majorana}}),\ }\href
  {\doibase 10.1103/PhysRevD.99.072004} {\bibfield  {journal} {\bibinfo
  {journal} {Phys. Rev. D}\ }\textbf {\bibinfo {volume} {99}},\ \bibinfo
  {pages} {072004} (\bibinfo {year} {2019}{\natexlab{a}})},\ \Eprint
  {http://arxiv.org/abs/1812.01090} {arXiv:1812.01090 [hep-ex]} \BibitemShut
  {NoStop}%
\bibitem [{\citenamefont {Alvis}\ \emph {et~al.}(2018)\citenamefont {Alvis}
  \emph {et~al.}}]{Alvis:2018yte}%
  \BibitemOpen
  \bibfield  {author} {\bibinfo {author} {\bibfnamefont {S.}~\bibnamefont
  {Alvis}} \emph {et~al.} (\bibinfo {collaboration} {{\sc Majorana}}),\ }\href
  {\doibase 10.1103/PhysRevLett.120.211804} {\bibfield  {journal} {\bibinfo
  {journal} {Phys. Rev. Lett.}\ }\textbf {\bibinfo {volume} {120}},\ \bibinfo
  {pages} {211804} (\bibinfo {year} {2018})},\ \Eprint
  {http://arxiv.org/abs/1801.10145} {arXiv:1801.10145 [hep-ex]} \BibitemShut
  {NoStop}%
\bibitem [{\citenamefont {Dror}\ \emph {et~al.}(2020)\citenamefont {Dror},
  \citenamefont {Elor},\ and\ \citenamefont {Mcgehee}}]{Dror:2019dib}%
  \BibitemOpen
  \bibfield  {author} {\bibinfo {author} {\bibfnamefont {J.~A.}\ \bibnamefont
  {Dror}}, \bibinfo {author} {\bibfnamefont {G.}~\bibnamefont {Elor}}, \ and\
  \bibinfo {author} {\bibfnamefont {R.}~\bibnamefont {Mcgehee}},\ }\href
  {\doibase 10.1007/JHEP02(2020)134} {\bibfield  {journal} {\bibinfo  {journal}
  {JHEP}\ }\textbf {\bibinfo {volume} {02}},\ \bibinfo {pages} {134} (\bibinfo
  {year} {2020})},\ \Eprint {http://arxiv.org/abs/1908.10861} {arXiv:1908.10861
  [hep-ph]} \BibitemShut {NoStop}%
\bibitem [{\citenamefont {Ha}\ \emph {et~al.}(2019)\citenamefont {Ha} \emph
  {et~al.}}]{COSINE-100:2018ged}%
  \BibitemOpen
  \bibfield  {author} {\bibinfo {author} {\bibfnamefont {C.}~\bibnamefont {Ha}}
  \emph {et~al.} (\bibinfo {collaboration} {COSINE-100}),\ }\href {\doibase
  10.1103/PhysRevLett.122.131802} {\bibfield  {journal} {\bibinfo  {journal}
  {Phys. Rev. Lett.}\ }\textbf {\bibinfo {volume} {122}},\ \bibinfo {pages}
  {131802} (\bibinfo {year} {2019})},\ \Eprint
  {http://arxiv.org/abs/1811.09344} {arXiv:1811.09344 [astro-ph.IM]}
  \BibitemShut {NoStop}%
\bibitem [{\citenamefont {Agostini}\ \emph
  {et~al.}(2016{\natexlab{a}})\citenamefont {Agostini} \emph
  {et~al.}}]{Agostini:2016rsa}%
  \BibitemOpen
  \bibfield  {author} {\bibinfo {author} {\bibfnamefont {M.}~\bibnamefont
  {Agostini}} \emph {et~al.} (\bibinfo {collaboration} {{\sc Gerda}}),\ }\href
  {\doibase 10.1140/epjc/s10052-016-4454-5} {\bibfield  {journal} {\bibinfo
  {journal} {Eur. Phys. J. C}\ }\textbf {\bibinfo {volume} {76}},\ \bibinfo
  {pages} {652} (\bibinfo {year} {2016}{\natexlab{a}})},\ \Eprint
  {http://arxiv.org/abs/1605.01756} {arXiv:1605.01756 [nucl-ex]} \BibitemShut
  {NoStop}%
\bibitem [{\citenamefont {Blum}\ \emph {et~al.}(2018)\citenamefont {Blum},
  \citenamefont {Nir},\ and\ \citenamefont {Shavit}}]{Blum:2018ljv}%
  \BibitemOpen
  \bibfield  {author} {\bibinfo {author} {\bibfnamefont {K.}~\bibnamefont
  {Blum}}, \bibinfo {author} {\bibfnamefont {Y.}~\bibnamefont {Nir}}, \ and\
  \bibinfo {author} {\bibfnamefont {M.}~\bibnamefont {Shavit}},\ }\href
  {\doibase 10.1016/j.physletb.2018.08.022} {\bibfield  {journal} {\bibinfo
  {journal} {Phys. Lett. B}\ }\textbf {\bibinfo {volume} {785}},\ \bibinfo
  {pages} {354} (\bibinfo {year} {2018})},\ \Eprint
  {http://arxiv.org/abs/1802.08019} {arXiv:1802.08019 [hep-ph]} \BibitemShut
  {NoStop}%
\bibitem [{\citenamefont {Piscicchia}\ \emph {et~al.}(2015)\citenamefont
  {Piscicchia} \emph {et~al.}}]{Piscicchia:2015beq}%
  \BibitemOpen
  \bibfield  {author} {\bibinfo {author} {\bibfnamefont {K.}~\bibnamefont
  {Piscicchia}} \emph {et~al.},\ }\href {\doibase 10.5506/APhysPolB.46.147}
  {\bibfield  {journal} {\bibinfo  {journal} {Acta Phys. Polon. B}\ }\textbf
  {\bibinfo {volume} {46}},\ \bibinfo {pages} {147} (\bibinfo {year} {2015})},\
  \Eprint {http://arxiv.org/abs/1501.04462} {arXiv:1501.04462 [quant-ph]}
  \BibitemShut {NoStop}%
\bibitem [{\citenamefont {Graesser}\ \emph {et~al.}(2021)\citenamefont
  {Graesser}, \citenamefont {Shoemaker},\ and\ \citenamefont
  {Arellano}}]{Graesser:2021vkr}%
  \BibitemOpen
  \bibfield  {author} {\bibinfo {author} {\bibfnamefont {M.~L.}\ \bibnamefont
  {Graesser}}, \bibinfo {author} {\bibfnamefont {I.~M.}\ \bibnamefont
  {Shoemaker}}, \ and\ \bibinfo {author} {\bibfnamefont {N.~T.}\ \bibnamefont
  {Arellano}},\ }\href@noop {} {\  (\bibinfo {year} {2021})},\ \Eprint
  {http://arxiv.org/abs/2105.05769} {arXiv:2105.05769 [hep-ph]} \BibitemShut
  {NoStop}%
\bibitem [{\citenamefont {Aprile}\ \emph {et~al.}(2020)\citenamefont {Aprile}
  \emph {et~al.}}]{Aprile:2020tmw}%
  \BibitemOpen
  \bibfield  {author} {\bibinfo {author} {\bibfnamefont {E.}~\bibnamefont
  {Aprile}} \emph {et~al.} (\bibinfo {collaboration} {XENON}),\ }\href
  {\doibase 10.1103/PhysRevD.102.072004} {\bibfield  {journal} {\bibinfo
  {journal} {Phys. Rev. D}\ }\textbf {\bibinfo {volume} {102}},\ \bibinfo
  {pages} {072004} (\bibinfo {year} {2020})},\ \Eprint
  {http://arxiv.org/abs/2006.09721} {arXiv:2006.09721 [hep-ex]} \BibitemShut
  {NoStop}%
\bibitem [{\citenamefont {Klapdor-Kleingrothaus}\ \emph
  {et~al.}(2001)\citenamefont {Klapdor-Kleingrothaus} \emph
  {et~al.}}]{KlapdorKleingrothaus:2000sn}%
  \BibitemOpen
  \bibfield  {author} {\bibinfo {author} {\bibfnamefont {H.~V.}\ \bibnamefont
  {Klapdor-Kleingrothaus}} \emph {et~al.},\ }\href {\doibase
  10.1007/s100500170022} {\bibfield  {journal} {\bibinfo  {journal} {Eur. Phys.
  J. A}\ }\textbf {\bibinfo {volume} {12}},\ \bibinfo {pages} {147} (\bibinfo
  {year} {2001})},\ \Eprint {http://arxiv.org/abs/hep-ph/0103062}
  {arXiv:hep-ph/0103062} \BibitemShut {NoStop}%
\bibitem [{\citenamefont {Aalseth}\ \emph {et~al.}(2002)\citenamefont {Aalseth}
  \emph {et~al.}}]{Aalseth:2002rf}%
  \BibitemOpen
  \bibfield  {author} {\bibinfo {author} {\bibfnamefont {C.~E.}\ \bibnamefont
  {Aalseth}} \emph {et~al.} (\bibinfo {collaboration} {IGEX}),\ }\href
  {\doibase 10.1103/PhysRevD.65.092007} {\bibfield  {journal} {\bibinfo
  {journal} {Phys. Rev. D}\ }\textbf {\bibinfo {volume} {65}},\ \bibinfo
  {pages} {092007} (\bibinfo {year} {2002})},\ \Eprint
  {http://arxiv.org/abs/hep-ex/0202026} {arXiv:hep-ex/0202026} \BibitemShut
  {NoStop}%
\bibitem [{\citenamefont {Klapdor-Kleingrothaus}\ \emph
  {et~al.}(2004)\citenamefont {Klapdor-Kleingrothaus}, \citenamefont
  {Krivosheina}, \citenamefont {Dietz},\ and\ \citenamefont
  {Chkvorets}}]{KlapdorKleingrothaus:2004wj}%
  \BibitemOpen
  \bibfield  {author} {\bibinfo {author} {\bibfnamefont {H.~V.}\ \bibnamefont
  {Klapdor-Kleingrothaus}}, \bibinfo {author} {\bibfnamefont {I.~V.}\
  \bibnamefont {Krivosheina}}, \bibinfo {author} {\bibfnamefont
  {A.}~\bibnamefont {Dietz}}, \ and\ \bibinfo {author} {\bibfnamefont
  {O.}~\bibnamefont {Chkvorets}},\ }\href {\doibase
  10.1016/j.physletb.2004.02.025} {\bibfield  {journal} {\bibinfo  {journal}
  {Phys. Lett. B}\ }\textbf {\bibinfo {volume} {586}},\ \bibinfo {pages} {198}
  (\bibinfo {year} {2004})},\ \Eprint {http://arxiv.org/abs/hep-ph/0404088}
  {arXiv:hep-ph/0404088} \BibitemShut {NoStop}%
\bibitem [{\citenamefont {Andreotti}\ \emph {et~al.}(2011)\citenamefont
  {Andreotti} \emph {et~al.}}]{Andreotti:2010vj}%
  \BibitemOpen
  \bibfield  {author} {\bibinfo {author} {\bibfnamefont {E.}~\bibnamefont
  {Andreotti}} \emph {et~al.},\ }\href {\doibase
  10.1016/j.astropartphys.2011.02.002} {\bibfield  {journal} {\bibinfo
  {journal} {Astropart. Phys.}\ }\textbf {\bibinfo {volume} {34}},\ \bibinfo
  {pages} {822} (\bibinfo {year} {2011})},\ \Eprint
  {http://arxiv.org/abs/1012.3266} {arXiv:1012.3266 [nucl-ex]} \BibitemShut
  {NoStop}%
\bibitem [{\citenamefont {Arnold}\ \emph {et~al.}(2015)\citenamefont {Arnold}
  \emph {et~al.}}]{Arnold:2015wpy}%
  \BibitemOpen
  \bibfield  {author} {\bibinfo {author} {\bibfnamefont {R.}~\bibnamefont
  {Arnold}} \emph {et~al.} (\bibinfo {collaboration} {NEMO-3}),\ }\href
  {\doibase 10.1103/PhysRevD.92.072011} {\bibfield  {journal} {\bibinfo
  {journal} {Phys. Rev. D}\ }\textbf {\bibinfo {volume} {92}},\ \bibinfo
  {pages} {072011} (\bibinfo {year} {2015})},\ \Eprint
  {http://arxiv.org/abs/1506.05825} {arXiv:1506.05825 [hep-ex]} \BibitemShut
  {NoStop}%
\bibitem [{\citenamefont {Agostini}\ \emph
  {et~al.}(2020{\natexlab{b}})\citenamefont {Agostini} \emph
  {et~al.}}]{Agostini:2020xta}%
  \BibitemOpen
  \bibfield  {author} {\bibinfo {author} {\bibfnamefont {M.}~\bibnamefont
  {Agostini}} \emph {et~al.} (\bibinfo {collaboration} {{\sc Gerda}}),\ }\href
  {\doibase 10.1103/PhysRevLett.125.252502} {\bibfield  {journal} {\bibinfo
  {journal} {Phys. Rev. Lett.}\ }\textbf {\bibinfo {volume} {125}},\ \bibinfo
  {pages} {252502} (\bibinfo {year} {2020}{\natexlab{b}})},\ \Eprint
  {http://arxiv.org/abs/2009.06079} {arXiv:2009.06079 [nucl-ex]} \BibitemShut
  {NoStop}%
\bibitem [{\citenamefont {Aalseth}\ \emph
  {et~al.}(2018{\natexlab{a}})\citenamefont {Aalseth} \emph
  {et~al.}}]{Aalseth:2017btx}%
  \BibitemOpen
  \bibfield  {author} {\bibinfo {author} {\bibfnamefont {C.}~\bibnamefont
  {Aalseth}} \emph {et~al.} (\bibinfo {collaboration} {{\sc Majorana}}),\
  }\href {\doibase 10.1103/PhysRevLett.120.132502} {\bibfield  {journal}
  {\bibinfo  {journal} {Phys. Rev. Lett.}\ }\textbf {\bibinfo {volume} {120}},\
  \bibinfo {pages} {132502} (\bibinfo {year} {2018}{\natexlab{a}})},\ \Eprint
  {http://arxiv.org/abs/1710.11608} {arXiv:1710.11608 [nucl-ex]} \BibitemShut
  {NoStop}%
\bibitem [{\citenamefont {Alvis}\ \emph
  {et~al.}(2019{\natexlab{b}})\citenamefont {Alvis} \emph
  {et~al.}}]{Alvis:2019sil}%
  \BibitemOpen
  \bibfield  {author} {\bibinfo {author} {\bibfnamefont {S.~I.}\ \bibnamefont
  {Alvis}} \emph {et~al.} (\bibinfo {collaboration} {{\sc Majorana}}),\ }\href
  {\doibase 10.1103/PhysRevC.100.025501} {\bibfield  {journal} {\bibinfo
  {journal} {Phys. Rev. C}\ }\textbf {\bibinfo {volume} {100}},\ \bibinfo
  {pages} {025501} (\bibinfo {year} {2019}{\natexlab{b}})},\ \Eprint
  {http://arxiv.org/abs/1902.02299} {arXiv:1902.02299 [nucl-ex]} \BibitemShut
  {NoStop}%
\bibitem [{\citenamefont {Anton}\ \emph {et~al.}(2019)\citenamefont {Anton}
  \emph {et~al.}}]{Anton:2019wmi}%
  \BibitemOpen
  \bibfield  {author} {\bibinfo {author} {\bibfnamefont {G.}~\bibnamefont
  {Anton}} \emph {et~al.} (\bibinfo {collaboration} {EXO-200}),\ }\href
  {\doibase 10.1103/PhysRevLett.123.161802} {\bibfield  {journal} {\bibinfo
  {journal} {Phys. Rev. Lett.}\ }\textbf {\bibinfo {volume} {123}},\ \bibinfo
  {pages} {161802} (\bibinfo {year} {2019})},\ \Eprint
  {http://arxiv.org/abs/1906.02723} {arXiv:1906.02723 [hep-ex]} \BibitemShut
  {NoStop}%
\bibitem [{\citenamefont {Gando}\ \emph {et~al.}(2016)\citenamefont {Gando}
  \emph {et~al.}}]{KamLAND-Zen:2016pfg}%
  \BibitemOpen
  \bibfield  {author} {\bibinfo {author} {\bibfnamefont {A.}~\bibnamefont
  {Gando}} \emph {et~al.} (\bibinfo {collaboration} {KamLAND-Zen}),\ }\href
  {\doibase 10.1103/PhysRevLett.117.109903, 10.1103/PhysRevLett.117.082503}
  {\bibfield  {journal} {\bibinfo  {journal} {Phys. Rev. Lett.}\ }\textbf
  {\bibinfo {volume} {117}},\ \bibinfo {pages} {082503} (\bibinfo {year}
  {2016})},\ \bibinfo {note} {[Addendum: Phys. Rev. Lett.117, no.10,
  109903(2016)]},\ \Eprint {http://arxiv.org/abs/1605.02889} {arXiv:1605.02889
  [hep-ex]} \BibitemShut {NoStop}%
\bibitem [{\citenamefont {Adams}\ \emph {et~al.}(2020)\citenamefont {Adams}
  \emph {et~al.}}]{Adams:2019jhp}%
  \BibitemOpen
  \bibfield  {author} {\bibinfo {author} {\bibfnamefont {D.~Q.}\ \bibnamefont
  {Adams}} \emph {et~al.} (\bibinfo {collaboration} {CUORE}),\ }\href {\doibase
  10.1103/PhysRevLett.124.122501} {\bibfield  {journal} {\bibinfo  {journal}
  {Phys. Rev. Lett.}\ }\textbf {\bibinfo {volume} {124}},\ \bibinfo {pages}
  {122501} (\bibinfo {year} {2020})},\ \Eprint
  {http://arxiv.org/abs/1912.10966} {arXiv:1912.10966 [nucl-ex]} \BibitemShut
  {NoStop}%
\bibitem [{\citenamefont {Adams}\ \emph {et~al.}(2021)\citenamefont {Adams}
  \emph {et~al.}}]{Adams:2021rbc}%
  \BibitemOpen
  \bibfield  {author} {\bibinfo {author} {\bibfnamefont {D.~Q.}\ \bibnamefont
  {Adams}} \emph {et~al.} (\bibinfo {collaboration} {CUORE}),\ }\href@noop {}
  {\  (\bibinfo {year} {2021})},\ \Eprint {http://arxiv.org/abs/2104.06906}
  {arXiv:2104.06906 [nucl-ex]} \BibitemShut {NoStop}%
\bibitem [{\citenamefont {Adhikari}\ \emph {et~al.}(2021)\citenamefont
  {Adhikari} \emph {et~al.}}]{nEXO:2021ujk}%
  \BibitemOpen
  \bibfield  {author} {\bibinfo {author} {\bibfnamefont {G.}~\bibnamefont
  {Adhikari}} \emph {et~al.} (\bibinfo {collaboration} {nEXO}),\ }\href@noop {}
  {\  (\bibinfo {year} {2021})},\ \Eprint {http://arxiv.org/abs/2106.16243}
  {arXiv:2106.16243 [nucl-ex]} \BibitemShut {NoStop}%
\bibitem [{\citenamefont {Armstrong}\ \emph {et~al.}(2019)\citenamefont
  {Armstrong} \emph {et~al.}}]{CUPID:2019imh}%
  \BibitemOpen
  \bibfield  {author} {\bibinfo {author} {\bibfnamefont {W.~R.}\ \bibnamefont
  {Armstrong}} \emph {et~al.} (\bibinfo {collaboration} {CUPID}),\ }\href@noop
  {} {\  (\bibinfo {year} {2019})},\ \Eprint {http://arxiv.org/abs/1907.09376}
  {arXiv:1907.09376 [physics.ins-det]} \BibitemShut {NoStop}%
\bibitem [{\citenamefont {Avignone}\ and\ \citenamefont
  {Elliott}(2019)}]{Avignone:2019phg}%
  \BibitemOpen
  \bibfield  {author} {\bibinfo {author} {\bibfnamefont {F.~T.}\ \bibnamefont
  {Avignone}}\ and\ \bibinfo {author} {\bibfnamefont {S.~R.}\ \bibnamefont
  {Elliott}},\ }\href {\doibase 10.3389/fphy.2019.00006} {\bibfield  {journal}
  {\bibinfo  {journal} {Front. in Phys.}\ }\textbf {\bibinfo {volume} {7}},\
  \bibinfo {pages} {6} (\bibinfo {year} {2019})},\ \Eprint
  {http://arxiv.org/abs/1901.02805} {arXiv:1901.02805 [nucl-ex]} \BibitemShut
  {NoStop}%
\bibitem [{\citenamefont {Luke}\ \emph {et~al.}(1989)\citenamefont {Luke},
  \citenamefont {Goulding}, \citenamefont {Madden},\ and\ \citenamefont
  {Pehl}}]{Luke1989}%
  \BibitemOpen
  \bibfield  {author} {\bibinfo {author} {\bibfnamefont {P.}~\bibnamefont
  {Luke}}, \bibinfo {author} {\bibfnamefont {F.}~\bibnamefont {Goulding}},
  \bibinfo {author} {\bibfnamefont {N.}~\bibnamefont {Madden}}, \ and\ \bibinfo
  {author} {\bibfnamefont {R.}~\bibnamefont {Pehl}},\ }\href {\doibase
  10.1109/23.34577} {\bibfield  {journal} {\bibinfo  {journal} {IEEE
  Transactions on Nuclear Science}\ }\textbf {\bibinfo {volume} {36}},\
  \bibinfo {pages} {926} (\bibinfo {year} {1989})}\BibitemShut {NoStop}%
\bibitem [{\citenamefont {Agostini}\ \emph
  {et~al.}(2015{\natexlab{b}})\citenamefont {Agostini} \emph
  {et~al.}}]{Agostini:2014hra}%
  \BibitemOpen
  \bibfield  {author} {\bibinfo {author} {\bibfnamefont {M.}~\bibnamefont
  {Agostini}} \emph {et~al.} (\bibinfo {collaboration} {{\sc Gerda}}),\ }\href
  {\doibase 10.1140/epjc/s10052-014-3253-0} {\bibfield  {journal} {\bibinfo
  {journal} {Eur. Phys. J. C}\ }\textbf {\bibinfo {volume} {75}},\ \bibinfo
  {pages} {39} (\bibinfo {year} {2015}{\natexlab{b}})},\ \Eprint
  {http://arxiv.org/abs/1410.0853} {arXiv:1410.0853 [physics.ins-det]}
  \BibitemShut {NoStop}%
\bibitem [{\citenamefont {Cooper}\ \emph {et~al.}(2011)\citenamefont {Cooper},
  \citenamefont {Radford}, \citenamefont {Hausladen},\ and\ \citenamefont
  {Lagergren}}]{Cooper2011}%
  \BibitemOpen
  \bibfield  {author} {\bibinfo {author} {\bibfnamefont {R.~J.}\ \bibnamefont
  {Cooper}}, \bibinfo {author} {\bibfnamefont {D.~C.}\ \bibnamefont {Radford}},
  \bibinfo {author} {\bibfnamefont {P.~A.}\ \bibnamefont {Hausladen}}, \ and\
  \bibinfo {author} {\bibfnamefont {K.}~\bibnamefont {Lagergren}},\ }\href
  {\doibase https://doi.org/10.1016/j.nima.2011.10.008} {\bibfield  {journal}
  {\bibinfo  {journal} {Nucl. Instrum. Methods Phys. Res., Sect. A}\ }\textbf
  {\bibinfo {volume} {665}},\ \bibinfo {pages} {25} (\bibinfo {year}
  {2011})}\BibitemShut {NoStop}%
\bibitem [{\citenamefont {Comellato}\ \emph {et~al.}(2021)\citenamefont
  {Comellato}, \citenamefont {Agostini},\ and\ \citenamefont
  {Sch\"onert}}]{Comellato:2020ljj}%
  \BibitemOpen
  \bibfield  {author} {\bibinfo {author} {\bibfnamefont {T.}~\bibnamefont
  {Comellato}}, \bibinfo {author} {\bibfnamefont {M.}~\bibnamefont {Agostini}},
  \ and\ \bibinfo {author} {\bibfnamefont {S.}~\bibnamefont {Sch\"onert}},\
  }\href {\doibase 10.1140/epjc/s10052-021-08889-0} {\bibfield  {journal}
  {\bibinfo  {journal} {Eur. Phys. J. C}\ }\textbf {\bibinfo {volume} {81}},\
  \bibinfo {pages} {76} (\bibinfo {year} {2021})},\ \Eprint
  {http://arxiv.org/abs/2007.12910} {arXiv:2007.12910 [physics.ins-det]}
  \BibitemShut {NoStop}%
\bibitem [{\citenamefont {Agostini}\ \emph
  {et~al.}(2021{\natexlab{d}})\citenamefont {Agostini} \emph
  {et~al.}}]{Agostini:2021wzn}%
  \BibitemOpen
  \bibfield  {author} {\bibinfo {author} {\bibfnamefont {M.}~\bibnamefont
  {Agostini}} \emph {et~al.} (\bibinfo {collaboration} {{\sc Gerda}}),\ }\href
  {\doibase 10.1140/epjc/s10052-021-09184-8} {\bibfield  {journal} {\bibinfo
  {journal} {Eur. Phys. J. C}\ }\textbf {\bibinfo {volume} {81}},\ \bibinfo
  {pages} {505} (\bibinfo {year} {2021}{\natexlab{d}})},\ \Eprint
  {http://arxiv.org/abs/2103.15111} {arXiv:2103.15111 [physics.ins-det]}
  \BibitemShut {NoStop}%
\bibitem [{\citenamefont {Agostini}\ \emph
  {et~al.}(2020{\natexlab{c}})\citenamefont {Agostini} \emph
  {et~al.}}]{GERDA:2019cav}%
  \BibitemOpen
  \bibfield  {author} {\bibinfo {author} {\bibfnamefont {M.}~\bibnamefont
  {Agostini}} \emph {et~al.} (\bibinfo {collaboration} {{\sc Gerda}}),\ }\href
  {\doibase 10.1007/JHEP03(2020)139} {\bibfield  {journal} {\bibinfo  {journal}
  {JHEP}\ }\textbf {\bibinfo {volume} {03}},\ \bibinfo {pages} {139} (\bibinfo
  {year} {2020}{\natexlab{c}})},\ \Eprint {http://arxiv.org/abs/1909.02522}
  {arXiv:1909.02522 [nucl-ex]} \BibitemShut {NoStop}%
\bibitem [{\citenamefont {Abgrall}\ \emph {et~al.}(2016)\citenamefont {Abgrall}
  \emph {et~al.}}]{Abgrall:2016cct}%
  \BibitemOpen
  \bibfield  {author} {\bibinfo {author} {\bibfnamefont {N.}~\bibnamefont
  {Abgrall}} \emph {et~al.} (\bibinfo {collaboration} {{\sc Majorana}}),\
  }\href {\doibase 10.1016/j.nima.2016.04.070} {\bibfield  {journal} {\bibinfo
  {journal} {Nucl. Instrum. Meth. A}\ }\textbf {\bibinfo {volume} {828}},\
  \bibinfo {pages} {22} (\bibinfo {year} {2016})},\ \Eprint
  {http://arxiv.org/abs/1601.03779} {arXiv:1601.03779 [physics.ins-det]}
  \BibitemShut {NoStop}%
\bibitem [{\citenamefont {Buuck}(2019)}]{Buuck2019}%
  \BibitemOpen
  \bibfield  {author} {\bibinfo {author} {\bibfnamefont {M.}~\bibnamefont
  {Buuck}},\ }\emph {\bibinfo {title} {A Radiogenic Background Model for the
  \textsc{Majorana Demonstrator}}},\ \href@noop {} {Ph.D. thesis},\ \bibinfo
  {school} {University of Washington} (\bibinfo {year} {2019})\BibitemShut
  {NoStop}%
\bibitem [{\citenamefont {Gilliss}(2019)}]{Gilliss2019}%
  \BibitemOpen
  \bibfield  {author} {\bibinfo {author} {\bibfnamefont {T.~F.}\ \bibnamefont
  {Gilliss}},\ }\emph {\bibinfo {title} {Statistical Modeling and Markov Chain
  Monte Carlo Inference of \textsc{Majorana Demonstrator} Background Data}},\
  \href@noop {} {Ph.D. thesis},\ \bibinfo  {school} {University of North
  Carolina, Chapel Hill} (\bibinfo {year} {2019})\BibitemShut {NoStop}%
\bibitem [{\citenamefont {Agostini}\ \emph
  {et~al.}(2021{\natexlab{e}})\citenamefont {Agostini} \emph
  {et~al.}}]{Agostini:2021duc}%
  \BibitemOpen
  \bibfield  {author} {\bibinfo {author} {\bibfnamefont {M.}~\bibnamefont
  {Agostini}} \emph {et~al.} (\bibinfo {collaboration} {{\sc Gerda}}),\
  }\href@noop {} {\  (\bibinfo {year} {2021}{\natexlab{e}})},\ \Eprint
  {http://arxiv.org/abs/2103.13777} {arXiv:2103.13777 [physics.ins-det]}
  \BibitemShut {NoStop}%
\bibitem [{\citenamefont {Arnquist}\ \emph
  {et~al.}(2020{\natexlab{a}})\citenamefont {Arnquist}, \citenamefont {Beck},
  \citenamefont {di~Vacri}, \citenamefont {Harouaka},\ and\ \citenamefont
  {Saldanha}}]{Arnquist:2019fkc}%
  \BibitemOpen
  \bibfield  {author} {\bibinfo {author} {\bibfnamefont {I.~J.}\ \bibnamefont
  {Arnquist}}, \bibinfo {author} {\bibfnamefont {C.}~\bibnamefont {Beck}},
  \bibinfo {author} {\bibfnamefont {M.~L.}\ \bibnamefont {di~Vacri}}, \bibinfo
  {author} {\bibfnamefont {K.}~\bibnamefont {Harouaka}}, \ and\ \bibinfo
  {author} {\bibfnamefont {R.}~\bibnamefont {Saldanha}},\ }\href {\doibase
  10.1016/j.nima.2020.163573} {\bibfield  {journal} {\bibinfo  {journal} {Nucl.
  Instrum. Meth. A}\ }\textbf {\bibinfo {volume} {959}},\ \bibinfo {pages}
  {163573} (\bibinfo {year} {2020}{\natexlab{a}})},\ \Eprint
  {http://arxiv.org/abs/1910.04317} {arXiv:1910.04317 [physics.ins-det]}
  \BibitemShut {NoStop}%
\bibitem [{\citenamefont {Edzards}\ \emph {et~al.}(2020)\citenamefont {Edzards}
  \emph {et~al.}}]{Edzards:2020wfg}%
  \BibitemOpen
  \bibfield  {author} {\bibinfo {author} {\bibfnamefont {F.}~\bibnamefont
  {Edzards}} \emph {et~al.},\ }\href {\doibase 10.1088/1748-0221/15/09/P09022}
  {\bibfield  {journal} {\bibinfo  {journal} {JINST}\ }\textbf {\bibinfo
  {volume} {15}},\ \bibinfo {pages} {P09022} (\bibinfo {year} {2020})},\
  \Eprint {http://arxiv.org/abs/2005.10366} {arXiv:2005.10366
  [physics.ins-det]} \BibitemShut {NoStop}%
\bibitem [{\citenamefont {Terasaki}(2020)}]{Terasaki:2020ndc}%
  \BibitemOpen
  \bibfield  {author} {\bibinfo {author} {\bibfnamefont {J.}~\bibnamefont
  {Terasaki}},\ }\href {\doibase 10.1103/PhysRevC.102.044303} {\bibfield
  {journal} {\bibinfo  {journal} {Phys. Rev. C}\ }\textbf {\bibinfo {volume}
  {102}},\ \bibinfo {pages} {044303} (\bibinfo {year} {2020})},\ \Eprint
  {http://arxiv.org/abs/2003.03542} {arXiv:2003.03542 [nucl-th]} \BibitemShut
  {NoStop}%
\bibitem [{\citenamefont {Deppisch}\ \emph {et~al.}(2015)\citenamefont
  {Deppisch}, \citenamefont {Bhupal~Dev},\ and\ \citenamefont
  {Pilaftsis}}]{Deppisch:2015qwa}%
  \BibitemOpen
  \bibfield  {author} {\bibinfo {author} {\bibfnamefont {F.~F.}\ \bibnamefont
  {Deppisch}}, \bibinfo {author} {\bibfnamefont {P.~S.}\ \bibnamefont
  {Bhupal~Dev}}, \ and\ \bibinfo {author} {\bibfnamefont {A.}~\bibnamefont
  {Pilaftsis}},\ }\href {\doibase 10.1088/1367-2630/17/7/075019} {\bibfield
  {journal} {\bibinfo  {journal} {New J. Phys.}\ }\textbf {\bibinfo {volume}
  {17}},\ \bibinfo {pages} {075019} (\bibinfo {year} {2015})},\ \Eprint
  {http://arxiv.org/abs/1502.06541} {arXiv:1502.06541 [hep-ph]} \BibitemShut
  {NoStop}%
\bibitem [{\citenamefont {Abgrall}\ \emph {et~al.}(2018)\citenamefont {Abgrall}
  \emph {et~al.}}]{Abgrall:2017acl}%
  \BibitemOpen
  \bibfield  {author} {\bibinfo {author} {\bibfnamefont {N.}~\bibnamefont
  {Abgrall}} \emph {et~al.} (\bibinfo {collaboration} {{\sc Majorana}}),\
  }\href {\doibase 10.1016/j.nima.2017.09.036} {\bibfield  {journal} {\bibinfo
  {journal} {Nucl. Instrum. Meth. A}\ }\textbf {\bibinfo {volume} {877}},\
  \bibinfo {pages} {314} (\bibinfo {year} {2018})},\ \Eprint
  {http://arxiv.org/abs/1707.06255} {arXiv:1707.06255 [physics.ins-det]}
  \BibitemShut {NoStop}%
\bibitem [{\citenamefont {Barabanov}\ \emph {et~al.}(2006)\citenamefont
  {Barabanov}, \citenamefont {Belogurov}, \citenamefont {Bezrukov},
  \citenamefont {Denisov}, \citenamefont {Kornoukhov},\ and\ \citenamefont
  {Sobolevsky}}]{Barabanov:2005cw}%
  \BibitemOpen
  \bibfield  {author} {\bibinfo {author} {\bibfnamefont {I.}~\bibnamefont
  {Barabanov}}, \bibinfo {author} {\bibfnamefont {S.}~\bibnamefont
  {Belogurov}}, \bibinfo {author} {\bibfnamefont {L.~B.}\ \bibnamefont
  {Bezrukov}}, \bibinfo {author} {\bibfnamefont {A.}~\bibnamefont {Denisov}},
  \bibinfo {author} {\bibfnamefont {V.}~\bibnamefont {Kornoukhov}}, \ and\
  \bibinfo {author} {\bibfnamefont {N.}~\bibnamefont {Sobolevsky}},\ }\href
  {\doibase 10.1016/j.nimb.2006.05.011} {\bibfield  {journal} {\bibinfo
  {journal} {Nucl. Instrum. Meth. B}\ }\textbf {\bibinfo {volume} {251}},\
  \bibinfo {pages} {115} (\bibinfo {year} {2006})},\ \Eprint
  {http://arxiv.org/abs/nucl-ex/0511049} {arXiv:nucl-ex/0511049} \BibitemShut
  {NoStop}%
\bibitem [{\citenamefont {Bruyneel}\ \emph {et~al.}(2016)\citenamefont
  {Bruyneel}, \citenamefont {Birkenbach},\ and\ \citenamefont
  {Reiter}}]{Bruyneel:2016zih}%
  \BibitemOpen
  \bibfield  {author} {\bibinfo {author} {\bibfnamefont {B.}~\bibnamefont
  {Bruyneel}}, \bibinfo {author} {\bibfnamefont {B.}~\bibnamefont
  {Birkenbach}}, \ and\ \bibinfo {author} {\bibfnamefont {P.}~\bibnamefont
  {Reiter}},\ }\href {\doibase 10.1140/epja/i2016-16070-9} {\bibfield
  {journal} {\bibinfo  {journal} {Eur. Phys. J. A}\ }\textbf {\bibinfo {volume}
  {52}},\ \bibinfo {pages} {70} (\bibinfo {year} {2016})}\BibitemShut {NoStop}%
\bibitem [{\citenamefont {Agostini}\ \emph {et~al.}(2019)\citenamefont
  {Agostini} \emph {et~al.}}]{Agostini:2019mwn}%
  \BibitemOpen
  \bibfield  {author} {\bibinfo {author} {\bibfnamefont {M.}~\bibnamefont
  {Agostini}} \emph {et~al.} (\bibinfo {collaboration} {{\sc Gerda}}),\ }\href
  {\doibase 10.1140/epjc/s10052-019-7353-8} {\bibfield  {journal} {\bibinfo
  {journal} {Eur. Phys. J. C}\ }\textbf {\bibinfo {volume} {79}},\ \bibinfo
  {pages} {978} (\bibinfo {year} {2019})},\ \Eprint
  {http://arxiv.org/abs/1901.06590} {arXiv:1901.06590 [physics.ins-det]}
  \BibitemShut {NoStop}%
\bibitem [{\citenamefont {Barabanov}\ \emph {et~al.}(2009)\citenamefont
  {Barabanov}, \citenamefont {Bezrukov}, \citenamefont {Demidova},
  \citenamefont {Gurentsov}, \citenamefont {Kianovsky}, \citenamefont
  {Knopfle}, \citenamefont {Kornouhkov}, \citenamefont {Schwingenheuer},\ and\
  \citenamefont {Vasenko}}]{Barabanov:2009zz}%
  \BibitemOpen
  \bibfield  {author} {\bibinfo {author} {\bibfnamefont {I.}~\bibnamefont
  {Barabanov}}, \bibinfo {author} {\bibfnamefont {L.}~\bibnamefont {Bezrukov}},
  \bibinfo {author} {\bibfnamefont {E.}~\bibnamefont {Demidova}}, \bibinfo
  {author} {\bibfnamefont {V.}~\bibnamefont {Gurentsov}}, \bibinfo {author}
  {\bibfnamefont {S.}~\bibnamefont {Kianovsky}}, \bibinfo {author}
  {\bibfnamefont {K.~T.}\ \bibnamefont {Knopfle}}, \bibinfo {author}
  {\bibfnamefont {V.}~\bibnamefont {Kornouhkov}}, \bibinfo {author}
  {\bibfnamefont {B.}~\bibnamefont {Schwingenheuer}}, \ and\ \bibinfo {author}
  {\bibfnamefont {A.}~\bibnamefont {Vasenko}},\ }\href {\doibase
  10.1016/j.nima.2009.04.006} {\bibfield  {journal} {\bibinfo  {journal} {Nucl.
  Instrum. Meth. A}\ }\textbf {\bibinfo {volume} {606}},\ \bibinfo {pages}
  {790} (\bibinfo {year} {2009})}\BibitemShut {NoStop}%
\bibitem [{\citenamefont {Acciarri}\ \emph {et~al.}(2009)\citenamefont
  {Acciarri} \emph {et~al.}}]{Acciarri:2009xj}%
  \BibitemOpen
  \bibfield  {author} {\bibinfo {author} {\bibfnamefont {R.}~\bibnamefont
  {Acciarri}} \emph {et~al.},\ }\href {\doibase
  10.1016/j.nuclphysbps.2009.10.037} {\bibfield  {journal} {\bibinfo  {journal}
  {Nucl. Phys. B Proc. Suppl.}\ }\textbf {\bibinfo {volume} {197}},\ \bibinfo
  {pages} {70} (\bibinfo {year} {2009})}\BibitemShut {NoStop}%
\bibitem [{\citenamefont {Acciarri}\ \emph {et~al.}(2010)\citenamefont
  {Acciarri} \emph {et~al.}}]{Acciarri:2008kx}%
  \BibitemOpen
  \bibfield  {author} {\bibinfo {author} {\bibfnamefont {R.}~\bibnamefont
  {Acciarri}} \emph {et~al.} (\bibinfo {collaboration} {WArP}),\ }\href
  {\doibase 10.1088/1748-0221/5/05/P05003} {\bibfield  {journal} {\bibinfo
  {journal} {JINST}\ }\textbf {\bibinfo {volume} {5}},\ \bibinfo {pages}
  {P05003} (\bibinfo {year} {2010})},\ \Eprint {http://arxiv.org/abs/0804.1222}
  {arXiv:0804.1222 [nucl-ex]} \BibitemShut {NoStop}%
\bibitem [{\citenamefont {Agnes}\ \emph {et~al.}(2018)\citenamefont {Agnes}
  \emph {et~al.}}]{Agnes:2018fwg}%
  \BibitemOpen
  \bibfield  {author} {\bibinfo {author} {\bibfnamefont {P.}~\bibnamefont
  {Agnes}} \emph {et~al.} (\bibinfo {collaboration} {DarkSide}),\ }\href
  {\doibase 10.1103/PhysRevD.98.102006} {\bibfield  {journal} {\bibinfo
  {journal} {Phys. Rev. D}\ }\textbf {\bibinfo {volume} {98}},\ \bibinfo
  {pages} {102006} (\bibinfo {year} {2018})},\ \Eprint
  {http://arxiv.org/abs/1802.07198} {arXiv:1802.07198 [astro-ph.CO]}
  \BibitemShut {NoStop}%
\bibitem [{\citenamefont {Aalseth}\ \emph
  {et~al.}(2018{\natexlab{b}})\citenamefont {Aalseth} \emph
  {et~al.}}]{Aalseth:2017fik}%
  \BibitemOpen
  \bibfield  {author} {\bibinfo {author} {\bibfnamefont {C.~E.}\ \bibnamefont
  {Aalseth}} \emph {et~al.} (\bibinfo {collaboration} {DarkSide-20k}),\ }\href
  {\doibase 10.1140/epjp/i2018-11973-4} {\bibfield  {journal} {\bibinfo
  {journal} {Eur. Phys. J. Plus}\ }\textbf {\bibinfo {volume} {133}},\ \bibinfo
  {pages} {131} (\bibinfo {year} {2018}{\natexlab{b}})},\ \Eprint
  {http://arxiv.org/abs/1707.08145} {arXiv:1707.08145 [physics.ins-det]}
  \BibitemShut {NoStop}%
\bibitem [{\citenamefont {Wang}\ \emph {et~al.}(2019)\citenamefont {Wang} \emph
  {et~al.}}]{Wang2019}%
  \BibitemOpen
  \bibfield  {author} {\bibinfo {author} {\bibfnamefont {Y.}~\bibnamefont
  {Wang}} \emph {et~al.},\ }\href@noop {} {\enquote {\bibinfo {title} {{A
  Global Liquid Dark Matter Search Program}},}\ } (\bibinfo {year} {2019}),\
  \bibinfo {note} {{Presentation at APS April Meeting 2019, Session H17: WIMP
  DARK matter II}}\BibitemShut {NoStop}%
\bibitem [{\citenamefont {Agnes}\ \emph
  {et~al.}(2021{\natexlab{a}})\citenamefont {Agnes} \emph
  {et~al.}}]{Agnes:2020pbw}%
  \BibitemOpen
  \bibfield  {author} {\bibinfo {author} {\bibfnamefont {P.}~\bibnamefont
  {Agnes}} \emph {et~al.} (\bibinfo {collaboration} {DarkSide 20k}),\ }\href
  {\doibase 10.1088/1475-7516/2021/03/043} {\bibfield  {journal} {\bibinfo
  {journal} {JCAP}\ }\textbf {\bibinfo {volume} {03}},\ \bibinfo {pages} {043}
  (\bibinfo {year} {2021}{\natexlab{a}})},\ \Eprint
  {http://arxiv.org/abs/2011.07819} {arXiv:2011.07819 [astro-ph.HE]}
  \BibitemShut {NoStop}%
\bibitem [{\citenamefont {Agostini}\ \emph
  {et~al.}(2015{\natexlab{c}})\citenamefont {Agostini} \emph
  {et~al.}}]{Agostini:2015boa}%
  \BibitemOpen
  \bibfield  {author} {\bibinfo {author} {\bibfnamefont {M.}~\bibnamefont
  {Agostini}} \emph {et~al.},\ }\href {\doibase 10.1140/epjc/s10052-015-3681-5}
  {\bibfield  {journal} {\bibinfo  {journal} {Eur. Phys. J. C}\ }\textbf
  {\bibinfo {volume} {75}},\ \bibinfo {pages} {506} (\bibinfo {year}
  {2015}{\natexlab{c}})},\ \Eprint {http://arxiv.org/abs/1501.05762}
  {arXiv:1501.05762 [physics.ins-det]} \BibitemShut {NoStop}%
\bibitem [{\citenamefont {Cs\'athy}\ \emph {et~al.}(2016)\citenamefont
  {Cs\'athy}, \citenamefont {Bode}, \citenamefont {Kratz}, \citenamefont
  {Sch\"onert},\ and\ \citenamefont {Wiesinger}}]{Csathy:2016wdy}%
  \BibitemOpen
  \bibfield  {author} {\bibinfo {author} {\bibfnamefont {J.~J.}\ \bibnamefont
  {Cs\'athy}}, \bibinfo {author} {\bibfnamefont {T.}~\bibnamefont {Bode}},
  \bibinfo {author} {\bibfnamefont {J.}~\bibnamefont {Kratz}}, \bibinfo
  {author} {\bibfnamefont {S.}~\bibnamefont {Sch\"onert}}, \ and\ \bibinfo
  {author} {\bibfnamefont {C.}~\bibnamefont {Wiesinger}},\ }\href@noop {} {\
  (\bibinfo {year} {2016})},\ \Eprint {http://arxiv.org/abs/1606.04254}
  {arXiv:1606.04254 [physics.ins-det]} \BibitemShut {NoStop}%
\bibitem [{\citenamefont {Agostini}\ \emph {et~al.}(2018)\citenamefont
  {Agostini} \emph {et~al.}}]{Agostini:2017hit}%
  \BibitemOpen
  \bibfield  {author} {\bibinfo {author} {\bibfnamefont {M.}~\bibnamefont
  {Agostini}} \emph {et~al.} (\bibinfo {collaboration} {{\sc Gerda}}),\ }\href
  {\doibase 10.1140/epjc/s10052-018-5812-2} {\bibfield  {journal} {\bibinfo
  {journal} {Eur. Phys. J. C}\ }\textbf {\bibinfo {volume} {78}},\ \bibinfo
  {pages} {388} (\bibinfo {year} {2018})},\ \Eprint
  {http://arxiv.org/abs/1711.01452} {arXiv:1711.01452 [physics.ins-det]}
  \BibitemShut {NoStop}%
\bibitem [{\citenamefont {Michael}\ \emph {et~al.}(2008)\citenamefont {Michael}
  \emph {et~al.}}]{Michael:2008bc}%
  \BibitemOpen
  \bibfield  {author} {\bibinfo {author} {\bibfnamefont {D.}~\bibnamefont
  {Michael}} \emph {et~al.} (\bibinfo {collaboration} {MINOS}),\ }\href
  {\doibase 10.1016/j.nima.2008.08.003} {\bibfield  {journal} {\bibinfo
  {journal} {Nucl. Instrum. Meth. A}\ }\textbf {\bibinfo {volume} {596}},\
  \bibinfo {pages} {190} (\bibinfo {year} {2008})},\ \Eprint
  {http://arxiv.org/abs/0805.3170} {arXiv:0805.3170 [physics.ins-det]}
  \BibitemShut {NoStop}%
\bibitem [{\citenamefont {Adamson}\ \emph {et~al.}(2004)\citenamefont {Adamson}
  \emph {et~al.}}]{Adamson:2004mh}%
  \BibitemOpen
  \bibfield  {author} {\bibinfo {author} {\bibfnamefont {P.}~\bibnamefont
  {Adamson}} \emph {et~al.},\ }\href {\doibase 10.1016/j.nima.2003.10.105}
  {\bibfield  {journal} {\bibinfo  {journal} {Nucl. Instrum. Meth. A}\ }\textbf
  {\bibinfo {volume} {521}},\ \bibinfo {pages} {361} (\bibinfo {year}
  {2004})}\BibitemShut {NoStop}%
\bibitem [{\citenamefont {Ayres}\ \emph {et~al.}(2007)\citenamefont {Ayres}
  \emph {et~al.}}]{Ayres:2007tu}%
  \BibitemOpen
  \bibfield  {author} {\bibinfo {author} {\bibfnamefont {D.}~\bibnamefont
  {Ayres}} \emph {et~al.} (\bibinfo {collaboration} {NOvA}),\ }\href {\doibase
  10.2172/935497} {\  (\bibinfo {year} {2007}),\ 10.2172/935497},\ \bibinfo
  {note} {\url{https://www.osti.gov/biblio/935497}}\BibitemShut {NoStop}%
\bibitem [{\citenamefont {Ayres}\ \emph {et~al.}(2004)\citenamefont {Ayres}
  \emph {et~al.}}]{Ayres:2004js}%
  \BibitemOpen
  \bibfield  {author} {\bibinfo {author} {\bibfnamefont {D.}~\bibnamefont
  {Ayres}} \emph {et~al.} (\bibinfo {collaboration} {NOvA}),\ }\href@noop {} {\
   (\bibinfo {year} {2004})},\ \Eprint {http://arxiv.org/abs/hep-ex/0503053}
  {arXiv:hep-ex/0503053} \BibitemShut {NoStop}%
\bibitem [{\citenamefont {Pahlka}\ \emph {et~al.}(2019)\citenamefont {Pahlka},
  \citenamefont {Elpers}, \citenamefont {Huang}, \citenamefont {Lang},\ and\
  \citenamefont {Proga}}]{Pahlka:2019bxr}%
  \BibitemOpen
  \bibfield  {author} {\bibinfo {author} {\bibfnamefont {R.}~\bibnamefont
  {Pahlka}}, \bibinfo {author} {\bibfnamefont {G.}~\bibnamefont {Elpers}},
  \bibinfo {author} {\bibfnamefont {J.}~\bibnamefont {Huang}}, \bibinfo
  {author} {\bibfnamefont {K.}~\bibnamefont {Lang}}, \ and\ \bibinfo {author}
  {\bibfnamefont {M.}~\bibnamefont {Proga}},\ }\href@noop {} {\  (\bibinfo
  {year} {2019})},\ \Eprint {http://arxiv.org/abs/1911.03790} {arXiv:1911.03790
  [physics.ins-det]} \BibitemShut {NoStop}%
\bibitem [{\citenamefont {Barabash}\ \emph {et~al.}(2017)\citenamefont
  {Barabash} \emph {et~al.}}]{Barabash:2017sxf}%
  \BibitemOpen
  \bibfield  {author} {\bibinfo {author} {\bibfnamefont {A.}~\bibnamefont
  {Barabash}} \emph {et~al.},\ }\href {\doibase 10.1016/j.nima.2017.06.044}
  {\bibfield  {journal} {\bibinfo  {journal} {Nucl. Instrum. Meth. A}\ }\textbf
  {\bibinfo {volume} {868}},\ \bibinfo {pages} {98} (\bibinfo {year} {2017})},\
  \Eprint {http://arxiv.org/abs/1707.06823} {arXiv:1707.06823
  [physics.ins-det]} \BibitemShut {NoStop}%
\bibitem [{\citenamefont {Barton}\ \emph {et~al.}(2016)\citenamefont {Barton},
  \citenamefont {Amman}, \citenamefont {Martin},\ and\ \citenamefont
  {Vetter}}]{Barton:2015skb}%
  \BibitemOpen
  \bibfield  {author} {\bibinfo {author} {\bibfnamefont {P.}~\bibnamefont
  {Barton}}, \bibinfo {author} {\bibfnamefont {M.}~\bibnamefont {Amman}},
  \bibinfo {author} {\bibfnamefont {R.}~\bibnamefont {Martin}}, \ and\ \bibinfo
  {author} {\bibfnamefont {K.}~\bibnamefont {Vetter}},\ }\href {\doibase
  10.1016/j.nima.2015.12.031} {\bibfield  {journal} {\bibinfo  {journal} {Nucl.
  Instrum. Meth. A}\ }\textbf {\bibinfo {volume} {812}},\ \bibinfo {pages} {17}
  (\bibinfo {year} {2016})},\ \Eprint {http://arxiv.org/abs/1512.00574}
  {arXiv:1512.00574 [physics.ins-det]} \BibitemShut {NoStop}%
\bibitem [{\citenamefont {Bombelli}\ \emph {et~al.}(2011)\citenamefont
  {Bombelli}, \citenamefont {Fiorini}, \citenamefont {Frizzi}, \citenamefont
  {Alberti},\ and\ \citenamefont {Longoni}}]{Bombelli:2011}%
  \BibitemOpen
  \bibfield  {author} {\bibinfo {author} {\bibfnamefont {L.}~\bibnamefont
  {Bombelli}}, \bibinfo {author} {\bibfnamefont {C.}~\bibnamefont {Fiorini}},
  \bibinfo {author} {\bibfnamefont {T.}~\bibnamefont {Frizzi}}, \bibinfo
  {author} {\bibfnamefont {R.}~\bibnamefont {Alberti}}, \ and\ \bibinfo
  {author} {\bibfnamefont {A.}~\bibnamefont {Longoni}},\ }in\ \href {\doibase
  10.1109/NSSMIC.2011.6154396} {\emph {\bibinfo {booktitle} {2011 IEEE Nuclear
  Science Symposium Conference Record}}}\ (\bibinfo {year} {2011})\ pp.\
  \bibinfo {pages} {1972--1975}\BibitemShut {NoStop}%
\bibitem [{\citenamefont {Abgrall}\ \emph
  {et~al.}(2017{\natexlab{b}})\citenamefont {Abgrall} \emph
  {et~al.}}]{Abgrall:2017gpr}%
  \BibitemOpen
  \bibfield  {author} {\bibinfo {author} {\bibfnamefont {N.}~\bibnamefont
  {Abgrall}} \emph {et~al.},\ }\href {\doibase 10.1016/j.nima.2017.08.005}
  {\bibfield  {journal} {\bibinfo  {journal} {Nucl. Instrum. Meth. A}\ }\textbf
  {\bibinfo {volume} {872}},\ \bibinfo {pages} {16} (\bibinfo {year}
  {2017}{\natexlab{b}})},\ \Eprint {http://arxiv.org/abs/1702.02466}
  {arXiv:1702.02466 [physics.ins-det]} \BibitemShut {NoStop}%
\bibitem [{\citenamefont {Baudis}\ \emph {et~al.}(2013)\citenamefont {Baudis},
  \citenamefont {Ferella}, \citenamefont {Froborg},\ and\ \citenamefont
  {Tarka}}]{Baudis:2013kaa}%
  \BibitemOpen
  \bibfield  {author} {\bibinfo {author} {\bibfnamefont {L.}~\bibnamefont
  {Baudis}}, \bibinfo {author} {\bibfnamefont {A.~D.}\ \bibnamefont {Ferella}},
  \bibinfo {author} {\bibfnamefont {F.}~\bibnamefont {Froborg}}, \ and\
  \bibinfo {author} {\bibfnamefont {M.}~\bibnamefont {Tarka}},\ }\href
  {\doibase 10.1016/j.nima.2013.08.003} {\bibfield  {journal} {\bibinfo
  {journal} {Nucl. Instrum. Meth.}\ }\textbf {\bibinfo {volume} {A729}},\
  \bibinfo {pages} {557} (\bibinfo {year} {2013})},\ \Eprint
  {http://arxiv.org/abs/1303.6679} {arXiv:1303.6679 [physics.ins-det]}
  \BibitemShut {NoStop}%
\bibitem [{\citenamefont {Baudis}\ \emph {et~al.}(2015)\citenamefont {Baudis},
  \citenamefont {Benato}, \citenamefont {Carconi}, \citenamefont {Cattadori},
  \citenamefont {De~Felice}, \citenamefont {Eberhardt}, \citenamefont
  {Eichler}, \citenamefont {Petrucci}, \citenamefont {Tarka},\ and\
  \citenamefont {Walter}}]{Baudis:2015sba}%
  \BibitemOpen
  \bibfield  {author} {\bibinfo {author} {\bibfnamefont {L.}~\bibnamefont
  {Baudis}}, \bibinfo {author} {\bibfnamefont {G.}~\bibnamefont {Benato}},
  \bibinfo {author} {\bibfnamefont {P.}~\bibnamefont {Carconi}}, \bibinfo
  {author} {\bibfnamefont {C.~M.}\ \bibnamefont {Cattadori}}, \bibinfo {author}
  {\bibfnamefont {P.}~\bibnamefont {De~Felice}}, \bibinfo {author}
  {\bibfnamefont {K.}~\bibnamefont {Eberhardt}}, \bibinfo {author}
  {\bibfnamefont {R.}~\bibnamefont {Eichler}}, \bibinfo {author} {\bibfnamefont
  {A.}~\bibnamefont {Petrucci}}, \bibinfo {author} {\bibfnamefont
  {M.}~\bibnamefont {Tarka}}, \ and\ \bibinfo {author} {\bibfnamefont
  {M.}~\bibnamefont {Walter}},\ }\href {\doibase
  10.1088/1748-0221/10/12/P12005} {\bibfield  {journal} {\bibinfo  {journal}
  {JINST}\ }\textbf {\bibinfo {volume} {10}},\ \bibinfo {pages} {P12005}
  (\bibinfo {year} {2015})},\ \Eprint {http://arxiv.org/abs/1508.05731}
  {arXiv:1508.05731 [physics.ins-det]} \BibitemShut {NoStop}%
\bibitem [{\citenamefont {Nakamura}\ \emph {et~al.}(2011)\citenamefont
  {Nakamura}, \citenamefont {Shirakawa}, \citenamefont {Takahashi},\ and\
  \citenamefont {Shimizu}}]{Nakamura_2011}%
  \BibitemOpen
  \bibfield  {author} {\bibinfo {author} {\bibfnamefont {H.}~\bibnamefont
  {Nakamura}}, \bibinfo {author} {\bibfnamefont {Y.}~\bibnamefont {Shirakawa}},
  \bibinfo {author} {\bibfnamefont {S.}~\bibnamefont {Takahashi}}, \ and\
  \bibinfo {author} {\bibfnamefont {H.}~\bibnamefont {Shimizu}},\ }\href
  {\doibase 10.1209/0295-5075/95/22001} {\bibfield  {journal} {\bibinfo
  {journal} {{EPL} (Europhysics Letters)}\ }\textbf {\bibinfo {volume} {95}},\
  \bibinfo {pages} {22001} (\bibinfo {year} {2011})}\BibitemShut {NoStop}%
\bibitem [{\citenamefont {Majorovits}\ \emph {et~al.}(2018)\citenamefont
  {Majorovits}, \citenamefont {Eck}, \citenamefont {Fischer}, \citenamefont
  {Gooch}, \citenamefont {Hayward}, \citenamefont {Kraetzschmar}, \citenamefont
  {van~der Kolk}, \citenamefont {Muenstermann}, \citenamefont {Schulz},\ and\
  \citenamefont {Simon}}]{Majorovits:2017cqj}%
  \BibitemOpen
  \bibfield  {author} {\bibinfo {author} {\bibfnamefont {B.}~\bibnamefont
  {Majorovits}}, \bibinfo {author} {\bibfnamefont {S.}~\bibnamefont {Eck}},
  \bibinfo {author} {\bibfnamefont {F.}~\bibnamefont {Fischer}}, \bibinfo
  {author} {\bibfnamefont {C.}~\bibnamefont {Gooch}}, \bibinfo {author}
  {\bibfnamefont {C.}~\bibnamefont {Hayward}}, \bibinfo {author} {\bibfnamefont
  {T.}~\bibnamefont {Kraetzschmar}}, \bibinfo {author} {\bibfnamefont
  {N.}~\bibnamefont {van~der Kolk}}, \bibinfo {author} {\bibfnamefont
  {D.}~\bibnamefont {Muenstermann}}, \bibinfo {author} {\bibfnamefont
  {O.}~\bibnamefont {Schulz}}, \ and\ \bibinfo {author} {\bibfnamefont
  {F.}~\bibnamefont {Simon}},\ }\href {\doibase 10.1063/1.5019011} {\bibfield
  {journal} {\bibinfo  {journal} {AIP Conf. Proc.}\ }\textbf {\bibinfo {volume}
  {1921}},\ \bibinfo {pages} {090001} (\bibinfo {year} {2018})},\ \Eprint
  {http://arxiv.org/abs/1708.09265} {arXiv:1708.09265 [physics.ins-det]}
  \BibitemShut {NoStop}%
\bibitem [{\citenamefont {Wetzel}\ \emph {et~al.}(2020)\citenamefont {Wetzel},
  \citenamefont {Bostan}, \citenamefont {K\"oseyan}, \citenamefont {Tiras},\
  and\ \citenamefont {Bilki}}]{Bilki:2019lep}%
  \BibitemOpen
  \bibfield  {author} {\bibinfo {author} {\bibfnamefont {J.}~\bibnamefont
  {Wetzel}}, \bibinfo {author} {\bibfnamefont {N.}~\bibnamefont {Bostan}},
  \bibinfo {author} {\bibfnamefont {O.~K.}\ \bibnamefont {K\"oseyan}}, \bibinfo
  {author} {\bibfnamefont {E.}~\bibnamefont {Tiras}}, \ and\ \bibinfo {author}
  {\bibfnamefont {B.}~\bibnamefont {Bilki}},\ }\href {\doibase
  10.3906/fiz-1912-9} {\bibfield  {journal} {\bibinfo  {journal} {Turk. J.
  Phys.}\ }\textbf {\bibinfo {volume} {44}},\ \bibinfo {pages} {437} (\bibinfo
  {year} {2020})},\ \Eprint {http://arxiv.org/abs/1912.11342} {arXiv:1912.11342
  [physics.ins-det]} \BibitemShut {NoStop}%
\bibitem [{\citenamefont {Ku\'zniak}\ \emph {et~al.}(2019)\citenamefont
  {Ku\'zniak}, \citenamefont {Broerman}, \citenamefont {Pollmann},\ and\
  \citenamefont {Araujo}}]{Kuzniak:2018dcf}%
  \BibitemOpen
  \bibfield  {author} {\bibinfo {author} {\bibfnamefont {M.}~\bibnamefont
  {Ku\'zniak}}, \bibinfo {author} {\bibfnamefont {B.}~\bibnamefont {Broerman}},
  \bibinfo {author} {\bibfnamefont {T.}~\bibnamefont {Pollmann}}, \ and\
  \bibinfo {author} {\bibfnamefont {G.~R.}\ \bibnamefont {Araujo}},\ }\href
  {\doibase 10.1140/epjc/s10052-019-6810-8} {\bibfield  {journal} {\bibinfo
  {journal} {Eur. Phys. J. C}\ }\textbf {\bibinfo {volume} {79}},\ \bibinfo
  {pages} {291} (\bibinfo {year} {2019})},\ \Eprint
  {http://arxiv.org/abs/1806.04020} {arXiv:1806.04020 [physics.ins-det]}
  \BibitemShut {NoStop}%
\bibitem [{\citenamefont {Boulay}\ \emph {et~al.}(2021)\citenamefont {Boulay}
  \emph {et~al.}}]{Boulay:2021njr}%
  \BibitemOpen
  \bibfield  {author} {\bibinfo {author} {\bibfnamefont {M.~G.}\ \bibnamefont
  {Boulay}} \emph {et~al.},\ }\href@noop {} {\  (\bibinfo {year} {2021})},\
  \Eprint {http://arxiv.org/abs/2106.15506} {arXiv:2106.15506
  [physics.ins-det]} \BibitemShut {NoStop}%
\bibitem [{\citenamefont {Garankin}\ \emph {et~al.}(2018)\citenamefont
  {Garankin}, \citenamefont {Plukis}, \citenamefont {Plukienė}, \citenamefont
  {Lagzdina},\ and\ \citenamefont {Remeikis}}]{Garankin2018}%
  \BibitemOpen
  \bibfield  {author} {\bibinfo {author} {\bibfnamefont {J.}~\bibnamefont
  {Garankin}}, \bibinfo {author} {\bibfnamefont {A.}~\bibnamefont {Plukis}},
  \bibinfo {author} {\bibfnamefont {R.}~\bibnamefont {Plukienė}}, \bibinfo
  {author} {\bibfnamefont {E.}~\bibnamefont {Lagzdina}}, \ and\ \bibinfo
  {author} {\bibfnamefont {V.}~\bibnamefont {Remeikis}},\ }\href {\doibase
  10.1109/TNS.2017.2785683} {\bibfield  {journal} {\bibinfo  {journal} {IEEE
  Transactions on Nuclear Science}\ }\textbf {\bibinfo {volume} {65}},\
  \bibinfo {pages} {739} (\bibinfo {year} {2018})}\BibitemShut {NoStop}%
\bibitem [{\citenamefont {Efremenko}\ \emph {et~al.}(2019)\citenamefont
  {Efremenko} \emph {et~al.}}]{Efremenko:2019xbs}%
  \BibitemOpen
  \bibfield  {author} {\bibinfo {author} {\bibfnamefont {Y.}~\bibnamefont
  {Efremenko}} \emph {et~al.},\ }\href {\doibase
  10.1088/1748-0221/14/07/P07006} {\bibfield  {journal} {\bibinfo  {journal}
  {JINST}\ }\textbf {\bibinfo {volume} {14}},\ \bibinfo {pages} {P07006}
  (\bibinfo {year} {2019})},\ \Eprint {http://arxiv.org/abs/1901.03579}
  {arXiv:1901.03579 [physics.ins-det]} \BibitemShut {NoStop}%
\bibitem [{\citenamefont {Heindl}\ \emph {et~al.}(2010)\citenamefont {Heindl},
  \citenamefont {Dandl}, \citenamefont {Hofmann}, \citenamefont {Krucken},
  \citenamefont {Oberauer}, \citenamefont {Potzel}, \citenamefont {Wieser},\
  and\ \citenamefont {Ulrich}}]{Heindl:2010zz}%
  \BibitemOpen
  \bibfield  {author} {\bibinfo {author} {\bibfnamefont {T.}~\bibnamefont
  {Heindl}}, \bibinfo {author} {\bibfnamefont {T.}~\bibnamefont {Dandl}},
  \bibinfo {author} {\bibfnamefont {M.}~\bibnamefont {Hofmann}}, \bibinfo
  {author} {\bibfnamefont {R.}~\bibnamefont {Krucken}}, \bibinfo {author}
  {\bibfnamefont {L.}~\bibnamefont {Oberauer}}, \bibinfo {author}
  {\bibfnamefont {W.}~\bibnamefont {Potzel}}, \bibinfo {author} {\bibfnamefont
  {J.}~\bibnamefont {Wieser}}, \ and\ \bibinfo {author} {\bibfnamefont
  {A.}~\bibnamefont {Ulrich}},\ }\href {\doibase 10.1209/0295-5075/91/62002}
  {\bibfield  {journal} {\bibinfo  {journal} {EPL}\ }\textbf {\bibinfo {volume}
  {91}},\ \bibinfo {pages} {62002} (\bibinfo {year} {2010})},\ \Eprint
  {http://arxiv.org/abs/1511.07718} {arXiv:1511.07718 [physics.ins-det]}
  \BibitemShut {NoStop}%
\bibitem [{\citenamefont {Christofferson}\ \emph {et~al.}(2018)\citenamefont
  {Christofferson} \emph {et~al.}}]{Christofferson:2017nih}%
  \BibitemOpen
  \bibfield  {author} {\bibinfo {author} {\bibfnamefont {C.~D.}\ \bibnamefont
  {Christofferson}} \emph {et~al.} (\bibinfo {collaboration} {{\sc
  Majorana}}),\ }\href {\doibase 10.1063/1.5019001} {\bibfield  {journal}
  {\bibinfo  {journal} {AIP Conf. Proc.}\ }\textbf {\bibinfo {volume} {1921}},\
  \bibinfo {pages} {060005} (\bibinfo {year} {2018})},\ \Eprint
  {http://arxiv.org/abs/1711.10361} {arXiv:1711.10361 [physics.ins-det]}
  \BibitemShut {NoStop}%
\bibitem [{\citenamefont {LaFerriere}\ \emph {et~al.}(2015)\citenamefont
  {LaFerriere}, \citenamefont {Maiti}, \citenamefont {Arnquist},\ and\
  \citenamefont {Hoppe}}]{LaFerriere:2014rva}%
  \BibitemOpen
  \bibfield  {author} {\bibinfo {author} {\bibfnamefont {B.}~\bibnamefont
  {LaFerriere}}, \bibinfo {author} {\bibfnamefont {T.}~\bibnamefont {Maiti}},
  \bibinfo {author} {\bibfnamefont {I.}~\bibnamefont {Arnquist}}, \ and\
  \bibinfo {author} {\bibfnamefont {E.}~\bibnamefont {Hoppe}},\ }\href
  {\doibase 10.1016/j.nima.2014.11.052} {\bibfield  {journal} {\bibinfo
  {journal} {Nucl. Instrum. Meth. A}\ }\textbf {\bibinfo {volume} {775}},\
  \bibinfo {pages} {93} (\bibinfo {year} {2015})}\BibitemShut {NoStop}%
\bibitem [{\citenamefont {Neder}\ \emph {et~al.}(2000)\citenamefont {Neder},
  \citenamefont {Heusser},\ and\ \citenamefont {Laubenstein}}]{NEDER:2000}%
  \BibitemOpen
  \bibfield  {author} {\bibinfo {author} {\bibfnamefont {H.}~\bibnamefont
  {Neder}}, \bibinfo {author} {\bibfnamefont {G.}~\bibnamefont {Heusser}}, \
  and\ \bibinfo {author} {\bibfnamefont {M.}~\bibnamefont {Laubenstein}},\
  }\href {\doibase https://doi.org/10.1016/S0969-8043(00)00132-9} {\bibfield
  {journal} {\bibinfo  {journal} {Applied Radiation and Isotopes}\ }\textbf
  {\bibinfo {volume} {53}},\ \bibinfo {pages} {191 } (\bibinfo {year}
  {2000})}\BibitemShut {NoStop}%
\bibitem [{\citenamefont {Rau}\ and\ \citenamefont {Heusser}(2000)}]{Rau:2000}%
  \BibitemOpen
  \bibfield  {author} {\bibinfo {author} {\bibfnamefont {W.}~\bibnamefont
  {Rau}}\ and\ \bibinfo {author} {\bibfnamefont {G.}~\bibnamefont {Heusser}},\
  }\href {\doibase https://doi.org/10.1016/S0969-8043(00)00155-X} {\bibfield
  {journal} {\bibinfo  {journal} {Applied Radiation and Isotopes}\ }\textbf
  {\bibinfo {volume} {53}},\ \bibinfo {pages} {371 } (\bibinfo {year}
  {2000})}\BibitemShut {NoStop}%
\bibitem [{\citenamefont {Zuzel}\ and\ \citenamefont
  {Simgen}(2009)}]{Zuzel:2009}%
  \BibitemOpen
  \bibfield  {author} {\bibinfo {author} {\bibfnamefont {G.}~\bibnamefont
  {Zuzel}}\ and\ \bibinfo {author} {\bibfnamefont {H.}~\bibnamefont {Simgen}},\
  }\href {\doibase https://doi.org/10.1016/j.apradiso.2009.01.052} {\bibfield
  {journal} {\bibinfo  {journal} {Applied Radiation and Isotopes}\ }\textbf
  {\bibinfo {volume} {67}},\ \bibinfo {pages} {889} (\bibinfo {year} {2009})},\
  \bibinfo {note} {5th International Conference on Radionuclide Metrology -
  Low-Level Radioactivity Measurement Techniques ICRM-LLRMT'08}\BibitemShut
  {NoStop}%
\bibitem [{\citenamefont {Zuzel}(2005)}]{Zuzel:2005hag}%
  \BibitemOpen
  \bibfield  {author} {\bibinfo {author} {\bibfnamefont {G.}~\bibnamefont
  {Zuzel}},\ }\href {\doibase 10.1063/1.2060465} {\bibfield  {journal}
  {\bibinfo  {journal} {AIP Conf. Proc.}\ }\textbf {\bibinfo {volume} {785}},\
  \bibinfo {pages} {142} (\bibinfo {year} {2005})}\BibitemShut {NoStop}%
\bibitem [{\citenamefont {Zuzel}\ \emph {et~al.}(2017)\citenamefont {Zuzel},
  \citenamefont {Pelczar},\ and\ \citenamefont {Wójcik}}]{Zuzel:2017}%
  \BibitemOpen
  \bibfield  {author} {\bibinfo {author} {\bibfnamefont {G.}~\bibnamefont
  {Zuzel}}, \bibinfo {author} {\bibfnamefont {K.}~\bibnamefont {Pelczar}}, \
  and\ \bibinfo {author} {\bibfnamefont {M.}~\bibnamefont {Wójcik}},\ }\href
  {\doibase https://doi.org/10.1016/j.apradiso.2017.01.030} {\bibfield
  {journal} {\bibinfo  {journal} {Applied Radiation and Isotopes}\ }\textbf
  {\bibinfo {volume} {126}},\ \bibinfo {pages} {165 } (\bibinfo {year}
  {2017})},\ \bibinfo {note} {proceedings of the 7th International Conference
  on Radionuclide Metrology – Low-Level Radioactivity Measurement
  Techniques}\BibitemShut {NoStop}%
\bibitem [{\citenamefont {Zuzel}(2019)}]{Zuzel:2019}%
  \BibitemOpen
  \bibfield  {author} {\bibinfo {author} {\bibfnamefont {G.}~\bibnamefont
  {Zuzel}},\ }in\ \href@noop {} {\emph {\bibinfo {booktitle} {7th Topical
  Workshop on Low Radioactivity Techniques (LRT 2019): Jaca, Spain, May 19 -
  23}}}\ (\bibinfo {year} {2019})\BibitemShut {NoStop}%
\bibitem [{\citenamefont {Zuzel}\ \emph {et~al.}(2018)\citenamefont {Zuzel},
  \citenamefont {Pelczar},\ and\ \citenamefont {W\'{o}jcik}}]{Zuzel:2018fzl}%
  \BibitemOpen
  \bibfield  {author} {\bibinfo {author} {\bibfnamefont {G.}~\bibnamefont
  {Zuzel}}, \bibinfo {author} {\bibfnamefont {K.}~\bibnamefont {Pelczar}}, \
  and\ \bibinfo {author} {\bibfnamefont {M.}~\bibnamefont {W\'{o}jcik}},\
  }\bibfield  {booktitle} {\emph {\bibinfo {booktitle} {{Proceedings, 6th
  Topical Workshop on Low Radioactivity Techniques (LRT 2017): Seoul, Korea,
  May 24-26, 2017}}},\ }\href {\doibase 10.1063/1.5019007} {\bibfield
  {journal} {\bibinfo  {journal} {AIP Conf. Proc.}\ }\textbf {\bibinfo {volume}
  {1921}},\ \bibinfo {pages} {070004} (\bibinfo {year} {2018})}\BibitemShut
  {NoStop}%
\bibitem [{\citenamefont {Mroz}(2019)}]{Mroz:2019}%
  \BibitemOpen
  \bibfield  {author} {\bibinfo {author} {\bibfnamefont {K.}~\bibnamefont
  {Mroz}},\ }in\ \href@noop {} {\emph {\bibinfo {booktitle} {7th Topical
  Workshop on Low Radioactivity Techniques (LRT 2019): Jaca, Spain, May 19 -
  23}}}\ (\bibinfo {year} {2019})\BibitemShut {NoStop}%
\bibitem [{\citenamefont {Heusser}(1995)}]{Heusser:1995wd}%
  \BibitemOpen
  \bibfield  {author} {\bibinfo {author} {\bibfnamefont {G.}~\bibnamefont
  {Heusser}},\ }\href {\doibase 10.1146/annurev.ns.45.120195.002551} {\bibfield
   {journal} {\bibinfo  {journal} {Ann. Rev. Nucl. Part. Sci.}\ }\textbf
  {\bibinfo {volume} {45}},\ \bibinfo {pages} {543} (\bibinfo {year}
  {1995})}\BibitemShut {NoStop}%
\bibitem [{\citenamefont {Peurrung}\ \emph {et~al.}(1997)\citenamefont
  {Peurrung}, \citenamefont {Bowyer}, \citenamefont {Craig},\ and\
  \citenamefont {Reeder}}]{Peurrung:1997wc}%
  \BibitemOpen
  \bibfield  {author} {\bibinfo {author} {\bibfnamefont {A.~J.}\ \bibnamefont
  {Peurrung}}, \bibinfo {author} {\bibfnamefont {T.~W.}\ \bibnamefont
  {Bowyer}}, \bibinfo {author} {\bibfnamefont {R.~A.}\ \bibnamefont {Craig}}, \
  and\ \bibinfo {author} {\bibfnamefont {P.~L.}\ \bibnamefont {Reeder}},\
  }\href {\doibase 10.1016/S0168-9002(97)00819-X} {\bibfield  {journal}
  {\bibinfo  {journal} {Nucl. Instrum. Meth. A}\ }\textbf {\bibinfo {volume}
  {396}},\ \bibinfo {pages} {425} (\bibinfo {year} {1997})}\BibitemShut
  {NoStop}%
\bibitem [{\citenamefont {Amsbaugh}\ \emph {et~al.}(2007)\citenamefont
  {Amsbaugh} \emph {et~al.}}]{Amsbaugh:2007ke}%
  \BibitemOpen
  \bibfield  {author} {\bibinfo {author} {\bibfnamefont {J.~F.}\ \bibnamefont
  {Amsbaugh}} \emph {et~al.},\ }\href {\doibase 10.1016/j.nima.2007.05.321}
  {\bibfield  {journal} {\bibinfo  {journal} {Nucl. Instrum. Meth. A}\ }\textbf
  {\bibinfo {volume} {579}},\ \bibinfo {pages} {1054} (\bibinfo {year}
  {2007})},\ \Eprint {http://arxiv.org/abs/0705.3665} {arXiv:0705.3665
  [nucl-ex]} \BibitemShut {NoStop}%
\bibitem [{\citenamefont {Agostini}\ \emph {et~al.}(2014)\citenamefont
  {Agostini} \emph {et~al.}}]{Agostini:2013tek}%
  \BibitemOpen
  \bibfield  {author} {\bibinfo {author} {\bibfnamefont {M.}~\bibnamefont
  {Agostini}} \emph {et~al.} (\bibinfo {collaboration} {{\sc Gerda}}),\ }\href
  {\doibase 10.1140/epjc/s10052-014-2764-z} {\bibfield  {journal} {\bibinfo
  {journal} {Eur. Phys. J. C}\ }\textbf {\bibinfo {volume} {74}},\ \bibinfo
  {pages} {2764} (\bibinfo {year} {2014})},\ \Eprint
  {http://arxiv.org/abs/1306.5084} {arXiv:1306.5084 [physics.ins-det]}
  \BibitemShut {NoStop}%
\bibitem [{\citenamefont {Agostini}\ \emph
  {et~al.}(2017{\natexlab{b}})\citenamefont {Agostini} \emph
  {et~al.}}]{Agostini:2016mof}%
  \BibitemOpen
  \bibfield  {author} {\bibinfo {author} {\bibfnamefont {M.}~\bibnamefont
  {Agostini}} \emph {et~al.} (\bibinfo {collaboration} {{\sc Gerda}}),\ }\href
  {\doibase 10.1016/j.astropartphys.2017.03.003} {\bibfield  {journal}
  {\bibinfo  {journal} {Astropart. Phys.}\ }\textbf {\bibinfo {volume} {91}},\
  \bibinfo {pages} {15} (\bibinfo {year} {2017}{\natexlab{b}})},\ \Eprint
  {http://arxiv.org/abs/1611.06884} {arXiv:1611.06884 [physics.ins-det]}
  \BibitemShut {NoStop}%
\bibitem [{\citenamefont {Freund}\ \emph {et~al.}(2016)\citenamefont {Freund}
  \emph {et~al.}}]{Freund:2016fhz}%
  \BibitemOpen
  \bibfield  {author} {\bibinfo {author} {\bibfnamefont {K.}~\bibnamefont
  {Freund}} \emph {et~al.},\ }\href {\doibase 10.1140/epjc/s10052-016-4140-7}
  {\bibfield  {journal} {\bibinfo  {journal} {Eur. Phys. J. C}\ }\textbf
  {\bibinfo {volume} {76}},\ \bibinfo {pages} {298} (\bibinfo {year} {2016})},\
  \Eprint {http://arxiv.org/abs/1601.05935} {arXiv:1601.05935
  [physics.ins-det]} \BibitemShut {NoStop}%
\bibitem [{\citenamefont {Lubashevskiy}\ \emph {et~al.}(2018)\citenamefont
  {Lubashevskiy} \emph {et~al.}}]{Lubashevskiy:2017lmf}%
  \BibitemOpen
  \bibfield  {author} {\bibinfo {author} {\bibfnamefont {A.}~\bibnamefont
  {Lubashevskiy}} \emph {et~al.},\ }\href {\doibase
  10.1140/epjc/s10052-017-5499-9} {\bibfield  {journal} {\bibinfo  {journal}
  {Eur. Phys. J. C}\ }\textbf {\bibinfo {volume} {78}},\ \bibinfo {pages} {15}
  (\bibinfo {year} {2018})},\ \Eprint {http://arxiv.org/abs/1708.00226}
  {arXiv:1708.00226 [physics.ins-det]} \BibitemShut {NoStop}%
\bibitem [{\citenamefont {Busch}\ \emph {et~al.}(2018)\citenamefont {Busch}
  \emph {et~al.}}]{Busch:2017kxq}%
  \BibitemOpen
  \bibfield  {author} {\bibinfo {author} {\bibfnamefont {M.}~\bibnamefont
  {Busch}} \emph {et~al.},\ }\bibfield  {booktitle} {\emph {\bibinfo
  {booktitle} {{Proceedings, 6th Topical Workshop on Low Radioactivity
  Techniques (LRT 2017): Seoul, Korea, May 24-26, 2017}}},\ }\href {\doibase
  10.1063/1.5019005} {\bibfield  {journal} {\bibinfo  {journal} {AIP Conf.
  Proc.}\ }\textbf {\bibinfo {volume} {1921}},\ \bibinfo {pages} {070002}
  (\bibinfo {year} {2018})},\ \Eprint {http://arxiv.org/abs/1712.04985}
  {arXiv:1712.04985 [physics.ins-det]} \BibitemShut {NoStop}%
\bibitem [{\citenamefont {Dobson}\ \emph {et~al.}(2018)\citenamefont {Dobson},
  \citenamefont {Ghag},\ and\ \citenamefont {Manenti}}]{Dobson:2017esw}%
  \BibitemOpen
  \bibfield  {author} {\bibinfo {author} {\bibfnamefont {J.}~\bibnamefont
  {Dobson}}, \bibinfo {author} {\bibfnamefont {C.}~\bibnamefont {Ghag}}, \ and\
  \bibinfo {author} {\bibfnamefont {L.}~\bibnamefont {Manenti}},\ }\href
  {\doibase 10.1016/j.nima.2017.10.014} {\bibfield  {journal} {\bibinfo
  {journal} {Nucl. Instrum. Meth.}\ }\textbf {\bibinfo {volume} {A879}},\
  \bibinfo {pages} {25} (\bibinfo {year} {2018})},\ \Eprint
  {http://arxiv.org/abs/1708.08860} {arXiv:1708.08860 [physics.ins-det]}
  \BibitemShut {NoStop}%
\bibitem [{\citenamefont {Lindemann}\ and\ \citenamefont
  {Simgen}(2014)}]{Lindemann:2013kna}%
  \BibitemOpen
  \bibfield  {author} {\bibinfo {author} {\bibfnamefont {S.}~\bibnamefont
  {Lindemann}}\ and\ \bibinfo {author} {\bibfnamefont {H.}~\bibnamefont
  {Simgen}},\ }\href {\doibase 10.1140/epjc/s10052-014-2746-1} {\bibfield
  {journal} {\bibinfo  {journal} {Eur. Phys. J. C}\ }\textbf {\bibinfo {volume}
  {74}},\ \bibinfo {pages} {2746} (\bibinfo {year} {2014})},\ \Eprint
  {http://arxiv.org/abs/1308.4806} {arXiv:1308.4806 [physics.ins-det]}
  \BibitemShut {NoStop}%
\bibitem [{\citenamefont {Christofferson}(2019)}]{Christofferson:2019}%
  \BibitemOpen
  \bibfield  {author} {\bibinfo {author} {\bibfnamefont {C.~D.}\ \bibnamefont
  {Christofferson}},\ }in\ \href@noop {} {\emph {\bibinfo {booktitle} {7th
  Topical Workshop on Low Radioactivity Techniques (LRT 2019): Jaca, Spain, May
  19 - 23}}}\ (\bibinfo {year} {2019})\BibitemShut {NoStop}%
\bibitem [{\citenamefont {Guiseppe}\ \emph {et~al.}(2018)\citenamefont
  {Guiseppe}, \citenamefont {Christofferson}, \citenamefont {Hair},\ and\
  \citenamefont {Adams}}]{Guiseppe:2017yah}%
  \BibitemOpen
  \bibfield  {author} {\bibinfo {author} {\bibfnamefont {V.~E.}\ \bibnamefont
  {Guiseppe}}, \bibinfo {author} {\bibfnamefont {C.~D.}\ \bibnamefont
  {Christofferson}}, \bibinfo {author} {\bibfnamefont {K.~R.}\ \bibnamefont
  {Hair}}, \ and\ \bibinfo {author} {\bibfnamefont {F.~M.}\ \bibnamefont
  {Adams}},\ }\href {\doibase 10.1063/1.5019006} {\bibfield  {journal}
  {\bibinfo  {journal} {AIP Conf. Proc.}\ }\textbf {\bibinfo {volume} {1921}},\
  \bibinfo {pages} {070003} (\bibinfo {year} {2018})},\ \Eprint
  {http://arxiv.org/abs/1712.08167} {arXiv:1712.08167 [physics.ins-det]}
  \BibitemShut {NoStop}%
\bibitem [{\citenamefont {Zuzel}\ and\ \citenamefont
  {Wójcik}(2012)}]{Zuzel:2012a}%
  \BibitemOpen
  \bibfield  {author} {\bibinfo {author} {\bibfnamefont {G.}~\bibnamefont
  {Zuzel}}\ and\ \bibinfo {author} {\bibfnamefont {M.}~\bibnamefont
  {Wójcik}},\ }\href {\doibase https://doi.org/10.1016/j.nima.2011.12.043}
  {\bibfield  {journal} {\bibinfo  {journal} {Nuclear Instruments and Methods
  in Physics Research Section A: Accelerators, Spectrometers, Detectors and
  Associated Equipment}\ }\textbf {\bibinfo {volume} {676}},\ \bibinfo {pages}
  {140 } (\bibinfo {year} {2012})}\BibitemShut {NoStop}%
\bibitem [{\citenamefont {Zuzel}\ \emph {et~al.}(2012)\citenamefont {Zuzel},
  \citenamefont {Wójcik}, \citenamefont {Majorovits}, \citenamefont
  {Lampert},\ and\ \citenamefont {Wendling}}]{Zuzel:2012b}%
  \BibitemOpen
  \bibfield  {author} {\bibinfo {author} {\bibfnamefont {G.}~\bibnamefont
  {Zuzel}}, \bibinfo {author} {\bibfnamefont {M.}~\bibnamefont {Wójcik}},
  \bibinfo {author} {\bibfnamefont {B.}~\bibnamefont {Majorovits}}, \bibinfo
  {author} {\bibfnamefont {M.}~\bibnamefont {Lampert}}, \ and\ \bibinfo
  {author} {\bibfnamefont {P.}~\bibnamefont {Wendling}},\ }\href {\doibase
  https://doi.org/10.1016/j.nima.2011.12.020} {\bibfield  {journal} {\bibinfo
  {journal} {Nuclear Instruments and Methods in Physics Research Section A:
  Accelerators, Spectrometers, Detectors and Associated Equipment}\ }\textbf
  {\bibinfo {volume} {676}},\ \bibinfo {pages} {149 } (\bibinfo {year}
  {2012})}\BibitemShut {NoStop}%
\bibitem [{\citenamefont {Abgrall}\ \emph {et~al.}(2021)\citenamefont {Abgrall}
  \emph {et~al.}}]{Abgrall:2020jto}%
  \BibitemOpen
  \bibfield  {author} {\bibinfo {author} {\bibfnamefont {N.}~\bibnamefont
  {Abgrall}} \emph {et~al.} (\bibinfo {collaboration} {{\sc Majorana}}),\
  }\href {\doibase 10.1109/TNS.2020.3043671} {\bibfield  {journal} {\bibinfo
  {journal} {IEEE Trans. Nucl. Sci.}\ }\textbf {\bibinfo {volume} {68}},\
  \bibinfo {pages} {359} (\bibinfo {year} {2021})},\ \Eprint
  {http://arxiv.org/abs/2003.04128} {arXiv:2003.04128 [physics.ins-det]}
  \BibitemShut {NoStop}%
\bibitem [{\citenamefont {{Jordanov}}\ and\ \citenamefont
  {{Knoll}}(1994)}]{1994NIMPA.345..337J}%
  \BibitemOpen
  \bibfield  {author} {\bibinfo {author} {\bibfnamefont {V.~T.}\ \bibnamefont
  {{Jordanov}}}\ and\ \bibinfo {author} {\bibfnamefont {G.~F.}\ \bibnamefont
  {{Knoll}}},\ }\href {\doibase 10.1016/0168-9002(94)91011-1} {\bibfield
  {journal} {\bibinfo  {journal} {Nuclear Instruments and Methods in Physics
  Research A}\ }\textbf {\bibinfo {volume} {345}},\ \bibinfo {pages} {337}
  (\bibinfo {year} {1994})}\BibitemShut {NoStop}%
\bibitem [{\citenamefont {Agostini}\ \emph
  {et~al.}(2015{\natexlab{d}})\citenamefont {Agostini} \emph
  {et~al.}}]{gerda:2015:zac}%
  \BibitemOpen
  \bibfield  {author} {\bibinfo {author} {\bibfnamefont {M.}~\bibnamefont
  {Agostini}} \emph {et~al.} (\bibinfo {collaboration} {{\sc Gerda}}),\
  }\href@noop {} {\bibfield  {journal} {\bibinfo  {journal} {Eur. Phys. J. C}\
  }\textbf {\bibinfo {volume} {75}},\ \bibinfo {pages} {255} (\bibinfo {year}
  {2015}{\natexlab{d}})}\BibitemShut {NoStop}%
\bibitem [{\citenamefont {Agostini}\ \emph
  {et~al.}(2015{\natexlab{e}})\citenamefont {Agostini} \emph
  {et~al.}}]{Agostini:2015pta}%
  \BibitemOpen
  \bibfield  {author} {\bibinfo {author} {\bibfnamefont {M.}~\bibnamefont
  {Agostini}} \emph {et~al.} (\bibinfo {collaboration} {{\sc Gerda}}),\ }\href
  {\doibase 10.1140/epjc/s10052-015-3409-6} {\bibfield  {journal} {\bibinfo
  {journal} {Eur. Phys. J. C}\ }\textbf {\bibinfo {volume} {75}},\ \bibinfo
  {pages} {255} (\bibinfo {year} {2015}{\natexlab{e}})},\ \Eprint
  {http://arxiv.org/abs/1502.04392} {arXiv:1502.04392 [physics.ins-det]}
  \BibitemShut {NoStop}%
\bibitem [{\citenamefont {Arnquist}\ \emph
  {et~al.}(2020{\natexlab{b}})\citenamefont {Arnquist} \emph
  {et~al.}}]{Arnquist:2020veq}%
  \BibitemOpen
  \bibfield  {author} {\bibinfo {author} {\bibfnamefont {I.}~\bibnamefont
  {Arnquist}} \emph {et~al.} (\bibinfo {collaboration} {{\sc Majorana}}),\
  }\href@noop {} {\  (\bibinfo {year} {2020}{\natexlab{b}})},\ \Eprint
  {http://arxiv.org/abs/2006.13179} {arXiv:2006.13179 [physics.ins-det]}
  \BibitemShut {NoStop}%
\bibitem [{\citenamefont {Budjas}\ \emph {et~al.}(2009)\citenamefont {Budjas},
  \citenamefont {Barnabe~Heider}, \citenamefont {Chkvorets}, \citenamefont
  {Khanbekov},\ and\ \citenamefont {Schonert}}]{Budjas:2009zu}%
  \BibitemOpen
  \bibfield  {author} {\bibinfo {author} {\bibfnamefont {D.}~\bibnamefont
  {Budjas}}, \bibinfo {author} {\bibfnamefont {M.}~\bibnamefont
  {Barnabe~Heider}}, \bibinfo {author} {\bibfnamefont {O.}~\bibnamefont
  {Chkvorets}}, \bibinfo {author} {\bibfnamefont {N.}~\bibnamefont
  {Khanbekov}}, \ and\ \bibinfo {author} {\bibfnamefont {S.}~\bibnamefont
  {Schonert}},\ }\href {\doibase 10.1088/1748-0221/4/10/P10007} {\bibfield
  {journal} {\bibinfo  {journal} {JINST}\ }\textbf {\bibinfo {volume} {4}},\
  \bibinfo {pages} {P10007} (\bibinfo {year} {2009})},\ \Eprint
  {http://arxiv.org/abs/0909.4044} {arXiv:0909.4044 [nucl-ex]} \BibitemShut
  {NoStop}%
\bibitem [{\citenamefont {Alvis}\ \emph
  {et~al.}(2019{\natexlab{c}})\citenamefont {Alvis} \emph
  {et~al.}}]{Alvis:2019dzt}%
  \BibitemOpen
  \bibfield  {author} {\bibinfo {author} {\bibfnamefont {S.}~\bibnamefont
  {Alvis}} \emph {et~al.} (\bibinfo {collaboration} {{\sc Majorana}}),\ }\href
  {\doibase 10.1103/PhysRevC.99.065501} {\bibfield  {journal} {\bibinfo
  {journal} {Phys. Rev. C}\ }\textbf {\bibinfo {volume} {99}},\ \bibinfo
  {pages} {065501} (\bibinfo {year} {2019}{\natexlab{c}})},\ \Eprint
  {http://arxiv.org/abs/1901.05388} {arXiv:1901.05388 [physics.ins-det]}
  \BibitemShut {NoStop}%
\bibitem [{\citenamefont {Agostini}\ \emph {et~al.}(2012)\citenamefont
  {Agostini}, \citenamefont {Pandola},\ and\ \citenamefont
  {Zavarise}}]{Agostini:2011mh}%
  \BibitemOpen
  \bibfield  {author} {\bibinfo {author} {\bibfnamefont {M.}~\bibnamefont
  {Agostini}}, \bibinfo {author} {\bibfnamefont {L.}~\bibnamefont {Pandola}}, \
  and\ \bibinfo {author} {\bibfnamefont {P.}~\bibnamefont {Zavarise}},\ }\href
  {\doibase 10.1088/1742-6596/368/1/012047} {\bibfield  {journal} {\bibinfo
  {journal} {J. Phys. Conf. Ser.}\ }\textbf {\bibinfo {volume} {368}},\
  \bibinfo {pages} {012047} (\bibinfo {year} {2012})},\ \Eprint
  {http://arxiv.org/abs/1111.3582} {arXiv:1111.3582 [physics.data-an]}
  \BibitemShut {NoStop}%
\bibitem [{\citenamefont {Wiesinger}\ \emph {et~al.}(2018)\citenamefont
  {Wiesinger}, \citenamefont {Pandola},\ and\ \citenamefont
  {Schönert}}]{Wiesinger:2018qxt}%
  \BibitemOpen
  \bibfield  {author} {\bibinfo {author} {\bibfnamefont {C.}~\bibnamefont
  {Wiesinger}}, \bibinfo {author} {\bibfnamefont {L.}~\bibnamefont {Pandola}},
  \ and\ \bibinfo {author} {\bibfnamefont {S.}~\bibnamefont {Schönert}},\
  }\href {\doibase 10.1140/epjc/s10052-018-6079-3} {\bibfield  {journal}
  {\bibinfo  {journal} {Eur. Phys. J. C}\ }\textbf {\bibinfo {volume} {78}},\
  \bibinfo {pages} {597} (\bibinfo {year} {2018})},\ \Eprint
  {http://arxiv.org/abs/1802.05040} {arXiv:1802.05040 [hep-ex]} \BibitemShut
  {NoStop}%
\bibitem [{\citenamefont {Agostini}\ \emph {et~al.}(2013)\citenamefont
  {Agostini} \emph {et~al.}}]{gerda:2013:psd}%
  \BibitemOpen
  \bibfield  {author} {\bibinfo {author} {\bibfnamefont {M.}~\bibnamefont
  {Agostini}} \emph {et~al.} (\bibinfo {collaboration} {{\sc Gerda}}),\
  }\href@noop {} {\bibfield  {journal} {\bibinfo  {journal} {Eur. Phys. J. C}\
  }\textbf {\bibinfo {volume} {73}},\ \bibinfo {pages} {2583} (\bibinfo {year}
  {2013})}\BibitemShut {NoStop}%
\bibitem [{\citenamefont {Boswell}\ \emph {et~al.}(2011)\citenamefont {Boswell}
  \emph {et~al.}}]{Boswell:2010mr}%
  \BibitemOpen
  \bibfield  {author} {\bibinfo {author} {\bibfnamefont {M.}~\bibnamefont
  {Boswell}} \emph {et~al.},\ }\href {\doibase 10.1109/TNS.2011.2144619}
  {\bibfield  {journal} {\bibinfo  {journal} {IEEE Trans. Nucl. Sci.}\ }\textbf
  {\bibinfo {volume} {58}},\ \bibinfo {pages} {1212} (\bibinfo {year}
  {2011})},\ \Eprint {http://arxiv.org/abs/1011.3827} {arXiv:1011.3827
  [nucl-ex]} \BibitemShut {NoStop}%
\bibitem [{\citenamefont {Agostinelli}\ \emph {et~al.}(2003)\citenamefont
  {Agostinelli} \emph {et~al.}}]{Agostinelli:2002hh}%
  \BibitemOpen
  \bibfield  {author} {\bibinfo {author} {\bibfnamefont {S.}~\bibnamefont
  {Agostinelli}} \emph {et~al.} (\bibinfo {collaboration} {GEANT4}),\ }\href
  {\doibase 10.1016/S0168-9002(03)01368-8} {\bibfield  {journal} {\bibinfo
  {journal} {Nucl. Instrum. Meth. A}\ }\textbf {\bibinfo {volume} {506}},\
  \bibinfo {pages} {250} (\bibinfo {year} {2003})}\BibitemShut {NoStop}%
\bibitem [{\citenamefont {Wiesinger}(2020)}]{Wiesinger2021}%
  \BibitemOpen
  \bibfield  {author} {\bibinfo {author} {\bibfnamefont {C.}~\bibnamefont
  {Wiesinger}},\ }\emph {\bibinfo {title} {No neutrinos not found}},\
  \href@noop {} {Ph.D. thesis},\ \bibinfo  {school} {Technische Universität
  München}, \bibinfo {address} {München} (\bibinfo {year} {2020})\BibitemShut
  {NoStop}%
\bibitem [{\citenamefont {Pertoldi}(2021)}]{Pertoldi2021}%
  \BibitemOpen
  \bibfield  {author} {\bibinfo {author} {\bibfnamefont {L.}~\bibnamefont
  {Pertoldi}},\ }\emph {\bibinfo {title} {{Search for new physics with
  two-neutrino double-beta decay in GERDA data}}},\ \href
  {https://github.com/gipert/phd-thesis/releases/download/v1.1/pertoldi-phd-thesis.pdf}
  {Ph.D. thesis},\ \bibinfo  {school} {Universit\'a degli Studi di Padova}
  (\bibinfo {year} {2021})\BibitemShut {NoStop}%
\bibitem [{\citenamefont {Elliott}\ \emph {et~al.}(2010)\citenamefont
  {Elliott}, \citenamefont {Guiseppe}, \citenamefont {LaRoque}, \citenamefont
  {Johnson},\ and\ \citenamefont {Mashnik}}]{Elliott:2009cw}%
  \BibitemOpen
  \bibfield  {author} {\bibinfo {author} {\bibfnamefont {S.}~\bibnamefont
  {Elliott}}, \bibinfo {author} {\bibfnamefont {V.}~\bibnamefont {Guiseppe}},
  \bibinfo {author} {\bibfnamefont {B.}~\bibnamefont {LaRoque}}, \bibinfo
  {author} {\bibfnamefont {R.}~\bibnamefont {Johnson}}, \ and\ \bibinfo
  {author} {\bibfnamefont {S.}~\bibnamefont {Mashnik}},\ }\href {\doibase
  10.1103/PhysRevC.82.054610} {\bibfield  {journal} {\bibinfo  {journal} {Phys.
  Rev. C}\ }\textbf {\bibinfo {volume} {82}},\ \bibinfo {pages} {054610}
  (\bibinfo {year} {2010})},\ \Eprint {http://arxiv.org/abs/0912.3748}
  {arXiv:0912.3748 [nucl-ex]} \BibitemShut {NoStop}%
\bibitem [{\citenamefont {Benziger}\ \emph {et~al.}(2007)\citenamefont
  {Benziger} \emph {et~al.}}]{Benziger:2007iv}%
  \BibitemOpen
  \bibfield  {author} {\bibinfo {author} {\bibfnamefont {J.}~\bibnamefont
  {Benziger}} \emph {et~al.},\ }\href {\doibase 10.1016/j.nima.2007.08.176}
  {\bibfield  {journal} {\bibinfo  {journal} {Nucl. Instrum. Meth. A}\ }\textbf
  {\bibinfo {volume} {582}},\ \bibinfo {pages} {509} (\bibinfo {year}
  {2007})},\ \Eprint {http://arxiv.org/abs/physics/0702162}
  {arXiv:physics/0702162} \BibitemShut {NoStop}%
\bibitem [{\citenamefont {Agnes}\ \emph
  {et~al.}(2021{\natexlab{b}})\citenamefont {Agnes} \emph
  {et~al.}}]{DarkSide:2021mpp}%
  \BibitemOpen
  \bibfield  {author} {\bibinfo {author} {\bibfnamefont {P.}~\bibnamefont
  {Agnes}} \emph {et~al.} (\bibinfo {collaboration} {DarkSide-20k}),\ }\href
  {\doibase 10.1140/epjc/s10052-021-09121-9} {\bibfield  {journal} {\bibinfo
  {journal} {Eur. Phys. J. C}\ }\textbf {\bibinfo {volume} {81}},\ \bibinfo
  {pages} {359} (\bibinfo {year} {2021}{\natexlab{b}})},\ \Eprint
  {http://arxiv.org/abs/2101.08686} {arXiv:2101.08686 [physics.ins-det]}
  \BibitemShut {NoStop}%
\bibitem [{\citenamefont {Lehnert}(2016)}]{Lehnert:2016phd}%
  \BibitemOpen
  \bibfield  {author} {\bibinfo {author} {\bibfnamefont {B.}~\bibnamefont
  {Lehnert}},\ }\emph {\bibinfo {title} {Search for 2$\nu\beta\beta$ Excited
  State Transitions and HPGe Characterization for Surface Events in GERDA Phase
  II}},\ \href@noop {} {Ph.D. thesis},\ \bibinfo  {school} {Technische
  Universit\"{a}t Dresden} (\bibinfo {year} {2016})\BibitemShut {NoStop}%
\bibitem [{\citenamefont {Barton}(2021)}]{Barton:2020fiz}%
  \BibitemOpen
  \bibfield  {author} {\bibinfo {author} {\bibfnamefont {C.}~\bibnamefont
  {Barton}} (\bibinfo {collaboration} {LEGEND}),\ }\href {\doibase
  10.22323/1.390.0195} {\bibfield  {journal} {\bibinfo  {journal} {PoS}\
  }\textbf {\bibinfo {volume} {ICHEP2020}},\ \bibinfo {pages} {195} (\bibinfo
  {year} {2021})}\BibitemShut {NoStop}%
\bibitem [{\citenamefont {Maneschg}\ \emph {et~al.}(2008)\citenamefont
  {Maneschg}, \citenamefont {Laubenstein}, \citenamefont {Budjas},
  \citenamefont {Hampel}, \citenamefont {Heusser}, \citenamefont {Knopfle},
  \citenamefont {Schwingenheuer},\ and\ \citenamefont
  {Simgen}}]{Maneschg:2008zz}%
  \BibitemOpen
  \bibfield  {author} {\bibinfo {author} {\bibfnamefont {W.}~\bibnamefont
  {Maneschg}}, \bibinfo {author} {\bibfnamefont {M.}~\bibnamefont
  {Laubenstein}}, \bibinfo {author} {\bibfnamefont {D.}~\bibnamefont {Budjas}},
  \bibinfo {author} {\bibfnamefont {W.}~\bibnamefont {Hampel}}, \bibinfo
  {author} {\bibfnamefont {G.}~\bibnamefont {Heusser}}, \bibinfo {author}
  {\bibfnamefont {K.~T.}\ \bibnamefont {Knopfle}}, \bibinfo {author}
  {\bibfnamefont {B.}~\bibnamefont {Schwingenheuer}}, \ and\ \bibinfo {author}
  {\bibfnamefont {H.}~\bibnamefont {Simgen}},\ }\href {\doibase
  10.1016/j.nima.2008.05.036} {\bibfield  {journal} {\bibinfo  {journal} {Nucl.
  Instrum. Meth. A}\ }\textbf {\bibinfo {volume} {593}},\ \bibinfo {pages}
  {448} (\bibinfo {year} {2008})}\BibitemShut {NoStop}%
\bibitem [{\citenamefont {Westerdale}\ and\ \citenamefont
  {Meyers}(2017)}]{Westerdale:2017kml}%
  \BibitemOpen
  \bibfield  {author} {\bibinfo {author} {\bibfnamefont {S.}~\bibnamefont
  {Westerdale}}\ and\ \bibinfo {author} {\bibfnamefont {P.~D.}\ \bibnamefont
  {Meyers}},\ }\href {\doibase 10.1016/j.nima.2017.09.007} {\bibfield
  {journal} {\bibinfo  {journal} {Nucl. Instrum. Meth. A}\ }\textbf {\bibinfo
  {volume} {875}},\ \bibinfo {pages} {57} (\bibinfo {year} {2017})},\ \Eprint
  {http://arxiv.org/abs/1702.02465} {arXiv:1702.02465 [physics.ins-det]}
  \BibitemShut {NoStop}%
\bibitem [{\citenamefont {Smith}(2012)}]{Smith:2012fq}%
  \BibitemOpen
  \bibfield  {author} {\bibinfo {author} {\bibfnamefont {N.~J.~T.}\
  \bibnamefont {Smith}},\ }\href {\doibase 10.1140/epjp/i2012-12108-9}
  {\bibfield  {journal} {\bibinfo  {journal} {Eur. Phys. J. Plus}\ }\textbf
  {\bibinfo {volume} {127}},\ \bibinfo {pages} {108} (\bibinfo {year}
  {2012})}\BibitemShut {NoStop}%
\bibitem [{\citenamefont {Heise}(2020)}]{Heise:2017rpu}%
  \BibitemOpen
  \bibfield  {author} {\bibinfo {author} {\bibfnamefont {J.}~\bibnamefont
  {Heise}},\ }\href {\doibase 10.5281/zenodo.1300395} {\bibfield  {journal}
  {\bibinfo  {journal} {J. Phys. Conf. Ser.}\ }\textbf {\bibinfo {volume}
  {1342}},\ \bibinfo {pages} {012085} (\bibinfo {year} {2020})},\ \Eprint
  {http://arxiv.org/abs/1710.11584} {arXiv:1710.11584 [physics.ins-det]}
  \BibitemShut {NoStop}%
\bibitem [{\citenamefont {Reichhart}\ \emph {et~al.}(2013)\citenamefont
  {Reichhart} \emph {et~al.}}]{Reichhart:2013xkd}%
  \BibitemOpen
  \bibfield  {author} {\bibinfo {author} {\bibfnamefont {L.}~\bibnamefont
  {Reichhart}} \emph {et~al.},\ }\href {\doibase
  10.1016/j.astropartphys.2013.06.002} {\bibfield  {journal} {\bibinfo
  {journal} {Astropart. Phys.}\ }\textbf {\bibinfo {volume} {47}},\ \bibinfo
  {pages} {67} (\bibinfo {year} {2013})},\ \Eprint
  {http://arxiv.org/abs/1302.4275} {arXiv:1302.4275 [physics.ins-det]}
  \BibitemShut {NoStop}%
\bibitem [{\citenamefont {{Boulby Feasibility Study
  Report}}(2021)}]{BFSR:2021}%
  \BibitemOpen
  \bibfield  {author} {\bibinfo {author} {\bibnamefont {{Boulby Feasibility
  Study Report}}},\ }\href@noop {} {\bibfield  {journal} {\bibinfo  {journal}
  {\emph{In preparation}}\ } (\bibinfo {year} {2021})}\BibitemShut {NoStop}%
\bibitem [{\citenamefont {Agostini}\ \emph
  {et~al.}(2016{\natexlab{b}})\citenamefont {Agostini} \emph
  {et~al.}}]{GERDA:2016lhn}%
  \BibitemOpen
  \bibfield  {author} {\bibinfo {author} {\bibfnamefont {M.}~\bibnamefont
  {Agostini}} \emph {et~al.} (\bibinfo {collaboration} {{\sc Gerda}}),\ }\href
  {\doibase 10.1016/j.astropartphys.2016.08.002} {\bibfield  {journal}
  {\bibinfo  {journal} {Astropart. Phys.}\ }\textbf {\bibinfo {volume} {84}},\
  \bibinfo {pages} {29} (\bibinfo {year} {2016}{\natexlab{b}})},\ \Eprint
  {http://arxiv.org/abs/1601.06007} {arXiv:1601.06007 [physics.ins-det]}
  \BibitemShut {NoStop}%
\bibitem [{\citenamefont {Abgrall}\ \emph
  {et~al.}(2017{\natexlab{c}})\citenamefont {Abgrall} \emph
  {et~al.}}]{MAJORANA:2016ifg}%
  \BibitemOpen
  \bibfield  {author} {\bibinfo {author} {\bibfnamefont {N.}~\bibnamefont
  {Abgrall}} \emph {et~al.} (\bibinfo {collaboration} {{\sc Majorana}}),\
  }\href {\doibase 10.1016/j.astropartphys.2017.01.013} {\bibfield  {journal}
  {\bibinfo  {journal} {Astropart. Phys.}\ }\textbf {\bibinfo {volume} {93}},\
  \bibinfo {pages} {70} (\bibinfo {year} {2017}{\natexlab{c}})},\ \Eprint
  {http://arxiv.org/abs/1602.07742} {arXiv:1602.07742 [nucl-ex]} \BibitemShut
  {NoStop}%
\bibitem [{\citenamefont {Duncan}\ \emph {et~al.}(2010)\citenamefont {Duncan},
  \citenamefont {Noble},\ and\ \citenamefont {Sinclair}}]{Duncan2010SNOLAB}%
  \BibitemOpen
  \bibfield  {author} {\bibinfo {author} {\bibfnamefont {F.}~\bibnamefont
  {Duncan}}, \bibinfo {author} {\bibfnamefont {A.}~\bibnamefont {Noble}}, \
  and\ \bibinfo {author} {\bibfnamefont {D.}~\bibnamefont {Sinclair}},\
  }\href@noop {} {\bibfield  {journal} {\bibinfo  {journal} {Annu. Rev. Nucl.
  Part. Sci.}\ }\textbf {\bibinfo {volume} {60}},\ \bibinfo {pages} {163}
  (\bibinfo {year} {2010})}\BibitemShut {NoStop}%
\bibitem [{\citenamefont {Aglietta}\ \emph {et~al.}(1992)\citenamefont
  {Aglietta} \emph {et~al.}}]{Aglietta:1992dy}%
  \BibitemOpen
  \bibfield  {author} {\bibinfo {author} {\bibfnamefont {M.}~\bibnamefont
  {Aglietta}} \emph {et~al.},\ }\href {\doibase 10.1007/BF02740929} {\bibfield
  {journal} {\bibinfo  {journal} {Nuovo Cim. A}\ }\textbf {\bibinfo {volume}
  {105}},\ \bibinfo {pages} {1793} (\bibinfo {year} {1992})}\BibitemShut
  {NoStop}%
\bibitem [{\citenamefont {Alimonti}\ \emph {et~al.}(2009)\citenamefont
  {Alimonti} \emph {et~al.}}]{Borexino:2008gab}%
  \BibitemOpen
  \bibfield  {author} {\bibinfo {author} {\bibfnamefont {G.}~\bibnamefont
  {Alimonti}} \emph {et~al.} (\bibinfo {collaboration} {Borexino}),\ }\href
  {\doibase 10.1016/j.nima.2008.11.076} {\bibfield  {journal} {\bibinfo
  {journal} {Nucl. Instrum. Meth. A}\ }\textbf {\bibinfo {volume} {600}},\
  \bibinfo {pages} {568} (\bibinfo {year} {2009})},\ \Eprint
  {http://arxiv.org/abs/0806.2400} {arXiv:0806.2400 [physics.ins-det]}
  \BibitemShut {NoStop}%
\bibitem [{\citenamefont {Pandola}\ \emph {et~al.}(2007)\citenamefont
  {Pandola}, \citenamefont {Bauer}, \citenamefont {Kroninger}, \citenamefont
  {Liu}, \citenamefont {Tomei}, \citenamefont {Belogurov}, \citenamefont
  {Franco}, \citenamefont {Klimenko},\ and\ \citenamefont
  {Knapp}}]{Pandola:2007hv}%
  \BibitemOpen
  \bibfield  {author} {\bibinfo {author} {\bibfnamefont {L.}~\bibnamefont
  {Pandola}}, \bibinfo {author} {\bibfnamefont {M.}~\bibnamefont {Bauer}},
  \bibinfo {author} {\bibfnamefont {K.}~\bibnamefont {Kroninger}}, \bibinfo
  {author} {\bibfnamefont {X.}~\bibnamefont {Liu}}, \bibinfo {author}
  {\bibfnamefont {C.}~\bibnamefont {Tomei}}, \bibinfo {author} {\bibfnamefont
  {S.}~\bibnamefont {Belogurov}}, \bibinfo {author} {\bibfnamefont
  {D.}~\bibnamefont {Franco}}, \bibinfo {author} {\bibfnamefont
  {A.}~\bibnamefont {Klimenko}}, \ and\ \bibinfo {author} {\bibfnamefont
  {M.}~\bibnamefont {Knapp}},\ }\href {\doibase 10.1016/j.nima.2006.10.103}
  {\bibfield  {journal} {\bibinfo  {journal} {Nucl. Instrum. Meth. A}\ }\textbf
  {\bibinfo {volume} {570}},\ \bibinfo {pages} {149} (\bibinfo {year}
  {2007})}\BibitemShut {NoStop}%
\end{thebibliography}%
